\documentclass[english,12pt, titlepage]{article}

\newcommand{\RR}{\mathbb{R}}
\newcommand{\CC}{\mathbb{C}}
\newcommand{\ZZ}{\mathbb{Z}}
 \newcommand{\N}{{\cal N}}

\renewcommand{\a}{\alpha}

  \newcommand{\s}{\sigma}

  \newcommand{\m}{\mu}

\newcommand{\eps}{\epsilon}

  \newcommand{\D}{\Delta}
  \newcommand{\G}{\Gamma}

\def\V{{\cal V}}

\newcommand{\Qsl}{Q\!\!\!\!\slash\,}
\newcommand{\Dsl}{D\!\!\!\!\slash\,}
\newcommand{\thetasl}{\theta\!\!\!\slash}

\def\pr{\partial}

\usepackage[latin1]{inputenc}
\usepackage[T1]{fontenc}
\usepackage[german,danish,english]{babel}

\usepackage{amsmath}
\usepackage{vmargin}
\usepackage{dsfont}
\usepackage{longtable}

\usepackage{ae,aecompl}
\usepackage{amssymb}
\usepackage{textcomp}
\usepackage{type1cm}
\usepackage{bm}
\usepackage[multiple]{footmisc}
\usepackage{multicol}
\usepackage{algorithmic}
\usepackage{algorithm}
\usepackage{ifthen}
\usepackage{layout}
\usepackage{acronym}

\usepackage{cite}
\usepackage{fancyhdr}
\usepackage{amssymb}
\usepackage{epic,eepic}
\usepackage{graphicx}
\usepackage{color}
\usepackage{epsfig}
\usepackage{rotating}

\usepackage{multirow}
\usepackage{hhline}
\usepackage{subfigure}
\usepackage{calc}
\usepackage{slashed}
\usepackage[bf,footnotesize]{caption}
\usepackage{babel}
\usepackage{ae,aecompl}
\usepackage{textcomp}
\usepackage{latexsym}
\usepackage{amsfonts}

\usepackage{amsmath,amsthm}
\allowdisplaybreaks[4]
\usepackage[multiple]{footmisc}
\usepackage{graphicx}
\usepackage{index} 

\usepackage{array}
\usepackage{blkarray}
\usepackage{acronym}
\usepackage{minitoc}
\usepackage{amsthm} 

\makeatletter

\setpapersize{A4}

\definecolor{darkblue}{cmyk}{0.95, 0.95, 0.0, 0.05}
\definecolor{darkmagenta}{cmyk}{0, 0.4, 0, 0.4}

\begin{document}

\thispagestyle{empty}

{\hfill {        HU-EP-04-36 \\
\mbox{}\bigskip

\bigskip

\begin{center} \noindent \Large \bf
Generalizations of the AdS/CFT correspondence
\end{center}

\bigskip\bigskip\bigskip

\centerline{ \normalsize \bf Ingo Kirsch\footnote[1]{\noindent \tt 
    PhD thesis, Humboldt University,
    Berlin.}\footnote[2]{\noindent \tt ik@physik.hu-berlin.de} }

\bigskip
\bigskip\bigskip

\centerline{\it Institut f\"ur Physik}
\centerline{\it Humboldt-Universit\"at zu Berlin}
\centerline{\it Newtonstra\ss e 15 }
\centerline{\it D-12489 Berlin, Germany}
\bigskip\bigskip\bigskip

\centerline{\bf \small Abstract}
\medskip\bigskip

\begin{small}
  We consider generalizations of the AdS/CFT correspondence in which probe
  branes are embedded in gravity backgrounds dual to either conformal or
  confining gauge theories. These correspond to defect conformal field
  theories (dCFT) or QCD-like theories with fundamental matter, respectively.
  Moreover, starting from the dCFT we discuss the deconstruction of
  intersecting M5-branes and M-theory.  We obtain the following results:

i) {\it Holography of defect conformal field theories.} We consider holography
for a general D3-D$p$ brane intersection in type IIB string theory ($p \in
\{3,5,7\}$). The corresponding near-horizon geometry is given by a probe
AdS-brane in \mbox{$AdS_5 \times S^5$}. The dual defect conformal field theory
describes $\N=4$ super Yang-Mills degrees of freedom coupled to
fundamental matter on a lower-dimensional space-time defect. We derive the
spectrum of fluctuations about the brane embedding and determine the behaviour
of correlation functions involving defect operators.  We also study the dual
conformal field theory in the case of intersecting D3-branes. To this end, we
develop a convenient superspace approach in which both two- and
four-dimensional fields are described in a two-dimensional $(2,2)$ superspace.
We show that quantum corrections vanish to all orders in perturbation theory,
such that the theory remains a (defect) conformal field theory when quantized.

ii) {\it Flavour in generalized AdS/CFT dualities.} We present a holographic
non-per\-tur\-ba\-tive description of QCD-like theories with a large number of
colours by embedding D7-brane probes into two non-supersymmetric gravity
backgrounds.  Both backgrounds exhibit confinement of fundamental matter and a
discrete glueball and meson spectrum.  We numerically compute the
$\bar\psi\psi$ quark condensate and meson spectrum associated with these
backgrounds. In the first background, we find some numerical evidence for a
first order phase transition at a critical quark mass where the D7 embedding
undergoes a geometric transition. In the second, we find a chiral
symmetry breaking condensate as well as the associated Goldstone boson.

iii) {\it Deconstruction of extra dimensions.} We apply the deconstruction
method to the dCFT of intersecting D3-branes to obtain a field theory
description for intersecting M5-branes.  The resulting theory corresponds to
two six-dimensional (2,0) superconformal field theories which we show to have
tensionless strings on their four-dimensional intersection.  Moreover,
we argue that the $SU(2)_L$ R-symmetry of the dCFT matches the manifest
$SU(2)$ R-symmetry of the M5-M5 intersection. We finally explore the
fascinating idea of deconstructing M-theory itself.  We give arguments for an
equivalence of M-theory on a certain background with the Higgs branch of a
four-dimensional non-supersymmetric (quiver) gauge theory: In addition to a
string theoretical motivation, we find wrapped M2-branes in the mass spectrum
of the quiver theory at low energies.
\end{small}



\pagestyle{plain}

\selectlanguage{english}

\tableofcontents



 \setcounter{equation}{0}\setcounter{figure}{0}\setcounter{table}{0}

\subsection*{}
\hfill
\begin{minipage}{8.3cm}
\begin{small}
{\it
\mbox{}\hspace{0.7cm} And since the portions of the great and of the small 
are equal in amount, for this reason, too, all things will be in
everything; nor is it possible for them to be apart, but all things
have a portion of everything. Since it is impossible for there to be a
least thing, they cannot be separated, nor come to be by themselves;
but they must be now, just as they were in the beginning, all
together. And in all things many things are contained, and an equal
number both in the greater and in the smaller of the things that are
separated off.} \\

\hfill Anaxagoras of Clazomenae (500-428 B.C.)
\end{small}
\end{minipage}

\bigskip

\section{Introduction} \label{introduction}

\noindent Holography is an ever fascinating concept since the early days 
of natural philo\-sophy in ancient Greece. In modern language, holography
(from the Greek word `holo', meaning `whole', and `graphy', meaning `(the form
of) writing') means that all the physics in a volume of arbitrary dimension
can be described in terms of the degrees of freedom of the surface or boundary
of the volume with one less dimension. This definition is in analogy to a
traditional hologram which stores a three-dimensional image in a
two-dimensional surface.  A philosophy which has a resemblance to holo\-graphy
is the ontology of Anaxagoras (500-428 B.C.). He was one of the pre-Socratics
who wanted to solve the problem of change posed by Parmenides (504-456 B.C.).
Anaxagoras suggested that each and every substance of the universe may be
divided infinitely into ever smaller parts, but even in the tiniest part of
the world there are fragments of all other things.  His notion of ``everything
in everything'' can most appropriately be illustrated by a
hologram.\footnote{I first encountered the comparison of Anaxagoras'
  philosophy with a hologram in \cite{JG}.  The above quotation of Anaxagoras
  is taken from the book \cite{Ralph} (Fr.~6).}  Unlike normal photographs,
every part of a hologram contains all the information possessed by the whole.
In other words, if a hologram is fragmented, each piece of the hologram
depicts a smaller version of the original picture and not just a part of
it.\footnote{Here we think of an idealized hologram. In a true hologram, the
  smaller image is less sharp than the whole hologram due to the finite
resolution.}

The philosophy of Anaxagoras fell behind due to the success of the ``atomic
{theory}'' of Democritus (ca.~460-370 B.C.), another pre-Socratic philosopher.
\mbox{Democritus'} idea of an indivisible object (the a-tom) entered physics
at the beginning of the 20th century when Rutherford and Bohr developed their
atom models.  Much later, these developments led to the concept of elementary
particles which are nowadays described by the Standard Model of Elementary
Particle Physics.  The ``Standard Model'' is a term used to describe the
quantum theory that includes the theory of strong interactions (quantum
chromodynamics or QCD) and the unified theory of weak and electromagnetic
interactions (electroweak).  Though the Standard Model is substantially
confirmed by experiment, it is incomplete in the sense that it does not
incorporate gravity. To overcome this shortcoming, physicists seek for a
unified theory which describes all forces including gravity. The most
promising candidate is currently string theory which describes nature by tiny
one-dimensional objects called ``strings''.

String theory has its origin as an attempted theory of strong interactions
\cite{Nambu:1997wf, Nielson, Susskind:1970xm} in 1970. However, in the years
1973 and 1974 an alternative theory of the strong interaction emerged in the
form of quantum chromodynamics (QCD) replacing the previous string models
(also called ``dual resonance models'').  It was then realised that dual models
contain spin~2 particles with gravity-like couplings \cite{Scherk:1974ca}.
The original motivation for string theory as a theory of hadrons gave way to
the view of string theory as a unified theory of fundamental forces. For this
reason the efforts to find a consistent and anomaly-free string theory
continued and culminated in a manifest supersymmetric formulation in the years
1981 and 1982 \cite{Green1, Green2, Green3}.  Much later (in 1995), string
theory was again revolutionised by the discovery of {\em D-branes}
\cite{Polchinski} which can be thought of as solitonic extended objects on
which open strings can end.


A revival of Anaxagoras' ontology came very recently in string theory with
\mbox{Maldacena's} discovery of the {\em AdS/CFT correspondence}
\cite{Juan,Gub,Wit} in 1997 which can be regarded as a modern realisation of
holography.  AdS/CFT is a conjectured holographic relation between a theory
with gravity in $d$ dimensions and a (local) quantum field theory in $d-1$
dimensions.  The field theory is invariant under angle-preserving
transformations (conformal transformations) and is located on the boundary of
an Anti-de~Sitter (AdS) space. An AdS space is a maximally symmetric Einstein
space with negative cosmological constant.  The interior (or the bulk) of the
AdS space is governed by string theory which includes (super-)gravity.
AdS/CFT states the equivalence or duality between both theories.

Such a holographic duality of two theories is quite remarkable for both
physicists and philosophers. First, we propose to consider gauge-gravity
dualities as a synthesis of both pre-Socratic philosophies: While the local
field theory on the boundary is the (preliminary) final stage of a development
which began with Democritus, the underlying philosophy of the bulk theory is
associated predominantly with \mbox{Anaxagoras'} notion of
holography.\footnote{The first indication for the holographic behaviour of
  gravity was found in the thermodynamics of black holes.  The entropy of the
  black holes is proportional to the area of the horizon. If gravity had
  similar local degrees of freedom as a field theory, one would have expected
  the entropy proportional to the volume.}  Second, AdS/CFT is a considerable
attempt to describe particle physics by a gravitational theory.  The hope is
that one day a generalization of AdS/CFT will contribute to the understanding
of some parameter regime of the Standard Model which is not accessible by
perturbation theory.

In this paper we consider generalizations of the correspondence in which
conformal symmetry and supersymmetry are broken. These are potentially useful
for describing realistic quantum field theories. In particular, it is hoped
that methods based on gauge-gravity duality will eventually be applicable to
QCD. Note that already in 1974 't~Hooft suggested that a large $N$ version of
QCD with $N$ the number of colours can be described by a string theory
\cite{'tHooft:1974jz}. The simplest generalizations involve deforming AdS by
the inclusion of relevant operators \cite{Girardello:1998pd}.  These
geometries are asymptotically AdS, with the deformations interpreted as
renormalization group (RG) flow from a super-conformal gauge theory in the
ultraviolet to a QCD-like theory in the infrared. Moreover a number of
non-supersymmetric ten-dimensional geometries of this or related form have
been found \cite{Witten2,Gubser:1999pk,Babington:2002ci,Babington:2002qt,CM,
  Kuperstein:2003yt} and have been shown to describe confining gauge dynamics.
There have been interesting calculations of the glueball spectrum in three and
four-dimensional QCD by solving classical supergravity equations in various
deformed AdS geometries
\cite{Csaki1,Koch,Zyskin,Russo,Minahan,Ooguri,Csaki2,Russo2,Csaki3,Constable2,Brower}.

A difficulty with describing QCD in this way arises due to the
asymptotic freedom of QCD. The vanishing of the 't~Hooft coupling
in the UV requires the dual geometry to be infinitely
curved in the region corresponding to the UV. In this case
classical supergravity is insufficient and one needs to use full
string theory. Formulating string theory in the relevant
backgrounds has thus far proven difficult.  The existing glueball
calculations involve geometries with small curvature that return
asymptotically to AdS (the field theory returns to the strongly
coupled ${\cal N}=4$ theory in the UV), and are in the same
coupling regime as strong coupling lattice calculations far from the
continuum limit. There is nevertheless optimism that the glueball
calculations are fairly accurate, based on comparisons with
lattice data \cite{Teper,Morningstar,CE}.

All AdS/CFT dualities considered so far conjecture the equivalence of a
particular string (or supergravity) theory and a pure Yang-Mills theory with
matter in the adjoint representation of the gauge group. For a more realistic
gauge-gravity duality the inclusion of matter in the {\em fundamental}
representation (``quarks'') is a mandatory requirement. The introduction of
quarks into the AdS/CFT correspondence is a prerequisite for studying a number
of non-perturbative phenomena in QCD in terms of a weakly coupled string
theory. Examples are the formation of hadrons, spontaneous chiral symmetry
breaking, pion scattering and decay, quark confinement, etc., to mention only
the most prominent among the strong coupling phenomena.
 
The main objectives of this paper are to lay the foundations for a
holographic study of Yang-Mills theories with flavour and to show that
some of the non-per\-tur\-ba\-tive phenomena can be understood in a string
theoretical framework at least in a qualitative way.

As an additional aspect we study several models which are based on a method
known as {\em deconstruction}. As we will explain in detail later,
deconstruction is a technique for generating extra dimensions in field
theories. The discussion of deconstruction is somewhat deviating from the
general discussion of flavour in AdS/CFT. Nevertheless, some of the models to
which deconstruction is applied arise out of the study of defect conformal
field theories (dCFT).  In particular, we will discuss intersecting M5-branes
the action of which is obtained from the dCFT of intersecting D3-branes.
Subsequently, we will investigate to some extend the exciting idea to
deconstruct a discrete action for M-theory itself.

The paper is organised as follows. In Chapter \ref{cha:dCFT} we discuss
holographic duals of defect conformal field theories in which fundamental
matter was first considered in the context of AdS/CFT. In
Chapter~\ref{cha:flavour} we consider gauge-gravity dualities with flavour in
four spacetime dimensions.  In particular, we will compute meson spectra in
large $N$ QCD-like theories via supergravity and demonstrate spontaneous
$U(1)_A$ chiral symmetry breaking. In Chapter~\ref{cha:deconstruction} we
consider the deconstruction of intersecting M5-branes and M-theory.

In the following, we give an introduction to each of these three topics.
After a brief review of standard AdS/CFT in Sec.~\ref{sec11}, we will give
an introduction to defect conformal field theories and their supergravity
duals in Sec.~\ref{sec12}.  This will lead us to the discussion of
flavours in four spacetime dimensions in Sec.~\ref{sec13}.  In
Sec.~\ref{sec14} we close the introduction by discussing the theory
of intersecting M5-branes which is related to a particular defect conformal
field theory via the deconstruction method.

\subsection{A brief introduction to the AdS/CFT correspondence} \label{sec11}
\fancyhead[RO]{\bfseries \ref{sec11} A brief introduction to the
  AdS/CFT correspondence \hfill \thepage}

Before introducing fundamental matter into confining gauge-gravity dualities,
we briefly review some basic aspects of the standard AdS/CFT correspondence.
For a more detailed introduction to AdS/CFT we refer the interested reader to
some excellent reviews on the subject \cite{D'Hoker2, Maldacena, Klebanov2,
  Bigatti, Bertolini, Petersen:1999zh, Douglas:1999ww, DiVecchia:1999yr}.

Essential for the AdS/CFT correspondence is the concept of D(irichlet)-branes
in string theory.  D$p$-branes are $p+1$-dimensional solitonic objects in
string theory which can be understood as hypersurfaces on which open strings
can end \cite{Polchinski}.  On the open strings attached to the D-branes one
imposes Dirichlet boundary conditions.  An interesting property of D-branes is
that they realise gauge theories on their world-volume. The low-energy
effective field theory of massless open string modes on $N$ coincident
D$p$-branes is a $p+1$-dimensional $U(N)$ super Yang-Mills theory with 16
supercharges.

D-branes also have an interpretation in terms of closed strings.  Polchinski
showed \cite{Polchinski} that D$p$-branes carry an elementary charge with
respect to the $p+1$-form potential from the Ramond-Ramond (closed string)
sector of the superstring. This implies that D-branes act as sources for
closed strings which induce a back-reaction on the background. Indeed, one can
show that massless closed string excitations of D-branes generate
Ramond-Ramond charged extremal $p$-brane solutions in supergravity.

There exists a limit in string theory (the Maldacena limit), in which the
string coupling $g_s$ and the number of D-branes are kept fixed, while the
string length $l_s$ goes to zero ($l_s \rightarrow 0$). In this limit open
strings and closed strings do not interact anymore leaving two decoupled
descriptions of the same system: one in terms of open strings, the other in
terms of closed strings.  Generally speaking, the AdS/CFT correspondence
conjectures both descriptions to be equivalent.
\begin{figure}[!ht]
\begin{center}
\includegraphics[scale=0.9]{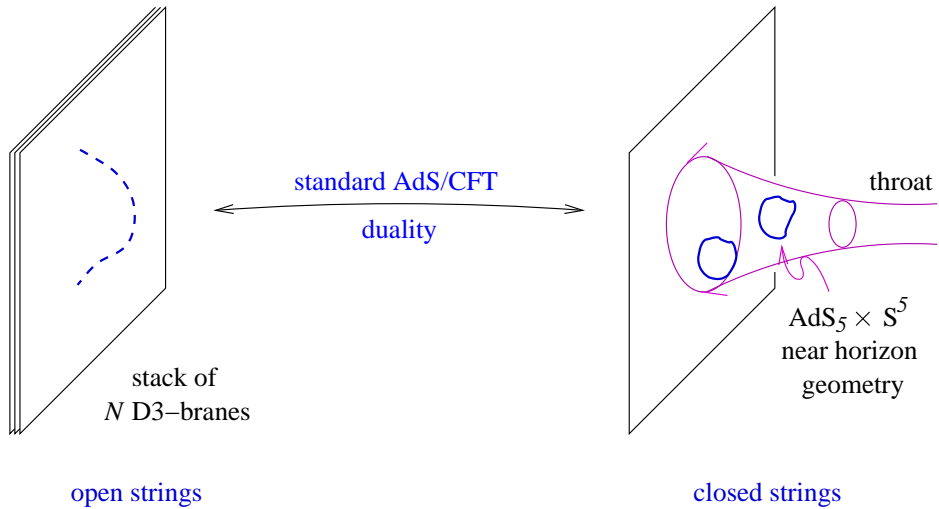}
\caption{Standard AdS/CFT: The left figure shows the description of a stack
  of D3-branes in terms of open strings (Yang-Mills description), the right
  figure in terms of closed strings (supergravity description).}
\label{figure1}
\end{center}
\end{figure}

In the standard AdS/CFT correspondence \cite{Juan,Gub,Wit}, a system of $N$
coincident D3-branes is considered within type IIB string theory.  Its
description in terms of open and closed strings is shown in
Fig.~\ref{figure1}. At low energies massive string modes decouple and the
effective theory generated by open string modes is $\N=4$ $SU(N)$ super
Yang-Mills theory in 3+1 dimensions which is known to be a conformal field
theory.\footnote{The diagonal $U(1)$ factor inside the group $U(N)$ decouples
  at low energies.}  This theory is located on the world-volume of the
D3-branes.

The holographic dual theory is generated by massless closed string modes.  In
the Maldacena limit, the metric of the D3-branes reduces to its near-horizon
(throat) region which is $AdS_5 \times S^5$. This is a product
space of a five-dimensional Anti-de-Sitter space and a five-sphere.
Closed strings in the asymptotic flat region decouple from the theory
inside the throat region.

So far only very little is known about string theory quantization on a curved
background including $AdS_5 \times S^5$.\footnote{Some progress has been made
  by considering the Penrose or plane-wave limit of $AdS_5 \times S^5$ on
  which string theory is exactly solvable \cite{Berenstein:2002jq,
    Metsaev:2001bj, Metsaev:2002re, Blau:2001ne, Blau:2002dy}.  String theory
  in this limit is conjectured to be dual to a sector of large $N$ $\N=4$
  super Yang-Mills theory with divergent R-charge $J \sim \sqrt{N}$. In this
  paper we will not consider this limit.}\footnote{Berkovits delevoped a
  formalism for the covariant quantization of string theory on a curved
  background.  For a review see \cite{Berkovits:2002zk}. It is however not
  (yet) possible to use this approach to compute the string excitation
  spectrum on these backgrounds.}  One therefore takes the 't~Hooft limit,
sending $N\rightarrow \infty$, while keeping the 't~Hooft coupling $\lambda =
4\pi g_s N$ fixed.  Taking also a large 't Hooft coupling, $\lambda \gg 1$,
the radius of curvature $L^4 = \lambda \alpha'{}^2$ of the AdS space becomes
large leading to a small curvature (there the string length is much smaller
than the size of AdS, so we see particles instead of strings).  The full
quantum string theory then reduces to classical type IIB supergravity on
$AdS_5 \times S^5$.

Thus, at low energies we have two different descriptions of a stack of $N$
D3-branes which are conjectured to be equivalent 
(at large $\lambda$):\footnote{To be precise, there are actually two
 decoupled systems on both sides of the correspondence. On both sides
 the additional system is supergravity in flat space. For a detailed
 discussion of this subtlety see e.g.~\cite{Maldacena}.}     
\begin{itemize}
\item one in terms of the $\N=4$ $SU(N)$ super Yang-Mills theory in 3+1
dimensions generated by the
massless {\em open} string modes,
\item the other in terms of type IIB supergravity on $AdS_5 \times S^5$ (with
  integer flux of the five-form Ramond-Ramond field strength, $N=\int_{S^5}
  F_5$) generated by the massless {\em closed} string degrees of freedom,
\end{itemize}
where the parameters of the two theories are related as
\begin{align}
 g_s = g_{YM}^2 \,,\qquad L^4 = \lambda \alpha'{}^2 \,.
\end{align}
Here $g_s$ is the string coupling, $g_{YM}$ the Yang-Mills coupling, $L$ the
radius of curvature of both AdS space and $S^5$, and $\alpha'=l_s^2$ is
related to the string tension by $T=1/2\pi\alpha'$.  

Although a strict proof of the AdS/CFT correspondence is still missing, there
is a lot of evidence that there is some truth in it. At least the above stated
weak form of the correspondence has been very well tested by now.

Note first that the map between AdS and CFT quantities is given by
\begin{align}
{\cal Z}_{sugra}(\phi_i)= \left\langle\exp\left(\int d^4x
\phi_i^0 {\cal O}_i \right) \right\rangle \,,
\end{align}
where the left-hand side is the supergravity partition function evaluated on
the classical solution given by $\phi_i$ (which satisfies $\phi_i
\vert_{\partial AdS} = \phi_i^0$) and the right-hand side is the generating
function for super Yang-Mills theory. $\phi_i^0$ denotes the value of the
supergravity field $\phi_i$ at the boundary, where it acts as a source for the
operator ${\cal O}_i$. We see that there is a one-to-one correspondence
between the operators ${\cal O}_i$ and fields $\phi_i$. It has been verified
\cite{Wit} that there is a precise match between supergravity fields and
so-called BPS operators in the gauge theory.  As a consistency check observe
that the isometries of the $AdS_5 \times S^5$ space correspond to the
symmetries of the conformal field theory.  The isometry group $SO(4,2)$ of
$AdS_5$ is the conformal group in four dimensions, while the isometry group
$SO(6) \simeq SU(4)$ of $S^5$ is the R-symmetry of $\N=4$ supersymmetry.  More
generally, both the supergravity fields as well as the Yang-Mills BPS
operators fall in the same multiplets of the supergroup $PSU(2,2 \vert 4)$.

Moreover, correlation functions of Yang-Mills operators have been computed via
supergravity and compared to field theory results (for a review see
\cite{D'Hoker2}). It is quite non-trivial that there is an agreement in the
general behaviour of the correlation functions in both computations.  Of
course, numerical factors usually differ since the field theory correlators
are computed at weak coupling, while the AdS computation
yields correlators at strong coupling.
 
It is also interesting to compare the vacuum expectation value of a Wilson
loop, which in field theory can be expanded in terms of local operators. It
became more and more clear that the fundamental string in AdS is the same as
the QCD string of large $N$ Yang-Mills theory.  For instance, open strings are
(dual to) spin chains of adjoint fields (``gluons'') with fundamental fields
(``quarks'') at their ends \cite{DeWolfe:2004zt}. We also encounter such
operators below (see Ch.\ \ref{moreD3D3}).

There are many other checks and tests of the correspondence, which we cannot
review here, all of them supporting the conjecture.

\subsection{Holography of defect conformal field theories}\label{sec12}
\fancyhead[RO]{\bfseries \ref{sec12} Holography of defect conformal field
  theories \hfill \thepage}

A generalization of the AdS/CFT correspondence is obtained by embedding an
additional probe brane into the $AdS_5 \times S^5$ background. Depending on
the dimension of the probe brane, the dual field theory of this supergravity
set-up is then a conformal field theory with a space-time {defect}. These
defect conformal field theories (dCFT) involve fields which are confined to a
lower-dimensional subspace of the original four-dimensional space-time.  For
these dCFT the four-dimensional conformal symmetry is broken to the
lower-dimensional conformal group of the defect. A typical \mbox{Feynman}
diagram corresponding to the interaction of four-dimensional bulk degrees of
freedom with defect fields is shown in Fig.~\ref{figdef}.  As a special case
one can also have a ``defect'' of codimension zero corresponding to flavour in
four spacetime dimensions \cite{Katz}. This will become important later when
we discuss mesons in QCD-like theories.

\begin{figure}[!ht]
\begin{center}
\includegraphics[scale=0.7]{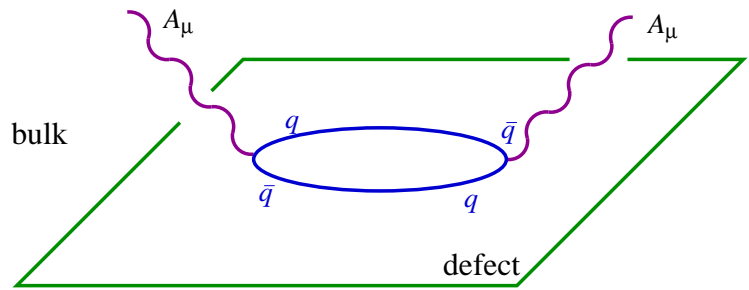}
\caption{Interaction of bulk gauge bosons and defect quarks.}  
\label{figdef}
\end{center}
\end{figure}

The general problem of introducing a spatial defect into a conformal field
theory has been studied in several contexts~\cite{Cardy,Osborn}.  Within
string theory such defect conformal field theories arise in various brane
constructions.  They were first studied in this context as matrix model
descriptions of compactified NS5-branes \cite{Sethi} and more generally as
effective field theories describing various D-brane
intersections~\cite{Ganor, KapustinSethi}.

The first AdS/CFT setup leading to a dCFT was considered by Karch and
\mbox{Randall} in \cite{Karchrandall1, Karchrandall2, Karch:2001cw}.  They
conjectured an AdS/CFT duality in which a D5-brane probe (along
$x^0,x^1,x^2,x^4,x^5,x^6$) orthogonally intersects a stack of $N$ D3-branes
(along $x^0,x^1,x^2,x^3$) on a three-dimensional subspace with coordinates
$x^0,x^1,x^2$, as shown on the left-hand side in Fig.~\ref{figure2}. The
near-horizon limit of this D3-D5 brane system is $AdS_5 \times S^5$ with the
D5-brane wrapping an $AdS_4 \times S^2$ submanifold.  The supersymmetry of the
$AdS_4 \times S^2$ embedding was demonstrated in \cite{Skenderis}. The $AdS_5$
geometry can be visualized as the interior of a disk as shown on the
right-hand side in Fig.~\ref{figure2}, while the $AdS_4$ brane ends on the
boundary of the disk.

There are various strings in the set-up: As usual, open string modes with both
endpoints on the D3-branes generate the $\N=4$ super Yang-Mills theory, while
closed string modes give rise to type IIB supergravity on $AdS_5 \times S^5$.
However, we have additional strings due to the embedding of a probe brane.
First, there are strings stretching between the D5-brane and the D3-branes.
They give rise to a {\em fundamental} hypermultiplet (``quarks'') in the
low-energy theory.  Due to the decoupling of open strings on the D5-brane in
the infrared, the $U(1)$ gauge group on the D5-brane, or $U(N_f)$ in case of
$N_f$ D5-branes, turns into the flavour group of the fundamental matter.
Second, there are open strings ending on the D5-brane wrapping $AdS_4
\times S^2$.  In the probe approximation, one neglects the back-reaction of the
D5-brane on the near-horizon background of the D3-branes.  Classically, the
fluctuation modes of the $AdS_4$-brane are then described by the Dirac-Born-Infeld
action of the D5-brane (plus Wess-Zumino term).

\begin{figure}[!ht]
\begin{center}
\includegraphics[scale=0.85]{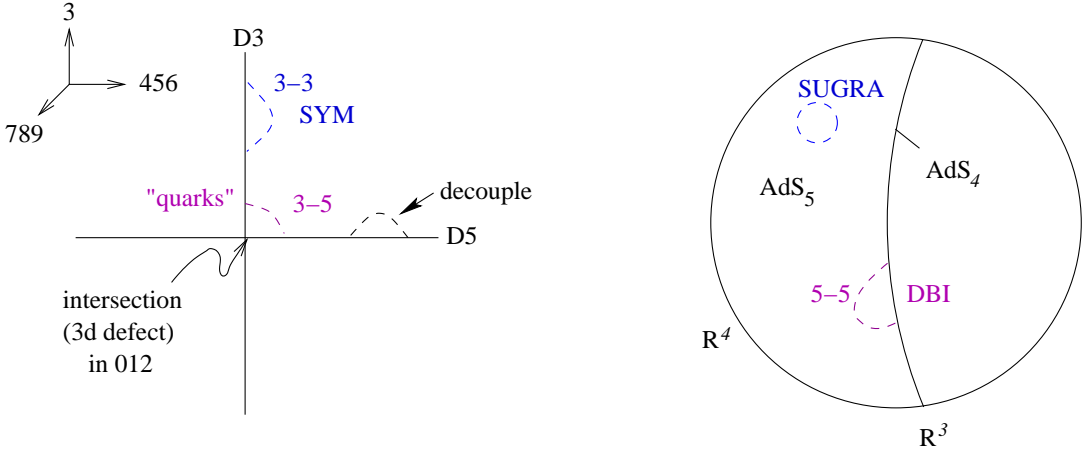}
\caption{Holography of the D3-D5 brane intersection.}  \label{figure2}
\end{center}  
\end{figure}

Karch and Randall conjecture the AdS/CFT duality to act `twice': First there
is the standard AdS/CFT duality between open strings ending on the D3-branes
(3-3 strings) and closed strings in type IIB string theory on
\mbox{$AdS_5\times S^5$}.  Secondly, they conjecture an additional duality
between open strings, stretching between the D5 and D3-branes (3-5 and 5-3
strings), and open strings ending on the D5-brane (5-5 strings) wrapping
$AdS_4 \times S^2$. 

The world-volume theory of this configuration is a four-dimensional conformal
field theory coupled to a codimension one defect. This defect conformal field
theory describes the decoupling limit of the D3-D5 intersection, and consists
of the ${\cal N} = 4$, $d=4$ super Yang-Mills theory coupled to an ${\cal
  N}=4$, $d=3$ hypermultiplet localized at the defect.  In \cite{DFO} DeWolfe,
Freedman and Ooguri constructed the action of the model and developed a
precise dictionary between composite operators in the field theory and
fluctuation modes on the $AdS_4$-brane.  In \cite{EGK}, we wrote the action
compactly in an ${\cal N} =2, d=3$ superspace and gave field theoretic
arguments for quantum conformal invariance.

In summary, the AdS/dCFT correspondence conjectures the equivalence of
the following two theories:
\begin{center}
\begin{minipage}{13cm}
\begin{center}
\hspace{0.6cm} {\color{darkblue}4d $\N=4$ $SU(N)$   \hfill type IIB supergravity\hspace{0.2cm}

\hspace{0.03cm} super Yang-Mills theory \hfill on $AdS_5 \times S^5$\hspace{1.0cm}}

              +  \hspace{4cm}  $\longleftrightarrow$ \hspace{3cm}  +

{\color{darkmagenta}
3d $\N=4$ hypermultiplet \hfill Dirac-Born-Infeld theory 

\hspace{1cm}on $\RR^3$ defect \hfill on $AdS_4 \times S^2$ \hspace{0.9cm}}

\end{center}
\end{minipage}
\end{center}

\subsubsection*{Generalization to a general D3-D$p$ intersection}

In Ch.~\ref{cha:dCFT} we generalize the above duality to the case of a D3-D$p$
brane intersection with $p \in \{(1),3,5,7 \}$. The ``monopole'' case $p=1$
will only be mentioned marginally. Since the D3-D9 intersection is
non-supersymmetric, we exclude the case $p=9$.  All other D3-D$p$
intersections are supersymmetric and related by T-duality. We also omit the
instanton case $p=-1$ since the D(-1)-D3 system does not have a defect
interpretation nor does it correspond to an $AdS$ embedding.

The near-horizon geometry is again $AdS_5 \times S^5$ with the D$p$-brane
wrapping an $AdS_{k+2} \times S^k$ submanifold. Here $k+1=(p+1)/2$ is the
dimension of the intersection which agrees with the dimension of the defect.
In general, the dual field theory describes the degrees of freedom of the
$\N=4$ $SU(N)$ super Yang-Mills theory coupling to a $k+1$-dimensional defect
hypermultiplet in the fundamental representation of the gauge group. The
defect breaks supersymmetry by a half. The theory is thus invariant
under eight supercharges, i.e.\ under 2d $(4,4)$, 3d $\N=4$, or
4d $\N=2$ supersymmetry in the case of a two-, three-, or 
four-dimensional defect, respectively.

In the following we give an overview over all D3-D$p$ brane intersections  and
their corresponding world-volume theories.  These configurations are:

{\bf D1-D3:} The case of $N'$ D1-branes ending on a D3-brane has an
interpretation in terms of an $SU(N')$ magnetic monopole
\cite{Diaconescu:1997rk}. From the point of view of the effective theory on
the D3-brane, the D1-branes act as a point source of magnetic charge for the
gauge field.  The near-horizon geometry in supergravity is given by an $AdS_2$
brane embedded in $AdS_5 \times S^5$. The dual field theory is a
four-dimensional defect CFT where the fundamental hypermultiplets are
localized on a one-dimensional defect \cite{KapustinSethi, Tsimpis:1998zh}.
Two-dimensional conformal field theories with a one-dimensional defect dual to
$AdS_2$ branes in $AdS_3$ have been studied in \cite{Bachas,Quella}. However,
holography of the D1-D3 system corresponding to an $AdS_2$ embedding
inside $AdS_5$ has not yet been discussed. Since abelian
monopoles do not exist, progress towards a holographic description requires
the discussion of a non-abelian Dirac-Born-Infeld action of the D1-branes. 
We will not discuss this case in detail.

In Ch.~\ref{cha:dCFT} we will mainly focus on the D3-D3 intersection, for
which reason we give a more detailed introduction and an overview over the
expected results.

{\bf D3-D3:} This system, which has first been studied in
\cite{Constable:2002xt}, consists of a stack of D3-branes spanning the $0123$
directions and an orthogonal stack (of D3$'$-branes) spanning the $0145$
directions such that eight supercharges are preserved, realising a $(4,4)$
supersymmetry on the common $1+1$ dimensional world-volume.  Unlike the D3-D5
intersection, open strings on both stacks of branes remain coupled as
$\alpha^{\prime} \rightarrow 0$. However, in the probe approximation a
holographic duality can be found relating fluctuations in an $AdS$ background
to operators in the dual field theory. One simply takes the number of
D3-branes, $N$, in the first stack to infinity, keeping $g_sN$ and the number
of D3-branes in the second stack, $N^{\prime}$, fixed. In this limit, the
't~Hooft coupling of the gauge theory on the second stack, $\lambda^{\prime} =
g_s N^{\prime}$, vanishes.  Thus the open strings with all endpoints on the
second stack decouple, and one is left with a four-dimensional CFT with a
codimension two defect.  The defect breaks half of the original \mbox{$\N=4$},
$d=4$ supersymmetry, leaving eight real supercharges realising a
two-dimensional $(4,4)$ supersymmetry algebra.  The conformal symmetry of the
theory is a global $SL(2,R) \times SL(2,R)$, corresponding to a subgroup of
the four-dimensional conformal symmetries.  The degrees of freedom at the
defect are a $(4,4)$ hypermultiplet arising from the open strings connecting
the orthogonal stacks of D3-branes.


The classical Higgs branch of this theory has an interpretation as a smooth
resolution of the intersection to the holomorphic curve $wy\sim
c\alpha^{\prime}$, where $w= X^2 + i X^3$ and $y = X^4 + iX^5$. However, due
to the two-dimensional nature of the fields which parameterize these curves,
the quantum vacuum spreads out over the entire classical Higgs branch.  It has
been argued that due to the spreading over the Higgs branch a fully localized
supergravity solution for this D3-brane intersection does not exist
\cite{marolfpeet,Gomberoff,peet}. Obtaining a closed string description of
this defect CFT would therefore seem to be difficult. These objections
do not hold in the probe approximation.

In the limit described above, the holographic dual is obtained by focusing on
the near horizon region for the first stack of D3-branes, while treating the
second stack as a probe.  The result is an $AdS_5 \times S^5$ background with
$N^{\prime}$ probe D3-branes wrapping an $AdS_3\times S^1$ subspace. This
embedding was shown to be supersymmetric in \cite{Skenderis}.  We will
demonstrate that there is a one complex parameter family of such
embeddings, corresponding to the holomorphic curves $wy\sim c$, all of which
preserve a set of isometries corresponding to the super-conformal group.  As
in the D3-D5 system, holographic duality is conjectured to act ``twice''.
First there is the standard AdS/CFT duality relating closed strings in $AdS_5
\times S^5$ to operators in ${\cal N}=4$ super Yang-Mills theory. Second,
there is a duality relating open strings on the probe D$3^{\prime}$ wrapping
$AdS_3\times S^1$ to operators localized on the $1+1$ dimensional defect.

One of the original motivations to search for holographic dualities for defect
conformal field theories \cite{Karchrandall1, Karchrandall2, Karch:2001cw} is
that such a duality might imply the localization of gravity on branes in
string theory.  In the context of a brane wrapping an $AdS_3$ geometry
embedded inside $AdS_5$, localization of gravity would indicate the existence
of a Virasoro algebra in the dual CFT, through a Brown-Henneaux mechanism
\cite{BrownHenneaux}. We do not find any evidence for the existence of a
Virasoro algebra in the conformal field theory. Although this theory has a
$(4,4)$ superconformal algebra, only the finite part of the algebra is
realised in any obvious way. Roughly speaking, the $(4,4)$ superconformal
algebra is the common intersection of two ${\cal N} = 4, d=4$ superconformal
algebras, both of which are finite. The even part of the superconformal group
is $SL(2,R) \times SL(2,R) \times SU(2)_L \times SU(2)_R \times U(1)$, which
is also realised as an isometry of the $AdS_5$ background which preserves the
probe embedding. Enhancement to the usual infinite dimensional algebra would
require the existence of a decoupled two-dimensional sector. Correctly
addressing this issue would require going beyond the probe limit and studying
the back-reaction of the D$3^{\prime}$-branes on the $AdS_5\times S^5$
geometry as well as gaining a deeper understanding of the dynamics of the
defect CFT.

The action for the D3-D3 intersection is most easily and elegantly
constructed in $(2,2)$ superspace.  Although it may seem unusual
to write the ${\cal N} =4, d=4$ components of the action in
$(2,2)$ superspace, this is actually quite natural because the
four-dimensional supersymmetries are broken by couplings to the
defect hypermultiplet.  In writing this action,  we will not take
the limit which decouples one stack of D3-branes. With the help of
the manifest chirality of $(2,2)$ superspace we are able to find
an argument for the absence of quantum corrections to the combined
2d/4d actions, which implies that the theory remains conformal
upon quantization.  Although this theory has two-dimensional
fields coupled to gauge fields, the gauge couplings are
exactly marginal due to the four-dimensional nature of the gauge
fields.

We give a detailed dictionary between Kaluza-Klein fluctuations on
the probe D3-brane and operators localized on the defect. Of
particular interest will be a certain subset of the fluctuations
which describe the embedding of the probe inside  $AdS_5$. This
subset is dual to operators containing defect scalar fields, which
appear without any derivative or vertex operator structure. Due to
strong infrared effects in two dimensions, these fields are not
conformal fields associated to states in the Hilbert space. From
the point of view of the probe-supergravity system, there is at
first sight nothing unusual about these fluctuations.  However,
upon applying the usual $AdS_3$/CFT$_2$ rules to compute the dual
two-point correlator,  one finds identically zero due to extra
surface terms in the probe action.  Thus there is no clear
interpretation of these fluctuations as sources for the generating
function of the CFT. We shall find however that the bottom of the
Kaluza-Klein tower for these fluctuations (with appropriate
boundary conditions) parameterizes the aforementioned holomorphic
embedding of the probe inside $AdS_5$. While the interpretation of
this fluctuation as a source is unclear, it nevertheless labels
points on the classical Higgs branch. Since the infrared dynamics
of two dimensions implies that the vacuum is spread out over the
entire Higgs branch, one should in principle sum over holomorphic
embeddings when performing computations in the $AdS$ background.

The fluctuations of the probe $S^1$ embedding inside $S^5$ satisfy
the Breitenlohner-Freedman bound despite the lack of topological
stability. These fluctuations are dual to a multiplet of scalar
operators with defect fermion pairs which we identify with BPS
superconformal primaries localized at the intersection. We also
find fluctuations of the probe embedding inside $AdS_5$ which are
dual to descendants of these operators. Remarkably, the AdS
computation of the corresponding correlators, which is valid for
large 't Hooft coupling $\lambda$, shows no dependence on
$\lambda$. We also study perturbative quantum corrections to the
two-point function of the BPS primary operators and find that such
corrections are absent at order $g_{YM}^2$. Together with the
$AdS$ strong coupling result, this suggests the existence of a
non-renormalization theorem.

{\bf D3-D5:} The world-volume theory of this system is a defect CFT with a
codimension one defect. Holography of this system was extensively studied in
\cite{DFO}.  Historically, it was the first set-up in which defect conformal
field theories were studied in AdS/CFT.  We reviewed this brane intersection
at the beginning of this section (see p.~\pageref{sec12}).

{\bf D3-D7:} In the D3-D7 brane configuration, first studied by Karch and Katz
\cite{Katz, Katz2}, a spacetime filling D7-brane was added to the $AdS_5/{\rm
  CFT}_4$ correspondence. The D7-brane completely fills the $AdS_5$ space and
wraps a maximal $S^3$ inside $S^5$. This supergravity configuration is dual to
a four-dimensional ${\mathcal N}=2$ Yang-Mills theory describing open strings
in the presence of one D7 and $N$ D3-branes sharing 3+1 dimensions.  The
degrees of freedom are those of the ${\mathcal N} =4$ super Yang-Mills theory,
coupled to an ${\mathcal N} =2$ hypermultiplet with fields in the fundamental
representation of $SU(N)$. The latter arise from strings stretched between the
D7 and D3-branes. This set-up becomes important in the next section, 
where we discuss flavour in confining supergravity backgrounds corresponding
to quarks in four-dimensional non-supersymmetric QCD-like theories.

Holography of defect conformal field theories and the embedding of branes in
AdS/CFT have also been considered in various other contexts: Gravitational
aspects were discussed in \cite{Porrati,Fayya,Liu,Giannakis}. The Penrose
limit of this background was studied in \cite{Skenderis, Lee}, wherein a map
between defect operators with large $R$-charge and open strings on a D3-brane
in a plane wave background was constructed.  RG flows related to defect
conformal field theories were discussed in \cite{Yama}. Finally, defect CFT's
were discussed in connection with the phenomenon of supertubes
in~\cite{Mateos}.  The dCFT on the D3-D5 intersection in connection to
integrable open spin chains was studied in \cite{DeWolfe:2004zt}. Further
related papers not mentioned so far are \cite{Aharony:2003qf,
  Yamaguchi:2003ay, Robert, Kumar:2002wc, Kumar:2003xi, Guralnik:2004ve}.

\subsection{Meson spectra in AdS/CFT and spontaneous chiral symmetry breaking} 
\label{sec13}
\fancyhead[RO]{\bfseries \ref{sec13} Meson spectra in AdS/CFT and 
chiral symmetry breaking \hfill \thepage}

In the previous section we have seen how matter in the fundamental
representation of the gauge group can be introduced in AdS/CFT via the
embedding of a probe brane in $AdS_5 \times S^5$. The corresponding D3-D$p$
brane intersection accommodates the holographic dual (defect) conformal field
theory on its world-volume. 

The D3-D7 configuration is special since fundamental fields are allowed to
propagate in all four space-time dimensions.  This opens up the possibility
for studying flavour in supersymmetric extensions of QCD. It is possible to
introduce mass for the fundamental matter by separating the D7-brane from the
D3-branes. The dual description involves a probe D7 on which the induced
metric is only asymptotically $AdS_5 \times S^3$.  In this case there is a
discrete spectrum of mesons.  This spectrum has been computed (exactly!) at
large 't~Hooft coupling \cite{MateosMyers} using an approach analogous to the
glueball calculations in deformed AdS backgrounds. The novel feature here is
that the ``quark'' bound states are described by the scalar fields in the
Dirac-Born-Infeld action of the D7-brane probe.

In view of a gravity description of Yang-Mills theory with confined quarks, it
is natural to attempt to generalize these calculations to probes of deformed
AdS spaces. For instance in \cite{Sonnenschein}, a way to embed D7-branes in
the Klebanov-Strassler (KS) background \cite{KS} was found, following the
suggestion of \cite{Katz}.  Moreover in \cite{Sonnenschein}, the spectrum of
mesons dual to fluctuations of the D7-brane probe in the KS-geometry was
calculated.  The underlying theory is an ${\mathcal N}=1$ gauge theory with
massive chiral superfields in the fundamental representation.  Calculations of
meson spectra for $\N=1$ supersymmetric gauge theories have also been
performed in \cite{Hu,Nunez1,Ouyang}.  Related work may also be found in
\cite{Nastase:2003dd, Strassler2}.

One of the most important features of QCD dynamics is chiral symmetry
breaking by a quark condensate, but since this is forbidden by unbroken
supersymmetry,\footnote{A quark bilinear $\Psi\tilde\Psi$, where $\Psi$ and
  $\tilde \Psi$ are fermionic components of chiral superfields $Q = q + \theta
  \Psi \cdots, \tilde Q =\tilde q + \theta \tilde \Psi + \cdots$, can be
  written as a SUSY variation of another operator (it is an F-term of the
  composite operator $\tilde{Q}Q$).} these constructions do not let us address
this issue. In Ch.~\ref{cha:flavour}, we
attempt to come somewhat closer to QCD by considering the embedding of
D7-branes in two {\em non-supersymmetric} backgrounds which exhibit
confinement.  Although neither of these backgrounds corresponds exactly to QCD
since they contain more degrees of freedom than just gluons and quarks, we
might nevertheless expect chiral symmetry breaking behaviour. The quark mass
$m$ and the quark condensate expectation value $c$ are given by the UV
asymptotic behaviour of the solutions to the supergravity equations of motion
in the standard holographic way (see \cite{PolStrass} for an example of this
methodology).  In the $\N=2$ supersymmetric Karch-Katz scenario \cite{Katz}
with a D7 probe in standard AdS space, we show that there cannot be any
regular solution which has $c\neq 0$; the supersymmetric theory does not allow
a quark condensate. We then find that for the deformed AdS backgrounds we
consider, there are regular solutions with $c \neq 0$. The case $c\neq 0$ with
$m=0$ corresponds to spontaneous chiral symmetry breaking.

The first supergravity background we consider is the {\em Schwarzschild black
  hole} in $AdS_5\times S^5$.  In the absence of D7-branes, this background is
dual to strongly coupled ${\mathcal N} =4$ super Yang-Mills at finite
temperature and is in the same universality class as three-dimensional pure
QCD \cite{Witten2}. Glueball spectra in this case were computed in
\cite{Csaki4,Zyskin}. We introduce D7-branes into this background and compute
the quark condensate as a function of the bare quark mass, as well as the
meson spectrum.  This background is dual to the finite temperature version of
the ${\mathcal N} = 2$ super Yang-Mills theory considered in
\cite{Katz,MateosMyers}. The finite temperature ${\mathcal N} =2$ theory is
not in the same universality class as three-dimensional QCD with light quarks
since the antiperiodic boundary conditions for fermions in the Euclidean time
direction give a non-zero mass to the quarks upon reduction to
three-dimensions, even if the hypermultiplet mass of the underlying ${\mathcal
  N}=2$ theory vanishes.  In fact these quarks decouple if one takes the
temperature to infinity in order to obtain a truly three-dimensional theory.
Nevertheless at finite temperature the geometry describes an interesting
four-dimensional strongly coupled gauge configuration with quarks.  The meson
spectrum we obtain has a mass gap of order of the glueball mass. Furthermore
we find that the $\bar\Psi\Psi$ condensate vanishes for zero hypermultiplet
mass, such that there is no spontaneous violation of parity in three
dimensions or chiral symmetry in four dimensions.  However for $m \neq 0$ we
find a condensate $c$ which at first grows linearly with $m$, and then shrinks
back towards zero. Increasing $m$ further, the D7 embedding undergoes a
geometric transition at a critical mass $m_c$. At sufficiently large $m \gg
m_c$ the condensate is negligible and the spectrum matches smoothly with the
one found in \cite{MateosMyers} for the ${\mathcal N}=2$ theory. Our numerics
give some evidence that the geometric transition corresponds to a first order
phase transition in the dual gauge theory, at which the condensate $c(m)$ is
discontinuous.

The second non-supersymmetric background which we consider was found by
Constable and Myers \cite{CM}.  This background is asymptotically $AdS_5
\times S^5$ but has a non-constant dilaton and $S^5$ radius. In the field
theory an operator of dimension four with zero R-charge has been introduced
(such as ${\rm tr}\, F^{\mu \nu}F_{\mu \nu}$).  This deformation does not give
mass to the adjoint fermions and scalars of the underlying ${\mathcal N}=4$
theory but does leave a non-supersymmetric gauge background.  Furthermore,
unlike the AdS black-hole background, the geometry has a naked singularity.
Nevertheless, in a certain parameter range, this background gives an area law
for the Wilson loop and a discrete spectrum of glueballs with a mass gap.

We obtain numerical solutions for the D7-brane equations of motion in
the Constable-Myers background
with asymptotic behaviour determined by a quark mass $m$ and
chiral condensate~$c$.   We compute the condensate $c$ as a function of
the quark mass $m$ subject to a regularity constraint. Remarkably,
our results are  not
sensitive to the singular behaviour of the metric in the IR.
For a given mass there are two regular solutions of which the physical, lowest
action solution corresponds to the D7-brane ``ending'' before reaching
the curvature singularity. Of course the D7-brane does
not really end, however the $S^3$ about which it is wrapped
contracts to zero size, similarly to the scenario discussed in
\cite{Katz}.  In our case the screening of the singularity is
related to the existence of the condensate. Furthermore we find numerical
evidence for a non-zero condensate in the limit $m\rightarrow 0$.
This corresponds to spontaneous breaking of the $U(1)$
chiral symmetry which is non-anomalous in the large $N$ limit
\cite{Wittenetaprime} (for a review see \cite{Donoghue}).

We also compute the meson spectrum by studying classical fluctuations about
the D7-embedding.  For zero quark mass, the meson spectrum contains a massless
mode, as it must due to the spontaneous chiral symmetry breaking.  Note that
since the spontaneously broken axial symmetry is $U(1)$ for a single D7-brane,
the associated Goldstone mode is a close cousin to the $\eta^{\prime}$ of QCD,
which is a Goldstone boson in the large $N$ limit.  We briefly comment on
generalizations to the case of more than one flavour or, equivalently, more
than one D7-brane. Moreover we give a holographic version of the Goldstone
theorem.

The main message of this part of the paper is that non-supersymmetric gravity
duals of gauge theories dynamically generate quark condensates and can break
chiral symmetries. We stress that the physical interpretation of naked
singularities is a delicate issue, for instance in the light of the analysis
of \cite{Gubser:2000nd}. This applies in particular to the discussion of light
quarks and mesons.  It is therefore an important part of our analysis that in
the presence of a condensate the physical solutions to the supergravity and
DBI equations of motion never reach the singularity in the IR. Of course it
would be interesting to understand this mechanism further and to see if it
occurs in other supergravity backgrounds as well.

Subsequent to \cite{Babington:2003vm, Babington:2003up}, chiral symmetry
breaking in a non-supersymmetric background obtained from type IIA string
theory has been found in \cite{MateosMyers2}. Chiral symmetry breaking
in the Constable-Myers background has been studied further in 
\cite{Evansnew}.

\subsection{Deconstruction of extra dimensions}\label{sec14}
\fancyhead[RO]{\bfseries \ref{sec14} Deconstruction of extra dimensions
\hfill \thepage}

In the last section we have seen that by relating a five-dimensional
gravitational theory to a four-dimensional quantum field theory, AdS/CFT
introduces an extra dimension in a natural way. Another possibility to
generate extra dimensions is given by the deconstruction technique.  In
Ch.~\ref{cha:deconstruction} we apply the deconstruction method to the dCFT of
intersecting D3-branes for which the AdS dual is discussed in
Ch.~\ref{cha:dCFT}. Generating two compact extra-dimensions in this system, we
obtain a {\em discrete} field theory description for intersecting M5-branes
wrapped on a two-torus. Due to the obstructions of finding a {\em continuous}
lagrangian description of M5-branes, the M5-M5 intersection is not very well
understood at present. We demonstrate that deconstruction is able to
contribute to the understanding of intersecting M5-branes.

Deconstruction is a method to generate (discrete) extra dimensions in theories
with internal gauge symmetries. This innovative method has been developed by
Arkani-Hamed, Cohen and Georgi \cite{Arkani-Hamed:2001} and, independently, by
Hill, Pokorski and Wang \cite{Hill:2000mu} (for early work on this subject,
see \cite{Georgi:au, Halpern:1975yj}). The deconstruction method has been used
in many fields in theoretical high energy physics and phenomenology.  For
instance, lattice gauge theorists make use of the discrete nature of the
generated extra dimensions to study supersymmetric theories on the lattice
\cite{Kaplan}. In phenomenology, deconstruction and the physics of so-called
theory spaces play an important role in stabilizing the electro-weak scale,
see \cite{Arkani-Hamed:2001} and references thereof. In these models the Higgs
boson appears as a pseudo-Goldstone boson protecting the Higgs mass.  In the
context of quantizing gravity, physicists also got interested in the
generation of discrete gravitational dimensions in Einstein's General
Relativity \cite{Arkani-Hamed:2003vb, Arkani-Hamed:2002sp, Schwartz:2003vj,
  Jejjala, Jejjala:2003qg}.  Amongst many other subjects these examples
demonstrate the broad application of deconstruction and the universality of
this method.

Subsequent to these developments Arkani-Hamed et al \cite{Arkani-Hamed}
published a further pioneering work, which made deconstruction highly
interesting for string theorists.  String theory predicts the existence of
non-gravitational theories in five and six dimensions, even though no
consistent Lagrangians are known for interacting theories in these dimensions.
These theories have been discovered by some limit of string theory
configurations involving five-branes. A particularly interesting example is
the six-dimensional theory with $(2,0)$ supersymmetry describing the
decoupling limit of multiple parallel M5-branes \cite{Strominger}.  Although
this theory is believed to be a local quantum field theory, obstructions to
finding a Lagrangian description arise because of difficulties in constructing
a non-abelian generalization of a chiral two-form (see for example
\cite{BHS}).  The spectrum includes tensionless BPS strings, which are in some
sense the ``off-diagonal'' excitations of the non-abelian chiral two-form.
Until recently, the only known formulation of this theory was in terms of a
matrix model describing its discrete light cone quantization \cite{Aharony}.

Recently, a formulation was found \cite{Arkani-Hamed, Csaki:2002fy} using the
deconstruction technique: In order to obtain a description for the
six-dimensional superconformal theory located on M5-branes, one considers the
low-energy effective theory of a stack of $kN$ D3-branes at an orbifold of the
type $\CC^2/\ZZ_k$. The resulting theory is a $\N=2$ four-dimensional super
Yang-Mills theory with a product gauge group $SU(N)^k$ whose field content is
given by the (quiver) diagram shown in Fig.~\ref{quiver1}.

\begin{figure}[!ht]
\begin{center}
{\includegraphics[scale=0.8]{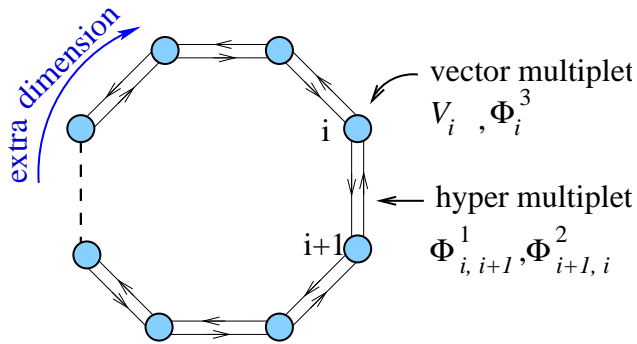}}
\caption{Theory space (quiver diagram) 
for parallel D3-branes at a $\CC^2/\ZZ_k$ orbifold. Each node
corresponds to an ${\cal N} =2$ vector multiplet, while double lines
between neighboring nodes correspond to an $\N=2$ hypermultiplet.}
\label{quiver1}
\end{center}
\end{figure} 

Each of the sites is associated with one of the $k$ $SU(N)$ gauge groups and
represents a vector multiplet $V_{i}$ as well as an adjoint chiral multiplet
$\Phi^3_i$ ($i=1,...,k$) which together form a $\N=2$ vector multiplet. Two
neighbouring sites are connected by two oppositely oriented links representing
the complex scalars $\Phi^1_{i,i+1}$ and $\Phi^2_{i+1,i}$ which together form
a $\N=2$ hyper multiplet $(\Phi^1, {\bar \Phi}^2)_{i,i+1}$. It can now be
argued that at low-energies (on the Higgs branch) this particular field theory
generates two extra dimensions. This means that the theory, which is
four-dimensional at high energies, behaves as a six-dimensional theory at low
energies, in which the two extra dimensions are compactified on a discrete
toroidal lattice. Such a lattice with sites and link fields is called a {\em
  theory space}.

It is not possible to make further statements about this
six-dimensional theory unless one considers the string theory in more
detail from which the above field theory descends in a low-energy
limit.  A string theoretical analysis shows that the Higgs branch of
the field theory corresponds to moving the D3-branes a finite distance
away from the orbifold singularity, where they become M5-branes after
an appropriate T-duality and lift to M-theory. The deconstructed
six-dimensional theory can therefore be identified with the
world-volume theory of M5-branes, which is the 6d (2, 0) superconformal
field theory. In Sec.~\ref{Ch43} we will review the deconstruction
of the non-abelian M5-brane action in much more detail.

In this paper we extend the above deconstruction to the case of intersecting
M5-branes, about which even less is known. For details of this deconstruction
please see Sec.~\ref{Ch44} which is based on \cite{Constable,
  Constable:2002fb}. We will study further the defect conformal field theory
associated with two stacks of D3-branes intersecting each other over a
1+1-dimensional subspace. This brane configuration is placed at a
$\CC^2/\ZZ_k$ orbifold, such that the supersymmetry of the defect conformal
field theory is broken to $(4,0)$.  Applying the method of deconstruction of
\cite{Arkani-Hamed} generates two extra compact dimensions in an appropriate
$k\rightarrow \infty$ limit.  In this way we generate a low-energy description
of two intersecting stacks of M5 branes.  By identifying moduli of the M5-M5
intersection in terms of those of the defect CFT, we will argue that the
$SU(2)_L$ R-symmetry of the $(4,0)$ defect CFT matches the $SU(2)$ R-symmetry
of the ${\cal N} =2, d=4$ theory of the M5-M5 intersection.  An amazing result
is that the intersection is described by a four-dimensional tensionless
string-theory.

In Sec.~\ref{Ch45} we finally elaborate on the fascinating idea to find a
four-dimensional Yang-Mills theory which generates seven extra-dimensions on
its Higgs branch such that it is able to describe M-theory itself.
This section is based on \cite{Oprisa}. We propose
to deconstruct M-theory from a four-dimensional non-super\-symmetric quiver
gauge theory with gauge group $SU(N)^{N_4N_6N_8}$ and $N_{4,6,8}$ three
positive integers. The corresponding orbifold realisation is given by a stack
of D3-branes in type IIB string theory placed at the origin of $\CC^3/\Gamma$,
where the orbifold group $\Gamma$ is the product of three cyclic groups
$\ZZ_{N_4} \times \ZZ_{N_6} \times \ZZ_{N_8}$. The quiver diagram is a
discretized three-torus with a body-centred cubic lattice structure.

At a certain point in the moduli space, each of the $\ZZ_{N}$ factors
generates a circular discretized extra dimension.  In an appropriate
$N_{4,6,8} \rightarrow \infty$ limit, the extra dimensions become continuous,
such that the theory appears to be seven-dimensional on the Higgs branch
(corresponding to the world-volume theory of D6-branes).
There is however a peculiarity in this deconstruction which suggests that the
strongly coupled Higgs branch theory is actually an eleven-dimensional
gravitational theory: The deconstructed seven-dimensional gauge theory breaks
down at a cut-off $\Lambda_{\rm 7d}$. It requires an ultra-violet completion
which by string theory arguments can be shown to be M-theory on an $A_{N-1}$
singularity (the M-theory lift of D6-branes). This suggests the equivalence of
M-theory on $A_{N-1}$ with the continuum limit of the Higgs branch of the
present quiver theory.

The equivalence is also supported by the following properties of the quiver
theory on the Higgs branch. We find Kaluza-Klein states in the spectrum of
massive gauge bosons which are responsible for the generation of three extra
dimensions. Since in M-theory on $A_{N-1}$ the gauge theory is localized at
the singularity which is a seven-dimensional submanifold, we can see three of
the seven extra dimensions in the gauge boson spectrum.  Moreover, we identify
states in the spectrum of massive dyons which are identical to M2-branes
wrapping two of the three compact extra dimensions.

For further applications of deconstruction in string theory see
Refs.~\cite{Mukhi:2002ck, Alishahiha:2002jj, Giedt:2003xr,
  Brax, Persson, Poppitz:2003uz, Iqbal,Dai:2003ak, Dai:2003dy,
  Dai:2004ke}.




\section{Defect conformal field theories and holography}
\label{cha:dCFT}

 \setcounter{equation}{0}\setcounter{figure}{0}\setcounter{table}{0}
 
In this chapter we study  holography of the D3-D$p$ brane intersection
introduced in Ch.~1.  The results on the probe-supergravity side of the
correspondence hold for all $p \in \{3,5,7 \}$. The dual (defect) conformal
field theory can most conveniently be described in terms of $k+1$-dimensional
superfields.  Without loosing to much generality, we specialize the field
theory discussion to the case of intersecting D3-branes (i.e.~\mbox{$p=3$} and
$k=1$) which is formulated in a two-dimensional superspace. Not only that this
theory is interesting by itself as an example of a defect conformal field
theory, in Ch.~\ref{cha:deconstruction} we will need it again for studying
intersecting M5-branes. The D3-D5 intersection can be described in an
analogous approach using a three-dimensional superspace formalism \cite{EGK}
or component notation \cite{DFO}.  The D3-D7 intersection is a degenerate
``defect'' conformal field theory with a codimension zero defect.  The
corresponding field theory can easily be formulated in an $\N=1$
four-dimensional superspace and will be discussed in Ch.\ \ref{cha:flavour}.

This chapter is organised as follows.  In Sec.~\ref{Ch21} we present the
D3-D$p$ brane setup and its dual description in terms of lower-dimensional
AdS-branes in $AdS_5$. In Sec.~\ref{flucsection} we obtain the spectrum of
low-energy fluctuations about the probe geometry.  In
Sec.~\ref{correlatorsection} we determine the dependence of general n-point
functions associated with these fluctuations on the 't~Hooft coupling.  We
compute one-point functions and bulk-defect two-point functions and show that
their scaling behaviour agrees with the general structure fixed by conformal
invariance. In Sec.~\ref{sec3} we study the field theory associated with the
intersecting D3-branes.  In Sec.~\ref{moreD3D3} we focus on some peculiarities
of the D3-D3 intersection which are due to the two-dimensional conformal
symmetry on the defect. For instance, we show that the two-point correlators
of a special class of defect operators do not have the usual power-law
behaviour. Moreover we discuss the classical Higgs branch of intersecting
D3-branes and derive the fluctuation-operator dictionary for the conjectured
AdS/CFT correspondence. We also demonstrate that two-point functions of the
BPS primary operators do not receive any radiative corrections to order $g^2$,
thus providing evidence for a non-renormalization theorem.

\subsection{Holography for the D3-D$p$ brane intersection}\label{Ch21}

\subsubsection{The D3-D$p$ brane configuration} 

We are interested in the conformal field theory describing the low
energy limit of a stack of $N$ D3-branes in the $x^0,x^1,x^2,x^3$
directions intersecting another stack of $N^{\prime}$ D$p$-branes,
where $p \in \{1,3,5,7\}$.  Depending on $p$ the stack of D$p$-branes
is aligned in the directions, as indicated in the following table:
\bigskip
\begin{table}[!h]
\begin{center}
\begin{tabular}{|l|c|c|c|c|c|c|c|c|c|c|}
\hline
D$p$ & 0 & 1 & 2 & 3 & 4 & 5 & 6 & 7 & 8 & 9 \\
\hline
D1   & X &   &   &   & X  &   &   &   &   &   \\
\hline
D$3$ & X & X &   &   & X & X &   &   &   &   \\
\hline
D$5$ & X & X & X  &   & X & X & X  &   &   &   \\
\hline
D$7$ & X & X & X  & X  & X & X & X  & X &   &   \\
\hline
\end{tabular}
\end{center}
\end{table}

\noindent Orthogonal D$p$--D$q$ brane intersections preserve 8~supercharges,
i.e.\ $1/4$ of the maximal supersymmetry, if $p$ and $q$ fulfill the
condition, see e.g.\ \cite{Vijay96},
\begin{align} \label{cond}
p+q -2k = 0\,\, {\rm mod}\,\, 4 \,,
\end{align}
with $k$ the number of intersecting (spatial) dimensions. The
D$3$--D$p$ intersections have $q=3$ and $k=(p-1)/2$ such that 
Eq.~(\ref{cond}) is automatically satisfied. 

The massless open string degrees of freedom of the D$3$--D$p$ intersection
correspond to a ${\cal N}=4$ super-Yang-Mills multiplet (generated by 3-3
strings) coupled to a fundamental hypermultiplet (3-$p$ and $p$-3 strings)
localized at the $k+1$-dimensional intersection. The decoupling of closed
strings is achieved by scaling $N\rightarrow\infty$ while keeping the 't~Hooft
coupling $\lambda\equiv g^2_{YM} N = 4\pi g_s N$ fixed.  This is the usual
't~Hooft limit for the gauge theory describing the $N$ D3-branes. The 't~Hooft
coupling for the $N^{\prime}$ orthogonal D$p$-branes is
\begin{align}
 \lambda' = 2 (2\pi)^{p-2} g_s
l_s^{p-3} N' = \lambda (2\pi l_s)^{p-3} N'/N
\end{align}
which vanishes in the above limit if $N'$ is kept fixed.  This implies that
the $SU(N')$ gauge theory on the D$p$-branes (generated by $p$-$p$ strings)
decouples and the group $SU(N')$ becomes the flavour symmetry of $N'$
flavours. For $\lambda \ll 1$ the appropriate description of this system is
given by a four-dimensional ${\cal N}=4$ $SU(N)$ gauge theory coupled to $N'$
hypermultiplets at a $k+1$ dimensional defect. The D3-D$p$ intersection and the
decoupling of strings is shown on the right hand side of Fig.~\ref{fig2}.

For $\lambda \gg 1$ one may replace the $N$ D3-branes by the geometry
$AdS_5\times S^5$, according to the usual AdS/CFT correspondence.  The
D$p$-branes may be treated as a probe of the $AdS_5\times S^5$
geometry. Comparing the tension of both stacks of branes,
\begin{align}
T_{Dp} = \frac{\nu}{(2\pi l_s)^{p-3}} \,T_{D3} \,\qquad (\nu =N'/N) \,,
\end{align}
we see that the tension $T_{Dp}$ and thus the backreaction of the
D$p$-branes can be neglected in the probe limit $\nu \rightarrow 0$
keeping $\nu/ l_s^{p-3} \ll 1$. As we will see shortly, the
D$p$-branes act as $AdS_{k+2} \times S^k$ probe branes. Consequently,
for large 't~Hooft coupling, the generating function for correlation
functions of the defect CFT should be given by the classical action of
IIB supergravity on $AdS_5 \times S^5$ coupled to a Dirac-Born-Infeld
theory on $AdS_{k+2} \times S^k$. The $AdS$ brane embedding in
$AdS_5 \times S^5$ is shown on the left hand side of Fig.~\ref{fig2}.

\begin{figure}[!ht]
\begin{center}
\includegraphics[scale=0.9]{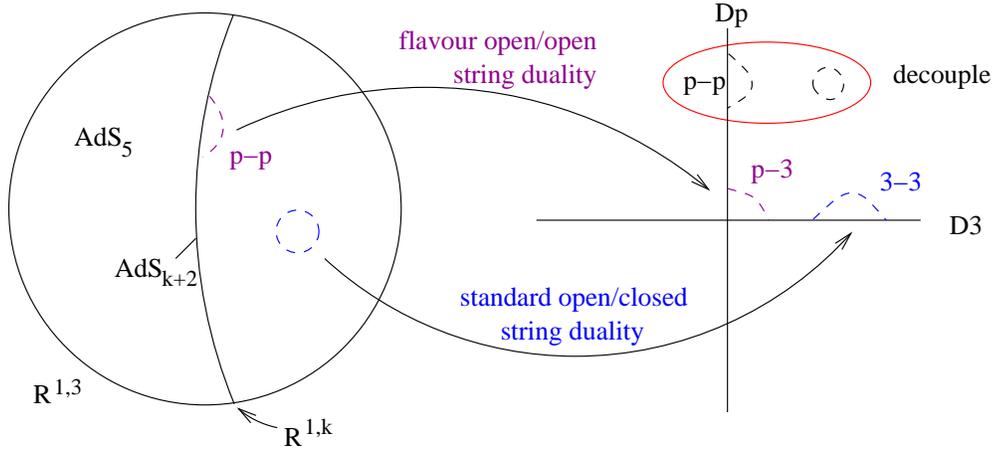}
\caption{AdS/CFT duality for an defect CFT. The duality acts
twice.  Once for the IIB supergravity on $AdS_5 \times S^5$, and
once for DBI theory on $AdS_{k+2} \times S^k$.}  \label{fig2}
\end{center}
\end{figure}

Following the arguments of \cite{Karchrandall1,Karchrandall2,
Karch:2001cw, DFO} we propose that the AdS/CFT
duality ``acts twice'' in the background with an $AdS_{k+2}$-brane
embedded in $AdS_5$. This means that the closed strings on $AdS_5$
should be dual to ${\cal N} =4$ $SU(N)$ super Yang-Mills theory on
$\mathbb{R}^{1,3}$, while open string modes on the probe $AdS_{k+2}$-brane 
should be dual to the fundamental hypermultiplet on
the $\mathbb{R}^{1,k}$ defect (see Fig.~\ref{fig2}). Interactions between the
defect hypermultiplet and the bulk $\N=4$ fields should correspond
to couplings between open strings on the probe D$p$-brane and closed
strings in  $AdS_5 \times S^5$. 

\subsubsection{AdS-branes in $AdS_5 \times S^5$}
We now demonstrate the existence of a one complex parameter family
of $AdS_{k+2} \times S^k$ embeddings of the probe D$p$-branes
in the $AdS_5\times S^5$ background. Consider first the geometry
of the stack of $N$ D3-branes before taking the near horizon
limit. The D3 metric is given by
\begin{align} \label{D3metric}
ds^2 = \left(1+\frac{L^4}{r^4}\right)^{-\frac{1}{2}}
(-dt^2 + dx_1^2+dx_2^2+dx_3^2)+
\left(1+\frac{L^4}{r^4}\right)^{\frac{1}{2}} (dx_4^2+\cdots
+dx_9^2) \,,
\end{align}
with $r^2=x_4^2+...+x_9^2$.  The probe sits at the origin of the space
transverse to its world-volume.  With this choice of embedding the induced
metric on the probe world-volume is
\begin{align}
 ds^2_{probe}= h^{-1/2}\left(-dx_0^2+...+dx_k^2 \right)+
 h^{1/2} d\vec y^2
\label{probemet}
\end{align}
where $h=1+L^4/|\vec y|^4$ ($y^2=x_4^2+...+x_{4+k}^2$) is the harmonic
function appearing in the background geometry evaluated at the
position of the probe.  In the near horizon limit, $L/r\gg 1$, the
D3-brane geometry becomes $AdS_5
\times S^5$,
\begin{align} \label{AdS5metric}
ds^2_{AdS_5 \times S^5} =&
\frac{L^2}{u^2}(-dt^2+dx_1^2+dx_2^2+dx_3^2+du^2) \hspace{1.6cm} \\ &+
L^2\left(d\phi^2_5+ s_{\phi_5}^2 d\phi_4^2 + s_{\phi_5}^2 s_{\phi_4}^2
d{\phi_3}^2 + s_{\phi_5}^2 s_{\phi_4}^2 s_{\phi_3}^2 d{\phi_2}^2 +
s_{\phi_5}^2 s_{\phi_4}^2 s_{\phi_3}^2 s_{\phi_2}^2 d{\phi_1}^2\right) ,
\nonumber
\end{align} where $u\equiv \frac{L^2}{r}$ and we have defined 
angular variables $\phi_1, {\phi_2}, {\phi_3}, {\phi_4}, {\phi_5}$ via
\begin{align}
x^4 &= r s_{\phi_5} s_{\phi_4} s_{\phi_3} s_{\phi_2} s_{\phi_1}
&&x^5 = r s_{\phi_5} s_{\phi_4} s_{\phi_3}
s_{\phi_2} c_{\phi_1} \,,\nonumber \\
x^6 &= r s_{\phi_5} s_{\phi_4}
s_{\phi_3} c_{\phi_2} \,,
&& x^7 = r s_{\phi_5} s_{\phi_4} c_{\phi_3} \,,\nonumber \\ 
x^8 &= r s_{\phi_5} c_{\phi_4}
\,,
&& x^9 = r c_{\phi_5}\,,
\label{sphere} 
\end{align}
where $s_{\phi_1} = \sin\phi_1$, $c_{\phi_1} =
\cos{\phi_1}$ etc. It is instructive to consider this limit from the
point of view of the probe metric. One can easily show that in the
near-horizon region the induced metric on the probe becomes 
\begin{align}
ds^2_{probe}
=\frac{{L}^2}{\tilde{u}^2}\left(-dx^2_0+...+dx_k^2+d\tilde{u}^2\right)
+{L^2} d\Omega_k^2 \label{nearprobe} \,, 
\end{align} where 
$\tilde{u} = u|_{x^{5+k},...,x^9=0}$. One immediately recognizes
Eq.~(\ref{nearprobe}) as the metric on $AdS_{k+2}\times S^k$ with radius
of curvature~${L}$. The probe wraps a $k$-sphere $S^k$ of maximal
radius inside the $S^5$. We summarize the AdS-branes in Tab.~\ref{tabAdS}.
\begin{table}[ht]
\begin{center}
\begin{tabular}{|c|c|c|}
\hline
D1: &$(0 \vert$ D1 $\perp$ D3) & $AdS_2$ \\
\hline 
D3: &$(1 \vert$ D3 $\perp$ D3) & $AdS_3 \times S^1$ \\
\hline
D5: &$(2 \vert$ D5 $\perp$ D3) & $AdS_4 \times S^2$ \\
\hline
D7: &$(3 \vert$ D7 $\perp$ D3) & $AdS_5 \times S^3$ \\
\hline
\end{tabular}
\caption{AdS-brane embeddings in $AdS_5 \times S^5$ which preserve
8 supercharges. \mbox{$(k \vert$ Dp $\perp$ D3)} denotes the D3-D$p$ brane
intersection on $k$ spatial dimensions.} \label{tabAdS}
\end{center}
\end{table}

The boundary of the embedded $AdS_{k+2}$-brane is a $k+1$-dimensional
Minkowski space $\mathbb{R}^{1,k}$ at $\tilde{u}=0$, and lies within the
$\mathbb{R}^{1,3}$ boundary of $AdS_5$. This embedding is indeed
supersymmetric, as was verified in
\cite{Skenderis}. Thus this configuration is stable despite the
fact that the $S^k$ is contractible within the $S^5$.  As we will see
in Sec.~\ref{ch222}, the naively unstable modes associated with contracting the
$S^k$ satisfy the Breitenlohner-Freedman bound \cite{BF} for scalars
in $AdS_{k+2}$, and therefore do not lead to an instability.

\subsubsection{Isometries}

In the absence of the (probe) D$p$-branes, the isometry group of the
$AdS_5 \times S^5$ background is $SO(2,4) \times SO(6)$.  The
$SO(2,4)$ component acts as conformal transformations on the
boundary of $AdS_5$, while the $SO(6) \simeq SU(4)$ isometry of
$S^5$ is the R-symmetry of four-dimensional ${\N =4}$ super
Yang-Mills theory,  under which the six real scalars
$X^{4,5,6,7,8,9}$ transform in the vector representation.

In the presence of the D$p$-branes, the $AdS_5 \times S^5$
isometries are broken to the subgroup which leaves the embedding
equations of the D$p$-branes invariant:
\begin{align} 
SO(2,4)  \times SO(6)  \rightarrow SO(2,k+1) \times SO(3-k) 
 \times SO(k+1) \times SO(5-k) 
\label{unb}
\end{align} 
Out of the $SO(2,4)$ isometry of $AdS_5$ only $SO(2,k+1) \times
SO(3-k)$ is preserved. The $SO(2,k+1)$ factor is the isometry group of
$AdS_{k+2}$, while the $SO(3-k)$ factor (non-trivial only for $k=0,1$)
acts as a rotation of the coordinates $X^{k+1},...,X^3$. Out of the
$SO(6) \simeq SU(4)$ isometry of $S^5$, only $SO(k+1) \times SO(5-k)$
is preserved.  The $SO(k+1)$ factor here rotates the $S^k$ of the
D$p$-brane world-volume.  The $SO(5-k)$ component acts on the
coordinates $X^{5+k,...,9}$.

\subsection{Fluctuations in the probe-supergravity background} \label{flucsection}

Following the conjecture put forth in \cite{Karchrandall1,Karchrandall2,
  Karch:2001cw} and elaborated upon in \cite{DFO}, we expect the holographic
duals of defect operators localized on the intersection are open strings on
the D$p$-branes, whose world-volume is an $AdS_{k+2} \times S^k$ submanifold
of $AdS_5 \times S^5$.  The operators with protected conformal dimensions
should be dual to probe Kaluza-Klein excitations at ``sub-stringy'' energies,
$m^2 \ll \lambda/L^2$. In this section we shall find the mass spectra of these
excitations.  Later we will find this spectrum to be consistent with the
dimensions of operators localized on the intersection.

\subsubsection{The probe-supergravity system}

The full action describing physics of the background as well as
the probe is given by 
\begin{align}
 S_{bulk}=S_{IIB}+S_{DBI}+S_{WZ} \,.
\label{totact} 
\end{align} 
The contribution of the bulk supergravity piece of the action in
Einstein frame is 
\begin{align}
 S_{IIB}=\frac{1}{2\kappa^2}\int
 d^{10}x\sqrt{-g}\left(R-\frac{1}{2}e^{2\Phi} (\partial\Phi)^2
 +\cdots\right) \label{IIB}  \,,
\end{align}
where $2\kappa^2=(2\pi)^7g_s^2l_s^8$. The dynamics of the probe
D$p$-brane is given by a Dirac-Born-Infeld term and a Wess-Zumino term
\cite{leigh}, 
\begin{align}
S&=S_{DBI} + S_{WZ} \,, 
\end{align}
where
\begin{align}
S_{DBI}&= -T_{Dp}\int d^{p+1} \sigma
e^{-\Phi}\sqrt{-\det\left(g_{ab}^{PB} + e^{-\Phi/2}{\cal
F}_{ab}\right) }  \,. \label{d3act}
\end{align}
The brane tension $T_{Dp}$ is given by $T_{Dp}=(g(2\pi)^p
\alpha'^{(p+1)/2})^{-1}$ with $g=\kappa/(8\pi^{7/2}\alpha'{}^2)$.  The metric
$g_{ab}^{PB}$ is the pull back of the bulk $AdS_5 \times S^5$ metric to the
world-volume of the probe. ${\cal F}_{ab} = B_{ab}+ 2\pi l_s^2F_{ab}$ is the
total world-volume field strength. The Wess-Zumino action $S_{WZ}$ will be
discussed in Sec.~\ref{wierd}.

We work in a static gauge where the world-volume coordinates of the
brane are identified with the space time coordinates by $\sigma^{a}
\sim x^0,...,x^k,u,\phi_{1},...,\phi_{k}$. With this identification the
DBI action is 
\begin{align}
 S_{DBI}=-T_{Dp} \int d^{p+1} \sigma\sqrt{-\det\left(g_{ab}+
\partial_aZ^i\partial_bZ^jg_{ij}+ {\cal F}_{ab} +
2g_{ai}\partial_bZ^i\right)} \label{actt}  \,,
\end{align} 
where $i,j$ label the transverse directions to the probe and the scalars $Z^i$
represent the fluctuations of the transverse scalars $X^{k+1},...,X^3,
\phi_{k+1},...,\phi_{5}$. Also, $e^{-\Phi}=g_s^{-1}=1/g_{YM}^2$ has been set
to one.  Henceforth we will only consider the open string fluctuations on the
probe and thus drop terms involving closed string fields\footnote{Such terms
  encode the physics of operators in the bulk of the dual ${\cal N} =4$ theory
  restricted to the defect.} $B_{ab}$ and $g_{ai}$. The embedding conditions
of the $AdS_{k+2}$-brane are given by
\begin{align} \label{embedcond}
X^{k+1} = ... = X^3= 0 \,,\qquad
\phi_{k+1} = ... = \phi_{5}= \frac{\pi}{2} \,,
\end{align}
where the conditions on the $\phi_i$ are equivalent to $X^{k+5}= ...  =X^9=0$,
as can be seen from Eq.~(\ref{sphere}). The $AdS$-brane wraps a maximal
$k$-sphere $S^k$ inside $S^5$.  To quadratic order in fluctuations, the action
takes the form\footnote{For a detailed computation see App.~\ref{appDBI}.}
\begin{align}
S_{DBI}=-T_{Dp}L^{p+1} \int d^{p+1} \sigma \sqrt{\bar g_{p+1}}
&\left[1+ \sum_{i=k+1}^{5} (\frac{1}{2}\partial_a\phi'_i\partial^a\phi'_i
  -\frac{k}{2} {\phi'}_i^2 ) \right.  \label{quadBI}\\
&\left. + \sum_{i=k+1}^3
\frac{1}{2u^2}\partial_a X^i \partial^a X^i
+\frac{1}{4}(2\pi l_s^2)^2
F_{ab} F^{ab} \right] \,, \nonumber
\end{align}
where $\phi'_i \equiv\phi_i-\frac{\pi}{2}$ and $\bar{g}_{p+1}$ is the
determinant of the rescaled $AdS_{k+2} \times S^k$ metric $\bar
g^{p+1}_{ab}$ given by
\begin{align} \label{AdSkmetric}
d\bar s^2
=\frac{1}{{u}^2}\left(-dt^2+...+dx_k^2+d{u}^2\right)
+ d\Omega_k^2 \,.
\end{align}

\subsubsection{$S^k$ fluctuations inside $S^5$} \label{ch222}

From Eq.~(\ref{quadBI}) we see that the angular fluctuations $\phi'_{k+1},
..., \phi'_{5}$ are minimally coupled scalars on $AdS_{k+2}\times S^k$
satisfying the equation of motion 
\begin{align}
(\Box+k) \phi' = 0 \,,\quad{\rm with}\quad \Box=\Box_{AdS} 
+ \Box_{S^k} \,.
\end{align}
  Interestingly
they have $m^2 =-k$ which, although negative, satisfies (saturates for
$k=1$) the Breitenlohner-Freedman bound $m^2
\ge -d^2/4$, where $d=k+1$ for $AdS_{k+2}$. We can separate variables
by means of the ansatz
\begin{align}
\phi'= \phi'_l(\vec{x},u) {\cal Y}^l(S^k) \,,
\end{align}
where the spherical harmonics on $S^k$ satisfy
\begin{align}
\Box_{S^k} {\cal Y}^l = - l(l+k-1) {\cal Y}^l \,
\end{align}
with $\Box_{S^k}$ the Laplacian on the $k$-sphere $S^k$. For the
Kaluza-Klein modes of these scalars we then find the masses $m^2 = -k
+ l (l+k-1)$. This leads to a spectrum of conformal dimensions of dual
defect operators given by
\begin{align}
\Delta_\pm = \frac{1}{2} \left(k+1 \pm (2l+k-1) \right) =
\left\{ \begin{matrix} k + l \\ 1 - l \end{matrix} \right. \,,
\end{align}
where we have used the standard $AdS_{d+1}/CFT_d$ relation
(\ref{massoprel}) for a scalar and $d=k+1$.  For $l>0$ one should
choose the positive branch for unitarity, while for $l<0$ one should
choose the negative branch. To leading order in fluctuations of the
$S^k$ embedding we see from (\ref{sphere}) that
\begin{align}
x^{k+5} &= -r \phi'_{k+1} \nonumber\\
\vdots  \\
x^{9} &= -r \phi'_{5} \nonumber
 \,. 
\end{align} 
Thus the angular variables $\phi'_{k+1}, ...,\phi'_{5}$ belong to
the vector representation of the group $SO(5-k)$ which is
part of the isometry group (\ref{unb}).

\subsubsection{Gauge field fluctuations \label{gff}}

Let us now turn to the fluctuations of the world-volume gauge
field. It is convenient to rescale fields according to $\hat{F}_{ab} =
2\pi l_s^2 F_{ab}$ so that the gauge field fluctuations have the same
normalization as the scalars in the previous subsection. We have
\begin{align}
S_{gauge}&=-T_{Dp}L^{p+1} \int d^{p+1}\sigma\sqrt{\bar{g}_{p+1}}
\frac{1}{4} \hat{F}_{ab}\hat{F}^{ab} \,
\nonumber \\
&=-T_{Dp} L^{p+1} \int d^{p+1}\sigma\sqrt{\bar{g}_{p+1}}\frac{1}{4}
\left(\hat{F}_{\mu\nu}\hat{F}^{\mu\nu} +
  2\hat{F}_{\mu\alpha}\hat{F}^{\mu\alpha} + 
 \hat{F}_{\alpha\beta}\hat{F}^{\alpha\beta}  \right) \,.
\label{gaugeact} 
\end{align}
In order to decouple the $AdS_{k+2}$ components ($\hat A_\mu$) of the gauge
field from that on the $S^k$ ($\hat A_\alpha$) it is convenient to work in
the gauge $D^\alpha \hat A_\alpha=0$. The last term in Eq.~(\ref{gaugeact})
vanishes in this gauge. Expanding the components $\hat A_\mu$ in spherical
harmonics on the $S^k$ so that $\hat{A}_{\mu}=\hat{A}^l_{\mu} {\cal
Y}^l(S^k)$, we find the equations of motion  
\begin{align}
D^{\mu}\hat{F}^l_{\mu\nu} + l(l+k-1) \hat{A}^l_{\nu} = 0
\label{procaeqn}
\end{align}
which are just the Maxwell-Proca equations for a vector field with
$m^2=l(l+k-1)$. Using the standard relation (\ref{massoprel}) with $p=1$ 
relating the mass of a one-form field to the
dimension of its dual operator we find the spectrum 
\begin{align}
 \Delta_\pm = \left\{ \begin{matrix} k + l \\ 1 - l \end{matrix}
 \right. \,,
\end{align}
which for $l>1$ requires us to choose the positive branch.

\subsubsection{$AdS_{k+2}$ fluctuations inside $AdS_5$}
\label{wierd}

Let us finally compute the conformal dimensions of the operators dual
to the scalars which describe the fluctuations of the probe inside of
$AdS_5$. Here we specialize to the case of intersecting D3-branes
\mbox{(1$\vert$D3 $\perp$ D3)}, i.e.\ we consider $AdS_{3}$
fluctuations inside $AdS_5$. In the \mbox{(2$\vert$D3 $\perp$ D5)}
configuration the computation of $AdS_4$ fluctuations $Z^x \equiv X^3$
inside $AdS_5$ is slightly more involved due to couplings
$F_{\alpha\beta} \partial_u Z^x \subset S_{WZ}$ to the gauge field
fluctuations. These $AdS_4$ fluctuations are discussed in
\cite{DFO}. The \mbox{(3$\vert$D3 $\perp$ D7)} does not have any $AdS$
fluctuations since the D7-branes are spacetime filling.

For the $AdS_3$ fluctuations represented by the scalars $X^2$ and $X^3$ 
we require the Wess-Zumino term 
\begin{align}
S_{WZ} = - \,T_{Dp}\int \left( 
C^{(p+1)}_{PB} +  C^{(p-1)}_{PB} \wedge {\cal F} + ... \right) \,,
\end{align}
where $C^{(q)}_{PB}$ is the pull-back of a bulk Ramond-Ramond
q-form to the D$p$-brane. In the $AdS_5 \times S^5$ background
only $C^{(4)}_{PB}$ given by   
\begin{eqnarray}\label{comp}
 C^{PB}_{abcd}&=& C_{abcd} + 4\partial_{[a}Z^iC_{bcd]i}
 +6\partial_{[a}Z^i\partial_{b}Z^{j}C_{cd]ij} \nonumber\\ &+&4
 \partial_{[a}Z^i\partial_{b}Z^{j}\partial_{c}Z^kC_{d]ijk} +
\partial_{[a}Z^i\partial_{b}Z^{j}\partial_{c}Z^k\partial_{d]}Z^lC_{ijkl} \, 
\end{eqnarray}
is a nonvanishing Ramond-Ramond field. We can choose a gauge in
which
\begin{align}
C_{0123} = \frac{L^4}{u^4} 
\end{align}
while the remaining components,  which are determined by the self
duality of $dC^{(4)}$, contribute only to terms in the pull back
with more than two $\partial Z$'s.  We do not need such terms to
obtain the fluctuation spectrum.  The quadratic term arising from
(\ref{comp}) is 
\begin{align}
C_{PB}^{(4)}=\left(\partial_{u}X^2\partial_{\phi_1}X^3-\partial_{u}X^3
\partial_{\phi_1}X^2\right)
C_{0123}dt \wedge dx^1\wedge du\wedge d\phi_1 \, .
\end{align}
The Wess-Zumino action is then
\begin{align}
S_{WZ} = - T_{D3}L^4 \int d^4\sigma \frac{1}{u^4}(\partial_u X^2
\partial_{\phi_1}X^3 - \partial_u X^3 \partial_{\phi_1} X^2) \,.
\label{quadWZ}
\end{align}
From Eqns.~(\ref{quadBI}) and (\ref{quadWZ}) the action
for $X^2$ and $X^3$ is 
\begin{align}
S_{2,3} = &-T_{D3}L^4\int d^4\sigma\sqrt{\bar{g}_4} 
\left(\frac{1}{2u^2}\partial_a
X^2\partial^aX^2 +\frac{1}{2u^2}\partial_aX^3\partial^aX^3\right)
\nonumber
\\
&+ T_{D3}L^4\int d^4\sigma\left(\frac{1}{u^4}\partial_{\phi_1} X^2
\partial_u  X^3 - \frac{1}{u^4}\partial_u X^2 \partial_{\phi_1}  X^3\right)
\label{x2x3act}               \,. 
\end{align}
Writing $\sqrt{2\pi}X^i = X^i_l \exp(il{\phi_1})$ for $i=2,3$ and doing the
integral over $\phi_1$ gives
\begin{align}
 S_{2,3}=&- T_{D3}L^4\int
d^3\sigma\sqrt{g_3}\left(1+\frac{1}{2u^2}(g_3^{ab}\partial_aX_{-l}^i
\partial_bX^i_l + l^2 X^i_{-l}X^i_l)\right) \nonumber\\ 
&+ T_{D3}L^4\int d^3\sigma
\frac{1}{u^4}\left(il X_l^3\partial_{u}X_{-l}^2 -
ilX^2_l\partial_{u}X_{-l}^3 \right) \,,
\end{align} where $g^3_{ab}$ is the
metric for the $AdS_3$ geometry
\begin{align}
ds^2=\frac{1}{u^2}\left(-dt^2+dx_1^2+du^2\right) \,.
\end{align}
The $X^2, X^3$ mixing in the Wess-Zumino term is diagonalized by
working with the field $w_l \equiv X^2_l + i X^3_l$, in terms of
which the action is
\begin{align}
S_w = - T_{D3}L^4 \int d^3 \sigma \left(\sqrt{g_3}\frac{1}{2u^2}(g^{ab}_3
\partial_aw_l^* \partial_b w_l + l^2 w^*_l w_l) 
+  \frac{1}{2u^4} \partial_u(lw^*_l
w_l) \right) \,.
\end{align}
The usual action for a scalar field in $AdS_3$ is obtained by defining
$\tilde w_l = w_l/u$, giving
\begin{align}
S_w = - T_{D3}L^4 \int d^3\sigma
\sqrt{g_3}\frac{1}{2}\left(g_3^{ab}\partial_a\tilde w^*_l\partial_b\tilde w_l + (l^2
-4l + 3) \tilde w^*_l \tilde
w_l\right) \label{reg}\\
+  T_{D3}L^4 (l-1) \int d^3 \sigma \frac{1}{2}\partial_u (\frac{1}{u^2}
\tilde w^*_l \tilde w_l) \label{surf} \,.
\end{align}
The surface term (\ref{surf}) does not effect the equations of motion,
but will be significant later when we compute correlation functions of
the dual operators. Inserting the spectrum $m^2 =l^2 - 4l+3$ into the
standard formula (\ref{massoprel}) for a scalar gives
\begin{align}
 \Delta = 1\pm |l-2|\,. \label{ads4series}
\end{align}
This gives two series of dimensions, $\Delta = l-1$ and $\Delta =
3-l$, which are possible in the ranges of $l$ for which $\Delta$ is
non-negative.  The entry in the AdS/CFT dictionary for the series
$\Delta = l-1$ holds several remarkable surprises which we will
encounter in Sec.~\ref{moreD3D3}.

\subsection{Correlators from strings on the probe-super\-gravity background}
\label{correlatorsection}

The rules for using classical supergravity in an AdS background to
compute CFT correlators have a natural generalization to defect CFT's
dual to AdS probe-supergravity backgrounds. The generating function
for correlators in the defect CFT is identified with the classical
action of the combined probe-supergravity system with boundary
conditions set by the sources. This approach was first used to compute
correlators in the dCFT describing the D3-D5 system in
\cite{DFO}. In the following we compute the general structure of 
bulk one-point and bulk-defect two-point correlators in a general
D3-D$p$ system using the holographic dual of the corresponding defect
CFT.  

\subsubsection{Dependence of the correlators on the 't\,\,Hooft \mbox{coupling} 
and the number of colours}
\label{sec41}
As in refs.~\cite{Boonstra, Bianchi, DFO} it is useful to work with a
Weyl rescaled metric
\begin{align}g_{MN}=L^2\bar{g}_{MN} \,,\end{align}
where $L^2 \sim \sqrt{g_sN}l_s^2$.
In terms of the rescaled metric, the supergravity action
(\ref{IIB}) becomes
\begin{align}
 \frac{L^8}{2\kappa^2}\int
d^{10}x\sqrt{-\bar{g}}\left(R-\frac{1}{2}e^{2\Phi}
(\partial\Phi)^2 +\cdots\right)\sim N^2\int d^{10}x\sqrt{-\bar{g}}
\left(R-\frac{1}{2}e^{2\Phi} (\partial\Phi)^2 +\cdots\right)
\label{scaledact} 
\end{align} 
As in the usual AdS/CFT correspondence correlation functions of gauge
invariant operators in the bulk of \mbox{$\N=4$} SYM at large 't~Hooft
coupling are calculated by expanding this action around the $AdS_5 \times S^5$
vacuum of type IIB. Here the presence of the probe D$p$-brane will make
additional contributions both through its world-volume fields but also through
the pull backs of the $AdS_5 \times\, S^5$ fields. Terms involving the pull
backs are dual to couplings between the bulk of the field theory and the $k+1$
dimensional defect.  After Weyl rescaling the metric as above, the DBI action
for the D$p$-brane becomes
\begin{align}
 S_{DBI}&=-L^{p+1}T_{Dp}\int
d^{p+1}{\sigma}\sqrt{\bar{g}}(1+ {\rm fluctuations})\nonumber \\ 
&\sim N \lambda^{(k-1)/2} \int
d^{p+1}{\sigma}\sqrt{\bar{g}}(1+ {\rm fluctuations}) \,,
\label{scaleBI} 
\end{align}
where we used $T_{Dp} \sim (g_s l_s^{p+1})^{-1}$ and $k=(p-1)/2$.  The
Wess-Zumino term scales identically in $N$ and~$\lambda$.  Generic correlation
functions involving $n$ fields $\psi$ living on the D$p$-brane probe and $m$
fields $\phi$ from the bulk of $AdS_5$ arise from
\begin{align}
 S_{DBI}&=N \lambda^{(k-1)/2}\int
d^4\sigma\left((\partial\psi)^2+\phi^m\psi^n\right) 
\nonumber \\
&=\int d^{p+1}\sigma\left((\partial\psi^{\prime})^2+
\frac{1}{N^{n/2+m-1} 
\lambda^{(n/4-1/2)(k-1)}}\psi^{\prime}{}^n\phi^{\prime}{}^m\right) \,,
\label{scales}
\end{align}
where $\psi^{\prime}=N^{1/2} \lambda^{(k-1)/4}\psi$ and
$\phi^{\prime}=N\phi$ are the canonically normalized probe and $AdS_5$
fields respectively. The DBI action (\ref{scales}) of the D$p$-brane
probe determines the scale dependence on $N$ and $\lambda$ of the
correlation functions of defect and bulk operators $\hat {{\cal
O}}_{\hat \D}$ and ${{\cal O}}_\D$. For the bulk one-point
function we have ($m=1$, $n=0$) and Eq.~(\ref{scales}) shows that the
one-point function scales like $\lambda^{(k-1)/2}$. The two-point
function of a bulk field and a defect field ($m=1$,
$n=1$) scales like $\lambda^{(k-1)/4}/N^{1/2}$. The two-point
function of two defect fields ($m=0$, $n=2$) is independent of $N$
and~$\lambda$.

The system $(1 \vert$ D3 $\perp$ D3) (for which $k=1$) is peculiar in
that the dependence on the 't~Hooft coupling $\lambda = g_sN$ drops
completely out of the normalization of the action!  In this case it is
interesting to observe that none of the correlation functions has any
dependence on $\lambda$, at least in the strong-coupling regime where
the $AdS$ probe-supergravity description is valid.

\subsubsection{SUGRA calculation of one-point functions and bulk-defect 
two-point functions}

We now compute the space-time dependence of the bulk one-point and the
bulk-defect two-point function using holographic methods and show that
their structure agrees with the general results obtained from
conformal invariance in App.~\ref{confs}.  The one-point function
of the bulk operator ${{\cal O}}_\D$ is the integral of the standard
bulk-boundary propagator (\ref{btobprop}) in $AdS_5$
over the $AdS_{k+2}$ subspace. We find
\begin{align}
\langle {{\cal O}}_\D(\vec x, \vec y) \rangle &=
\lambda^{(k-1)/2} \int \frac{dw
d\vec w^{k+1}}{w^{k+2}} \frac{\G(\D)}{\pi^2 \G(\D-2)}
\left(\frac{w}{w^2+\vec x^2 + (\vec w-\vec y)^2} \right)^{\D}
\nonumber\\
&=\lambda^{(k-1)/2} \frac{1}{\vert \vec x \vert^{\D}}
\frac{\G(\frac{\D}{2}) \G(\frac{\D}{2}-\frac{k+1}{2})}{2 \G(\D-2)} \pi^{(k+1)/2-2} \,
\end{align}
which converges for $\D > k+1$ (for details of the computation 
see App.~\ref{appA2}). The scaling
behaviour $\vert \vec x
\vert^{-\D}$ has been expected from the structure of the one-point
function (\ref{onepoint}) on the CFT side. 

\begin{figure}[!ht]
\begin{center}
\includegraphics[scale=0.9]{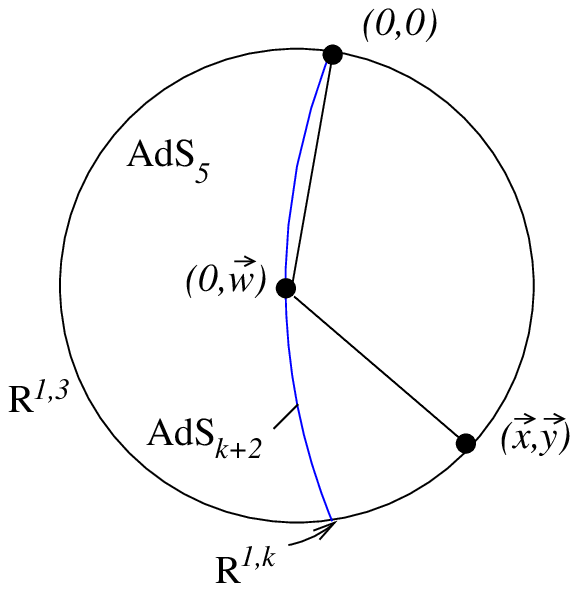}
\caption{Witten diagram for bulk-defect two-point functions.}  
\label{Figwitten}
\end{center}
\end{figure}
\vspace{-7.3cm} \hspace{6.5cm} $\hat {{\cal O}}_{\hat \D}$

\vspace{1.1cm} \hspace{7cm} $K_{\hat \D}$

\vspace{1.0cm} \hspace{7.6cm} $K_{\D}$

\vspace{0.8cm} \hspace{8.2cm} ${{\cal O}}_\D$
\vspace{2.6cm}

The two-point function $\langle {{\cal O}}_\D (\vec x, \vec y)
\hat {{\cal O}}_{\hat \D} (0) \rangle$ is the integral over the
product of the bulk-boundary propagators $K_{\D}\big(w,(\vec
x,\vec y),(\vec 0,\vec w)\big)$ and $K_{\hat \D}\big(w,(\vec
0,\vec w),(\vec 0,\vec 0)\big)$,
\begin{align}
  \langle {{\cal O}}_\D(\vec x, \vec y) \hat {{\cal O}}_{\hat \D} (\vec 0)
\rangle =
  \frac{\lambda^{(k-1)/4}}{N^{1/2}} \frac{\G(\D)}{\pi^2 \G(\D-2)}
  \frac{\G(\hat \D)}{\pi^{(k+1)/2} \G(\hat \D-\frac{k+1}{2})} J( \vec x, \vec
  y; \D, \hat \D)
\end{align}
with the integral
\begin{align}
  J( \vec x, \vec y; \D, \hat \D)&=\int \frac{dw d\vec w^{k+1}}{w^{k+2}}
  \left(\frac{w}{w^2+\vec x^2 + (\vec w-\vec y)^2} \right)^{\D}
  \left(\frac{w}{w^2 + \vec w^2} \right)^{\hat \D} \,\\
  &= \frac{1}{(\vec x^2 + \vec y^2)^{\D}} \int dw' d\vec w'^{k+1}
     \, w'^{\hat \D-(k+2)}
     \left(\frac{w'}{w'^2+\vec x'^2 + (\vec w'-\vec y')^2} \right)^{\D}\,.
\nonumber
\end{align}
In the last line we made use of the inversion trick
\cite{FreedmanMathur} by defining
\begin{align}
(w',0,\vec w') = \frac{1}{w^2+\vec w^2}(w,0,\vec w)  ,\qquad (\vec
x',\vec y') =\frac{1}{\vec x^2 + \vec y^2}(\vec x,\vec y) \,.
\end{align}
The corresponding Witten diagram is shown in Fig.~\ref{Figwitten}.

As in \cite{DFO}, we rescale $\vec w' =\vec y' + \sqrt{\vec x'^2 +
w'^2} \vec v$ and $w'=\vert \vec x' \vert u$ and find
\begin{align}
J( \vec x, \vec y; \D, \hat \D)&= \frac{1}{(\vec x^2 + \vec
y^2)^{\hat \D} \vert \vec x \vert^{\D -\hat \D}} \int du
\frac{u^{\hat \D-(k+2)+\D}}{(1+u^2)^{\D-(k+1)/2}}
\int d\vec v^{k+1} \frac{1}{(1+\vec v^2)^{\D}} \nonumber\\
&=\frac{1}{(\vec x^2 + \vec y^2)^{\hat \D} \vert \vec x \vert^{\D
-\hat \D}} \frac{ \G(\frac{\D-\hat \D}{2}) \G(\frac{\D+\hat \D}{2}-\frac{k+1}{2})}
{2 \G(\D)} \,.
\end{align}
This converges if $\D > \hat \D$ and $\D+\hat \D>k+1$.  The scaling $
1/( (\vec x^2 +
\vec y^2)^{\hat \D} \vert \vec x \vert^{\D -\hat \D} )$ agrees
with the behaviour of the two-point function fixed by conformal
invariance, cf.\ Eq.\ (\ref{twopoint}).

The defect-defect correlator $\langle \hat {{\cal O}}_{\hat \D}(
\vec y) \hat {{\cal O}}_{\hat \D} (\vec 0) \rangle$ for a defect operator 
$\hat {{\cal O}}_{\hat \D}$ is given by \cite{FreedmanMathur}
\begin{align}
\langle \hat {{\cal O}}_{\hat \D}(
\vec y) \hat {{\cal O}}_{\hat \D} (\vec 0) \rangle
= \eta \epsilon^{2(\hat \Delta-d)} \frac{2\hat\Delta-d}{\hat \Delta}
\frac{\Gamma(\hat \Delta+1)}{\pi^\frac{d}{2} \Gamma (\hat\Delta - \frac{d}{2})}
\frac{1}{\vert \vec y \vert^{2\hat \Delta}}
\end{align}
with $d=k+1$, which holds for $\hat\Delta > d/2$.\footnote{As will see
in Sec.~\ref{vanish}, two-point functions of defect operators may
deviate from the power-law behaviour $1/\vert \vec
y\vert^{2\hat\Delta}$ in the case of intersecting D3-branes.}


\subsection{The conformal field theory of the D3-D3 brane intersection}
\label{sec3}

Thus far we have only studied the defect CFT on the D3-D$p$ intersection with
\mbox{$p \in \{3,5,7 \}$} in terms of its holographic dual, without ever
writing the action. In the following we demonstrate the construction of its
action specialising to $p=3$.  We will find the low-energy effective action
of $N$ D3-branes orthogonally intersecting $N'$ D$3^{\prime}$-branes
over two common dimensions. In the discussion of holo\-graphy it was assumed
that $N\rightarrow \infty$ with $\lambda=g_{YM}^2N$ and $N'$ fixed, such that
the open strings with both endpoints on the D3$'$-brane decoupled. We will not
make this assumption in constructing the action.

The $\N=4$ SYM $SU(N)$ theory located on the D3-branes and the $\N=4$
SYM $SU(N')$ theory located on the D$3^{\prime}$-branes couple to a
$(4,4)$ hypermultiplet at a two-dimensional defect.  Although $(4,4)$
supersymmetry with $8$ supercharges is preserved, it is convenient to
work with $(2,2)$ superspace.\footnote{$(2,2)$ and $(4,4)$
superymmetry in two dimensions can be obtained from $\N=1$ and $\N=2$
superspace formalism in four dimensions upon dimensional reduction.}
The world-volume of both stacks of D3-branes can be viewed as two
$\N=2, d=4$ superspaces, intersecting over a two-dimensional $(2,2)$
superspace.  One of the $\N=2,d=4$ superspaces is spanned by
\begin{align} {\cal X} \sim (z^+,z^-, w, \bar w,
\theta_{(i)}^{\alpha},\bar\theta^{(i)}_{\dot \alpha}) \,,
\end{align} with $z^\pm = X^0\pm X^1$ and $w=X^2+iX^3$. The index
$\alpha$ is a spinor index with values $1,2$, while the index $i$
accounts for the $\N=2$ supersymmetry and has values $1,2$. The
other $\N=2,d=4$ superspace is spanned by \begin{align}{\cal X}'
\sim (z^+, z^-,y,\bar y, \Theta_{(i)}^{\alpha},
\bar\Theta^{(i)}_{\dot\alpha})\,,\end{align} where $y= X^4 + iX^5$ and
one makes the identification\footnote{We put brackets around the
indices 1 and 2, which label the two Grassmann coordinates, in
order to distinguish these indices from spinor indices $\a,
\dot\a=1,2$.}
\begin{align}
\theta^1_{(1)} = \Theta^1_{(1)} \equiv \theta^+ \,, \\
\theta^2_{(2)} = \Theta^2_{(2)} \equiv \bar\theta^- \,.
\end{align} This is not the unique choice.  For instance
one could have written $\theta^2_{(2)} = \Theta^2_{(2)} \equiv
\theta^-$ which is related to the first choice by mirror symmetry
\cite{Hori}. The intersection is the $(2,2), d=2$ superspace
spanned by \begin{align} {\cal X} \cap {\cal X}' \sim (z^+,z^-,
\theta^+, \theta^-, \bar\theta^+, \bar\theta^-)\,.\end{align} All
the degrees of freedom describing the D3-D3$'$ intersection can be
written in $(2,2)$ superspace. For instance the D3-D3 strings,
which are not restricted to the intersection, can be described by
$(2,2)$ superfields carrying extra (continuous) labels $w,\bar w$.
Similiarly superfields associated to the
D$3^{\prime}$-D$3^{\prime}$ strings carry the extra labels $y,\bar
y$. Fields associated to D3-D$3^{\prime}$ strings are localized on
the intersection and have no extra continuous labels.

Due to the breaking of four-dimensional supersymmetry by the couplings to the
degrees of freedom localized at the intersection, it is convenient to write
the action in a language in which the unbroken $(2,2)$ symmetry is manifest.
This leads to a somewhat unusual form for the four-dimensional parts of the
action.  One way to obtain this action is somewhat akin to
deconstruction~\cite{Arkani-Hamed:2001}. The basic idea is to start with a
conventional $(4,4)$ two-dimensional action in $(2,2)$ superspace, add an
extra continuous label $w,\bar w$ to all the fields, and then try to add terms
preserving $(4,4)$ supersymmetry such that there is a (non-manifest)
four-dimensional Lorentz invariance. A four-dimensional Lorentz invariant
theory which has a two-dimensional $(4,4)$ supersymmetry must also have ${\cal
  N} =4$ supersymmetry in four dimensions.  The procedure of constructing a
supersymmetric D-dimensional theory using a lower dimensional superspace has
been employed in several contexts
\cite{Arkani-Hamed:2001tb,EGK,Hebecker,Guralnik:2003di}. The reader wishing to
skip directly to the action of the D3-D3 intersection in $(2,2)$ superspace
may proceed to Sec.~\ref{intersectaction}.

\subsubsection{Four-dimensional
actions in lower dimensional superspaces} \label{decomp}

The approach of building four-dimensional Lorentz invariance
starting with a conventional $(4,4)$ supersymmetric theory is an
indirect but effective way to obtain the ${\cal N} =4, d=4$ super
Yang-Mills action in a two-dimensional superspace.  There is also
a more direct approach which gives a $(2,2)$ superspace
representation for the part of the ${\cal N} =4, d=4$ action
containing only the ${\cal N} = 2, d=4$ vector multiplet. The
${\cal N} =2, d=4$ vector multiplet has a straightforward
decomposition under two-dimensional $(2,2)$ supersymmetry.  On the
other hand, there is no off-shell ${\cal N} =2, d=4$ formalism for
the hypermultiplet, unless one uses harmonic superspace. We
demonstrate the decomposition of the vector multiplet below.  This
provides a useful check of at least part of the action appearing
in Sec.~\ref{intersectaction}.

\subsubsection*{Embedding $(2,2)$, $d=2$ in $\N=2$, $d=4$}

We begin by showing how to embed $(2,2)$, $d=2$ superspace into
$\N=2$, $d=4$ superspace. The $\N=2, d=4$ superspace is
parameterized by ($z^+, z^-, w, \bar w$, $\theta^{(i)}_\a$, $\bar
\theta^{(i)}_{\dot\a}$). For the embedding let us redefine these
coordinates as
\begin{align} \label{coordinates}
\theta^+ &\equiv \theta_{(1)}^1, \quad
\thetasl^+ \equiv \theta_{(2)}^1\,, \nonumber\\
\bar \theta^- &\equiv \theta_{(2)}^2,\quad \thetasl^- \equiv
\theta_{(1)}^2 \,.
\end{align}
In the absence of central charges, the $\N=2$, $d=4$ supersymmetry
algebra is
\begin{align}
  \{Q_{(i)\a}, \bar Q^{(j)}{}_{\dot\beta}\} &= 2 \rho^\m_{\a\dot\beta}
 P_\m \delta_{i}^j,
  \qquad i,j=1,2 \,, \nonumber\\
  \{ Q_{(i)\a}, Q_{(j)\beta} \} &=\{\bar Q^{(i)}{}_{\dot\a}, \bar
  Q^{(j)}{}_{\dot\beta}\}=0 \,
\label{d4algebra}
\end{align}
with Pauli matrices $\rho^\m$ given by Eq.~(\ref{Pauli}).  We define
supersymmetry charges $Q_+ \equiv Q_{(1)1}$, $\bar Q_- \equiv Q_{(2)2}$,
$\Qsl_+ \equiv Q_{(2)1}$, and ${{\Qsl_-} \equiv Q_{(1)2}}$.  Following the
methods of refs.~\cite{hori2, Hellerman}, we introduce a superspace defect
at
\begin{align}
w = 0 \nonumber, \quad \thetasl^+ =\thetasl^-= 0 \,,
\end{align}
which implies that the generators $P_2$, $P_3$, $\Qsl_\pm $, and $\bar
\Qsl_\pm$  are
broken. The unbroken subalgebra of (\ref{d4algebra}) is generated
by $Q_\pm$ and $\bar Q_\pm$ and turns out to be the $(2,2)$, $d=2$
supersymmetry algebra given by
\begin{align}
\{Q_\pm, \bar Q_\pm \} = 2 (P_0 \pm P_1) \,.
\end{align}
Other anticommutators of the $Q$'s vanish due to the absence of
central charges.

\subsubsection*{$\N=2$, $d=4$ Super Yang-Mills action in $(2,2)$,
$d=2$ language} \label{sec2.2} In order to derive the $\N=2$
Yang-Mills action in $(2,2)$ language, we decompose the
four-dimensional $\N=2$ abelian vector superfield $\Psi$ in terms
of a two-dimensional $(2,2)$ chiral superfield $\Phi$, a twisted
chiral superfield $\Sigma$, and a vector superfield $V$. In the
abelian case, the twisted chiral superfield (see e.g.\
Ref.~\cite{Hori,Witten}) is related to the vector multiplet by
\begin{align}
\Sigma \equiv \bar D_+ D_- V
\end{align}
and satisfies $\bar D_+ \Sigma = D_- \Sigma=0$.  The $(2,2)$
vector and chiral superfields can be obtained by dimensional
reduction of their ${\cal N} =1, d=4$ counterparts.

In App.~\ref{newappendix} we show that the $\N=2$, $d=4$
vector supermultiplet $\Psi$ decomposes into
\begin{align} \label{decomposition}
  \Psi = -i \Sigma + \thetasl^+ \bar D_+ \left( \bar \Phi -
    \pr_{\bar w}  V \right) + \thetasl^- D_- \left( \Phi -
    \pr_w V \right) + \thetasl^+ \thetasl^- G  \,,
\end{align}
where $\pr_w$ is the transverse derivative and $G$ an auxiliary
$(2,2)$ superfield. An interesting result of the decomposition is
that the auxiliary field $D$ of the twisted chiral superfield
$\Sigma$ is related to the component $D'$ and transverse
derivatives of the components $v'_2$ and $v'_3$ of the
four-dimensional vector superfield,
\begin{align}
D=\frac{1}{\sqrt{2}} \left(D'+f'_{32} \right) \,, \label{Dterm}
\end{align}
where $f'_{32}=\partial_3 v'_2 -\partial_2 v'_3$.
Note that in distinction to the conformal field theory dual to the
$(2\vert \rm D3\perp D5)$ intersection studied in \cite{DFO, EGK}
there are no transverse derivatives like $\pr_w \phi'$ in
the auxiliary fields $F$ of the (2,2) superfield $\Phi$.

With the above decomposition of $\Psi$, we can now write down the
$\N=2$, $d=4$ (abelian) Yang-Mills action in (2,2) language.
Substituting Eq.\ (\ref{decomposition}) with $G=\bar D_+ D_- (i
\Sigma^\dagger + ...)$ into the usual form of the YM action, we
find
\begin{align}
  &\frac{1}{4\pi}{\rm\,Im\,}\tau \int d^4x d^2\theta_{(1)} d^2\theta_{(2)}
  \, \frac{1}{2} \Psi^2 \\
  &=\frac{1}{4\pi}{\rm\,Im\,} \tau \int d^4x d^4\theta \, \left(
    \bar \Sigma \Sigma + \bar \Phi \Phi +  \pr_{\bar w} V \Phi -
    \bar \Phi \pr_w V -  \pr_{\bar w} V \pr_w V \right) ,\nonumber
\end{align}
with $d^4\theta= \frac{1}{4} d\theta^+ d\theta^- d\bar\theta^+
d\bar\theta^- $. From this one can easily deduce the corresponding
non-abelian Yang-Mills action for vanishing $\theta$ angle,
\begin{align} \label{vectoraction}
  S^{\rm nonab}_{\rm YM}=\frac{1}{g^2} \int d^4x d^4\theta {\rm\,tr} \left(
    \Sigma^{\dagger}\Sigma + (\partial_{\bar w} + \bar \Phi) e^{V}
    (\partial_w + \Phi) e^{- V} \right) \,.
\end{align}

\subsubsection{The D3-D3 action in $(2,2)$ superspace} \label{intersectaction}

We now present the full action for the $(4,4)$ supersymmetric
theory describing the intersecting stacks of D3-branes. The action
has the form
\begin{align}
S= S_{\rm D3} + S_{\rm D3^{\prime}} + S_{\rm D3-D3^{\prime}} \,.
\end{align}
For each stack of parallel D3-branes we have separate actions,
$S_{\rm D3}$ and $S_{\rm D3^{\prime}}$, each of which correspond
to an $\N=4, d=4$ SYM theory with gauge groups $SU(N)$ and
$SU(N')$, respectively. The term $S_{\rm D3-D3^{\prime}}$
describes the coupling of these theories to matter on the
two-dimensional intersection.

In $(2,2)$ superspace, the field content of $S_{\rm D3}$ is as
follows. First, there is a vector multiplet $V(z^\pm, \theta^\pm,
\bar \theta^\pm; w, \bar w)$ or, more precisely, a continuous set
of vector multiplets labeled by $w, \bar w$ which are functions on
the $(2,2)$ superspace spanned by $(z^\pm, \theta^\pm,\bar
\theta^\pm)$.  The label $w = X^2+ i X^3$ parameterizes the
directions of the D3 world-volume transverse to the intersection,
while $z^\pm = X^0 \pm X^1$ parameterizes the remaining
directions.  Under gauge transformations $V$ transforms as
\begin{align}
e^V \rightarrow e^{-i\Lambda^{\dagger}} e^V e^{i\Lambda} \,, \qquad
e^{-V} \rightarrow e^{-i\Lambda} e^{-V} e^{i\Lambda^{\dagger}} \,,
\end{align}
where $\Lambda$ is a $(2,2)$ chiral superfield which also depends
on $w, \bar w$. From $V$ one can build a twisted chiral (or field
strength) multiplet  as
\begin{align}
\Sigma = \frac{1}{2} \{ \bar {\cal D}_+, {\cal D}_- \} \,,
\end{align}
where ${\cal D}_\pm=e^{-V} D_\pm e^V$, ${\cal \bar D}_\pm=e^V \bar
D_\pm e^{-V}$. Additionally one has a pair of adjoint chirals
$Q_1$ and $Q_2$, transforming as
\begin{align} Q_i \rightarrow
  e^{-i\Lambda}Q_ie^{i\Lambda} \,.
\end{align}
Finally there is a $(2,2)$ chiral field $\Phi$ which transforms
such that $\partial_{\bar w} + \Phi$ is a covariant derivative:
\begin{align}
\partial_{\bar w} + \Phi \rightarrow e^{-i\Lambda}(\partial_{\bar w} + \Phi)
e^{i\Lambda} \,.
\end{align}
The complex scalar which is the lowest component of $\Phi$ is
equivalent to the gauge connection $v_2 + i v_3$ of the
four-dimensional SYM theory described by $S_{\rm D3}$.  This structure
was also seen in the explicit decomposition of the ambient $\N=2$, $d=4$
vector field $\Psi$ under $(2,2), d=2$ supersymmetry discussed in
Sec.~\ref{decomp}, cf.\ Eq.~(\ref{redefinition}).

The action of the second D3-brane (D$3^{\prime}$) is identical to
that of the first D3-brane with the replacements
\begin{align}
w \rightarrow y \,,\quad V \rightarrow {\cal V}\,,\quad \Sigma
\rightarrow \Omega\,,\quad Q_i \rightarrow S_i\,,\quad \Phi
\rightarrow \Upsilon\,,
\end{align}
and is invariant under gauge transformations $\Lambda^{\prime}$.

The fields corresponding to D3-D$3^{\prime}$ strings are the
chiral multiplets $B$ and $\tilde B$,  which are bifundamental and
anti-bifundamental respectively with respect to \linebreak \mbox{$SU(N) \times
SU(N')$} gauge transformations;
\begin{align}
B\rightarrow e^{-i\Lambda} B e^{i \Lambda^{\prime}} \,, \qquad
\tilde B \rightarrow e^{-i\Lambda^{\prime}} \tilde B e^{i\Lambda} \,.
\end{align}

Using a canonical normalization ($V \rightarrow g V$ etc.), the
components of the action are as follows:
\begin{align} \label{action1}
S_{\rm D3} = \,&\frac{1}{g^2} \int d^2z  d^2w d^4\theta {\rm\, tr}
\left(\Sigma^{\dagger}\Sigma + (\partial_{w} +  g\bar \Phi) e^{gV}
(\partial_{\bar w} + g\Phi) e^{-gV}
+ \sum_{i=1,2} e^{-gV} \bar Q_i e^{gV} Q_i \right) \nonumber\\
+ & \int d^2 z d^2w  d^2 \theta  \epsilon_{ij} {\rm\, tr\,} Q_i
[\partial_{\bar w} + g \Phi, Q_j] + c.c
\end{align}
\begin{align} 
S_{\rm D3^{\prime}}=\,&\frac{1}{g^2}\int d^2z  d^2y
d^4\theta {\rm\,tr} \left( \Omega^{\dagger}\Omega +
(\partial_{y} + g \bar \Upsilon) e^{g{\cal V}} (\partial_{\bar y} +
g\Upsilon)
e^{-g{\cal V}}
+ \sum_{i=1,2} e^{-g{\cal V}} \bar S_i e^{g{\cal V}} S_i \right) \nonumber\\
+ & \int d^2 z d^2y  d^2 \theta \epsilon_{ij} {\rm
\,tr\,} S_i [\partial_{\bar y} + g\Upsilon, S_j] + c.c \label{action2}
\end{align}
\begin{align}
S_{\rm D3- D3^{\prime}} = &\int d^2z d^4 \theta {\rm\, tr}
\left(e^{-g \cal V}\bar B e^{gV} B + e^{-gV} \bar {\tilde B}
e^{g \cal V} \tilde B \right) \hspace{5.5cm} \nonumber \\
+ & \frac{ig}{{2}} \int d^2z d^2\theta {\,\rm tr}\left(
  B \tilde B Q_1 - \tilde B B S_1\right) + c.c.  \label{defectaction}
\end{align}
with $d^4\theta= \frac{1}{4} d\theta^+ d\theta^- d\bar\theta^+
d\bar\theta^-$ and $d^2\theta = \frac{1}{2} d \theta^+ d
\theta^-$.

Some comments about $S_{\rm D3}$ are in order.  We have already
presented part of this action, as the first two terms in the
$S_{\rm D3}$ are given by Eq.\ (\ref{vectoraction}). Upon
integrating out auxiliary fields, $S_{\rm D3}$ can be seen to
describe the ${\cal N} =4$ SYM theory. To illustrate how
four-dimensional Lorentz invariance arises, consider the
superpotential $\epsilon_{ij} {\rm\, tr\,} Q_i [\partial_{\bar w}
+ \Phi, Q_j]$.  Upon integrating out the F-terms of $Q_1$ and
$Q_2$, one gets kinetic terms in the $X^2, X^3$ directions which
are the four-dimensional Lorentz completion of the kinetic terms
in the $X^0, X^1$ directions arising from $e^{-V} \bar Q_i e^V
Q_i$.

The form of $S_{\rm D3-D3'}$ is dictated by gauge invariance and
$(4,4)$ supersymmetry.  The geometric interpretation of various
fields can be seen from this part of the action.  The vacuum
expectation values for the scalar components of $Q_1$ and $S_1$
give rise to mass terms for the fields $B$ and $\tilde B$
localized at the intersection. There are also ``twisted'' mass
terms for $B$ and $\tilde B$ which arise when the scalar
components of the twisted chiral fields $\Sigma$ and $\Omega$ (or
equivalently of $V$ and ${\cal V}$) get expectation values.  One
expects  $B$ and $\tilde B$ fields to become massive when the
D3-branes are separated from the D3$'$-branes in the $X^{6,7,8,9}$
directions transverse to both. Thus we associate the scalar
components of $(Q_1, \Sigma)$ or $(S_1, \Omega)$ with fluctuations
in $(X^6+ iX^7, X^8+iX^9)$.

Note that in $(2,2)$ superspace, $Q_2$ and $S_2$ are not directly
coupled to the fields $B$ and $\tilde B$, although derivative
couplings arise after integrating out the F-terms of $Q_1$ and
$S_1$.  The scalar component of $Q_2$ describes fluctuations of
the D3-branes in the $y=X^4+iX^5$ plane parallel to the
D3$'$-branes. Similiarly the scalar components of $S_2$ describe
fluctuations of the D3$'$-branes in the $w=X^2+i X^3$ plane
parallel to the D3-branes. 

\subsubsection{R-symmetries}

The isometries of the AdS background are $SL(2,R) \times SL(2,R) \times U(1)
\times SU(2)_L \times SU(2)_R \times U(1)$, as can be seen from
Eq.~(\ref{unb}) for $k=1$.  The $SU(2)_L \times SU(2)_R$ component is an
R-symmetry which acts as rotations in the $6,7,8$ and $9$ directions
transverse to all the D3-branes. The first $U(1)$ R-symmetry acts as a
rotation in the $w$ (or $23$) plane, while the second $U(1)$ acts as a
rotation in the $y$ (or $45$) plane.  In the near horizon geometry, the probe
Kaluza-Klein momentum on $S^1$ is a contribution to $J_{45}$.  The charge
$J_{23}$ generates a rotation in $AdS_5$ directions orthogonal to the probe.

Below we summarize the R-charges and engineering dimensions of the
fields of the D3-D3 intersection.
\begin{table}[ht]
\begin{center}
\begin{tabular}{ccllccc}
(4,4)& (2,2) &  components  & $(j_L, j_R)$ & $J_{23}$ & $J_{45}$ &
$\Delta$ \\
\hline & & $\sigma, q_1$ &
$(\frac{1}{2},\frac{1}{2})$ &$0$&$0$& $1$ \\
Vector& $Q_1, \Sigma$ &  $\psi_{q_1}^+, \bar\lambda_{\sigma}^+$ &
$(0,\frac{1}{2})$ &$\frac{1}{2}$& $-\frac{1}{2}$& $\frac{3}{2}$ \\
& &  $\psi_{q_1}^-, \bar\lambda_{\sigma}^-$ &
$(\frac{1}{2},0)$ &$ \frac{1}{2}$&$-\frac{1}{2}$& $\frac{3}{2}$ \\
& & $v_0, v_1$ & $(0,0)$ &$0$&$0$& $1$ \\
\hline
  & & $\phi$ & $(0,0)$ &$-1$&$0$& $1$ \\
Hyper& $Q_2, \Phi$ & $ q_2$ & $(0,0)$&$0$&$1$ & $1$ \\
& & $\psi_\phi^+, \bar\psi_{q_2}^+$ & $(\frac{1}{2},0)$ &
$-\frac{1}{2}$ & $-\frac{1}{2}$ &
$\frac{3}{2}$ \\
& & $\psi_\phi^-, \bar\psi_{q_2}^-$ & $(0,\frac{1}{2})$ &
$-\frac{1}{2}$&$-\frac{1}{2}$&
$\frac{3}{2}$\\
\hline
&  & $b$ & $(0,0)$ & $-\frac{1}{2}$ & $\frac{1}{2}$ & 0 \\
Hyper& $B, \tilde B$ & $\tilde b$ & $(0,0)$ & $-\frac{1}{2}$ &
$\frac{1}{2}$ & $0$ \\
& & $\psi_b^+, \bar\psi_{\tilde b}^+$  & $(\frac{1}{2},0)$ & $0$ &
$0$ &
$\frac{1}{2}$ \\
& & $\psi_b^-, \bar\psi_{\tilde b}^-$ & $(0,\frac{1}{2})$ & $0$ &
$0$ &
$\frac{1}{2}$ \\
\hline & &$\omega, s_1$ &
$(\frac{1}{2},\frac{1}{2})$ & $0$ &$0$& $1$ \\
Vector& $S_1, \Omega$ &  $\psi_{s_1}^+, \bar\psi_{\omega}^+$ &
$(0,\frac{1}{2})$ & $\frac{1}{2}$ & $-\frac{1}{2}$ & $\frac{3}{2}$ \\
& &  $\psi_{s_1}^-, \bar\psi_{\omega}^-$ &
$(\frac{1}{2},0)$ & $\frac{1}{2}$ & $-\frac{1}{2}$ & $\frac{3}{2}$ \\
& & $\tilde v_0, \tilde v_1$ & $(0,0)$ & $0$ &$0$ & $1$ \\
\hline
  & & $\upsilon$ & $(0,0)$ & $0$ & $1$ & 1 \\
Hyper& $S_2, \Upsilon$ & $ s_2$ & $(0,0)$ & $-1$ & $0$ & $1$ \\
& & $\lambda_\upsilon^+, \bar\psi_{s_2}^+$ & $(\frac{1}{2},0)$ &
$\frac{1}{2}$ & $\frac{1}{2}$ &
$\frac{3}{2}$ \\
& & $\lambda_\upsilon^-, \bar\psi_{s_2}^-$ & $(0,\frac{1}{2})$ &
$\frac{1}{2}$ & $\frac{1}{2}$ &
$\frac{3}{2}$\\
\end{tabular}
\caption{Field content of the D3-D3 intersection.}\label{rsym}
\end{center}
\end{table}

The $U(1)$ symmetries generated by $J_{45}$ and $J_{23}$ are
manifest in $(2,2)$ superspace.  The $U(1)$ generated by $J_{45}$
has the following action:
\begin{align}
&\theta^{+} \rightarrow e^{i\alpha/2} \theta^{+}\,,
&&B\rightarrow e^{i \alpha/2} B\,, &&
Q_2 \rightarrow e^{i\alpha} Q_2 \,,&&&&\nonumber\\
&\theta^{-} \rightarrow e^{i\alpha/2} \theta^{-}\,,
&&\tilde B \rightarrow e^{i \alpha/2} \tilde B\,,  &&
\Upsilon \rightarrow  e^{+i\alpha} \Upsilon \,, \nonumber\\
&y\rightarrow e^{i\alpha} y\,,
\end{align}
with all remaining fields being singlets. The $U(1)$ generated by
$J_{23}$ acts as
\begin{align}
&\theta^{+} \rightarrow e^{-i\alpha/2} \theta^{+}\,,
&&B\rightarrow e^{-i \alpha/2} B\,, &&
S_2 \rightarrow e^{-i\alpha} S_2\,,&&&& \nonumber \\
&\theta^{-} \rightarrow e^{-i\alpha/2} \theta^{-}\,,
&&\tilde B \rightarrow e^{-i \alpha/2} \tilde B\,,  &&
\Phi\rightarrow e^{-i\alpha}\Phi\,, \nonumber \\
&w\rightarrow  e^{-i\alpha} w\,.
\end{align}
The reader may be surprised that these R-symmetries act on the coordinates 
$w$
and~$y$.\footnote{Upon toroidal compactification of $w$ and $y$ the $U(1)$
  R-symmetry generated by $J_{23} + J_{45}$ is enhanced to $SU(2)$. Note 
that
  the $(4,4)$ supersymmetry algebra admits an $SU(2)_L \times SU(2)_R \times
  SU(2)$ automorphism \cite{DiaconescuSeiberg} which in the compactified 
case
  is also realised as a symmetry.}  However in the language of
two-dimensional superspace, these are continuous labels rather than 
space-time
coordinates. Recall also that $J_{23}$ (or $J_{45}$) is an R-symmetry of the
${\cal N} = 4$ algebra associated with one stack of D3-branes, but a Lorentz
symmetry for the orthogonal stack.

\subsubsection{Quantum conformal invariance}

Here we give an argument that the action given by (\ref{action1}),
(\ref{action2}) and (\ref{defectaction}) does not receive quantum corrections,
such that it remains conformal to all orders in perturbation theory.  We will
not attempt a rigorous proof here, but we give an argument for conformal
invariance using our $(2, 2)$, $d=2$ formulation of the model. The argument
relies on the assumption that the classical $(2, 2)$, $d=2$ supersymmetry is
unbroken by quantum corrections. This argument is analogous to the discussion
of the 3d/4d case in \cite{EGK}, where more details on the renormalization
procedure may be found.

The argument for excluding possible quantum breakings of conformal symmetry by
defect operators relies on considering the $(2, 2)$ supercurrent and its
possible anomalies, and by making the assumption that $(4, 4)$ supersymmetry
is preserved by the quantum corrections. 

We begin by recalling the situation in $\N=1$, \mbox{$d=4$} theories, see for
instance \cite{West:1990tg}. In this case there is a supermultiplet $J_{\dot
  \alpha \beta} = \sigma^\mu{}_{\dot \alpha \beta} J_\mu$, which has the
R-current $R_\mu$ (sometimes also denoted as $j_\mu^{(5)}$), the supersymmetry
currents $Q_{\mu\alpha}$ and the energy-momentum tensor $T_{\mu\nu}$ among its
components. These are the Noether currents corresponding to R transformations,
supersymmetry transformations and translations. The supermultiplet $J_\mu$ can
be expanded as
\begin{align}
J_\mu (x,\theta,\bar \theta) = R_\mu (x) - i \theta^\alpha Q_{\mu\alpha}(x)
+ \bar \theta_{\dot \alpha} \bar Q_{\mu}^{\dot \alpha}(x)
-2 (\theta \sigma^\nu \bar \theta) T_{\mu\nu} (x) + ...
\end{align}
Potential superconformal anomalies may be
written in the form
\begin{equation}
\bar D^{\dot \alpha} J_{\dot \alpha \beta} \, = \, D_\beta S \, ,
\label{fourj}
\end{equation}
with $S$ a chiral superfield. When $S=0$, superconformal symmetry is
conserved. In $\N=1$, \mbox{$d=4$} the superfield $S$ is proportional to the
operator $W_\alpha W^\alpha$. When written in components, Eq.~(\ref{fourj})
contains both the trace anomaly and the anomalous divergences of the
R-symmetry and supersymmetry currents.

By standard dimensional reduction to $(2,2)$ supersymmetry in two
dimensions, we obtain from (\ref{fourj}), as shown in \cite{West89},
\begin{equation}
(\gamma^M)_A{}^B \bar D_B {\cal J}_M \, = D_A \, {\cal S} \, ,
\end{equation}
where $M=\{1,2 \}$, $A,B=\{+,-\}$, $\gamma^M=\{\sigma^1, i
\sigma^2\}$ are the two-dimensional gamma matrices, ${\cal J}_M$
is the two-dimensional $(2,2)$ supercurrent and the possible
conformal anomaly is given by the $(2,2)$ chiral superfield ${\cal
S}$. ${\cal J}_M$ contains the 2d R-current, the four $(2,2)$
supersymmetry currents and the 2d energy-momentum tensor.

For 2d/4d models like the one given by (\ref{action1}),
(\ref{action2}) and (\ref{defectaction}), the classically
conserved two-dimensional supercurrent is given by
\begin{equation}
{\cal J}_M (z)\,=\, {\cal J}^{\rm def}_M (z) \, + \, \int\! d^2w
\; {\cal J}^{\rm bulk, 1}_{M} (w,z) \, + \, \int\! d^2y \; {\cal
J}^{\rm bulk, 2}_{M} (y,z)  \, .
\end{equation}
Let us first consider possible defect operator contributions to
the anomaly ${\cal S}$, which have to be gauge invariant and of
dimension 1. The possible defect contributions to the anomaly
${\cal S}$ are given by
\begin{gather} \label{twoj}
{\cal S}_D \, = \, {\rm Tr}\,   [ u\, \bar D^+ \bar D^- ( e^{-\cal
V}\bar B e^{V} B + e^{-V} \bar {\tilde B} e^{\cal V} \tilde B) \,
+ \, \, v \, (B \tilde B Q_1 - \tilde B B S_1) \, ] \, .
\end{gather}
It is important to note that there is no gauge anomaly term
contributing to this equation, since ${\rm Tr\,} \Sigma$ or $ {\rm
Tr\,} \Omega$, which would have the right dimension, are twisted
chiral and not chiral. $u$ and $v$ are coefficients which may be
calculated perturbatively. They are related to the $\beta$ and
$\gamma$ functions. From the standard supersymmetric
non-renormalization theorem we know that $v=0$ since the
corresponding operator is chiral. $u$ may be non-zero in a general
$(2,2)$ supersymmetric gauge theory. However $u$ and $v$ are
related by $(4,4)$ supersymmetry. Therefore if we assume that
$(4,4)$ supersymmetry is preserved upon quantization, $v=0$ also
implies $u=0$. Thus there are no defect contributions breaking
conformal symmetry.

We may also show that there are no contributions from four-dimensional
operators to the conformal anomaly ${\cal S}$. Such terms would have to
originate from bulk action counter\-terms. Consider correlation functions of
bulk fields in the limit of large $\vert w \vert$ (or $\vert  y
\vert$) but fixed momenta $\vec p$ parallel to the defect.  These receive the
usual contributions from diagrams involving only bulk fields.  Such
contributions are finite due to the finiteness of the ${\cal N}=4$, $d=4$
theory.  Contributions from diagrams which involve bulk-defect interactions
(see Fig.~\ref{zdep}) are $w$ (or $y$) dependent and fall off with
distance from the defect.  Therefore local counterterms generating the
anomalies would be of the form
\begin{equation}
{\cal S}_B \, \sim \, \int\!  d^2 w  \,|w|^{-s_1} \Lambda^{t_1}
{\cal O}_1 (w, z) \, + 
\, \int\!  d^2 y \,|y|^{-s_2} \Lambda^{t_2}
\, {\cal O}_2 (y, z)  \, ,
\end{equation}
with $\Lambda$ a regulator scale, and $s_i \geq 2$, $t_i\geq 0$
for $i=1,2$. However there are no such operators available in the
theory. From dimensional analysis, only ${\rm Tr\,} \Sigma$ or
${\rm Tr\,} \Omega$ would be possible for ${\cal O}_1$ or ${\cal
O}_2$, respectively, but again these are twisted chiral and not
chiral. Therefore we conclude that there a no
terms breaking $SO(2,2)$ conformal invariance, such that the
theory is conformal to all orders in perturbation theory.

\begin{figure}[!ht]
\begin{center}
 \scalebox{.75}{\includegraphics{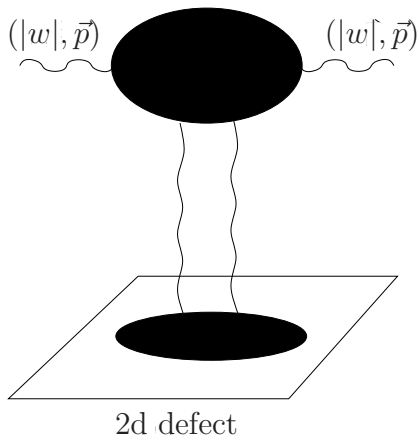}}
\caption{A $w$ dependent contribution to a bulk-bulk propagator.}
\label{zdep}
\end{center}
\vspace{-7.1cm} \hspace{4.9cm}$( |w|, \vec p)$ \hspace{2.7cm} 
$(|w|, \vec p)$
\vspace{6.6cm}

\vspace{-1.9cm}  \hspace{6.2cm} 2d defect
\vspace{1.4cm}
\end{figure}

\subsection{More on the D3-D3 intersection} \label{moreD3D3}

In this section we consider some interesting properties of the D3-D3
intersection which do not arise for the D3-D5 or the D3-D7 brane
configuration. Most of the pecularities are due to the two-dimensional
conformal symmetry preserved by the defect which differs in many
aspects from conformal invariance in higher dimensions. A natural
question to ask is whether the theory satisfies an
infinite-dimensional Virasoro algebra.  We will see however that only
the finite part of the $(4,4)$ superconformal algebra, whose even part
is $SL(2, R) \times SL(2, R)$, is realised in an obvious way.  Roughly
speaking, the $(4,4)$ superconformal algebra is the common
intersection of two ${\cal N} = 4, d=4$ superconformal algebras, both
of which are finite. Enhancement to the usual infinite-dimensional
algebra would require the existence of a decoupled two-dimensional
sector which does not exist.

We then give a detailed dictionary between Kaluza-Klein fluctuations
on the probe D3-brane and operators localized on the defect.  Of
particular interest will be a certain subset of the fluctuations which
describe the embedding of the probe inside $AdS_5$. This subset is
dual to operators containing defect scalar fields, which appear
without any derivative or vertex operator structure. Due to strong
infrared effects in two dimensions, these fields are not conformal
fields associated to states in the Hilbert space. From the point of
view of the probe-supergravity system, there is at first sight nothing
unusual about these fluctuations.  However upon applying the usual
$AdS_3$/CFT$_2$ rules, we will find that the dual two-point correlator
does {\em not} show a power law behaviour.  Thus there is no clear
interpretation of these fluctuations as sources for the generating
function of the CFT. We shall find however that the bottom of the
Kaluza-Klein tower for these fluctuations (with appropriate boundary
conditions) parameterizes the holomorphic embedding $w y \sim
c\alpha'$ of the $AdS_3$ probe inside $AdS_5$ which we discussed in the last
section. While the interpretation of this fluctuation as a source is
unclear, it nevertheless labels points on the classical Higgs branch.

We finally compute perturbative quantum corrections to the two-point function
of the BPS primary operators and find that such corrections are absent at
order $g_{YM}^2$. In Sec.~\ref{correlatorsection} we have seen in an AdS
computation that correlation functions are independent of the 't~Hooft
coupling. Both results suggest the existence of a non-renormalization theorem.

\subsubsection{The $(4,4)$ superconformal algebra} \label{superalg}

The D3-D3 intersection has a $(4,4)$ superconformal group whose
even part is 
\begin{align}
SL(2,R) \times SL(2,R) \times SU(2)_L \times SU(2)_R 
\times U(1)\,.
\end{align}  
A comparison with the isometry group given by Eq.~(\ref{unb}) for $k=1$,
$SO(2,2) \times SO(2) \times SO(4) \times SO(2)$, shows that only a
certain combination of the two $U(1)\simeq SO(2)$ factors enters the
superconformal algebra.  We emphasize that this system does not give a
standard $(4,4)$ superconformal algebra.  Because of the couplings
between two and four-dimensional fields, the algebra does not
factorize into left and right moving parts. Neither an infinite
Virasoro algebra nor an affine Kac-Moody algebra are realised in any
obvious way. The superconformal algebra for the D3-D3 system should be
thought of as a common ``intersection'' of two ${\cal N}=4, d=4$
superconformal algebras, both of which are finite. If there is a
hidden affine algebra, it should arise via some dynamics which gives a
decoupled two-dimensional sector, for which we presently have no
evidence.

For comparative purposes,  it is helpful to first review the
situation for more familiar two-dimensional $(4,4)$ theories with
vector multiplets and hypermultiplets,  such as those considered
in \cite{wittenhiggs}. These theories may have classical Higgs and
Coulomb branches which meet at a singularity of the moduli space.
For finite coupling, quantum states spread out over both the Higgs
and Coulomb branches. However in the infrared (or strong coupling)
limit, one obtains a separate $(4,4)$ CFT on the Higgs and Coulomb
branches \cite{wittenhiggs}. One argument for the decoupling of
the Higgs and Coulomb branches is that the $(4,4)$ superconformal
algebra contains an $SU(2)_l \times SU(2)_r$ R-symmetry with a
different origin in the original $SU(2)_L \times SU(2)_R \times
SU(2)$ R-symmetry depending on whether one is on the Higgs branch
or Coulomb branch. The CFT scalars must be uncharged under the
R-symmetries. This means for example that the original $SU(2)_L
\times SU(2)_R$ factor may be the R-symmetry of the CFT on the
Higgs branch but not the Coulomb branch.

For the linear sigma model describing the D3-D3 intersection, a
$(4,4)$ super-conformal theory arises only on the Higgs branch,
which is parameterized by the two-dimensional scalars of the
defect hypermultiplet.  On the Coulomb branch,  the orthogonal
D3-branes are separated by amounts characterised by the VEV's of
four-dimensional scalar fields.  One obtains a CFT on the Higgs
branch without flowing to the IR, since the gauge fields propagate
in four dimensions and the gauge coupling is exactly
marginal. Furthermore scalar degrees
of freedom of the CFT may carry R-charges, since the R-currents do
not break up into purely left and right moving parts.  Of course
the scalars of the defect hypermultiplet must still be uncharged
under the R-symmetries, since for $g_{YM}=0$ the free $(4,4)$
hypermultiplet realises a conventional two-dimensional $(4,4)$
CFT.  However the four-dimensional scalar fields, which are not
decoupled at finite $g_{YM}$, transform non-trivially under the
R-symmetries of the defect CFT.

In more familiar considerations of the $AdS_3/CFT_2$ duality, the full
Virasoro algebra is realised in terms of diffeomorphisms that leave the form
of the metric invariant asymptotically, near the boundary of $AdS_3$
\cite{BrownHenneaux}.  Of these diffeomorphisms, the finite $SL(2,R) \times
SL(2,R)$ subalgebra is realised as an exact isometry.  However the
three-dimensional diffeomorphisms which are asymptotic isometries of $AdS_3$,
and correspond to higher-order Virasoro generators, do not have an extension
into the bulk which leave the $AdS_5$ metric asymptotically invariant.  The
existence of a Virasoro algebra seems to require localized gravity on $AdS_3$.
This could only be seen through a consideration of the back-reaction. Note
also that a necessary condition for the existence of a Virasoro algebra is the
existence of a two-dimensional conserved local energy-momentum tensor. This
requirement is not satisfied as shown in App.~\ref{confs}. In the defect CFT,
the two-dimensional conformal algebra contains only those generators which can
be extended to conformal transformations of the four-dimensional parts of the
world-volume, namely $L_{-1}, L_0, L_1,\tilde L_{-1},\tilde L_0$ and $\tilde
L_1$.

The `global' $(4,4)$ superconformal algebra of defect CFT  gives
relations between the dimensions and R-charges of BPS operators.
We will later find that these relations are consistent with the
spectrum of fluctuations in the probe-$AdS$ background. To
construct the relevant part of the algebra, it is helpful to note
that the algebra should be a subgroup of an ${\cal N} =4, d=4$
superconformal algebra (or actually an unbroken intersection of
two such algebras).

Let us start by writing down the relevant part of the ${\cal N}
=4, d=4$ superconformal algebra for the D3-branes in the $0123$
directions. The supersymmetry generators are $Q_{\alpha}^a$, where
$\alpha =1,2$ is a spinor index and $a=1, \cdots,4$ is an index in
the representation ${\bf 4}$ of the $SU(4)$ R-symmetry.  The
special superconformal generators are $S_{\beta b}$ which are in
the ${\bf 4}^*$ representation of $SU(4)$. The relevant part of
the ${\cal N} =4, d=4$ algebra is then
\begin{align}
\{Q^a_{\alpha},S_{\beta b} \} =
\epsilon_{\alpha\beta}({\delta_b}^a D + 4 J^A (T_A)^a_b) +
\frac{1}{2}{\delta_b}^a L_{\mu\nu}\sigma^{\mu\nu}_{\alpha\beta}\,
, \label{alg}
\end{align}
where $D$ is the dilation operator,  $J^A$ are the operators
generating $SU(4)$, and $L_{\mu\nu}$ are the generators of
four-dimensional Lorentz transformations.  The matrices $(T_
A)^a_b$ generate the fundamental representation of $SU(4)$, and
are normalized such that ${\rm Tr}(T^AT^B) = \frac{1}{2} \delta^{AB}$.

A $(4,4)$ supersymmetry sub-algebra is generated by the
supercharges $Q^a_1 \equiv Q^a_+$ with $a=1,2$, and $Q^a_2\equiv
Q^a_-$ with $a = 3,4$, on which an $SU(2)_L \times SU(2)_R \times
U(1)$ subgroup of the orginal $SU(4)$ R-symmetry acts.  The
embedding of the $SU(2)_L \times SU(2)_R \times U(1)$ generators
in $SU(4)$ is as follows:
\begin{align}SU(2)_L: \begin{pmatrix} \frac{1}{2}\sigma^A & 0\cr 0 & 0
\end{pmatrix}, \quad
SU(2)_R: \begin{pmatrix} 0 & 0 \cr 0 & \frac{1}{2}\sigma ^B
\end{pmatrix},\quad U(1): \frac{1}{\sqrt{8}} \begin{pmatrix}-I & 0 \cr 0 & I
\end{pmatrix} \,.\end{align}
The unbroken $SU(2)_L \times SU(2)_R$ R-symmetry corresponds to
rotations in the directions $6,7,8,9$ transverse to both stacks of
D3-branes, while the unbroken $U(1)$ describes rotation in the
$45$ plane. These symmetries act on adjoint scalars.  Since the
R-currents of the CFT do not break up into left and right moving
parts, there is no requirement that four-dimensional scalars are
uncharged under R-symmetries.  We shall call the generator of
rotations in the $45$ plane $J_{45}$, and normalize it such that
the supercharges $Q^a_\pm$ have $J_{45}$ eigenvalue $\pm 1/2$. The
special superconformal generators of the $(4,4)$ sub-algebra are
$S_{b2} \equiv S_{b-}$ with $b=1,2$ and $S_{b1} \equiv S_{b+}$
with $b=3,4$.    The term in the $(4,4)$ algebra inherited from
(\ref{alg}) is then
\begin{align}
\{Q^a_+,S_{b-} \} &= {\delta_b}^a D + 2J^L_A(\sigma^A)^a_b +
\delta^a_b J_{45} + \delta^a_b L_{01} + \delta^a_b L_{23}\, , \label{one} \\
\{Q^a_-,S_{b+} \} &=  -\delta_b^a D - 2J^R_A(\sigma^A)^a_b +
\delta^a_b J_{45} + \delta^a_b L_{01} + \delta^a_b L_{23} \,.
\label{two}
\end{align}
The unbroken Lorentz generators are $L_{01}$ and $L_{23}$. Note
that from a two-dimensional point of view,  the Lorentz
transformations are generated by $L_{01}$, whereas $L_{23}$ is an
R-symmetry.

For the orthogonal D3-branes spanning $0,1,4,5$, rotations in the
$45$ plane are Lorentz generators $L_{45}$ rather than a subgroup
of $SU(4)$. The rotations in the $23$ plane are an unbroken $U(1)$
part of the $SU(4)$ R-symmetry rather than a Lorentz
transformation.  This distinction is illustrated in Fig.~\ref{decompo}.

\begin{figure}[!ht]
\begin{center}
\includegraphics{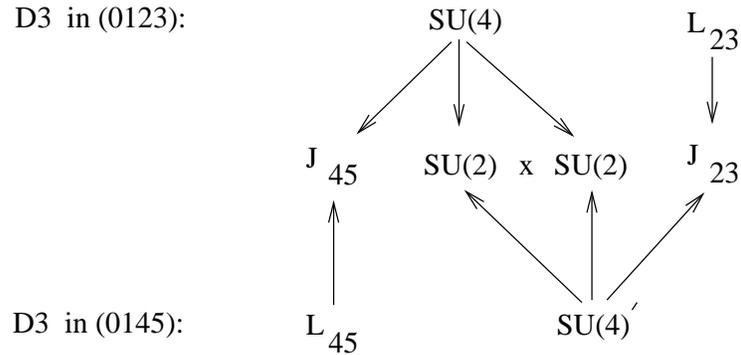}
\caption{Decomposition of the two $SU(4)$ R-symmetries.}  \label{decompo}
\end{center}
\end{figure}

\noindent From the two-dimensional point of view, both $23$ and
$45$ rotations are $U(1)$ R-symmetries. If we write $L_{23} =
J_{23}, \, L_{45} = J_{45}$ and define ${\cal J} = J_{23}+J_{45}$,
then the terms (\ref{one}) and (\ref{two}) become
\begin{align}
\{Q^a_+,S_{b-} \} &= {\delta_b}^a (L_{01}+D) + 2J^L_A(\sigma^A)^a_b
+
\delta^a_b {\cal J}\, , \label{thealg1} \\
\{Q^a_-,S_{b+} \} &=  \delta_b^a (L_{01}-D)  -
2J^R_A(\sigma^A)^a_b + \delta^a_b {\cal J}
\,.\label{thealg}\end{align} which are applicable to {\it both}
stacks of D3-branes.  This forms part of the $(4,4)$
superconformal algebra of the full D3-D3 system. The charge ${\cal
J}$ plays a somewhat unusual role. From the point of view of the
bulk four-dimensional fields, ${\cal J}$ is a combination of an
R-symmetry and a Lorentz symmetry, under which the preserved
supercharges are invariant. As we will see later, the fields
localized at the two-dimensional intersection are not charged
under ${\cal J}$. Upon decoupling the four-dimensional fields by
taking $g = 0$, the two-dimensional sector becomes a free $(4,4)$
superconformal theory with an affine $SU(2)_L \times SU(2)_R$
R-symmetry. However, for $g_{YM} \ne 0$, the algebra does not
factorize into left and right moving parts.

The algebra (\ref{thealg1},~\ref{thealg}) determines the dimensions of
the BPS superconformal primary operators, which are annihilated by
all the $S$'s and some of the $Q$'s.  The bounds on dimensions due
to the superconformal algebra are best obtained in Euclidean
space.  The Euclidean $(4,4)$ algebra of the defect CFT  contains
the terms
\begin{align}
\{ {\cal Q}^a_{1/2}, {\cal Q}_{1/2}^{b\dagger} \} = 2\delta^a_b
L_0 + 2J^L_A ({\sigma^A})^a_b + \delta^a_b {\cal
J}\, , \label{left} \\
\{ \tilde {\cal Q}^a_{1/2}, \tilde {\cal Q}_{1/2}^{b\dagger} \} =
2\delta^a_b \tilde L_0 + 2J^R_A  ({\sigma^A})^a_b  - \delta^a_b
{\cal J}\, . \label{right}
\end{align}
For $a=b$, the left hand side of (\ref{left}) and (\ref{right})
are positive operators, leading to the bounds
\begin{align}
h + j^L_3 + \textstyle\frac{1}{2}{\cal J} \geq 0 \,,\label{first}\\
h - j^L_3 + \textstyle\frac{1}{2}{\cal J} \geq 0 \,,\label{second}\\
\tilde h + j^R_3 - \textstyle\frac{1}{2}{\cal J} \geq 0\,, \label{third}\\
\tilde h - j^R_3 - \textstyle\frac{1}{2}{\cal J} \geq 0\,, \label{fourth}
\end{align}
some of which are saturated by the BPS super-conformal primaries.
As always, the dimensions are $\Delta = h + \tilde h$,  with
$h=\tilde h$ for scalar operators.

\subsubsection{Fluctuation--operator dictionary}

In the following we find the map between fluctuations on the probe
D3-brane and operators localized at the defect.  The single particle
states on the probe correspond to meson-like operators with strings of
adjoint fields sandwiched between pairs of defect fields in the
fundamental representation. The fluctuations can be devided into three
classes: $S^1$ fluctuations, gauge field fluctuations, and $AdS_3$
fluctuations. These fluctuations follow from the analysis in
Sec.~\ref{flucsection} for $k=1$.

\subsubsection*{$S^1$ fluctuations inside $S^5$}
\label{chiralprimaries}

The fluctuations of the probe $S^1$ embedding inside $S^5$ are
characterised by the mode $V^m_l$ where $m=6,7,8,9$. As shown in
Sec.~\ref{ch222} these fluctuations are scalars in the
$(\frac{1}{2},\frac{1}{2})$ (vector) representation of $SO(4) \simeq
SU(2)_L \times SU(2)_R$. Moreover, these fluctuations have $J_{23}=0$
and $J_{45}=l$ such that the $U(1)$ charge appearing in the algebra
(\ref{left}),~(\ref{right}) is ${\cal J}=l$. The possible series of
dimensions are $\Delta = 1 \pm l$.  We need only consider $l\ge 0$
since ${V^m_l}^*= V^m_{-l}$. In this case the sensible series of
dimensions is $\Delta = 1+l$. The only gauge invariant defect operator
consistent with this is
\begin{align}
{\cal C}^{\mu l} \equiv \sigma^\m_{ij} \left( \epsilon_{ik}\bar
\Psi^+_k q_2^l \Psi^-_j + \epsilon_{jk}\bar \Psi^-_k q_2^l
\Psi^+_i \right) \qquad (\m=0,...,3) \label{gurk}
\end{align}
where $\Psi^+_i$ and $\Psi^-_i$ are $SU(2)_L$ and $SU(2)_R$
doublets respectively,  given by
\begin{align}
\Psi^+_i = \begin{pmatrix} \psi_b^+ \cr \bar \psi_{\tilde b}^+
\end{pmatrix}
\qquad \Psi^-_i = \begin{pmatrix} \psi_b^- \cr \bar \psi_{\tilde
b}^-
\end{pmatrix}\,.\label{Psi}\end{align}
The index $\mu$ is an $SO(4)$ index and should not be confused with a
spacetime Lorentz index.  Note that (\ref{gurk}) is invariant under
parity, which exchanges the $SU(2)_L$ index $i$ with the $SU(2)_R$
index $j$, as well as $+$ with $-$.  This operator saturates the bound
(\ref{fourth}), i.e.~only one of the bounds in
(\ref{first})-(\ref{fourth}), so it is actually $1/4$ BPS. For $l=0$,
the operator is a pure defect operator which satisfies both the bounds
(\ref{second}) and (\ref{fourth}) and thus is $1/2$ BPS. This operator
will be shown to satisfy a non-renormalization theorem to order $g^2$
in Sec.~\ref{nonrenormal}.  

\subsubsection*{Gauge field fluctuations}

The gauge field fluctuations as derived in Sec.~\ref{gff} transform
trivially under $SU(2)_L \times SU(2)_R$ and have $J_{23}=0$ and $J_{45}=l$.
If we pick the positive branch, the dimension of this operator is
$\Delta=l+1$.  On the field theory side,
the operator at the bottom of the tower with the same quantum
numbers is the current associated with a global $U(1)_B$ under which the
defect fields transform,
\begin{align}
  {\cal J}^M_B \equiv \bar\Psi^\a_i \rho^M_{\a\beta} \Psi_i^\beta + i \bar b
  \overleftrightarrow{D}^M b + i \tilde b\overleftrightarrow{D}^M
\bar{\tilde
    b} \qquad(M=0,1) \, \label{gaugeoperator},
\end{align}
with Pauli matrices $\rho^M$ defined by Eq.~(\ref{Pauli}), $\Psi$
as in (\ref{Psi}), and $\alpha, \beta\in \left\{+,-\right\}$.
Although this current is conserved and satisfies the BPS bound of the
superconformal algebra, it is not a quasi-primary of the $SO(2,2)$
global conformal symmetry. This is essentially due to the fact
that it is in the same (short) supersymmetry multiplet as the
dimensionless field $\bar b b + \tilde b \bar{\tilde b} $.

The contributions to (\ref{gaugeoperator}) involving $b$, $\tilde
b$ lead to logarithms in the correlation functions. These are
actually present even in the purely two-dimensional free field
theory obtained by setting $g=0$ and thus decoupling the 2d from
the 4d theory. In this case we have a bosonic current contribution
of the form
\begin{align}
  J_M^{\rm 2d} = i \bar b \partial_M b
  - i (\partial_M \bar b) b \,, \label{bosonicgaugeop}
\end{align}
which is conserved. For Euclidean signature, this current has a
correlator of the form
\begin{align}
  \langle J_M^{\rm 2d} (x) J_N^{\rm 2d} (0) \rangle \propto
  {\textstyle\frac{1}{2}} \ln (x^2\m^2) \frac{I_{MN}(x)}{x^2} + \frac{x_M
    x_N}{x^2} \,,\quad I_{MN}(x) = \delta_{MN} - 2 \frac{x_Mx_N}{x^2}\,,
\end{align}
where $I_{MN}(x)$ is the inversion tensor. (\ref{bosonicgaugeop})
satisfies $\partial^x_M \langle J_M^{\rm 2d} (x) J_N^{\rm 2d} (0)
\rangle =0$ for $x \neq 0$. Note that in complex coordinates we
have $\partial_{\bar z} J_z^{\rm
  2d} + \partial_{z} J_{\bar z}^{\rm 2d} =0$, where only the sum vanishes,
not each term separately, such that there is no holomorphic -
antiholomorphic
splitting.

On the supergravity side, it is not quite clear if the current-current
correlator obtained from the gauge field fluctuations in Sec.~\ref{gff} is
well-defined. In $AdS_3$, the equation of motion for the gauge field leads
formally to a logarithmic propagator. This however does not satisfy the
required boundary condition to be identified as a bulk to boundary propagator.
A better understanding of the role played by two-dimensional scalars in this
model will be left for future work.

\subsubsection*{$AdS_3$ fluctuations inside AdS$_5$}

The fluctuations of the probe D3-brane wrapping $AdS_3$ inside
$AdS_5$ are characterised by $w_l$, which is the Fourier transform
of $w=X^2+ iX^3$ on $S^1$. The associated R-symmetry charges are
$J_{23} = -1$ and $J_{45} = l$, while there are no charges with
respect to $SU(2)_L \times SU(2)_R$.  Recall that the possible
series of dimensions for operators dual to these fluctuations are
$\Delta=l-1$ and $\Delta = 3-l$, cf.\ Eq.~(\ref{ads4series}).

\underline{$\Delta = l-1$ series}: 
Let us  determine the operators dual to this series. In the
free field limit, a gauge invariant scalar operator which is
localized on the defect and has $\Delta = l-1, J_{23} =-1$ and
$J_{45} = l$ with no $SU(2)_L \times SU(2)_R$ charges is
\begin{align}
{\cal B}^l \equiv \tilde b q_2^{l-1} b \,. \label{wierdops}
\end{align}
This operator has dimension $\Delta = {\cal J}$, which saturates the
bounds (\ref{third}, \ref{fourth}) due to the superconformal algebra.
An inspection of the supersymmetry variations of the fundamental
fields of the defect CFT also suggests that ${\cal B}^l$ is a chiral
primary. However this conclusion is erroneous.  In fact, ${\cal B}^l$
is not even a quasi-primary conformal field due to the presence of the
dimensionless scalars $b,\tilde{b}$. In other examples for probe brane
holography were the branes intersect over more than two dimensions
(for instance for the D3-D5 intersection), similar operators are in
fact chiral primaries. Here however, massless scalar fields in two
dimensions have strong infrared fluctuations and logarithmic
correlation functions. In a unitary two-dimensional CFT, it is
generally mandatory to take derivatives of massless scalars or
construct vertex operators from them in order to obtain operators
associated with states in the Hilbert space.\footnote{In our case, due
to the fact that $b$ and $\tilde b$ transform in the fundamental and
anti-fundamental representations, it is not clear how to build a gauge
covariant vertex operator with power law correlation functions.} It
may therefore seem remarkable that operators such as (\ref{wierdops})
appear at all in the AdS/CFT dictionary.  Note that even though the
apparent dimension of ${\cal B}^l$ is greater than zero for $l>1$, the
two-point functions do not have a standard power law behaviour.  This
can be readily seen in perturbation theory, where the scalars $b$ and
$\tilde b$ give rise to logarithmic terms in the two-point functions
for ${\cal B}^l$.  As we will discuss in Sec.~\ref{vanish}, this
behaviour of the two-point functions for this series is also seen in an
$AdS$ computation of the correlator. In Sec.~\ref{higgsbranch} we find
that the fluctuations $w_1$ are dual to the vacuum expectation value
of the operator ${\cal B}^1$ which parametrizes the classical Higgs
branch.

We note that the operators ${\cal B}^l$ have been proposed as
duals of the light-cone open string vacuum for D3-branes in a
plane-wave background \cite{Skenderis}.  The Penrose limit
giving rise to this background isolates a sector with large
$J_{45}$ in the defect CFT.  The light-cone energy in the plane
wave background corresponds to $\Delta - J_{45}$. For the
operators ${\cal B}^l$,  this quantity is negative: $\Delta -
J_{45} = -1$. Moreover we have seen that these operators are not
really chiral primaries (or even conformal fields).  Thus it is
not clear that they should be dual to the light-cone open string
vacuum. In fact it is not clear what the open string vacuum is,
due to the quantum mechanical spreading over the classical Higgs
branch, which corresponds different embeddings in the plane-wave
(or AdS) background.


\underline{$\Delta = 3-l$ series}:
Next let us consider the series $\Delta = 3-l$ with $l \leq 1$. A
gauge invariant scalar operator on the defect having $\Delta = 3-l,
J_{23} =-1, J_{45} = l$ with no $SU(2)_L \times SU(2)_R$ charges is
\begin{align} \label{Gl}
{\cal G}^l \equiv D_- \tilde b {q_2^{\dagger}}^{1-l} D_+ b
+ D_+ \tilde b {q_2^{\dagger}}^{1-l} D_- b
\end{align}
with the gauge covariant derivatives $D_\pm \equiv D_0 \pm D_1$.
These operators are obtained as two supercharge descendants of ${\cal
C}^{\mu l}$ defined in Eq.~(\ref{gurk}). Note that the two separate
terms are necessary for parity invariance under $z^+ \leftrightarrow
z^-$. The fluctuations modes $w_l$ are scalars rather than
pseudoscalars. These operators satisfy the bounds (\ref{first}) -
(\ref{fourth}).

\subsubsection{Correlators from probe fluctuations inside
AdS$_5$}\label{vanish}

Let us now compute the correlation functions of the operator ${\cal B}^l$
associated to the fluctuations $w_l$ of the probe brane inside $AdS_5$. For a
classical solution of the equation of motion, the action given by the sum of
(\ref{reg}) and (\ref{surf}) is given by the surface term
\begin{align}
S_{cl} = - T_{D3}L^4 \int d^3 \sigma \frac{1}{2}\partial_u\left[\frac{1}{u}
\tilde w_l^* \partial_u \tilde w_l - (l-1)\frac{1}{u^2} \tilde
w_l^* \tilde w_l \right] \,.\label{clact}
\end{align}
The first term in this expression is of the standard form obtained in
AdS/CFT, for instance in \cite{FreedmanMathur}.
The new feature here which does not appear in standard AdS
computations is the extra surface term with coefficient $(l-1)$.
This term has dramatic consequences.
To see this we compute the two-point function of the operator dual to $w_l$
following the procedure of \cite{FreedmanMathur}. 
We introduce an $AdS_3$ boundary at $u=\epsilon$ and
evaluate the action (\ref{clact}) for a solution of the form
\begin{align}
w_l(u,\vec k) = K^{(l)} (u,\vec k) \, w_l{}^b (\vec k)
\end{align}
in momentum space satisfying the boundary conditions
\begin{align}
\lim_{u\rightarrow \epsilon} K^{(l)}(u,\vec k) =1, \qquad
\lim_{u\rightarrow \infty} K^{(l)}(u, \vec k) = 0 \,.
\end{align}
The solution of the wave equation with these boundary conditions is
\begin{align}
K^{(l)}(u,\vec k) = \frac{u}{\epsilon} \frac{{\cal K}_{\nu}(u |\vec
k|)}{{\cal K}_{\nu}(\epsilon |\vec k|)} \,,
\end{align}
where $\nu = \Delta -1$ and ${\cal K}_\nu(x)$ is the modified
Bessel function which vanishes at $x\rightarrow \infty$. Note that
this coincides with the calculation of \cite{FreedmanMathur} where in
this case $d=2$. The two-point
function is given by
\begin{align}
  \langle {\cal B}^l(\vec k) {\cal B}^l(\vec k') \rangle &\equiv
  - \frac{\delta^2}{\delta w_l{}^b(\vec k) \delta w_l{}^b(\vec k')} S_{cl}
 \Big|_{w_l{}^b = 0}   \nonumber \\
  &=-\frac{1}{\epsilon} \delta(\vec k + \vec k') \lim_{u\rightarrow \epsilon}
   \left[ \partial_u K(u,\vec k) - (l-1) \frac{1}{u} K(u,\vec k) \right] 
 \,,\label{fe}
\end{align}
with $S_{cl}$ the Fourier transform of (\ref{clact}). 

The non-local part of the two-point function is obtained by expanding ${\cal
  K}_\nu(x)$ in a power series for small argument,\footnote{Expansion of the
  modified Bessel function ${\cal K}_\nu(x)$ for small argument:
\begin{align} \label{expandBessel}
 {\cal K}_\nu (x) = 2^{\nu-1} \Gamma(\nu) x^{-\nu} [1+ ...]
 - 2^{-\nu-1} \frac{\Gamma(1-\nu)}{\nu} x^\nu [1 + ...] \,.
\end{align}  
} keeping only the term
which scales like $\varepsilon^{ 2(\Delta-2)}$. The more singular
terms give rise to local contact terms of the form $\square^2
\delta(x-y)$ and are dropped. The non-local contribution to the
two-point function is given by
\begin{align}  \label{second2}
  \langle {\cal B}^l(\vec k) {\cal B}^l(\vec k') \rangle &=
\delta(\vec k + \vec k') \lim_{u\rightarrow
\epsilon}\left[-\epsilon^{-1}(\epsilon k)^{-1}\partial_u
\left(\frac{2^{-2(\Delta
-1)}\frac{\Gamma(2-\Delta)}{\Gamma(\Delta)}
(ku)^{\Delta}}{(k\epsilon)^{1-\Delta}}\right)\right.  \\
 &\left. \hspace{3cm} +\,  (l-1)\,
\epsilon^{-2}(\epsilon
k)^{-1}
\frac{2^{-2(\Delta-1)}\frac{\Gamma(2-\Delta)}{\Gamma(\Delta)}
(ku)^{\Delta}}{(k\epsilon)^{1-\Delta}} \right] + \cdots \,,\nonumber
\end{align}
where the dots indicate possible logarithmic terms.

The first of the two terms coincides exactly with the standard 
AdS calculation of \cite{FreedmanMathur}, whereas the second term
is an additional feature due to the presence of the probe brane.
Remarkably, there is an exact cancellation between the first and
the second term in (\ref{second2}) for the series
$\Delta = l-1$.  Thus for these fluctuations the usual calculation
does {\it not} give a power law correlation function of the form
$1/x^{2\Delta}$.  When we obtain the operators dual to these
fluctuations, it will become clear that one should not find a
power law. In particular, the lowest mode in this series is
the operator which parameterizes the classical Higgs branch.

\subsubsection{The classical Higgs
branch of intersecting D3-branes} \label{higgsbranch}

The classical Higgs branch of the D3-D3 intersection is parametrized
by vacuum expectation values of the scalar components of the defect
chiral fields $B$ and $\tilde B$. The vanishing of the F-terms of
the bulk chiral fields $S^1$ and $Q^1$ gives
\begin{align}
F_{q_1} = \partial_{\bar w} q_2 + [\phi,q_2] - g\delta^2(w) b
\tilde b &=0 \,, \nonumber \\
F_{s_1}= \partial_{\bar y} s_2 + [\upsilon, s_2] -
g\delta^2(y) \tilde b b &= 0\,, \label{holomeq}
\end{align} 
where $w=X^2+iX^3$ and $y=X^4+iX^5$.  In looking for solutions of
these equations, we shall take the gauge fields to vanish:
\mbox{$\phi=\upsilon=0$}.  With boundary conditions at infinity
($w\rightarrow \infty$ and $y \rightarrow \infty$) corresponding to
the original configuration of orthogonal intersecting branes, the
unique solution of (\ref{holomeq}) is
\begin{align}
q_2(w)=\frac{g b \tilde b}{2\pi i w}\,,\qquad 
s_2(y)=\frac{g \tilde b b}{2\pi i y}\,,
\end{align}
where we made use of the identity $\partial_{\bar w} \frac{1}{w} =2\pi
i \delta^2(w)$.  Because of the geometric identifications $q_2 \sim
y/\alpha^{\prime}$ and $s_2 \sim w/\alpha^{\prime}$, the solutions
give rise to holomorphic curves of the form
\begin{align}
 w y = c\alpha^{\prime} \,,
\end{align}
where $2\pi i c= g b\tilde b = g \tilde b b$. In other words, on the
Higgs branch the probe brane merges with one of the $N$ D3-branes as
shown in Fig.~\ref{figcurve}. Such brane recombinations have been studied
further in \cite{Robert}.

\begin{figure}[!ht]
\begin{center}
\includegraphics[scale=0.9]{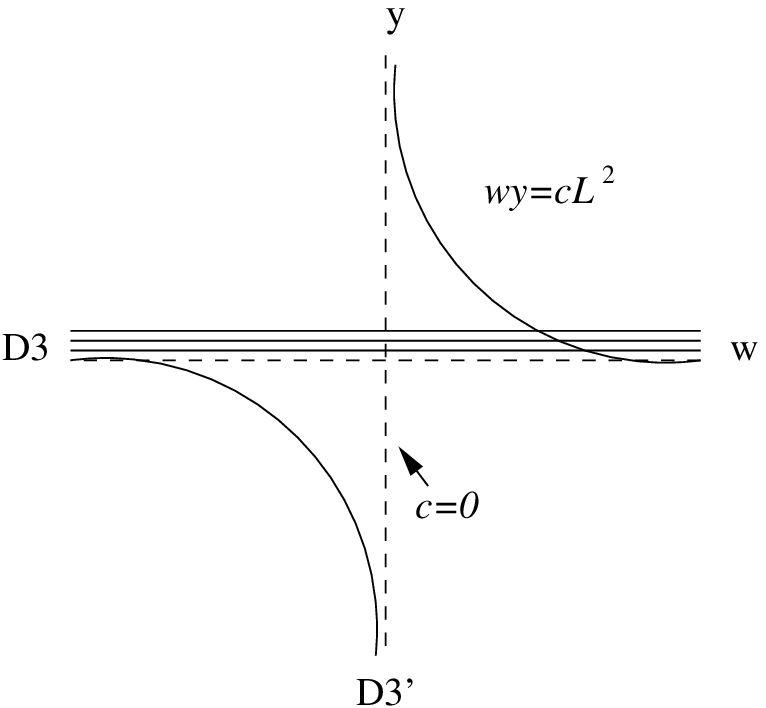}
\caption{Holomorphic curve $w y = c\alpha^{\prime}$.}  \label{figcurve}
\end{center}
\end{figure}

Let us come back to the fluctuation series $\Delta = l-1$.  There is a
very simple interpretation for the fluctuation $w_1$, the lowest mode
in this series, in the AdS background. According to
Eq.~(\ref{wierdops}) the dual operator is ${\cal B}^1 = \tilde b b$
whose vacuum value parametrizes the holomorphic curves $wy\sim\langle
\tilde{b}b\rangle L^2 = c L^2$. Furthermore, the probe brane can be embedded 
in $AdS_5 \times S^5$ so as to sit on such a holomorphic curve. In
this case the induced metric on the probe world-volume is
\begin{equation}
ds^2_{probe}= h^{-1/2}\left(-dt^2+dx_1^2\right)+
h^{1/2}\left(1+\frac{|c|^2L^4}{(|y|^2)^2}h^{-1}\right)dyd\bar{y}
\label{probemetnew}
\end{equation}
which reduces to (\ref{probemet}) for $c=0$ (note $k=1$ here). The
only effect on the near-horizon metric (\ref{nearprobe}) is the
modification of the radius of curvature $L$ which gets replaced by
$\tilde L=L\sqrt{1+|c|^2}$.  The near-horizon geometry is still
$AdS_3\times S^1$ but now with radius of curvature~$\tilde{L}$. Thus
it is natural to expect that these holomorphic embeddings correspond
to the classical fluctuations $w_1$ about the $c=0$ embedding.

To see this is in more detail let us elaborate on the relation
between the fluctuations $\tilde w_{1}$ and the classical Higgs branch.
Scalar fields in $AdS_3$ have the following behaviour near the
$u\rightarrow 0$ boundary of $AdS_3$:
\begin{align}
\phi \sim u^\Delta f(z^\pm) + u^{2-\Delta} g(z^\pm) \,.
\end{align}
As is standard in the AdS/CFT duality (with Lorentzian signature)
non-normalizable classical solutions are to be interpreted as sources for the
corresponding operators, while the normalizable solutions can be interpreted
as specifying a particular state in the Hilbert space
\cite{Balasubramanian,Klebanov}.  Only the VEV interpretation seems to make
sense for the fluctuations $\tilde w_l$ since, as shown in Sec.~\ref{vanish},
the two-point functions calculated in the usual way with source boundary
conditions do not have a power-law behaviour. Let us examine the $l=1$
fluctuation for which $\Delta = l-1= 0$, and consider the solutions $\tilde
w_1 = c$ where $c$ is a complex number. Naively one might conclude that this
amounts to choosing $\langle \tilde b b \rangle \sim c$. However since $\Delta
=0$, this solution is not normalizable, although it sits right at the border
of normalizability\footnote{Note that such solutions have as much right to be
  considered in Euclidean signature, since they are non-singular at the
  ``origin'' of $AdS$, $u = \infty$.}.  This is a reflection of the fact that
the quantum mechanical vacuum must spread out over the entire classical Higgs
branch, since the latter is parameterized by dimensionless scalars whose
correlators grow logarithmically with distance\footnote{This is the same
  spreading which accounts for the ``Coleman-Mermin-Wagner'' theorem
  \cite{CMW, CMW2} preventing spontaneously broken continuous symmetries in
  two dimensions.}.

Despite the lack of normalizability of the fluctuations $w_1 = c$,
the identification $c \sim \langle \tilde b b \rangle$ makes sense
at the classical level. This follows from the fact that the
solution $\tilde w_{1} = c$ corresponds to a holomorphic
embedding. To see this it is convenient to recall the following
coordinate definitions (with $L^2=1$):
\begin{align}
r= 1/u, \qquad z^{\pm} = X^0 \pm X^1, \quad w = u \tilde w = X^2 +
i X^3, \quad y = x^4+i x^5,
\end{align}
and define $\vec v = X^{6,7,8,9}$, in terms of which the D3-brane
metric is
\begin{align}
ds^2 = (1+ \frac{1}{r^4} )^ {-1/2}(-dz^+ dz^- + dw d\bar w) + (1 +
\frac{1}{r^4})^{1/2}(dy d\bar y + d\vec v^2) \,.
\end{align}
In the simplest case, the embedding of the probe D3$'$-brane is given
by $w=0, \vec v =0$ which agrees with the embedding conditions
(\ref{embedcond}) for $k=1$. On the probe, \mbox{$y= r\exp(-i\phi_1)$} where
$\phi_1$ is defined in (\ref{sphere}). Therefore $\tilde w_{1} = c$
implies
\begin{align}
w = u \tilde w_{1}e^{i\phi_1} = \frac{c}{r e^{-i\phi_1}} = \frac{c}{y} \,.
\end{align}
The holomorphic curve $w y =c$ is precisely that which arises from
(\ref{holomeq}), provided that
\begin{align}
b = \begin{pmatrix} v \cr 0 \cr  \vdots\end{pmatrix}  \qquad
\tilde b = \begin{pmatrix} v & 0 & \cdots \end{pmatrix}
\end{align}
with $gv^2 = c/(2\pi i)$.  In this background, the probe D3$'$-brane
combines with one of the $N$ D3-branes to form a single D3 on the
curve $w y =c$. In this sense the AdS field $w_1$ parameterizes the possible
embeddings of the probe brane within $AdS_5$ and the dual operator
$\tilde{b}b$ parameterizes the classical Higgs branch of the CFT.

As was noted earlier the curve $w y =c$ does not break the
superconformal symmetries. To see this, it is convenient to represent
$AdS_5$ by the
hyperboloid,
\begin{align}\label{hyperbol}
{\cal X}_0^2 + {\cal X}_5^2 - {\cal X}_1^2 - {\cal X}_2^2 - {\cal
X}_3^2 - {\cal X}_4^2 = 1
\end{align}
where
\begin{align}
ds^2 = -d{\cal X}_0^2 -d{\cal X}_5^2 + d{\cal X}_1^2 +d {\cal
X}_2^2 +d {\cal X}_3^2 +d {\cal X}_4^2 \,.
\end{align}
The coordinates on the Poincar\'{e} patch, $t , \vec x =
x^{1,2,3}$ and $r$, are related to these by
\begin{align}
{\cal X}_5 &= \frac{1}{2r} \left(1+r^2(1 + \vec x^2 -
t^2)\right), \qquad {\cal X}_0 = rt\,, \qquad
{\cal X}_{1,2,3} = r x^{1,2,3} \,, \\
{\cal X}_4 &= \frac{1}{2r} \left(1-r^2(1 + \vec x^2 - t^2)\right) \,.
\end{align}
The embedding $w y =c$,  or $x^2 + i x^3 = \frac{c}{r e^{i\phi_1}}$
can then be written as
\begin{align}
{\cal X}_2+ i{\cal X}_3 = ce^{-i\phi_1}  \,.
\end{align}
which when combined with eqn.~(\ref{hyperbol}) gives,
\begin{align}
{\cal X}_0^2 + {\cal X}_5^2 - {\cal X}_1^2  - {\cal X}_4^2 = 1+|c|^2.
\end{align}
This is exactly the hyperboloid which defines an $AdS_3$ spacetime
with radius of curvature $1+|c|^2$.
Further, this embedding is manifestly invariant under the isometry $SO(2,2)
\times SU(2)_L \times SU(2)_R \times U(1)'$.  The $U(1)'$ factor
is precisely that which appears in the superconformal algebra as a
combination of rotations in the $23$ and $45$ planes generated by
$J_{23} + J_{45}$.  This $U(1)'$ factor phase rotates $w$ and
shifts $\phi_1$ such that $we^{-i\phi_1}$ is invariant.

Similarly, we can understand the fluctuations $w_l$ with $l>1$. The
fluctuations $\tilde w_l$ behave as $\tilde w_l = c_l u^\Delta$ with
$\Delta=l-1$ and $c_l=\langle {\cal B}^l \rangle$ the expectation
value of the operator ${\cal B}^l$. If we ignore all other
fluctuations, i.e.\ if we set $w_k=0$ for all $k \neq l$, then we
obtain the holomorphic curve
\begin{align} \label{othercurves}
w = u \sum_k \tilde w_k e^{i\phi_1k} = u \tilde w_{l}e^{i\phi_1 l} 
  = \frac{c_l}{(r e^{-i\phi_1})^l} = \frac{c_l}{y^l} \,.
\end{align}
Such embeddings are still supersymmetric \cite{calibrate, calibrate2}, but
only for $l=1$ is the $AdS_3$ geometry preserved. So $w_l$ fluctuations
generate excited states which are not conformal, except for $l=1$. Note that
in a conformal quantum field theory only the vacuum must be invariant under
conformal symmetry. In particular, we do not expect to get the curves
(\ref{othercurves}) from the F-terms.  The flatness conditions on the F-terms
give only conformally invariant vacuum solutions.

Quantum mechanically we expect the vacuum to spread out over the
entire classical Higgs branch, since it is parameterized by massless
two-dimensional fields.  This differs from the situation on the
Coulomb branch, on which the orthogonal branes are separated in the
$X^{6,7,8,9}$ directions by giving VEV's to {\it four}-dimensional
fields $q_1, \sigma, s_1$ and $\omega$. Note that on the Higgs branch
one also has non-zero four-dimensional fields, of the form $q_2 = c/w,
s_2 = c/y$, however since the asymptotic values of the fields are
independent of $c$ in all but two of the four world-volume directions,
we expect that there is no obstruction to the wavefunction spreading
out as a function of $c$.  This suggests that the AdS/CFT prescription
for computing correlators should be modified to sum over embeddings of
holomorphic curves parameterized by $c$. A natural conjecture is that
the map between the generating function for correlators in the CFT and
the probe-supergravity action should have the form
\begin{align}
\langle e^{-J \hat O} \rangle = \int {\cal D} c\, e^{-S_{cl}(\phi, c)}
\end{align}
where, as usual,  the probe-supergravity fields $\phi$ have
boundary behaviour determined by the sources $J$.  Note that the
classical Higgs branch is non-compact, and it is unclear to us
what the measure ${\cal D} c$ should be.\footnote{We expect that
one contribution to the measure should arise from the fact that
the $AdS_3$ metric induced on the curve $wy =c$ has effective
curvature radius $\sqrt{1 + c^*c}$.}

\subsubsection{Nonrenormalization of the two-point function involving
${\cal C}^{\m l}$} \label{nonrenormal}

In Sec.~\ref{sec41} we found from considering strings on the
probe-supergravity background that correlators of both probe and bulk fields
should be independent of the 't~Hooft coupling $\lambda=g_{YM}^2N$. In
general, the weak and strong coupling behaviour do not have to be related.
Nevertheless, the remarkable result of complete 't~Hooft coupling independence
of the correlators at strong coupling suggests that nonrenormalization
theorems may be present in the defect conformal field theory.  In this section
we study the nonrenormalization behaviour of the correlators at weak coupling.
By showing the absence of order $g_{YM}^2$ radiative corrections to some of
the correlators, we give some field-theoretical evidence for the existence of
nonrenormalization theorems.  In particular, we consider the two-point
function of the chiral primary operator ${\cal C}^{\mu l}$ which is the lowest
component of a short representation of the $(4, 4)$ supersymmetry algebra
derived in Sec.~\ref{superalg}.

Let us consider the two-point correlator of the chiral primary ${\cal
C}^{\mu l}$. In the following we show that $\langle {\cal C}^{\mu l}
(x) \bar {\cal C}^{\mu l} (y) \rangle$ does not receive any
corrections at order $g_{YM}^2$ in perturbation theory.  It is
sufficient to show this for the component ${\cal C}^l \equiv {\cal
C}^{1l}$ given by
\begin{align}
  {\cal C}^l &\equiv \psi^-_{\tilde b} q_2^l \bar
    {\psi}^+_{\tilde b} - \bar \psi^+_b q_2^l \psi^-_b + \bar \psi^-_{\tilde
      b} q_2^l {\psi}^+_{\tilde b} - \psi^+_b q_2^l \bar \psi^-_b
  \,.\label{C2}
\end{align}
The nonrenormalization of the other components is guaranteed by
the $SO(4)$ R-symmetry.  The tree-level graph of the two-point
function $\langle {\cal
  C}^l(x) \bar {\cal C}^l(y) \rangle$ is depicted in Fig.~\ref{rainbow}.
There are three other graphs contributing to this propagator
corresponding to the remaining three terms in Eq.~(\ref{C2}).

\begin{figure}[!ht]
\begin{center}
\includegraphics{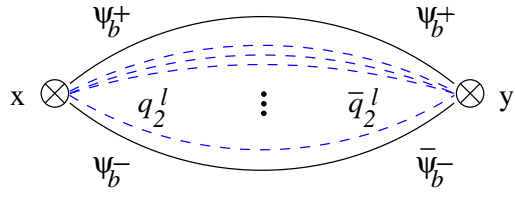}
\caption{One of the four graphs of the correlator $\langle {\cal
C}^l(x) \bar {\cal C}^l(y) \rangle$.}\label{rainbow}
\end{center}
\end{figure}

We show ${\cal O}(g^2)$ nonrenormalization for ${\cal C}^l$ with $l=0$ for
which $q_2$ exchanges are absent. The relevant propagators are
\begin{align}
&\langle v_M(x) v_N(y) \rangle = \frac{ \eta_{MN} }{ (2\pi)^2
(x-y)^2}\,, \qquad
\langle q_1(x) \bar q_1(y) \rangle = \frac{1}{(2\pi)^2 (x-y)^2}
\,,\label{bosonprop} \\
&\langle \psi_\alpha (x) \bar \psi_\beta (y) \rangle =
\frac{i}{2\pi} \frac{\rho^M_{\alpha\beta}  (x-y)_M}{(x-y)^2}
\,,
\end{align}
with $\eta_{MN}={\rm diag(+1,-1)}$, Pauli matrices $\rho^M
(M=0,1)$ defined in App.~\ref{appA}, and defect coordinates $x, y$.
The four-dimensional propagators in Eq.~(\ref{bosonprop}) are
pinned to the defect. The Feynman rules for the vertices can be
read off from the defect action in component form derived in
App.~\ref{appendixC}.

First we note that, similar as in $\N=4$, $d=4$ SYM theory \cite{Kovacs},
there are no one-loop self-energy corrections to the defect fermionic
propagator $\langle \bar \psi_b \psi_b \rangle$.  Self-energy corrections
involving a gaugino propagator are cancelled by those involving a
$\psi^{q_1}$
propagator which is the fermion of the superfield $Q_1$. There are also
self-energy graphs with $q_1$ and $\sigma$ propagators which arise from the
ambient scalars coupling to the defect. These cancel each other, too.

\begin{figure} [!ht]
\begin{center}
  \includegraphics{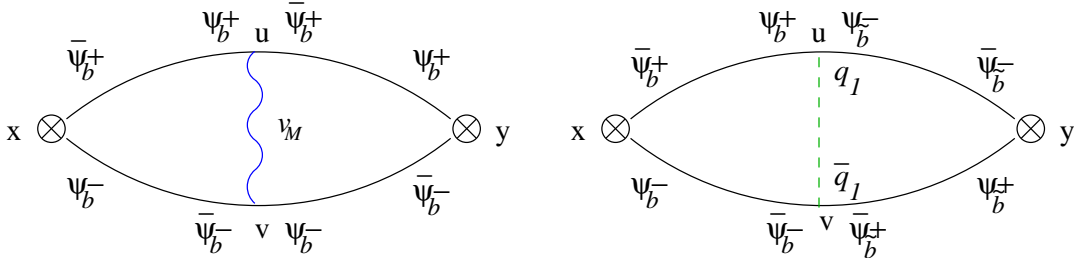}
\caption{First order corrections to the correlator $\langle  {\cal
C}^l(x) \bar {\cal C}^l(y) \rangle$ for $l=0$.}\label{rainbow2}
\end{center}
\end{figure}
However, we have two possible corrections from exchange graphs as shown in
Fig.~\ref{rainbow2}. Note that in Fig.~\ref{rainbow2}, two different
contributions to ${\cal C}^l$ ($l=0$) are depicted at the point $y$, which
originate from different terms in the sum (\ref{C2}).  These graphs include an
ambient gauge boson exchange and an ambient scalar exchange. There is no
$\sigma$ exchange contributing to the correlator $\langle {\cal C}^l(x) {\cal
  C}^l(y) \rangle$ (for $l=0$). In fact, it may be shown that for each of the
components of ${\cal C}^{\m l}$, there is either a $\sigma$ or a $q_1$
exchange. For all of the components, the vector exchange is cancelled by one
of these scalar exchanges while the other one vanishes.

For the gauge boson exchange in Fig.~\ref{rainbow2}a we find the
contribution
\begin{align}
  \frac{1}{2} \int d^2u d^2v \,& \frac{\rho_{++}\cdot(x-u)}{2\pi
    (x-u)^2} (-\frac{1}{2} g \rho^M_{++}) \frac{ \eta_{MN} }{ (2\pi)^2
    (u-v)^2} (-\frac{1}{2} g \rho^N_{--}) \frac{\rho_{++}\cdot(u-y)}{2\pi
    (u-y)^2}
  \nonumber\\
  &\times \frac{\rho_{--}\cdot(x-v)}{2\pi (x-v)^2}
  \frac{\rho_{--}\cdot(v-y)}{2\pi (v-y)^2} \,.
\end{align}
The overall factor $\frac{1}{2}$ comes from the definition $v_M=
\frac{1}{\sqrt{2}} v'_M$.

Let us now consider the contribution from the $q_1$
exchange in Fig.~\ref{rainbow2}b which is given by
\begin{align}
- \int d^2u d^2v \,& \frac{\rho_{++}\cdot(x-u)}{2\pi (x-u)^2}
(\frac{1}{2} ig) \frac{1 }{ (2\pi)^2 (u-v)^2} (-\frac{1}{2} ig)
\frac{\rho_{--}\cdot(u-y)}{2\pi (u-y)^2}
\nonumber\\
&\times \frac{\rho_{--}\cdot(x-v)}{2\pi (x-v)^2}
\frac{\rho_{++}\cdot(v-y)}{2\pi (v-y)^2}   \,. \label{C3}
\end{align}
Note that the operator at the external point $y$ in the graph of
Fig.~\ref{rainbow2}b is the conjugate of the first term in Eq.~(\ref{C2})
which leads to the minus sign in front of the integral in Eq.~(\ref{C3}).
In Fig.~\ref{rainbow2}a both external vertices have a minus sign, whereas
in Fig.~\ref{rainbow2}b the vertices have opposite signs.
Since $\eta_{MN} \rho^M_{++}\rho^N_{--}=2$, the vector exchange exactly
cancels the contribution from the scalar exchange.

Nonrenormalization of correlators of ${\cal C}^{\m l}$ with $l
\geq 1$ is more difficult to show.  As was shown for the operators
${\rm Tr\,}X^k$ in $\N=4$ super Yang-Mills theory \cite{Kovacs, D'Hoker},
there are no exchanges between the ambient propagators $\langle
q_2(x) \bar q_2(y) \rangle$ within the correlator $\langle {\cal
C}^l (x) \bar {\cal C}^l (y) \rangle$.  However, one could think
of a gauge boson exchange between a fermionic defect and a bosonic
ambient propagator.  If we do {\em not} work in Wess-Zumino gauge
then there is an additional interaction of the defect fermions
with a scalar $C$ which is the lowest component of the gauge
superfield $V$.  Keeping this in mind we expect that a CD exchange
\cite{Kovacs} cancels the above gauge boson exchange. 

\subsubsection{Vanishing of odd correlators of the BPS primaries 
${\cal C}^{\mu l}$}

Another property of the BPS primaries ${\cal C}^{\mu l}$ is the
vanishing of all $(2k+1)$-point functions ($k \in \mathbb{N}$).
Only even $n$-point functions may differ from zero. On the gravity
side this can be seen by studying once more the DBI action of the
probe D3-brane.  Due to the expansion of the cosines of the
angular fluctuations $\theta, \phi, \rho$, and $\chi$ in the
determinant, the BI action contains only even powers of the
fluctuations, see Eq.\ (\ref{quadBI}).  This implies vanishing odd
couplings for the Kaluza-Klein modes which, via the AdS/CFT
correspondence, implies vanishing odd $n$-point functions on the
field-theory side. In the dual conformal field theory these
Kaluza-Klein modes correspond to the BPS primary operators ${\cal
C}^{\mu l}$. Again we restrict to the component ${\cal C}^l \equiv
{\cal C}^{1 l}$.

On the field theory side too,  one finds for instance that the
three-point function $\langle {\cal
  C}^{l_1} (x) {\cal C}^{l_2} (y) {\cal C}^{l_3} (z) \rangle $ is absent.
This is due to a global $U(1)$ symmetry of the action,
\begin{align}
B &\rightarrow e^{i\frac{\phi}{2}} B \,,\qquad
\tilde B \rightarrow e^{i\frac{\phi}{2}} \tilde B\,,\qquad
Q_1 \rightarrow e^{-i\phi} Q_1 \,,\qquad
Q_2 \rightarrow e^{i\phi} Q_2  \,,
\end{align}
with all other fields being singlets under this symmetry. If we
choose $\phi=\pi$ then \mbox{${\cal C}^l \rightarrow (e^{i\pi})^{l+1}
{\cal C}^l$} and the three-point function transforms as
\begin{align}
  \langle {\cal C}^{l_1} (x) {\cal C}^{l_2} (y) {\cal C}^{l_3} (z) \rangle
  \rightarrow (-1)^{l_1+l_2+l_3+1} \langle {\cal C}^{l_1} (x) {\cal C}^{l_2}
  (y) {\cal C}^{l_3} (z) \rangle \,.
\end{align}
Since $l_1+l_2+l_3$ must be even, $(-1)^{l_1+l_2+l_3+1}=-1$ and $\langle
{\cal
  C}^{l_1} (x) {\cal C}^{l_2} (y) {\cal C}^{l_3} (z) \rangle$ vanishes.
Though we have restricted the discussion on ${\cal C}^{1l}$, the statement
also holds for the other components. This is guaranteed by the fact that
${\cal C}^{\m l}$ transforms as a vector under the $SO(4)$ R-symmetry group.

\subsubsection{Summary and discussion of the AdS/CFT dictionary}

Table \ref{table2} summarizes the fluctuations of the KK modes and
their dual operators.\footnote{The conformal dimensions of the
dual operators are lowered by one in comparison with the
corresponding series in the D3-D5 system studied in \cite{DFO}.
This is simply because the operators are bilinears of defect
fundamental fields, whose conformal dimensions are lowered by 1/2
in comparison with corresponding defect fields in the D3-D5 case.}
The angular fluctuations of the probe $S^1$ embedding inside $S^5$
are dual to $1/4$ BPS primaries ${\cal C}^{\m l}$. The $\Delta =
3-l$ fluctuations of the embedding of $AdS_3$ inside $AdS_5$ are
dual to ${\cal G}^l$ which are two-supercharge descendants of
these primaries. The $\Delta = l-1$ fluctuations of the embedding
of $AdS_3$ inside $AdS_5$ are not dual to conformal operators
which correspond to states in the Hilbert space. Naively the dual
operators ${\cal B}^l$ look like $1/2$ BPS (chiral) primaries, but
in fact they contain massless defect scalars which do not give
rise to power law correlation functions. These massless scalars
and their dual fluctuations include an entry ${\cal B}^1$ which
parameterizes the classical Higgs branch.  The fluctuations ${\cal
B}^l$ for $l>1$ correspond to other holomorphic curves $w =
c_l/y^{l}$, however we do not (as yet) have a clear interpretation
for these in the defect CFT. Lastly, the operator ${\cal J}^M_B$
which is dual to the gauge field fluctuations on $AdS_3$ is a
descendant of the dimensionless operator $\bar b b + \tilde b
\bar{\tilde b}$, which has a logarithmic two-point function and is
not a primary operator although formally it trivially satisfies
the BPS bounds.

\begin{table}[!h]
\begin{center}
\begin{tabular}{c|c|c|l|c|c}
fluctuations & $\Delta$ & $l$ & $(j_1,j_2)_{\cal J} $  & operator
&
interpretation\\
\hline
$S^1 \subset S^5$ & $l+1$ & $l \geq 0$ &
$(\frac{1}{2},\frac{1}{2})_l$
& ${\cal C}^{\m l}$ & $1/4$ BPS primary \\
$AdS_3 \subset AdS_5$ & $3-l$ & $l \leq 1$ &  $(0,0)_{l+1}$ &
${\cal G}^l$&
descendant \\
& $l-1$ & $l \geq 1$&$(0,0)_{l+1}$ &  ${\cal B}^l$ &  classical
Higgs branch\\
gauge field & $l+1$ & $l \geq 0$ &  $(0,0)_l$ & ${\cal
J}_B^{M l}$ &
---
\end{tabular}
\caption{Summary of fluctuation modes and field theory operators
with coincident quantum numbers.} \label{table2}
\end{center}
\end{table}



\section{Fundamental matter in the AdS/CFT correspondence}
\label{cha:flavour}

 \setcounter{equation}{0}\setcounter{figure}{0}\setcounter{table}{0}

One of the main objectives of this paper is the inclusion of particles in the
fundamental representation of the gauge group into the AdS/CFT correspondence.
This is an indispensable necessity to study QCD in terms of its holographic
dual.  As we have seen in the last chapter, fundamental representations can be
introduced by embedding a probe brane on an $AdS_d$ subspace of the full
$AdS_D$ geometry, where $d \leq D$ and the boundary of $AdS_d$ is part of the
boundary of the $AdS_D$. Embeddings for which $d=D$ give rise to a dual field
theory with ``quarks'' that are free to move in all spacetime dimensions.  The
D3-D7 brane intersection realises such an embedding in its near-horizon
geometry, where it wraps an $AdS_5 \times S^3$ submanifold. In order to study
non-perturbative effects in large $N$ QCD-like theories, we now embed a D7
probe in two non-supersymmetric supergravity backgrounds, both exhibiting
confinement of fundamental matter. This allows us to numerically compute quark
condensates and meson spectra.

The organisation of this chapter is as follows.  In Sec.~\ref{D3D7} we discuss
explicitly the holography of the D3-D7 intersection. Here we specialize the
general discussion of the D3-D$p$ brane system of the previous chapter to
$p=7$, i.e.~we study D7-probes in the AdS/CFT Correspondence.  We
demonstrate our numerical techniques by a supergravity computation of the
meson spectrum in the standard AdS/CFT correspondence.  In
Sec.~\ref{adsbhsection} we consider D7-branes in the AdS Schwarzschild black
hole background. We find a first order phase transition in the dual field
theory. We compute the quark condensate and the meson spectrum.  In
Sec.~\ref{mcsection} we study D7-brane probes in the Constable-Myers
background \cite{CM}. We demonstrate spontaneous $U(1)_A$ chiral symmetry
breaking and identify the corresponding Goldstone boson in the meson spectrum.
We summarize our results in Ch.~\ref{conclusion}.


\subsection{AdS/CFT duality for an ${\mathcal N} =2$ gauge theory with 
fundamental matter} \label{D3D7}
\lhead[\fancyplain{}{\bfseries\thepage}]%
{\fancyplain{}{\bfseries \ref{D3D7} AdS/CFT
 for an ${\mathcal N} =2$ gauge theory with
fundamental matter}}
\rhead[\fancyplain{}{\bfseries\leftmark}]%
{\fancyplain{}{\bfseries\thepage}} 

\subsubsection{The D3-D7 brane configuration}
It was first observed in \cite{Katz} that one can obtain a
holographic dual of a four-dimensional Yang-Mills theory with
fundamental matter by taking the near-horizon limit of a system of
intersecting D3 and D7-branes. We first review some of the features of
this duality, and then test numerical techniques which we will use
later to study deformations of this duality.

\begin{figure}[!h]
\begin{center}
\includegraphics[scale=0.76,clip=true,keepaspectratio=true]{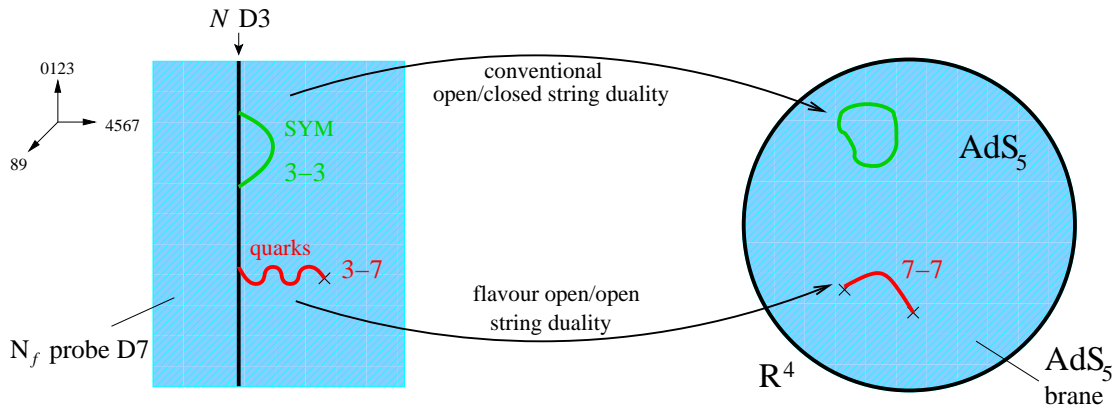}
\caption{Holography of the D3--D7 brane configuration.} \label{mapD7}
\end{center}
\end{figure}
According to the general description of a D3-D$p$ brane intersection,
see Sec.~\ref{Ch21}, we consider a stack of $N$ D3-branes spanning the
directions $x^{0},x^1,x^2,x^3$ and another stack of $N_f$ D7-branes
spanning the directions $x^0, ..., x^7$. This is shown in the left
picture of Fig.~\ref{mapD7}. The low-energy dynamics of open strings
in this setting is described by a ${\mathcal N} =2$ super Yang-Mills
theory.  This theory contains the degrees of freedom of the ${\mathcal
N} =4$ theory, namely an ${\mathcal N} =2$ vector multiplet and an
${\mathcal N}=2$ adjoint hypermultiplet, as well as $N_f$ ${\mathcal
N} =2$ hypermultiplets in the fundamental representation of
$SU(N)$. The theory is conformal in the limit $N \rightarrow \infty$
with $N_f$ fixed.  There is an $SU(2) \times U(1)$ R-symmetry.  The
$U(1)$ R-symmetry acts as a chiral rotation on the ``quarks'', which
are the fermionic components of the ${\mathcal N} =2$ hypermultiplet
composed of fundamental and anti-fundamental chiral superfields $Q$
and $\tilde Q$.  This symmetry also acts as a phase rotation on the
scalar component of one of the adjoint chiral superfields.  When the
D7-branes are separated from the D3-branes in the two mutually
transverse directions $X^8$ and $X^9$, the fields $Q,
\tilde Q$ become massive,  explicitly breaking the $U(1)$
R-symmetry and conformal invariance.  As shown in
\cite{Katz},  the ${\mathcal N} =2$ theory as well as
its renormalization group flow have an elegant holographic
description.

This holographic description is obtained as follows.  In the limit
of large $N$ at fixed but large 't~Hooft coupling $\lambda = g^2N
\gg 1$, the D3-branes may be replaced with their near-horizon $AdS_5
\times S^5$ geometry given by
\begin{align} \label{ads5}
ds^2 = \frac{r^2}{L^2} (-dt^2 + dx_1^2+dx_2^2+dx_3^2) + 
\frac{L^2}{r^2} d \vec y^2 \,,
\end{align}
where $\vec y = (X^4, \cdots, X^9)$, $r^2 \equiv \vec y^2$, and $L$
the radius of curvature. It will be convenient to write the transverse
metric $d\vec y^2$ in the following way
\begin{equation}
 d\vec y^2 = d
\rho^2 +\rho^2 d \Omega_3^2 + dy_5^2 + dy_6^2 \,, 
\end{equation}
where $d\Omega_3^2$ is a three-sphere metric and $r^2=\rho^2+y_5^2 +
y_6^2$. Since the number $N_f$ of D7-branes is finite, their
back-reaction on the geometry can be effectively ignored.

For massless flavours, the D7-brane embedding in the D3-metric
(\ref{D3metric}) is given by $y^5 = y^6 =0$ (corresponding to $X^8=X^9
=0$).  In the $AdS_5 \times S^5$ geometry, the induced geometry on
the D7-brane is given by the metric (\ref{nearprobe}) (for $k=3$)
which is $AdS_5 \times S^3$.  The D7-brane fills $AdS_5$, while
wrapping a great three-sphere of the $S^5$. The isometries of the
$AdS_5 \times S^5$ metric which preserve the embedding correspond to
the conformal group and R-symmetries of the ${\mathcal N} =2$ gauge
theory.  The conformal group $SO(2,4)$ is the isometry group of
$AdS_5$, while the $SU(2) \times U(1)$ R-symmetry corresponds to the
rotations of the $S^3$ inside $S^5$ and rotations of the $y^5, y^6$
coordinates.

The holographic description for massive flavours is found by
considering the D7-brane embedding $y^5 = X^8 =0$, $y^6 = X^9 = m$.
In this case the D7-geometry is still $AdS_5 \times S^3$ in the
$r\rightarrow \infty$ region corresponding to the ultraviolet.
However, from the induced metric on the
D7-brane,
\begin{align}
ds^2_{probe} = \frac{\rho^2+m^2}{L^2} (-dt^2 + dx_1^2+dx_2^2+dx_3^2)
+  \frac{L^2}{\rho^2+m^2}(d\rho^2+\rho^2d\Omega_3^2) \,,
\end{align}
we see that the radius of the three-sphere vanishes at $\rho=0$ or,
equivalently, at $r=m$. This is possible because the $S^3$ is contractible
within the $S^5$ of the full ten dimensional geometry. The D7-brane ``ends''
at the value of $r$ at which the $S^3$ collapses, meaning that it does not
fill all of $AdS_5$, but only a region outside a core of radius $r=m$.  This
is consistent with the fact that the fundamental degrees of freedom decouple
at energies below $m$.  Note that although the D7-brane ends at $r=m$, the
D7-geometry is perfectly smooth, as is illustrated in Fig.~\ref{flow}.  In the
massive case, the conformal and $U(1)$ symmetries are broken, and the D7
embedding is no longer invariant under the corresponding isometries.
\begin{figure}[!h]
\begin{center}
\includegraphics[height=6cm,clip=true,keepaspectratio=true]{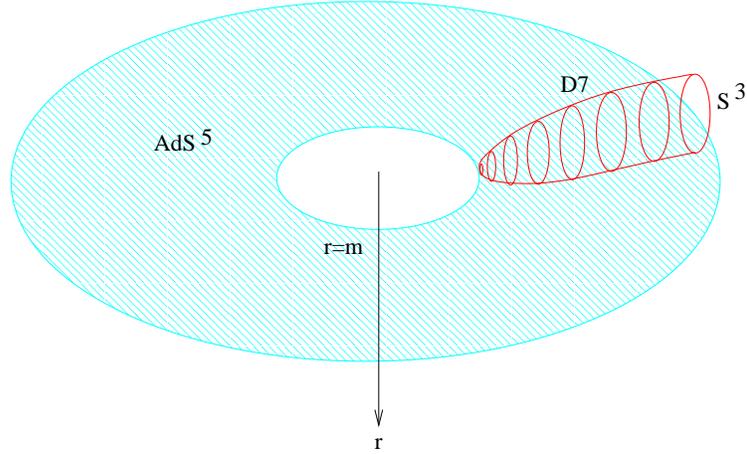}
\caption{The D7 embedding in $AdS_5 \times S^5$ for $m \neq 0$. The figure
shows the modification of the $AdS_5$ brane embedding on the r.h.s.~of
Fig.~\ref{mapD7} in the case of massive quarks.}\label{flow}
\end{center}
\end{figure}

\subsubsection{Testing numerical methods: Mesons in the D3-D7 
\mbox{intersection}}

In this chapter we will numerically compute condensates
and meson spectra in deformations of the duality discussed above.
Therefore we  first test these numerical techniques against
some exact results in the undeformed case.

To study the implications of the classical D7 probe dynamics for the dual
field theory, we now evaluate the scalar contributions to the
Dirac-Born-Infeld (DBI) action for the D7-brane in the $AdS_5 \times S^5$
background. We work in static gauge where the world-volume coordinates of the
brane are identified with the spacetime coordinates by $\xi^a \sim t,x_1,x_2,
x_3, y_1, ..., y_4$. The DBI action (\ref{d3act}) for $p=7$ is then
\begin{eqnarray}
S_{D7} & = & - T_{D7} \int d^8\xi \sqrt{- {\rm det}(g^{PB}_{ab}) }
=- T_{D7} \int d^8\xi\sqrt{-\det g_{ab}}\sqrt{1 + g^{ab} \partial_aZ^i\partial_bZ^jg_{ij}}
\nonumber \\ & = & - T_{D7} \int d^8 \xi ~ \epsilon_3 ~ \rho^3
\sqrt{1 + {\frac{g^{ab}}{  \rho^2 + y_5^2 + y_6^2}} (\partial_a y_5
\partial_b y_5 +
\partial_a y_6 \partial_b y_6)}\,,
\end{eqnarray}
where $Z^i$ are the transverse coordinates $y_5, y_6$. The metrics $g_{ab}$
and $g_{ij}$ are the $AdS_5 \times S^5$ metric restricted to eight and two
dimensions, respectively (set either $dZ^i=0$ or $d\xi^a=0$). The factor
$\epsilon_3$ is the determinant from the three sphere.  The ground state
configuration of the D7-brane is given by the solution of the Euler-Lagrange
equation with dependence only on the $\rho$ variable.  In this case the
equations of motion become (use $g^{\rho\rho}=r^2=\rho^2+y_5^2+y_6^2$)
\begin{equation} \label{d7eom} {\frac{d}
{d \rho}} \left[ {\rho^3 \over  \sqrt{1 + \left({d y_6 \over d
\rho}\right)^2}}  { d y_6 \over d \rho}\right] = 0 \, , 
\end{equation}
where we consider solutions with $y_5=0$ only.  Recall that the $U(1)$
R-symmetry corresponds to rotations in the $y^5$--$y^6$-plane.

The equations of motion have asymptotic ($\rho\rightarrow \infty$)
solutions of the form 
\begin{equation}
 y_6 = m + { \frac{c}{ \rho^2}} \,. \label{identif}
\end{equation} 
The identification of these constants as field theory operators requires a
coordinate transformation because the scalar kinetic term is not of the usual
canonical AdS form. Transforming to the coordinates of \cite{Katz} in which
the kinetic term has canonical form, we see that $m$ has dimension 1 and $c$
has dimension 3. The scalars are then identified \cite{PolStrass} with the
quark mass $m_q$ and condensate $\langle \bar{\psi} \psi \rangle$,
respectively, in agreement with the usual AdS/CFT dictionary. The dimension
three operator ${\cal C}^0=\bar \psi \psi$ is the higher-dimensional analog of
the operator ${\cal C}^{\mu 0}$ as given by Eq.~(\ref{gurk}) and is dual to
$S^3$ fluctuations inside $S^5$.

\begin{figure}[!h]
\begin{center}
\includegraphics[height=6cm,clip=true,keepaspectratio=true]{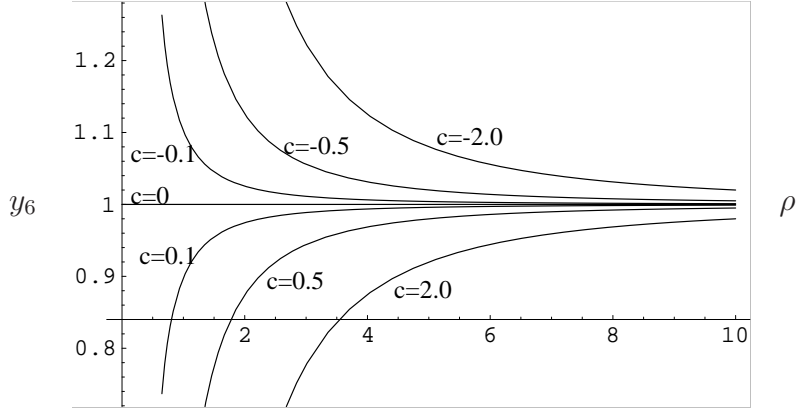}
\caption{Numerical solutions of the equations of motion in AdS showing 
that in the presence of a condensate asymptotically the solutions are
divergent.  The regular solution is the mass only
solution.}\label{numsol1}
\end{center}
\vspace{-5.5cm} \hspace{2cm} $y_6$ \hspace{9.6cm} $\rho$ \vspace{5cm}
\end{figure}
Note that $y_6(\rho) = m$ is an exact solution of the equations of
motion, corresponding to the embedding \cite{Katz} reviewed
above.  On the other hand there should be something ill-behaved about
the solutions with non-zero $c$, since a quark condensate is forbidden
by supersymmetry.  In Fig.~\ref{numsol1} we plot numerical solutions
of the equations of motion (obtained by a shooting technique using
Mathematica) for solutions with non-zero $c$, and find that they are
divergent.  The divergence of these solutions is not, by itself,
pathological because the variable $y_6$ is just the location of the
D7-brane.  However the AdS radius $r^2 = y_6(\rho)^2 + \rho^2$ is not
monotonically increasing as a function of $\rho$ for the divergent
solutions. This means that these solutions have no interpretation as a
renormalization group flow, or as a vacuum of the dual field
theory. As expected, the mass only solution is the only well-behaved
solution.

The other exact result which we wish to test numerically is the
meson spectrum. In order to find the states with zero spin on the
$S^3$, one looks for normalizable solutions of the equations of
motion of the form 
\begin{equation}
 y_6 + i y_5 = m + f (\rho) e^{i k
x}, \hspace{1cm} M^2 = - k^2 \,,
\end{equation}
where one linearizes in the small fluctuation $f(\rho)$. The
linearized equation of motion is
\begin{equation}
\partial_\rho^2 f(\rho)  + { \frac{3}{  \rho}} \partial_\rho 
f(\rho) + {\frac{M^2}{ (\rho^2+m^2)^2}} f(\rho) = 0 \,.
\end{equation}
This was solved exactly in~\cite{MateosMyers}, where it was shown that
the solutions can be written as hypergeometric functions 
\begin{equation}
f(\rho) = {\frac{A}{   (\rho^2 + 1)^{n+1}}} F(-n-1,-n;2,-\rho^2)
\end{equation} 
with $A$ a constant. The exact mass spectrum is then given by
\begin{equation} \label{formula} M = 2 m \sqrt{(n+1)(n+2)}, \hspace{1cm}
n=0,1,2, \dots \, . 
\end{equation}

We are interested in whether we can reproduce this result
numerically, via a shooting technique. The equation of motion can
be solved numerically subject to boundary conditions $f(\rho)
\sim c/\rho^2$ at large $\rho$ indicating that the meson is a
quark bilinear of dimension 3 in the UV. Solutions of the equation
must be regular at all $\rho$ so the allowed $M^2$ solutions can
be found by tuning to these regular forms. The result
(\ref{formula}) is easily reproduced to 2 significant figures. We
show an example of the method in action in Fig.~\ref{numsol}.
\begin{figure}[!ht]
\begin{center}
\includegraphics[height=6cm,clip=true,keepaspectratio=true]{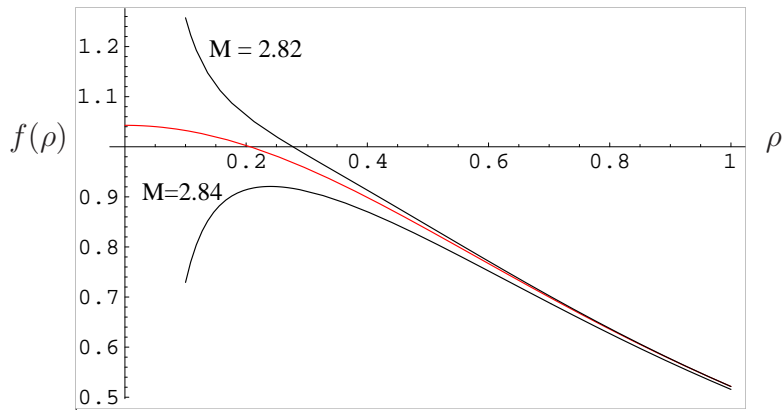}
\caption{Numerical solutions of the meson equation of motion for 
  different values of $M$ showing the identification of the first bound state
  mass. The exact regular solution is plotted between the two numerical
  flows.}\label{numsol}
\end{center}
\vspace{-6.4cm} \hspace{2.0cm} $f(\rho)$ \hspace{9cm} $\rho$  \vspace{5.9cm} 
\end{figure}

\subsection{The AdS-Schwarzschild Solution} \label{adsbhsection}
\lhead[\fancyplain{}{\bfseries\thepage}]%
{\fancyplain{}{\bfseries \rightmark}}
\rhead[\fancyplain{}{\bfseries\leftmark}]%
{\fancyplain{}{\bfseries\thepage}} 

\subsubsection{The background}

We now move on to study quark condensates and mesons in a
non-supersymmetric deformation of the AdS/CFT correspondence and
study the AdS-Schwarzschild black hole solution. This
geometry is dual to the ${\cal N}=4$ gauge theory at finite
temperature \cite{Witten2},  which is in the same universality
class as pure three-dimensional QCD.

The Euclidean AdS-Schwarzschild solution is given by 
\begin{equation}
\label{bh}ds^2 = K(r) d\tau^2 + {\frac{dr^2}{ K(r)}} + r^2
dx_{\parallel}^2 + d\Omega_5^2 \, , \end{equation} where 
\begin{equation} K(r) = r^2 - {\frac{b^4}
{r^2}} \,. 
\end{equation} 
This space-time is smooth and complete if $\tau$ is periodic with period $\pi
b$. Note that the $S^1$ parameterized by $\tau$ collapses at $r= b$.  The fact
that the geometry ``ends'' at $r=b$ is responsible for the existence of an
area law for the Wilson loop and a mass gap in the dual field theory (see
\cite{Witten2}).  The period of $\tau$ is equivalent to the inverse temperature
in the dual ${\mathcal N} =4$ gauge theory. The parameter $b$ sets the scale
of the deformation and for convenience in the numerical work below we shall
set it equal to 1.  At finite temperature, the fermions have anti-periodic
boundary conditions in the Euclidean time direction and become massive upon
dimensional reduction to three dimensions. The adjoint scalars also become
massive at one loop. Thus in the high-temperature limit, the adjoint fermions
and scalars decouple, leaving pure three-dimensional QCD.

We  now introduce a D7-brane into this background,  which
corresponds to the addition of matter in the fundamental
representation.  The dual gauge theory is the ${\mathcal N}=2$
gauge theory of Karch and Katz at finite temperature.  Note that
the fermions in the fundamental representation also have
anti-periodic boundary conditions in the Euclidean time direction.
Thus these also decouple in the high-temperature limit,  as do the
fundamental scalars which get masses at one loop,  leaving pure
QCD$_3$ as before.  Thus in this particular case we are not
interested in the high-temperature limit, but only the region
accessible to supergravity and Dirac-Born-Infeld theory. Although
the dual field theory cannot be viewed as a three-dimensional
gauge theory with light quarks, it is nevertheless a 
four-dimensional non-supersymmetric gauge theory with confined degrees
of freedom in the fundamental representation. This provides an
interesting, if exotic, setting to compute quark condensates and
meson spectra using Dirac-Born-Infeld theory.  The Constable-Myers
background which we will consider later turns out to have more
realistic properties.

\subsubsection{Embedding of a D7-brane}

To embed a D7-brane in the AdS black-hole background it is
useful to recast the metric (\ref{bh}) to a form with an explicit
flat 6-plane. To this end, we change variables from $r$ to $w$,
such that
\begin{equation}
{\frac{dw}{ w}}\equiv {\frac{r dr}{ (r^4-b^4)^{1/2}}} \, , 
\end{equation}
which is  solved by
\begin{equation}
2 w^2=r^2+\sqrt{r^4-b^4} \,.
\end{equation}
The metric is then
\begin{equation} \label{BHmetric}
ds^2=\left(w^2+\frac{b^4}{4w^2}\right)d\vec
x^2+\frac{(4w^4-b^4)^2}{4w^2(4w^4 +b^4)}dt^2+
\frac{1}{w^2}(\sum_{i=1}^6 dw_i^2) \, , 
\end{equation}
where $\sum_i dw_i^2 = dw^2 + w^2 d\Omega_5^2$,  which for reasons
of convenience will also be written as $d\rho^2 +
\rho^2d\Omega_3^2 + dw_5^2 + dw_6^2$ where $d\Omega_3^2$ is the
unit three-sphere metric.
The AdS black hole geometry asymptotically approaches $AdS_5
\times S^5$ at large $w$.  Here the background becomes
supersymmetric,  and the D7 embedding should approach that
discussed in \cite{Katz}.  The asymptotic solution has the
form $w_6 = m, w_5 = 0$,  where $m$ should be interpreted as a
bare quark mass. To take into account the deformation,  we will
consider a more general ansatz for the embedding of the form $w_6
= w_6(\rho), w_5=0$,  with the function $w_6(\rho)$ to be
determined numerically.
The DBI action for the orthogonal directions $w_5, w_6$ is
\begin{align} \label{DBIBH}
S_{D7} &=-\mu_7\int d^8\xi ~\epsilon_3 ~ {\cal G}(\rho,w_5,w_6) \\
&\times \left( 1 + { g^{ab} \over (\rho^2 + w_5^2 + w_6^2)} \partial_a w_5
\partial_b
w_5 + { g^{ab} \over (\rho^2 + w_5^2 + w_6^2)} \partial_a w_6 \partial_b
w_6 \right)^{1/2} \,,\nonumber
\end{align}
where the determinant of the metric is given by
\begin{align}
{\cal G}(\rho,w_5,w_6) 
&= \sqrt{ g_{tt} g_{xx}^3 \rho^6 \over (\rho^2 + w_5^2
+ w_6^2)^4} \nonumber\\
&= \rho^3 { ( 4(\rho^2 + w_5^2 + w_6^2)^2 + b^4) ( 4( \rho^2
+ w_5^2 + w_6^2)^2 - b^4) \over 16 (\rho^2 + w_5^2 + w_6^2)^4} \,.
\end{align}

With the ansatz $w_5=0$ and $w_6 = w_6(\rho)$, the equation of
motion becomes
\begin{equation}
\label{eqnmot} {d \over d\rho} \left[ {\cal G}(\rho,w_6)
\sqrt{ 1 \over 1 + \left({  d w_6 \over d \rho}\right)^2} { d w_6
\over d \rho} \right] - \sqrt{1 + \left({  d w_6 \over d
\rho}\right)^2} {b^8 \rho^3 w_6 \over 2 ( \rho^2 + w_6^2)^5} = 0
\,.
\end{equation}
The solutions of this equation determine the induced metric on the D7
brane which is given by
\begin{equation} ds^2 = \left( \tilde w^2 +
  \frac{b^4}{4 \tilde w^2} \right) d\vec x^2 + \frac{(4 \tilde w^4 -
  b^4)^2}{4\tilde w^2 (4\tilde w^4 +b^4)} dt^2 + \frac{1+(\partial_\rho
w_6)^2}{\tilde w^2} d
\rho^2
+ \frac{\rho^2}{\tilde w^2} d\Omega_3^2 \, , \label{D7metric}
\end{equation}
with $\tilde w^2=\rho^2+w_6^2(\rho)$.
The D7-brane metric becomes $AdS_5 \times S^3$ for $\rho \gg b,m$.

\subsubsection{Karch-Katz solutions versus condensate solutions}

Before computing the explicit D7-brane solutions, we remark that
there are several possibilities for the topology of the D7-brane
embedding which we illustrate in Fig.~\ref{quiv}. 

\begin{figure}[!h]
\begin{center}
\includegraphics[height=7cm,clip=true,keepaspectratio=true]{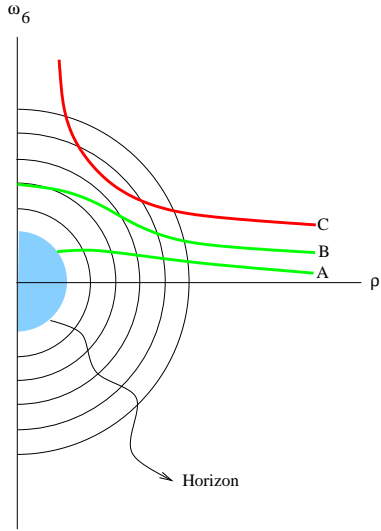}
\caption{\small{Different possibilities for solutions of the
D7-brane equations of motion.  The semicircles are lines of
constant $r$, which should be interpreted as a scale in the dual
Yang-Mills theory. The curves of type $A,B$ have an
interpretations as an RG flow, while the curve $C$ does not.}
}\label{quiv}
\end{center}
\end{figure}

The UV asymptotic (large $\rho$) solution, where the geometry returns
to $AdS_5\times S^5$,  is of the form 
\begin{equation}
\label{asymp} w_6(\rho) \sim m + \frac{c}{\rho^2} \,.
\end{equation}
The parameters $m$ and $c$ have the interpretation as a quark mass and
bilinear quark condensate respectively, as discussed below equation
(\ref{identif}).  These parameters can be taken as the boundary
conditions for the second order differential equation (\ref{eqnmot}),
which we solve using a numerical shooting technique.  Of course the
physical solutions should not have arbitrary $m$ and $c$. The
condition which we use to identify physical solutions is that the
D7-brane embedding should have an interpretation as a RG flow.  This
implies for instance that if one slices the D7-brane geometry at a
fixed value of $w^2$, or equivalently at a fixed value of $w_6^2 +
\rho^2$, one should obtain at most one copy of the geometry $R^4
\times S^3$.  In other words, $w^2 = \rho^2 + w_6(\rho)^2$ should be a
monotonically increasing function of $\rho$.  This is certainly not
the case for divergent solutions.  Such solutions are not in
correspondence with a vacuum of the dual gauge theory and are
discarded.

There are then two possible forms of regular solutions.  The geometry in which
the D7-brane is embedded has the boundary topology $R^3 \times S^1 \times
S^5$, which contains the D7-brane boundary $R^3 \times S^1 \times S^3$. Recall
that the $S^3$ is contractible within~$S^5$. Furthermore the $S^1$ is
contractible within the bulk geometry and shrinks to zero as one approaches
the horizon $r=b$ (i.e.~$w=b/\sqrt{2}$).  The D7-brane may either ``end'' at
some $r > b$ if the $S^3$ collapses, or it may continue all the way to the
horizon where the $S^1$ collapses but the $S^3$ has finite size. In other
words, the D7-topology may be either $R^3 \times B^4 \times S^1$ or $R^3
\times S^3 \times B^2$. The former is the type found in \cite{Katz}.  As one
might expect, this topology occurs when the quark mass $m$ is sufficiently
large compared to $b$. In this case, the $S^3$ of the D7-brane contracts to
zero size in the asymptotic region where the deformation of $AdS_5 \times S^5$
is negligible. We shall find the other topology for sufficiently small $m$.

\begin{figure}[!h]
\begin{center}
\includegraphics[height=6cm,clip=true,keepaspectratio=true]{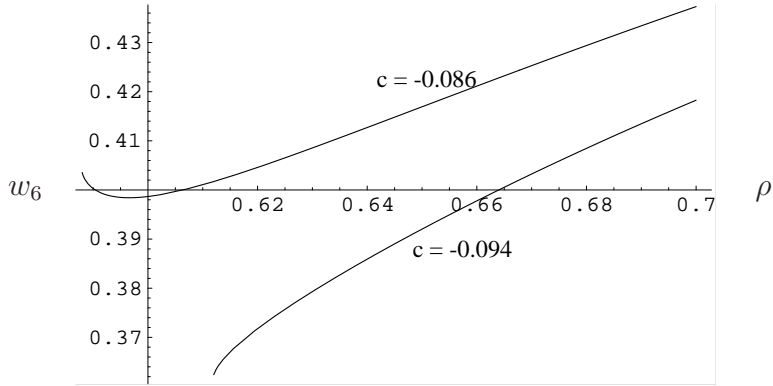}
\caption{An example (for
$m=0.6$) of the different flow behaviour around the regular
(physical) solution.}\label{quiv4}
\end{center}
\vspace{-5.2cm} \hspace{2.2cm} $w_6$ \hspace{9.2cm} $\rho$  \vspace{4.7cm} 
\end{figure}
For any chosen value of $m$, we find only a discrete choice of $c$
which gives a regular solution that can be interpreted as an RG
flow. In Fig.~\ref{quiv4} we show sample numerical flows used to
identify a regular solution.

For the regular  solutions the D7-brane either ends at the
horizon, 
\begin{equation} 
w_6^2 + \rho^2 = \frac{1}{2}b^2\,,
\end{equation} 
at which the $S^1$ collapses, or ends at a point outside the horizon,
\begin{equation}
 \rho =0\,,\qquad w_6^2 \ge \frac{1}{2}b^2,
\end{equation} 
at which the $S^3$ collapses (see (\ref{D7metric})).  Both types of solution
are illustrated in Fig.~\ref{quiv2} for several choices of $m$.  We choose
units such that $b=1$. We refer to solutions with collapsing $S^3$ as
Karch-Katz solutions, and solutions with collapsing $S^1$ as ``condensate''
solutions, for reasons that will become apparent shortly.  Note that the
boundary between the Karch-Katz and condensate solutions is at a critical
value of the mass $m=m_{\rm crit}$ such that $w_6(\rho=0)=\sqrt{1/2} b$.  In
this case both the $S^1$ and $S^3$ collapse simultaneously. Numerically, we
find $m_{\rm crit} \approx 0.92$. While the Karch-Katz solutions are
(approximately) constant for all values of $\rho$, the condensate solutions
bend towards the horizon for small $\rho$. As expected, the black hole exerts
an {\it attractive} force on the D7-brane.  

\begin{figure}[!h]
\begin{center}
\includegraphics[height=6cm,clip=true,keepaspectratio=true]{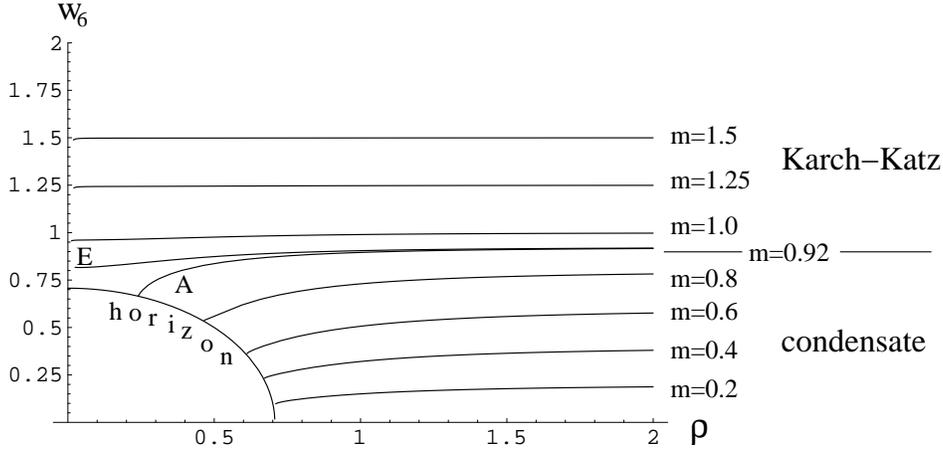}
\caption{Two classes of regular solutions in the AdS black hole
  background. The quark mass $m_q$ is the parameter $m$ in units of $\Lambda
  \equiv \frac{b/\sqrt{2}}{2\pi\alpha'}$: $m_q=m \Lambda$. Numerically we set
  $\Lambda=1/\sqrt{2}$.}\label{quiv2}
\end{center}
\end{figure}
As mentioned above, the dual gauge theory is considered at finite temperature.
The same gauge theory at zero temperature is dual to supergravity on the
near-horizon geometry of solitonic D3-branes instead of thermal D3-branes
(from which the AdS black hole descends in the case $T>0$).  Solitonic
D3-branes exert a repulsive force on the D7-branes and one would obtain a plot
similar to Fig.~\ref{quiv7}. This has been studied in detail in
\cite{MateosMyers2} for thermal and solitonic D4-branes.

There is an exact solution of the equation of motion $w_6 = 0$ which
is regular and corresponds to $m=c=0$. Thus there is no condensate
when the quarks are massless (from the four-dimensional point of
view). This should not be disappointing, since the theory is not in
the same universality class as QCD$_3$ with light quarks. The quarks
obtain a mass of order the temperature which will tend to suppress the
formation of a condensate.  For non-zero $m$ we obtain the solutions
numerically. 

\subsubsection*{Phase transitions in QCD-like theories at finite temperature}

The dependence of the condensate on the mass is illustrated in
Fig.~\ref{quiv3}a.  We find that as $m$ increases, the condensate $c$
initially increases and then decreases again. At sufficiently large $m$, the
condensate becomes negligible, which is to be expected as the D7-brane ends in
the region where the deformation of AdS is small.  Recall that there is no
condensate in the Yang-Mills theory with unbroken ${\mathcal N} =2$
supersymmetry described by D7-branes in un-deformed AdS.

\begin{figure}[!h]
\hspace{0.3cm}
\begin{minipage}{8cm}
\includegraphics[height=5cm,clip=true,keepaspectratio=true]{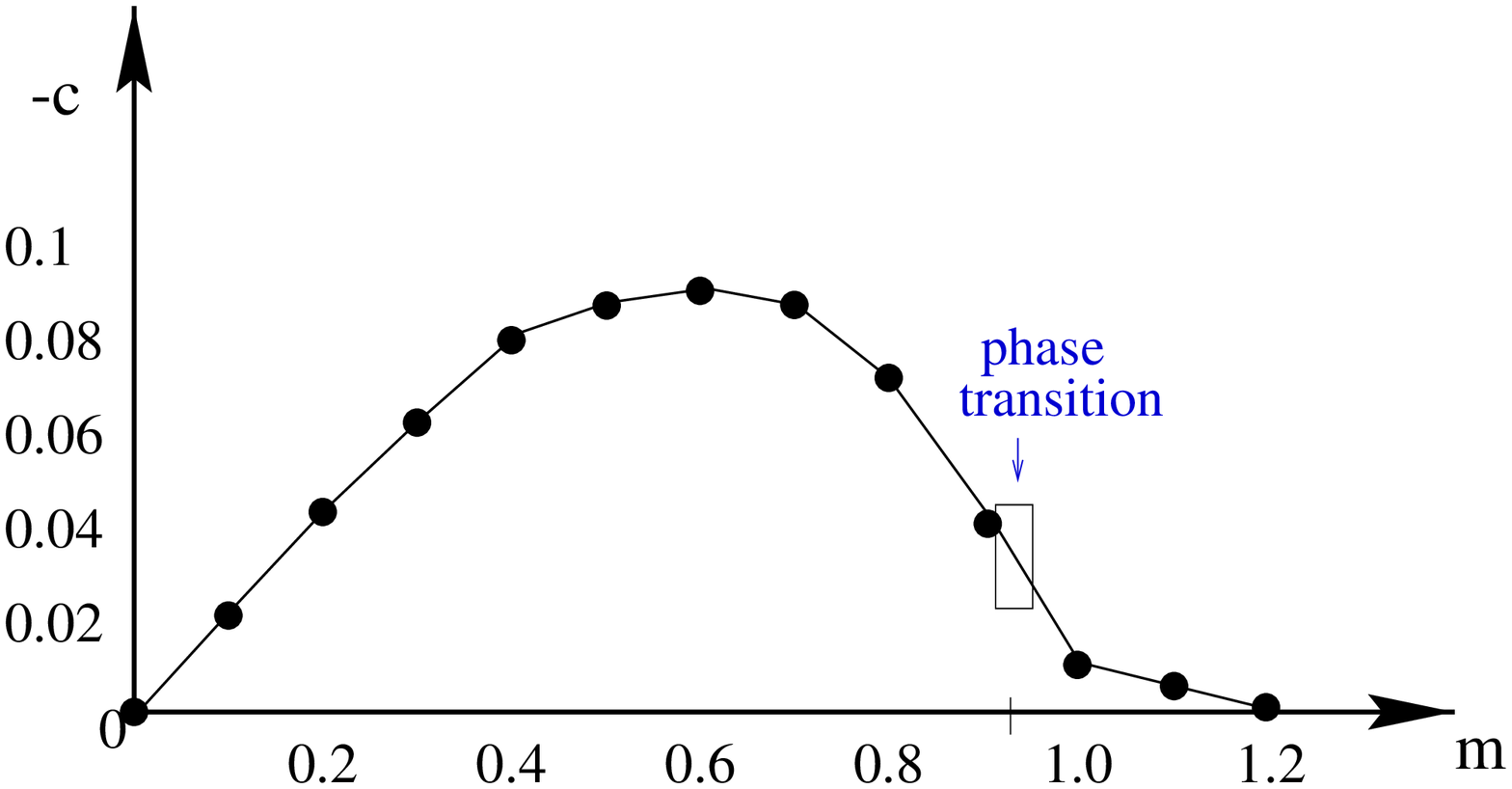}
\end{minipage}
\begin{minipage}{3.5cm}
\includegraphics[height=3.5cm,clip=true,keepaspectratio=true]{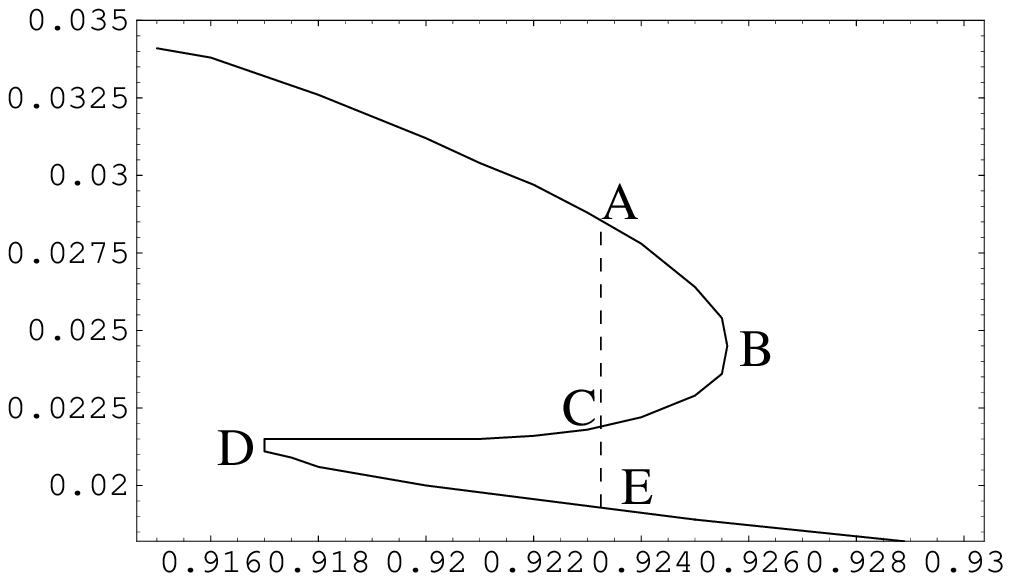}
\end{minipage}
\caption{A plot of the parameter $c$ vs $m$ for the regular
solutions in AdS Schwarzschild. The linear fit between points is just
to guide the eye. The right plot zooms in around $m_{\rm crit}
\approx 0.92$.}\label{quiv3}
\end{figure}

Since the D7-brane topology changes as one crosses $m_{\rm crit}$, one might
expect a phase transition to occur at this point.\footnote{For $m>m_{\rm
    crit}$ there is the interesting possibility of introducing an even spin
  structure on the $S^1$ of the D7-brane, since this $S^1$ is no longer
  contractible on the D7.  If this is a sensible (i.e stable) background, it
  would correspond to a different field theory, in which the fundamental
  fermions are periodic on $S^1$ and do not have a Kaluza-Klein mass. We have
  not analyzed this possibility.}  Zooming in around $m_{\rm crit}$, we see in
Fig.~\ref{quiv3}b that $c$ is multi-valued around the critical mass $m_{\rm
  crit}$. This means that two graphs in Fig.~\ref{quiv2} can have the same
asymptotic behaviour but a different behaviour at small values of~$\rho$. For
instance, the value of $c$ at the point $A$ in Fig.~\ref{quiv3}b corresponds
to the graph $A$ in Fig.~\ref{quiv2} which ends at the horizon, while $c$ at
the point $E$ is associated with the graph $E$ which ends above the horizon.
The dashed line between $A$ and $E$ is obtained by the Maxwell construction
and corresponds to the equilibrium transition between the two phases. In this
regime both phases are present. We can reformulate this result by keeping the
quark mass $m_q$ fixed and vary the horizon $b/\sqrt{2}$, i.e.~the temperature
$T \sim (\pi b)^{-1}$. Then the $c$-$m$ plot can also be considered as a plot
of the condensate $c$ in dependence of the temperature $T$.\footnote{Note that
  the quark mass $m_q$ is $m_q \sim m b$.  The temperature is thus related to
  $m$ by $T \sim b^{-1} = m/m_q$.} During the phase transition between the
points $A$ and $E$ the condensate decreases while the temperature remains
constant. The graph in Fig.~\ref{quiv3}b looks quite similar to an isotherm of
a fluid in a pressure-volume diagram.  As in the case of a fluid the phase
transition is discontinuous and thus of first order. A similar behaviour has
been found in the thermal D4-brane background studied in \cite{MateosMyers2}.
Such phase transitions seem to be a universal feature of QCD-like theories at
{\rm finite} temperature.


\subsubsection{Meson spectrum in the AdS black-hole background}

The meson spectrum can be found by solving the linearized
equations of motion for small fluctuations about the D7-embeddings
found above.  Let us consider the fluctuations of the variable
$w_5$ about the embedding,  which has $w_5=0$.  We take 
\begin{equation}
 w_5 = f(\rho)\sin( \vec k \cdot \vec x) \,. 
\end{equation} 
In App.~\ref{appB1} we compute the linearized (in
 $w_5$) equation of motion. We find
\begin{equation} \label{meep} \begin{array}{c} {d \over
 d\rho} \left[ {\cal G}(\rho, w_6) \sqrt{ 1 \over 1 + \left({ d w_6
 \over d \rho}\right)^2} { d f(\rho) \over d \rho} \right] + {\cal
 G}(\rho, w_6) \sqrt{ 1 \over 1 + \left({ d w_6 \over d
\rho}\right)^2} \left({ 4 \over 4 (\rho^2 +w_6^2)^2 + b^4}\right)
M^2 f(\rho)
\\\\
-  \sqrt{1 + \left({  d w_6 \over d \rho}\right)^2} { b^8 \rho^3
f(\rho) \over 2 ( \rho^2 + w_6^2)^5} = 0 \,.
\end{array}
\end{equation} 
where $M^2 = \vec k^2$. The allowed values of $\vec k^2$ are
determined by requiring the solution to be normalizable and
regular. Note that if the $U(1)$ symmetry which rotates $w_5$ and
$w_6$ were spontaneously broken by an embedding of the asymptotic form
$w_6 \sim c/\rho^2$ with non-zero $c$, there would be a massless state
in the spectrum associated with $w_5$ fluctuations.  This is not the
case in the present setting, since the condensate is only non-zero for
non-zero quark mass $m$. Instead we find a mass gap in the meson
spectrum.  We have computed the meson spectrum by solving (\ref{meep})
by a numerical shooting technique.  As in the Karch-Katz geometry, we
seek regular solutions for $w_5$ which are asymptotically of the form
$c/\rho^2$ in the presence of the background $w_6$ solution. The
results for the meson masses are plotted in Fig.~\ref{quiv5}. Of course the
meson mass gap here can be largely attributed to the Kaluza-Klein
masses of the constituent quarks, which are of the same order as the
temperature ($T \sim \pi^{-1}$ in units with $b=1$).

\begin{figure}[!h]
\begin{center}
\includegraphics[height=6.2cm,clip=true,keepaspectratio=true]{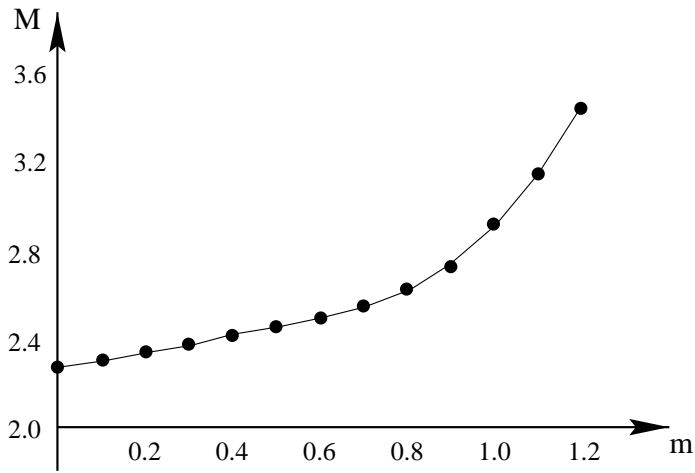}
\caption{A plot of the $w_5$ meson mass vs $m$ in AdS
Schwarzschild. The linear fit between points is just to guide the
eye.}\label{quiv5}
\end{center}
\end{figure}

Thus we have seen that, while the thermal gauge background allows
a quark condensate when it does not spontaneously break any
symmetries, there is no chiral (parity) symmetry breaking at zero quark
mass. The meson spectrum reflects this by having a mass gap - the
fermions have an induced mass from the presence of finite
temperature. In the subsequent discussion we will consider another
background which admits light constituent quarks and has
properties much closer to QCD.


\subsection{The Constable-Myers Deformation} \label{mcsection}

\subsubsection{The background}

We consider the non-supersymmetric deformed AdS geometry
originally constructed in \cite{CM}. This geometry
corresponds to the ${\mathcal N}=4$ super Yang-Mills theory
deformed by the presence of a vacuum expectation value for an
R-singlet operator with dimension four (such as $tr F^{\mu
\nu}F_{\mu \nu}$). The supergravity background has a dilaton and
$S^5$ volume factor depending on the radial direction. In a
certain parameter range, this background implies an area law for
the Wilson loop and a mass-gap in the glueball spectrum. Whether
the geometry,  which has a naked singularity, actually describes
the stable non-supersymmetric vacuum of a field theory is not well
understood \cite{CM}.  This is not so important from our
point of view though since the geometry is a well defined gravity
description of a non-supersymmetric gauge configuration. We can
just ask about the behaviour of quarks in that background.

The geometry in {\it Einstein frame} is given by
\begin{align}
 ds^2 = &\,H^{-1/2} \left( 1 + { 2 b^4 \over r^4}
\right)^{\delta/4} dx_{4}^2 \nonumber \\
&+ H^{1/2} \left( 1 + { 2 b^4 \over
r^4} \right)^{(2-\delta)/4} {r^2 \over \left( 1 + {b^4 \over
r^4}\right)^{1/2}} \left[ {r^6 \over (r^4 + b^4)^2} dr^2 + d
\Omega^2_5 \right], 
\end{align} 
where
\begin{equation}
 H =  \left( 1 + { 2 b^4 \over
r^4} \right)^{\delta} - 1
\end{equation}
and with the dilaton and four-form given by
\begin{equation}
e^{2 \phi} = e^{2 \phi_0} \left( 1 + { 2 b^4 \over r^4}
\right)^{\Delta}, \hspace{1cm} C_{(4)} = - {1 \over 4} H^{-1} dt
\wedge dx \wedge dy \wedge dz \,.
\end{equation}
The parameter $b$ corresponds
to the vev of the dimension 4 operator. The parameters $\Delta$
and  $\delta$ are constrained by
\begin{equation}
 \Delta^2 + \delta^2 = 10 \,. 
\end{equation}
Asymptotically the AdS curvature is given by $L^4 = 2 \delta b^4$, so
it makes sense to set (with $L=1$)
\begin{equation}
 \delta = {1 \over 2 b^4}. 
\end{equation}

As in \cite{Evansnew} we define the fundamental energy scale $\Lambda_b$ as
\begin{align}
\Lambda_b=\frac{b}{2\pi\alpha'} \,.
\end{align}
$\Lambda_b$ is the only free parameter in the geometry and its value sets the
scale of the conformal symmetry and supersymmetry breaking $\Lambda_{\rm
  susy}$ below which we have confinement and a discrete glueball spectrum.
Since this is also approximately the scale of chiral symmetry breaking, we
assume that $\Lambda_b \simeq \Lambda_{\rm susy} \simeq
\Lambda_{\rm QCD}$.  Then the quark mass is given by
\begin{align}
m_q = \frac{b}{2\pi\alpha'} m = m \Lambda_b\,.
\end{align}
We will numerically set $\Lambda_b$ equal to 1 below.

To embed a D7-brane in this background it will again be convenient
to recast the metric in a form containing an explicit flat
6-plane. To this end, we change variables from $r$ to $w$, such that
\begin{equation}
\frac{dw}{w}\equiv \frac{r^3 dr}{r^4+b^4},
\end{equation}
which is solved by
\begin{equation}
\ln (w/w_0)^4 =\ln (r^4+b^4)
\end{equation}
or
\begin{equation}
(w/w_0)^4=r^4+b^4.
\end{equation}
So for the case of $b=0$ we should set the integration constant
$w_0=1$. The full metric is now
\begin{equation} \label{mcmetric}
 ds^2 = H^{-1/2} \left( { w^4 + b^4 \over w^4-b^4}
\right)^{\delta/4} dx_{4}^2 + H^{1/2} \left( {w^4 + b^4 \over w^4-
b^4}\right)^{(2-\delta)/4} {w^4 - b^4 \over w^4 } \sum_{i=1}^6
dw_i^2, 
\end{equation} where
\begin{equation} H =  \left(  { w^4
+ b^4 \over w^4 - b^4}\right)^{\delta} - 1
\end{equation}
and the dilaton and four-form become
\begin{equation} e^{2 \phi} = e^{2 \phi_0} \left( { w^4 + b^4 \over
w^4 - b^4} \right)^{\Delta}, \hspace{1cm} C_{(4)} = - {1 \over 4}
H^{-1} dt \wedge dx \wedge dy \wedge dz.
\end{equation}
We now consider the D7-brane action in the static gauge with
world-volume coordinates identified with the four Minkowski
coordinates - denoted by $x_{4}$ - and with
$w_{1,2,3,4}$. The transverse fluctuations are parameterized by
$w_5$ and $w_6$.  It is again convenient to define a coordinate
$\rho$ such that $\sum_{i=1}^4 dw_i^2 = d\rho^2 + \rho^2
d\Omega_3^2$. The DBI action in Einstein frame
\begin{equation}
 S_{D7} = - T_{D7} \int d^8 \xi e^{\phi} \sqrt{- {\rm
det} (g^{PB}_{ab})} 
\end{equation} can then be written as 
\begin{equation} 
 S_{D7} = - T_{D7}
\int d^8 \xi~ \epsilon_3 ~  e^{\phi} { \cal G}(\rho,w_5,w_6)
\left( 1 + g^{ab} g_{55} \partial_a w_5
\partial_b w_5 + g^{ab} g_{66} \partial_a w_6
\partial_b w_6 \right)^{1/2},
\end{equation}
where
\begin{equation}
 {\cal G}(\rho,w_5,w_6) = \rho^3 {( (\rho^2 + w_5^2
+ w_6^2)^2 + b^4) ( (\rho^2 + w_5^2 + w_6^2)^2 - b^4) \over
(\rho^2 + w_5^2 + w_6^2)^4}.
\end{equation}
We  again look for classical solutions to the equation of motion of the form
\begin{align}
w_6 = w_6(\rho)\,,\qquad w_5 =0 \,,
\end{align} that define the ground state. They satisfy
\begin{equation}
 \label{eommc}{ d \over d \rho} \left[ {e^{\phi}  { \cal
G}(\rho,w_6) \over \sqrt{ 1 + (\partial_\rho w_6)^2}}
(\partial_\rho w_6)\right] - \sqrt{ 1 + (\partial_\rho w_6)^2} { d
\over d w_6} \left[ e^{\phi} { \cal G}(\rho,w_6) \right] = 0\,.
\end{equation}
The last term in the above equation is a ``potential'' like term that is
evaluated to be 
\begin{equation} \label{potliketerm}
 { d \over d w_6} \left[ e^{\phi} { \cal G}(\rho,w_6)
\right] = { 4 b^4 \rho^3 w_6 \over (\rho^2 + w_6^2)^5} \left( {
(\rho^2 + w_6^2)^2 + b^4 \over  (\rho^2 + w_6^2)^2 - b^4}
\right)^{\Delta/2} (2 b^4  - \Delta (\rho^2 + w_6^2)^2)\,. 
\end{equation}
\begin{figure}[!h]
\begin{center}
\includegraphics[height=6cm,clip=true,keepaspectratio=true]{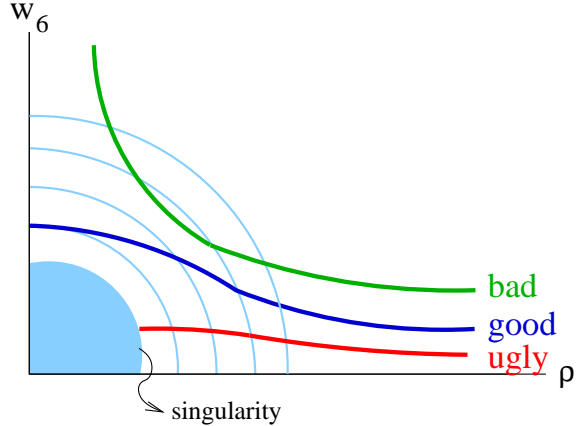}
\caption{\small{Different possibilities for solutions of the
D7-brane equations of motion.  The semicircles are lines of
constant $r$, which should be interpreted as a scale in the dual
Yang-Mills theory. The ``Bad'' curve cannot be interpreted as an
RG flow.  The other curves have an RG flow interpretation, however
the infrared (small r) region of the ``Ugly'' curve is outside the
range of validity of DBI/supergravity.}
}\label{quiv6}
\end{center}
\end{figure}

\noindent We now consider numerical solutions with the asymptotic behavior
\mbox{$w_6 \sim m + c/\rho^2$}, and find the physical solutions by
imposing a regularity constraint as discussed in the previous
section. Note that unlike the Euclidean AdS black hole,  the
Constable-Myers background has a naked singularity at $r=0$ or
$\rho^2 + w_6^2 = b^2$. Thus there are  two possibilities for a
solution with an interpretation as an RG flow, which are as
follows. Either the D7-brane terminates at a value of $r$ away
from the naked singularity via a collapse of the $S^3$, or the
D7-brane goes all the way to the singularity.  In the latter case
we would have little control over the physics without a better
understanding of string theory in such highly curved backgrounds.
Different possibilities for solutions of the D7-brane equations of
motion are illustrated in Fig.~\ref{quiv6}.

Fortunately something remarkable happens.  For positive values of $m$
we find that there is a discrete regular solution for each value of
the mass that terminates at $w\geq 1.3$ before reaching the
singularity at $w=1$.  Some of these regular solutions are plotted in
Fig.~\ref{quiv7}.  For $m=0$, the solution $w_6 =0$ is exact, which
naively seems to indicate the absence of a chiral condensate
$(c=0)$. However, this solution reaches the singularity, and therefore
cannot be trusted.  On the other hand for a very small but non-zero
mass, the regular solutions require a non-vanishing $c$ and terminate
before reaching the singularity! The numerical evidence (see
Fig.~\ref{quiv7}) suggests that there is a non-zero condensate in the
limit $m\rightarrow 0$.

We can study this phenomenon further in the deep infrared, in particular in
view of gaining further understanding of the behaviour of the solutions shown
in Fig.~\ref{quiv7}. Consider (\ref{eommc}) as $\rho \rightarrow 0$ with $w_6
\neq 0$ - the dilaton becomes $\rho$ independent whilst ${\cal G} \sim
\rho^3$. Thus the potential term vanishes as $\rho^3$ whilst the derivative
piece contains a term that behaves like $\rho^2 \partial_\rho w_6$ and
dominates.  Clearly there is a solution where $w_6$ is just a constant. This
is the regular behaviour we are numerically tuning to. It is now easy to find
the flows in reverse by setting the infrared constant value of $w_6$ and
numerically solving out to the UV. This method ensures that the flow is always
regular. We have checked that the asymptotic values of the condensate as a
function of mass match our previous computation at the level of a percent,
showing that the numerics are under control.
\begin{figure}[!h]
\begin{center}
\includegraphics[height=6cm,clip=true,keepaspectratio=true]{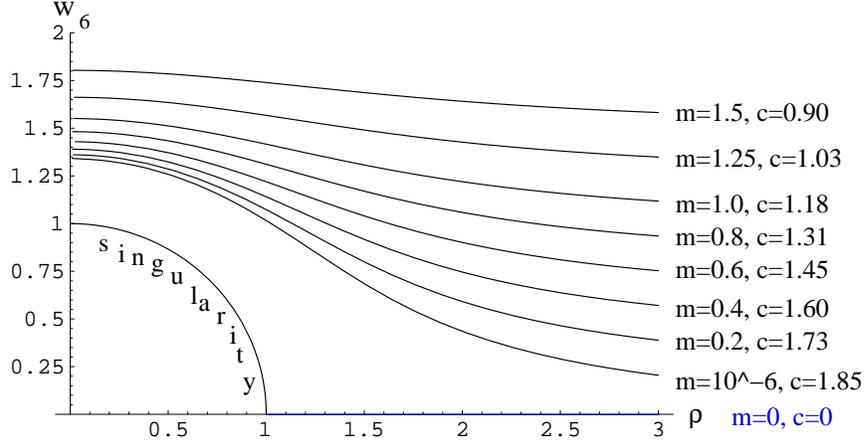}
\caption{Regular solutions in the Constable-Myers background. The quark mass
is measured in units of $\Lambda_b=1$.}
\label{quiv7}
\end{center}
\end{figure}

\begin{figure}[!h]
\begin{center}
\includegraphics[height=6cm,clip=true,keepaspectratio=true]{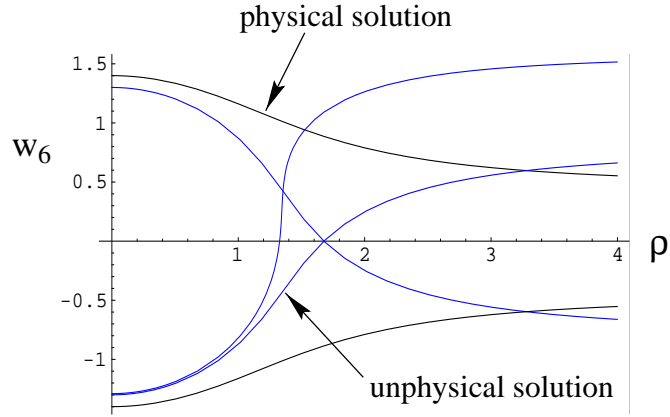}
\caption{Regular D7 embedding solutions in the Constable-Myers geometry
which lie close to the singularity in the infra-red.}
\label{new1}
\end{center}
\end{figure}

We now realise though that there are infrared solutions where
$w_6(\rho=0) < 1.3$.  These flows lie close to the singularity at
$w_6(\rho=0) = 1 $ which we had hoped to exclude. Solving these flows
numerically we find that they flow to negative masses in the
ultraviolet. We show these flows in Fig.~\ref{new1}. Since there is a
$w_6 \rightarrow - w_6$ symmetry of the solution though this means
there is a second regular flow for each positive mass which in the
infra-red flows to negative values, also shown in Fig.~\ref{new1}.
The flows that begin closer to the singularity in the infrared flow
out to larger masses in the ultraviolet. This strongly suggests that
these flows are not physical. When the quarks have a large mass,
relative to the scale of the deformation, we do not expect the
infra-red dynamics to have a large influence on the physics. Thus the
flows shown in Fig.~\ref{quiv7} that match onto the Karch-Katz type
solution for large mass are the expected physical solutions. To find
some analytic support for this conjecture we have calculated the
action of two of the solutions for each mass value.  The action is
formally infinite if we let the flow cover the whole space.  To see
the difference in action we have calculated the contribution for $0<
\rho<3$ only which covers the infrared part of the solution (varying
the upper limit does not change the conclusions). We plot the action
of the two solutions versus the quark mass in Fig.~\ref{new2}, from
which it can be seen that the action for the solution lying closer to
the singularity is larger. Therefore the corresponding solution is not
relevant for the physics.
\begin{figure}[!h]
\begin{center}
\includegraphics[height=5.5cm,clip=true,keepaspectratio=true]{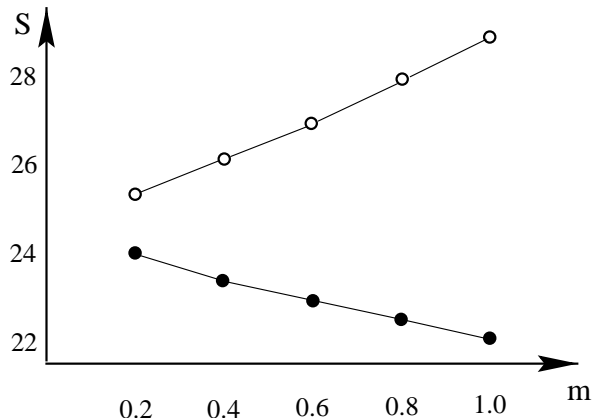}
\caption{A plot of action vs mass for the two regular D7 embedding
solutions in the Constable-Myers geometry. The higher action solutions
correspond to the flows that end at $|w_6| < 1.3$.}
\label{new2}
\end{center}
\end{figure}

In a certain sense, the condensate screens the probe physics from the naked
singularity. This is due to the centre of the Constable-Myers background which
exerts a repulsive force on the embedded D7-brane and triggers chiral symmetry
breaking.  In the limit of small explicit symmetry breaking parameter $m$, any
solution (or vacuum) which did not break the $U(1)$ symmetry would have
$w_6\rightarrow 0$ for all $\rho$.  If this were the case, the solution would
reach the singularity (see Fig.~\ref{quiv6}).  The screening of the
singularity is reminiscent of the enhan\c con \cite{Johnson:1999qt} found in
\mbox{${\cal N}=2$ } gravity duals - an important part of that analysis was
understanding that the singularity of the geometry was screened from the
physics of a D3 brane probe which led to an understanding of how the
singularity could be removed. It is possible we are seeing hints of something
similar, if more complicated, here, although at this stage we can not see how
to remove the singular behaviour. However, since the D7-brane avoids
the singularity, the existence of a singularity in the Constable-Myers
background is not too bad.

\subsubsection{Spontaneous $U(1)_A$ symmetry breaking and holographic
\mbox{version} of the Goldstone theorem}

One of the most important features of QCD dynamics is {\em spontaneous} chiral
symmetry breaking by a quark condensate $\psi\tilde\psi$.  The $U(1)$ chiral
symmetry $\psi \rightarrow e^{i\theta} \psi$, \mbox{$\tilde \psi \rightarrow
  e^{i\theta} \tilde \psi$} in the QCD-like field theory corresponds to a
$U(1)$ isometry in the $w_5-w_6$ plane transverse to the D7-brane. This
symmetry can be explicitly broken by a non-vanishing quark mass. In the
holographic dual this corresponds to the breaking of the $U(1)$ isometry due
to the separation of the D7-brane from the modified (``hairy'') D3-branes in
the $w_5 + iw_6$ direction, which give rise to the Constable-Myers geometry.
More interesting however than an explicit breaking by a mass term is the
question whether there is a spontaneous $U(1)$ chiral symmetry breaking in the
case of massless quarks.

Before we can answer this question, we have to verify that the $U(1)_A$
symmetry is a real symmetry in a QCD-like theory with a large number of
colours. Note that the $U(1)$ chiral symmetry is non-anomalous only in the
limit of a large number of colours, $N_c \rightarrow\infty$. This can be seen
in the anomaly equation of the axial $U(1)$ current $J_\mu=\bar \psi
\gamma_\mu\gamma_5 \psi$
\begin{align}
\partial_\mu J^\mu = \frac{2N_f}{N_c} \frac{g^2}{16 \pi^2} F \tilde F\,
\,,
\end{align}
where $N_f$ is the number of flavours.  For $N_f/N_c$ fixed the $U(1)$
symmetry is explicitly broken by instantons. 't Hooft showed \cite{tHooft}
that though the instanton term $i\theta F\tilde F$ is a total divergence it
does not fall off fast enough at infinity to allow neglect of surface terms.
The effect of the instantons is to give a mass to the $\eta'$ meson.

The $\eta'$ mass is given by the Witten-Veneziano
formula~\cite{Wittenetaprime, Veneziano:1979ec}
\begin{align} \label{witvene}
M^2_{\eta'} = \frac{4 N_f}{f_{\pi}^2} \chi_T  \,, 
\end{align}
where the pion decay constant, defined by $\langle 0 \vert \bar \psi \gamma_0
\gamma_5 \psi \vert \pi \rangle=m_\pi f_\pi$, has experimental value 132 MeV.
The topological susceptibility of the pure gauge theory,
defined by
\begin{align}
\chi_T = \frac{1}{(16\pi^2)^2} \int d^4x \langle {\rm Tr}(F\tilde F(x))
{\rm Tr}(F\tilde F(0)) \rangle \,,
\end{align} 
measures the fluctuation of the topological charge of the vacuum.  Since
$\chi_T$ is of order one and $f_{\pi}^2 \sim N_c$ for large $N_c$, the
eta-prime mass scales as $N_f/N_c$. The $\eta'$ meson is massless in a large
$N_c$ limit if the number of flavours $N_f$ is fixed. However, if $N_f$ scales
like $N_c$, i.e.~if $\nu=N_f/N_c$ is a fixed quantity, the $\eta'$ meson
becomes massive. This explains the experimental fact that the mass of the
$\eta'$ particle ($M_{\eta'}=958$ MeV) is much higher than the mass of the
pions (e.g.\ $M_{\pi^0}=135$ MeV), which are the Goldstone bosons of the
symmetry breaking $SU(N_f) \times SU(N_f) \rightarrow SU(N_f)$. However, for
$N_c\rightarrow\infty$ with $N_f$ fixed, the anomaly vanishes and the $\eta'$
becomes a real Goldstone boson \cite{Wittenetaprime} of a spontaneously broken
$U(1)_A$ symmetry.

\begin{figure}[!h]
\begin{center}
\includegraphics[height=6cm,clip=true,keepaspectratio=true]{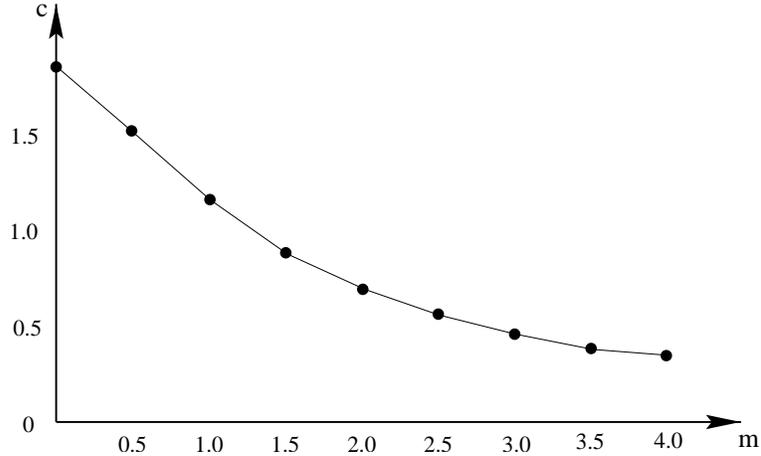}
\caption{A plot of the condensate parameter $c$ vs quark mass $m$ for
the regular solutions of the equation of motion in the Constable-Myers
background.}
\label{quiv8}
\end{center}
\end{figure}
In Fig.~\ref{quiv8} we plot the condensate $c$ of the solutions in
Fig.~\ref{quiv7} as a function of the quark mass $m$. The numerics
show that $c \neq 0$ for $m\rightarrow0$.  In other words the geometry
spontaneously breaks the $U(1)$ chiral symmetry. This seems to be
analogous to the situation in field theory, in which the path integral
{\it formally} gives no spontaneous symmetry breaking, which is found
only in the limit that a small explicit symmetry breaking parameter is
taken to zero.

Since there is $U(1)$ chiral symmetry breaking via a condensate in the
$m\rightarrow 0$ limit, we also expect there to be a Goldstone boson
in the meson spectrum, which is the analogon of the $\eta'$ particle
in QCD. Such a Goldstone mode must exist as a solution to the DBI
equation of motion, as the following holographic version of the
Goldstone theorem shows.

Assume a D7-embedding with $w_5 = 0$ and $w_6
\sim c/\rho^2$ asymptotically.  A small $U(1)$ rotation
$\exp(-i\epsilon)$ of $w_5 +i w_6$ generates a solution\footnote{Note:
$\tilde w_5 + i \tilde w_6 \approx (1-i\epsilon) (w_5 + iw_6) =
\epsilon w_6 + i w_6$.} with $\tilde w_6=w_6$ (to order $(\epsilon^2)$)
and $\tilde w_5 = \epsilon c/\rho^2$.  Thus a small fluctuation with
$\delta w_6=0$ and $\delta w_5 = \tilde w_5 \sin( k \cdot x)$ is a
normalizable solution of the {\it linearized} equations of motion,
\begin{align}
f(\delta w_5, \partial_\rho \delta w_5, \partial_x \delta w_5) = f(
\tilde w_5, \partial_\rho \tilde w_5) + M^2 \tilde w_5 c(\rho)  = 0 \,
, \label{diffeqn}
\end{align} 
provided $M^2=-k^2 = 0$. Here the differential equation of the
fluctuations splits into the equation of motion for $\tilde w_5$,
$f(\tilde w_5, \partial_\rho \tilde w_5) = \epsilon f(w_6,
\partial_\rho w_6)=0$, and a mass term with some coefficient
$c(\rho)$, cf.~Eq.~(\ref{diffeqn}) with the structure of
Eqs.~(\ref{meep}) or (\ref{lin}).  In other words there must be a
Goldstone boson associated with $w_5$ fluctuations. Note that if the
embedding were asymptotically $w_6 \sim m + c(m)/\rho^2$ for non-zero
$m$, a $U(1)$ rotation of $w_5 + i w_6$ would still generate another
solution. However this solution is no longer a {\it normalizable}
small fluctuation about the original embedding - asymptotically the
mass will acquire a different phase moving us to a different theory.
Thus if the mass is kept fixed asymptotically one does not find a
massless particle in the spectrum, which reflects the explicit
symmetry breaking by the quark mass $m$.


\subsubsection{Meson spectrum in the Constable-Myers background}

In the solutions discussed where $w_6$ has a background value,
fluctuations in $w_5$ should contain the Goldstone mode. Let
us turn to the numerical study of these fluctuations in the background
of the $w_6$ solutions we have obtained above.
The linearized equation of motion for small fluctuations of the
form $w_5 = f(\rho)\sin( k \cdot x)$, with $x$ the four Minkowski
coordinates, are 
\begin{equation} \begin{array}{c} {d
\over d \rho} \left[ {e^{\phi} {\cal G}(\rho,w_6) \over \sqrt{ 1
+ (\partial_\rho w_6)^2}} \partial_\rho f(\rho) \right] + M^2
{e^{ \phi} {\cal G} (\rho,w_6) \over \sqrt{ 1 + (\partial_\rho
w_6)^2}} H \left( { (\rho^2 + w_6^2)^2 + b^4 \over (\rho^2 +
w_6^2)^2 - b^4} \right)^{(1- \delta)/2}  {  (\rho^2 + w_6^2)^2 -
b^4 \over (\rho^2 + w_6^2)^2} f(\rho) \\
\\-\sqrt{ 1 + (\partial_\rho w_6)^2} { 4 b^4 \rho^3 \over (\rho^2 +
w_6^2)^5} \left( { (\rho^2 + w_6^2)^2 + b^4 \over  (\rho^2 +
w_6^2)^2 - b^4} \right)^{\Delta/2} ( 2 b^4 - \Delta (\rho^2 +
w_6^2)^2) f(\rho) = 0 \, , \label{lin}
\end{array}
\end{equation}
see App.~\ref{appB2} for details.  The meson mass as a function of quark mass
for the regular solutions for $w_5$ are plotted in Fig.~\ref{quiv9}.  The
meson mass indeed falls to zero as the quark mass is taken to zero providing
further evidence of chiral symmetry breaking.

At small $m$,  the meson mass associated to the $w_5$ fluctuations
scales like $\sqrt{m}$ in agreement with the Gell-Mann-Oakes-Renner
(GMOR) relation \cite{GMOR}:
\begin{align}
M_{\pi}^2= -\frac{m \langle \bar\psi \psi \rangle}{N_f f_\pi^2} \,.
\end{align}
A well-known field theory argument for this scaling is as follows. The
low-energy effective Lagrangian depends on a field $\eta^{\prime}$ where
$\exp(i\eta^{\prime}/f)$ parameterizes the vacuum manifold and transforms by a
phase under chiral $U(1)$ rotations. A quark mass term transforms by the same
phase under $U(1)$ rotations, and thus breaks the $U(1)$ explicitly.  A chiral
Lagrangian consistent with this breaking has a term $\mu^3 Re(m \exp( i
\eta^{\prime}/f))$ where $\mu$ is some parameter with dimensions of mass. For
real $m$, expanding this term to quadratic order gives a mass term
$\frac{\mu^3}{f^2}m {\eta^{\prime}}^2$. It would be very interesting to
demonstrate this scaling with $m$ analytically in the DBI/supergravity
setting, along with other low-energy ``theorems''. At large $m$, the meson
masses scale linearly in the quark mass. This differs from the behaviour of
the meson spectrum computed in the (non-supersymmetric) solitonic D4-brane
background \cite{MateosMyers2} which satisfies the GMOR relation to arbitrary
high quark masses.

For comparison it is interesting to study $w_6$ fluctuations as well,
which we expect to have a mass gap.  Analytically linearizing the
$w_6$ equation of motion is straightforward but the result is
unrevealingly messy. Since we must eventually solve the equation
numerically, we can use a simple numerical trick to obtain the
solutions.  We solve equation (\ref{eommc}) for $w_6$ above but write
$w_6 = w_{6}^0 + \delta w_6(r)$, where numerically we enforce $\delta
w_6$ to be very small relative to the background configuration
$w_6^0$.  With this ansatz we retain the field equation in its
non-linear form, but it is numerically equivalent to standard
linearization.
We must also add a term to the l.h.s.~of (\ref{eommc})
which takes into account the $x$ dependence of
$\delta w_6$. This dependence
 takes the same form as that in the linearized
$w_5$ equation (\ref{lin}), i.e.~$\delta w_6 =h(\rho)\, {\rm sin}(k
\cdot x)$. The extra term to be added to (\ref{eommc}) is
\begin{equation} \label{extraterm2}
\Delta V = M^2 {e^{ \phi} {\cal G}
(\rho,w_6) \over \sqrt{ 1 + (\partial_\rho w_6)^2}}
H \left( {
(\rho^2 + w_6^2)^2 + b^4 \over (\rho^2 + w_6^2)^2 - b^4}
\right)^{(1- \delta)/2}  {  (\rho^2 + w_6^2)^2 - b^4 \over
(\rho^2 + w_6^2)^2}
\delta w_6 \, .
\end{equation}

The numerical solutions for the $w_6$ fluctuations are plotted in
Fig.~\ref{quiv9}.  The $w_6$ fluctuations have a mass gap, as expected
since they are transverse to the vacuum manifold.

\begin{figure}[!h]
\begin{center}
\includegraphics[height=6.5cm,clip=true,keepaspectratio=true]{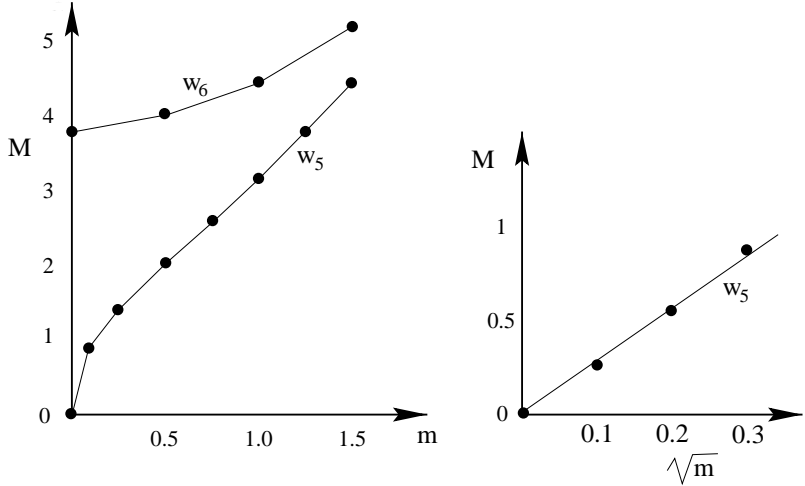}
\caption{A plot of the $w_5$ and $w_6$
meson mass vs quark mass $m$ associated with the fluctuations about
the regular solutions of the equation of motion for the
Constable-Myers flow. The Goldstone mass is also plotted vs $\sqrt{m}$ 
with a linear fit.}
\label{quiv9}
\end{center}
\end{figure}

\subsubsection{Comparison with pions in QCD}

Thus far we have found a particle in the spectrum which is similar
to the $\eta^{\prime}$ in the large $N$ limit of QCD
where it becomes a Goldstone boson.  
In order to obtain true
pions one must have a non-abelian flavour symmetry.
Unfortunately in the background which we consider,  taking 
$N_f >1$ D7-branes does not give rise to a $U(N_f)_L \times U(N_f)_R$ 
chiral symmetry.  Instead one gets only diagonal $U(N_f)$ times an 
axial $U(1)$.  The reason is that the theory contains a coupling 
$\tilde \psi_i X \psi_i$,  where $i$ is a flavour index and $X$ is an 
adjoint scalar without any flavour indices.  The coupling to $X$ explicitly
breaks the  $U(N_f)_L \times U(N_f)_R$ chiral symmetry 
to the diagonal subgroup,  but preserves
an axial $U(1)$ which acts as 
\begin{align}
\psi_i \rightarrow e^{i\theta} 
\psi_i, \qquad \tilde\psi_i \rightarrow e^{i\theta} 
\tilde \psi_i, \qquad X \rightarrow e^{-2i\theta} X.
\end{align}
Thus a $\tilde \psi_i \psi_i$ condensate will only give rise to  
one Goldstone boson, even if 
$N_f > 1$.  If $X$ were massive, there would be an approximate $U(N_f)_L
\times U(N_f)_R$ symmetry at low energies,  but this is not the case in the 
Constable-Myers background. 

Note that $N_f >1$ coincident D7-branes 
may be embedded in the same way as a single
 D7-brane.  In this case there are $N_f$
independent solutions to the linearized equations of motion for
 small fluctuations about this embedding,   
corresponding to fluctuations of the diagonal entries in a diagonal 
$N_f \times N_f$
matrix.  These fluctuations would naively
give rise to at least $N_f$ Goldstone bosons,  rather than one.  
These extra states will not remain massless though since the 
interaction with the scalar fields which breaks the symmetry will 
induce a mass.

Nevertheless we can still make a rough comparison between our $\eta'$ and QCD
pions. In a two-flavour large $N$ QCD model where the quarks are degenerate
one would expect four degenerate Goldstone bosons. As $N$ is decreased,
instanton effects will enter to raise the $\eta'$ mass.  However since we are
at large $N$, the mass formula for our Goldstone boson is applicable to the
pions.  It is therefore amusing to compare the Goldstone mass we predict to
that of the QCD pion. There is considerable uncertainty in matching the strong
coupling scale of our theory to that of QCD.  In QCD the bare up or down quark
mass is roughly $0.01 \Lambda$ and the pion mass of order $0.5 \Lambda$ (it is
of course hard to know precisely what value one should pick for $\Lambda$).
The comparison to our theory is a little hard to make but if we assume that
$\Lambda \simeq \frac{b}{2\pi \alpha'} = 1$ then for this quark mass we find
$m_\pi \simeq 0.25 \Lambda$. The gravity dual is correctly predicting the pion
mass at the level of a factor of two. Of course we cannot expect a perfect
match given the additional degrees of freedom in the deformed ${\cal N}=4$
theory relative to real QCD.



\section{Deconstructing extra dimensions}
\label{cha:deconstruction}

 \setcounter{equation}{0}\setcounter{figure}{0}\setcounter{table}{0}
 
In the first part of this paper we have used intersecting brane
configurations to study field theories with fundamental matter in the context
of the AdS/CFT correspondence. In this chapter we will make use of such brane
intersections in order to study some aspects of higher-dimensional field
theories which describe the low-energy dynamics of intersecting M5-branes in
M-theory. A description of this theory can be obtained from the low-energy
effective field theory of intersecting D3-branes studied in Sec.~\ref{sec3} by
a method known as {\em deconstruction}.

Deconstruction is a method to generate extra dimensions in theories
with internal gauge symmetries \cite{Arkani-Hamed:2001tb, Hill:2000mu}
(for early work on this subject, see \cite{Georgi:au,
Halpern:1975yj}).  Non-renormalizable theories in $D>4$ are
ill-defined above a certain cut-off at which they become strongly
coupled due to a coupling constant with negative dimension.  At high
energies these theories require an ultraviolet completion which can
frequently be provided by the deconstruction technique.  This UV
completion is generically a {\em quiver} theory \cite{Douglas:1996sw}
characterised by a discrete theory space, the {\em moose} or {\em
quiver} diagram, which represents the field content of the theory by a
lattice of sites and links. In a certain low-energy limit the quiver
theory develops one or more extra dimensions and reproduces the
higher-dimensional non-renormalizable theory. A peculiarity of the
deconstruction approach is that the ultraviolet theory has less
dimensions than the infrared theory.  This is different from
compactified theories which reveal their higher-dimensional behaviour
at energies above the inverse radius of the compactified dimension.

In Sec.~\ref{Ch41} we will review the low-energy effective field theory of
D3-branes located at an orbifold singularity. This will lead us to the
notion of a quiver gauge theory, an example of which with $\N=2$
supersymmetry is discussed in Sec.~\ref{Ch42}. In Sec.~\ref{Ch43} we
study the generation of extra dimensions on the Higgs branch of quiver
gauge theories. We demonstrate the deconstruction of the low-energy
theory of coincident M5-branes which is the six-dimensional (2, 0)
superconformal theory. In Sec.~\ref{Ch44} we generalize the procedure
to the case of intersecting M5-branes, a brane system which is not
very well understood so far.  In Sec.~\ref{Ch45} we will follow to some
extend the fascinating idea to deconstruct M-theory itself.

\subsection{D3-branes at orbifolds of the type $\CC^3/\Gamma$} \label{Ch41}

In this section we briefly review orbifolds of the type $\CC^3/\Gamma$
with $\Gamma$ a discrete subgroup of the group $SU(4)$. We are 
interested in the world-volume field theory of $N \vert
\Gamma \vert$ coincident D3-branes which are located at the orbifold 
singularity.  The effect of the orbifold is to project out degrees of
freedom of the four-dimensional $\N=4$ $U(N \vert \Gamma \vert)$ super
Yang-Mills theory which are not invariant under the orbifold group.
In general, this leads to a particular class of gauge theories known as quiver gauge theories which we will discuss in the next section.

For simplicity, let us choose for $\Gamma$ the cyclic group $\ZZ_k$ which
acts on the three complex coordinates $z_i$ ($i=1,2,3$) of the transverse 
space $\CC^3$ as 
\begin{align}
z_i \rightarrow \xi^{b_i} z_i \,,\quad
\end{align}
where $\xi=e^{\frac{2\pi i}{k}}$ is the generator of the group $\ZZ_k$ 
and $(b_1,b_2,b_3)$ is a triple of indices.

Since the orbifold group acts on the D-brane (Chan-Paton) indices, the
orbifold group $\ZZ_k$ is embedded into the gauge group $U(N k)$. The invariant
components of the gauge field satisfy
\begin{align}
  A_\mu = g(\xi) A_\mu g(\xi)^{-1} \,,      \label{transf1}
\end{align}
where $g(\xi)={\rm diag}({\bf 1},\xi {\bf 1}, \xi^2 {\bf 1}, ...,
\xi^{k-1}{\bf 1})$ is the regular representation of the generator $\xi$
and ${\bf 1}$ is the unit matrix of dimension $N \times N$. The gauge
field $(A_\mu)_{ab}$ ($a,b=1,...,Nk$ color indices) is a matrix in the
adjoint representation of $U(N k)$. Since only block-diagonal matrices
$A_\mu$ survive the projection, the gauge group is broken down to
$U(N)^{k}$.

The four Weyl fermions of the $\N=4$ super Yang-Mills theory transform in the
{\bf 4} of $SU(4)$. Those fermions which are invariant under the orbifold must
satisfy
\begin{align} \label{transf2}
  \psi^i = \xi^{-a_i} g(\xi) \psi^i g(\xi)^{-1} \,,
\end{align}
where $i=1,...,4$ and
\begin{align}
a_1+a_2+a_3+a_4 \equiv 0 \!\! \mod k \,. \label{condition}
\end{align}
The condition (\ref{condition}) guarantees that $\Gamma$ is a subgroup
of $SU(4)$ and not only of $U(4)$.

The complex scalars $\phi^i$, $i=1,2,3$, transform in the {\bf 6} of $SU(4)
\simeq \overline{SO}(6)$, which can be obtained from the anti-symmetric tensor
product of two {\bf 4}'s. Invariant scalars fulfill the conditions
\begin{align}
  \phi^i = \xi^{b_i} g(\xi) \phi^i g(\xi)^{-1} \,,     \label{transf3}
\end{align}
where $b_1 = a_2 + a_3$, $b_2 = a_3 + a_1$, and $b_3 = a_1 + a_2$. If
in addition the condition 
\begin{align}
b_1+b_2+b_3 \equiv 0 \!\! \mod k \, \label{conditionsusy}
\end{align}
or, equivalently, $a_4 \equiv 0$ is satisfied, the group $\ZZ_k$
is embedded into $SU(3)$ and at least $\N=1$ supersymmetry is preserved. 

In general, if we choose $\Gamma \subset SU(3)$ or $\Gamma \subset
SU(2)$ then the $\N=4$ parent theory reduces to an $\N=1$ or $\N=2$
supersymmetric theory, respectively.\footnote{The holonomy group
$\Gamma$ of the space $\CC^3/ \Gamma$ is a subgroup of either $SU(2)$
or $SU(3)$ which themselves are subgroups of the maximal holonomy
group of a six-fold $\overline{SO}(6) \simeq SU(4)$.  The $\mathbf{4}$
of $SU(4)$ decomposes into ${\mathbf 3+1}$ under $SU(3)$ or
$\mathbf{2+1+1}$ under $SU(2)$ guaranteeing one or two constant
spinors, respectively.} If
\mbox{$\Gamma \subset \!\!\!\!\!\!\!~/ \,\, SU(3)\subset SU(4)$} then no
supersymmetry is preserved.

A simple example of a non-supersymmetric orbifold, which will be of use
in Sec.~\ref{Ch45} below, is $\CC/\ZZ_{k}$ defined by the vectors
\begin{align}
a_i=(-1,1,1,-1)\,, \quad b_i=(2,0,0) \,.  \label{nonsusy}
\end{align}
Since $b_2=b_3=0$ the orbifold acts only in one of the three complex planes
transverse to the D3-branes while the other two planes remain flat.

\subsection{Quiver gauge theories} \label{Ch42}

An orbifold of the type $\CC^2/\ZZ_{k}$ with $\N=2$ supersymmetry is given by
the vectors
\begin{align}
a_i=(-1,1,0,0)\,, \quad b_i=(1,-1,0) \,.  \label{susy}
\end{align}
In this case the four fermions can be paired with the three scalars and the
gauge boson to form $\N=1$ chiral and vector multiplets,
\begin{align}
 (\phi^i, \psi^i) \rightarrow \Phi^i \quad(i=1,2,3)\,,\quad (A_\mu, \psi^4)
 \rightarrow V \,.
\end{align}
In terms of these superfields the conditions (\ref{transf1})--(\ref{transf3})
are given by  
\begin{align} 
\Phi^i_{ab} &= \xi^{a-b+b_i} \Phi^i_{ab} \,,\\
 V_{ab} &= \xi^{a-b} V_{ab} \,,
\end{align}
($a, b=1,...,k$) which have the solutions
\begin{align} \label{XY} 
\Phi^1=\!
\begin{pmatrix}
0 & \Phi^1_{12} & 0 & ...& 0 \\
0 & 0    & \Phi^1_{23} & ...&0 \\
\vdots & &       & \ddots &\vdots\\
\vdots & &       & & \Phi^1_{k-1,k} \\
\Phi^1_{k,1}&0& 0 &... & 0\\
\end{pmatrix}&, \,
\Phi^2=\!\begin{pmatrix}
0 &  0 & ... &... & \Phi^2_{1,k} \\
\Phi^2_{21} & 0     & &&0 \\
0 &  \Phi^2_{23} &       & &\vdots\\
\vdots &\vdots & \ddots&  &\vdots \\
0 &0& ... &  \Phi^2_{k,k-1} &0\\
\end{pmatrix} 
\end{align}
and
\begin{align}
\Phi^3 ={\rm diag}(\Phi^3_{11}, \Phi^3_{22},..., \Phi^3_{kk})\,&,\quad
V={\rm diag}(V_{11}, V_{22},..., V_{kk}) \,. \label{V}
\end{align} 
The bifundamentals $\Phi^1_{i,i+1}$ and $\Phi^2_{i+1,i}$ transform
in the
\begin{align}
(1,1,...,N,\bar N,...,1)\quad{\rm and}\quad (1,1,...,\bar N,N,...,1) 
\end{align}
of the gauge group $SU(N)^k$, while the fields $V_{ii}$ and
$\Phi^3_{ii}$ transform in the adjoint representation.

In summary, we have the bifundamentals $\Phi^1_{i,i+1}$, $\Phi^2_{i+1,i}$, the
adjoint chiral superfields $\Phi^3_i \equiv \Phi^3_{ii}$, and the vector
superfields $V_i\equiv V_{ii}$.  These fields can now be encoded in an
oriented diagram of sites and links known as a {\em quiver} or {\em moose}
diagram, as explained in the introduction, see Fig.~\ref{quiver1}. Each of the
sites is associated with one of the $k$ $SU(N)$ gauge groups and represents a
vector multiplet $V_{i}$ as well as an adjoint chiral multiplet $\Phi^3_i$
which together form a $\N=2$ vector multiplet. Two neighbouring sites are
connected by two oppositely oriented links representing the complex scalars
$\Phi^1_{i,i+1}$ and $\Phi^2_{i+1,i}$ which together form a $\N=2$ hyper
multiplet $(\Phi^1, {\bar \Phi}^2)_{i,i+1}$.

The low-energy effective action for this quiver model is given by
\begin{align} \label{quiveraction}
  S=\sum_{i=1}^k \,{\rm Tr} \int d^4x &\left[ \int d^4 \theta\, \big(
  e^{-gV_{i+1}} \bar \Phi^1_{i+1,i} e^{gV_i} \Phi^1_{i,i+1} +
  e^{gV_{i+1}} \Phi^2_{i+1,i} e^{-gV_i} \bar \Phi^2_{i,i+1}
  \right.\nonumber\\ & +\, e^{-gV_i} \bar\Phi^3_i e^{gV_i} \Phi^3_i
  \big) +\frac{1}{4g^2} \left( \int d^2\theta\,\frac{1}{4} W_i^\a
  W^i_\a + {\rm h.c.} \right) \nonumber\\ &+ i g \frac{ \sqrt{2}
  }{3}\left. \int d^2 \theta ( \Phi^3_i \Phi^1_{i,i+1} \Phi^2_{i+1,i}-
  \Phi^3_{i+1} \Phi^2_{i+1,i} \Phi^1_{i,i+1} )+{\rm h.c.}  \right]
\end{align}
This action is obtained by substituting the solutions (\ref{XY}) and
(\ref{V}) into the $\N=4$, $d=4$ parent super Yang-Mills action. The
embedding of the group $\ZZ_k$ into say $SU(2)_L$ of $SU(2)_R \times
SU(2)_L \times U(1) \subset SU(4)$ breaks the R-symmetry of the parent
theory down to $SU(2)_L \times U(1)$.  The quiver theory thus describes
$\N=2, d=4$ super Yang-Mills theory with gauge group $SU(N)^k$ which
couples to $k$ bifundamental hyper multiplets.


\subsection{Deconstruction of the M5-brane action} \label{Ch43}

In the last section we have derived the low-energy theory of a stack
of $kN$ D3-branes at an orbifold of the type $\CC^2/\ZZ_{k}$. The action
(\ref{quiveraction}) of this four-dimensional quiver theory is the
starting point for the deconstruction of the M5-brane theory.  We will
show that on the Higgs branch of the quiver gauge theory two extra
dimensions are generated and the theory becomes equivalent to the
six-dimensional (2, 0) theory.

\subsubsection{Deconstruction of extra dimensions} 

We start by considering a particular Higgs branch of the quiver theory
(\ref{quiveraction}) on which the scalars in all hyper multiplets have the
same expectation value, i.e.
\begin{align}
 \langle \phi^1_{i,i+1} \rangle = \langle \phi^2_{i+1,i} \rangle= v \bf{1}
 \,,
\end{align}
independent of $i$. This causes the breakdown of the gauge group
$SU(N)^k$ to its diagonal subgroup $SU(N)$ leading to massive gauge bosons
$A_i^\m$. To see this, we consider the kinetic term for
the scalars $\phi^1_{i,i+1}$ of the bifundamentals $\Phi^1_{i,i+1}$
(analogous for $\Phi^2_{i+1,i}$) which is given by
\begin{align} \label{kineticterm}
  {\rm Tr} \left( (D_\mu
  \phi^1_{i,i+1} )^\dagger D^\mu \phi^1_{i,i+1} \right) \subset {\rm Tr}
  \left(e^{-gV_{i+1}} \bar \Phi^1_{i+1,i} e^{gV_i} \Phi^1_{i,i+1}
  \right)\,,
\end{align}
where the covariant derivative is $D_\m \phi^1_{i,i+1} \equiv
\partial_\m \phi^1_{i,i+1} - g A^i_\m \phi^1_{i,i+1} - g \phi^1_{i,i+1} 
A^{i+1}_\m$.  Substituting the VEVs $\langle \phi^1_{i,i+1} \rangle=v
{\bf 1}$ into the Lagrangian, the term (\ref{kineticterm}) reduces to
\begin{align}
 g^2 v^2 (A_i^\m - A_{i+1}^\m)^2 &= g^2 v^2 \left[ 2 (A_i^\m)^2 -
 A_{i-1}^\m A_{i}^\m - A_{i+1}^\m A_{i}^\m \right] = \frac{1}{2} A_{i
 \m} M^2 A_j{}^\m \,,
\end{align}
where we defined the mass matrix $M^2$ by\footnote{The discrete subgroups of
  $SU(2)$ can be classified by the simply laced Lie groups (ADE
  classification). For instance, the cyclic group $\ZZ_k$ is associated with
  the group $A_{k-1}$. The McKay correspondence \cite{McKay} now states that
  there is a one-to-one correspondence between the vertices of the (extended)
  Dynkin diagram of the ADE group and equivalence classes of irreducible
  representations of the group $\Gamma \subset SU(2)$.  Since the regular
  representation is the sum of irreducible representations, the quiver diagram
  is equivalent to the (extended) Dynkin diagram. The information of a Dynkin
  diagram is encoded in the Cartan matrix.  It is therefore not surprising
  that the mass matrix (\ref{massmatrix}) agrees with the Cartan matrix of
  the $A_{k-1}$ group.}

\begin{align} \label{massmatrix}
M^2 \equiv 2 g^2 v^2 
\begin{pmatrix}
2  &-1 & & &-1\\
-1 & 2 & -1 & && \\
   & \ddots & \ddots & \ddots \\
   & & -1 & 2 & -1 \\
-1 & &    & -1& 2
\end{pmatrix} \,.
\end{align} 
The eigenvalues of $M$ turn out to be \cite{Csaki1}
\begin{align} \label{KKmass}
m_l = 2 \sqrt{2} g v \sin \frac{l \pi}{k} \,, \quad 0 \leq l \leq k-1\,.
\end{align}
For large $k$ (and small $l$) the mass spectrum becomes linear, $m_l
\approx 2 \sqrt{2} g v \frac{l \pi}{k}$ and approximates a
Kaluza-Klein tower of states corresponding to the compactification of
a fifth dimension. Comparing (\ref{KKmass}) with the conventional
expression of a Kaluza-Klein spectrum, $m_l=l/R_5$, we find
\begin{align}
2\pi R_5= \frac{k}{\sqrt{2} gv} \,
\end{align}   
for the radius $R_5$ of the compact dimension.

This spectrum was expected since the action (\ref{quiveraction}) looks like a
latticized five-dimen\-sional theory, where only the fifth dimension is
discritized on a spatial circle. The quiver diagram in Fig.~\ref{quiver1},
which encodes the field content of the theory, visualizes the discretization
of the extra dimension. As $k \rightarrow \infty$ the indices of the fields
turn into continuous labels parameterizing the extra dimension. For instance,
the discrete index $i$ of the four-dimensional gauge bosons $A_\mu^i(x)$
becomes a continuous label~$z$, i.e.\ $A_\mu^i \rightarrow A_\mu(x, z)$
yielding four components of a five-dimensional gauge boson $A_M(x, z)$
($M=0,1,2,3,4$).  Its fifth component $A_4$ is given by the imaginary part of
$\phi_{i,i+1}$.\footnote{Expand $\phi_{i,i+1}$ around the VEV $v$:
  $\phi_{i,i+1} = v + i A_{4,i} + \tilde \phi_{i,i+1}$.}

\subsubsection*{Dyonic excitations}
The Kaluza-Klein spectrum of massive gauge bosons shows the occurrence
of a fifth dimension on the Higgs branch. The deconstructed theory is
however not just a five-dimensional gauge theory. Surprisingly, there
exists a further Kaluza-Klein spectrum of magnetic excitations
indicating the deconstruction of a second extra dimension.  Due to the
conformal invariance of the quiver theory, which is inherited from the
$\N=4$ parent theory, the theory possesses an $SL(2, \ZZ)$
S-duality. Substituting $g \rightarrow k/g$ in
(\ref{KKmass}),\footnote{The complex coupling of the quiver theory
can be expressed in terms of the complex coupling of the parent
theory, $\tau=\tau_{\rm par}/k$. Then the S-duality of the parent
theory $\tau_{\rm par} \rightarrow 1/\tau_{\rm par}$ translates into
$g \rightarrow k/g$.} we obtain the magnetic mass spectrum
\begin{align}
M_l = 2 \sqrt{2} \frac{kv}{g} \sin \frac{l \pi}{k} \,,
\quad 0 \leq l \leq k-1\,.
\end{align}
As above, for small $l$ this approximates a KK spectrum 
which generates a sixth dimension of circumference   
\begin{align}
2\pi R_6 = \frac{g}{\sqrt{2} v} \,.
\end{align}

\begin{figure}[!ht]
\begin{center}
{\includegraphics[scale=0.91]{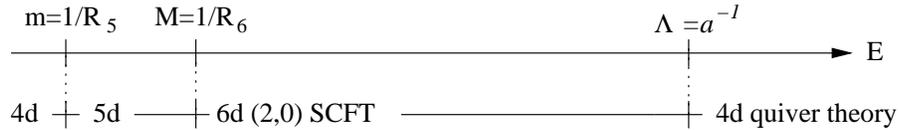}}
\caption{Deconstruction of extra dimensions at low energies.}
\label{energy2}
\end{center}
\end{figure}

In summary, on the Higgs branch the quiver theory develops two extra
dimensions corresponding to the two cycles of a torus with radii $R_5$ and
$R_6$. Kaluza-Klein modes with momenta in both directions are dyons in the
quiver theory. The higher-dimensional behaviour of the quiver theory at low
energies is shown schematically in Fig.~\ref{energy2}. The inverse of the
lattice spacing $a=1/(\sqrt{2}g v)$ plays the role of a cut-off.  For finite
$a$ the theory is six-dimensional up to the cut-off $a^{-1}$ above which the
theory is still well-defined but lacks six-dimensional Lorentz invariance. In
the continuum limit $g, v \rightarrow \infty$ keeping $R_5$ and $R_6$ fixed
the lattice spacing $a$ and the theory becomes six-dimensional to arbitrary
high dimensions.

\subsubsection{String theory analysis} \label{secString}

We now show that the Higgs branch theory of the quiver model
(\ref{quiveraction}) is the six-dimen\-sional (2, 0) theory compactified on a
torus which is the world-volume theory of wrapped M5-branes. To see this we
will make use of string theory dualities in the orbifold realisation of the
quiver theory.

\begin{figure}[!ht]
\begin{center}
{\includegraphics[scale=0.9]{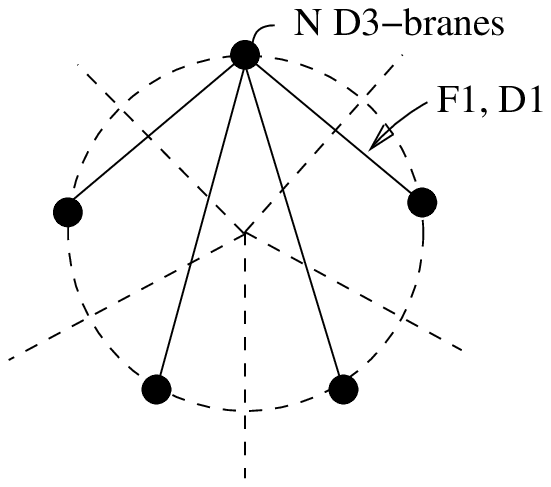}}
\caption{The orbifold $\CC^2/\ZZ_k$ for $k=5$.}
\label{quiver3}
\end{center}
\end{figure}

Fig.\ \ref{quiver3} shows the covering space of the orbifold
$\CC^2/\ZZ_k$ (for $k=5$) and the location of a stack of $N$ D3-branes
and their mirror branes.  The Higgs branch of the quiver gauge theory
is parameterized by the VEVs $v$ of the bifundamentals, i.e.\ the
D3-branes are located some distance $d \sim v$ away from the orbifold
singularity.  The KK states in the field theory descend from
fundamental strings and D-strings stretching between the D-branes and
its mirror branes.

The first issue to verify is the doubling of supersymmetry on the Higgs
branch. While the $\N=2$ quiver gauge theory has 8 supercharges at high
energies, we expect it to have 16 supercharges at low energies, where it is
supposed to turn into the M5-brane theory. The amount of supersymmetry is
effectively enhanced by a factor two.
\begin{figure}[!ht]
\begin{center}
{\includegraphics{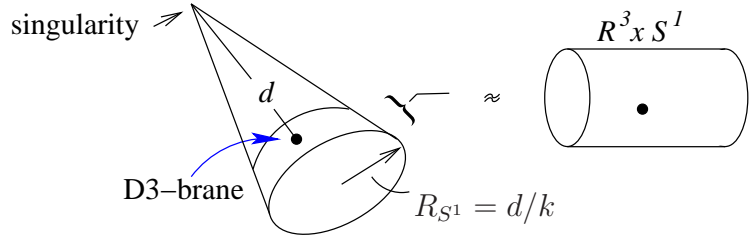}}
\caption{Far away from the orbifold singularity the cone can be
approximated locally by a cylinder.}
\label{cone}
\end{center}
\end{figure}

 \vspace{-3cm} \hspace{7.2cm} $R_{S^1}=d/k$ \vspace{2.5cm}

This can be understood from a geometrical point of view as follows. The
orbifold $\CC^2/\ZZ_k$ can be introduced in the six-dimensional transverse 
space of the D3-branes with flat metric $ds^2=dz_i d\bar z_i$ ($i=1,2,3$)
by redefining the coordinates,
\begin{align}
 z_1 =r_1 exp (i\theta + i\phi/k) \,,\quad
 z_2 =r_2 exp (i\theta - i\phi/k) \,,\quad
 z_3 \,\,\textmd{unchanged} \,
\end{align}
with ranges $\phi, \theta \in (0,2\pi)$. Expanding $r_1 = d/\sqrt{2}
+x_1$, $r_2 = d/\sqrt{2} +x_2$, $\theta=x_3/d$ the metric
becomes  
\begin{align}
 ds^2 = dz_3 d\bar z_3 + dx_1^2 + dx_2^2 + dx_3^2 + \frac{d^2}{k^2}
d\phi^2 
\end{align}
plus corrections proportional to $1/k$ which vanish in the large $k$ limit.
The metric shows that away from the singularity the orbifold geometry
$\CC^2/\ZZ_k$ degenerates into a cylinder, $\RR^3 \times S^1$, where the
radius of the $S^1$ is given by $R_{S^1}=l^2_s/R_5$ with $R_5 \equiv
kl^2_s/d$, see Fig.\ \ref{cone}. The D3-branes are effectively located in a
flat geometry, far away from the highly curved region close to the orbifold
singularity at the tip of the cone. Recall that the effect of the orbifold was
to break supersymmetry by a half. Now in the locally flat region of the
orbifold the supersymmetry of the effective D-brane theory is doubled.

In the decoupling limit $g_s$ fixed, $l_s \rightarrow 0$ the radius $R_{S^1}$
becomes sub-stringy. A more appropriate description is obtained by T-dualizing
the D3-branes to D4-branes which wrap around the T-dual $S^1$ with large
radius $R_5$. However, the type IIA string coupling $g_s'=g_s R_5/l_s$ becomes
very large for $l_s \rightarrow 0$ such that we have to lift the D4-branes to
M5-branes in M-theory. These M5-branes wrap a two-torus with radii $R_5$ and
$R_6=g_s' l_s=g_s R_5$. The eleven-dimensional Planck length
$l_p^3=l_s^3g_s'=l_s^2R_6$ goes to zero for $l_s \rightarrow 0$.  The
world-volume theory of the D3-branes at $\CC^2/\ZZ_k$ therefore turns on the
Higgs branch into the world-volume theory of M5-branes wrapped around a
two-torus.

To conclude, taking the number $k$ of nodes in the quiver diagram to
infinity, the Higgs branch of the 4D $\N=2$ orbifold theory  is equivalent
to the 6D $\N=(2,0)$ gauge theory compactified on a torus. The gauge group
is thereby broken down to its diagonal subgroup, $SU(N)^k \rightarrow 
SU(N)$ and supersymmetry is enhanced from eight to sixteen supercharges.



\subsection{Intersecting M5-branes} \label{Ch44}

In this section we discuss the low energy dynamics of orthogonally
intersecting M5-branes which are not very well understood by now.  In
addition to a non-abelian chiral two-form, this theory has tensionless
strings localized at the intersection corresponding to M2-branes
stretched between the M5-branes \cite{Hanany}.  These tensionless
strings are in some sense fundamental, as they are not excitations of
a chiral two-form.  The only known formulation of the M5-M5
intersection is the DLCQ M(atrix) description proposed in
\cite{Kachru2}.

Here we shall present the (de)construction of the M5-M5 intersection,
which is a natural extension of the (de)construction of parallel
M5-branes discussed in the last section.  This will be accomplished by
taking a $k \rightarrow \infty$ limit of the theory describing
intersecting D3-branes at a $\CC^2/\ZZ_k$ orbifold. At a certain point
in the moduli space, two compact latticized extra dimensions are
generated. In an appropriate \mbox{$k\rightarrow \infty$} limit, we
expect that the extra directions become continuous, such that the
intersection of four-dimensional world-volumes over $1+1$ dimensions
becomes an intersection of six-dimensional world-volumes over $1+3$
dimensions.

The infrared dynamics of the D3-D3 intersection at a $\CC^2/\ZZ_k$ orbifold is
described by a defect conformal field theory with two-dimensional $(4,0)$
supersymmetry.  This theory belongs to the class of conformal field theories
with defects which we studied in Ch.~\ref{cha:dCFT}.  The action of this
$(4,0)$ theory is readily constructed in $(2, 0)$ superspace, starting from
the action for the D3-D3 intersection in flat space which we constructed in
Sec.~\ref{intersectaction}.  The field content of the $(4,0)$ theory is
summarized by a quiver (or ``moose'') diagram consisting of two concentric
rings, and spokes stretching between the inner and outer rings. For large $k$
this gives rise to a discretized version of the field theory corresponding to
the low-energy limit of the M5-M5 intersection.  The spokes in the quiver
diagram will be seen to correspond to strings localized at the M5-brane
intersection.

Moreover, we examine the relation between the moduli space of
vacua of the $(4,0)$ defect conformal field theory and that of the
M5-M5 intersection. On a particular part of the Higgs branch of
the defect CFT, the resolution of the intersection to a
holomorphic curve $xy =c$ can be seen very explicitly from
F-flatness conditions.  This point in the Higgs branch corresponds
to a vacuum of the M5-M5 theory in which tensionless strings have
condensed. By going to another point on the Higgs branch of the
defect CFT for which the string tension in the M5-M5 theory is
non-zero, we will be able to match the $SU(2)_L$ R-symmetry of the
$(4,0)$ theory with the $SU(2)$ R-symmetry of the M5-M5
intersection, which has ${\cal N} =2, d=4$ supersymmetry.

\subsubsection{D3-D3 intersection at a $\CC^2/\ZZ_k$ orbifold}

To (de)construct the theory of the M5-M5 intersection, we shall
consider a pair of intersecting stacks of D3-branes at an
$\CC^2/\ZZ_k$ orbifold point. One set of D3-branes is located at
$X^{4,5,6,7,8,9} = 0$ while the other set of ${\rm
D3}^{\prime}$-branes is located at \mbox{$X^{2,3,6,7,8,9}=0$}. The
$\CC^2/\ZZ_k$ is spanned by the coordinates $u = X^6 + iX^7$ and $w=
X^8 + iX^9$ subject to the orbifold condition $u \sim \xi u, w \sim
\xi^{-1} w$ where $\xi = \exp(2\pi i/k)$.  Before orbifolding, the
theory of intersecting D3-branes has $(4,4)$ supersymmetry with an
$SU(2)_L \times SU(2)_R$ R-symmetry. The $SU(2)_L \times SU(2)_R$
component of the R-symmetry acts as an $SO(4)$ transformation on the
real components of $u$ and $w$, which are the coordinates
$X^{6,7,8,9}$. The orbifold breaks $SU(2)_L \times SU(2)_R$ to
$SU(2)_L$, under which the pair ($u,w^*$) transforms as a doublet.
Moreover, the supersymmetry is broken from $(4,4)$ to $(4,0)$.  It is
interesting to observe that the theory is chiral.

\subsubsection*{Orbifold projection}
The Lagrangian describing the D3-D3 intersection in the $\CC^2/\ZZ_k$ orbifold
can be obtained from the action of the D3-D3 intersection in flat space given
in (\ref{action1}) - (\ref{defectaction}). Following
Refs.~\cite{Douglas:1996sw,JM} we start with $Nk$ D3-branes intersecting
$N^\prime k$ D3-branes in a flat background and project out the degrees of
freedom which are not invariant under the $\ZZ_k$ orbifold group, which is
generated by a combination of a gauge symmetry and an R-symmetry. An important
constraint on the orbifold action is that the theory on each stack of D3
branes (ignoring strings connected to the other stack) should be the ${\cal N}
=2, d=4$ super Yang-Mills theory described by the quiver in Fig.~\ref{quiver1}
(in Ch.~1), with gauge group $SU(N)^k$ or $SU(N^{\prime})^k$.

The orbifold action which gives the quiver of Fig.~\ref{quiver1} for
both the D3 and the D3$^{\prime}$ degrees of freedom separately, and
breaks the $SU(2)_L \times SU(2)_R$ R-symmetry to $SU(2)_L$ is as
follows. The embedding of the $\ZZ_k$ orbifold group in the $U(Nk)$
and $U(N^{\prime}k)$ gauge groups is given by
\begin{align}
g(\xi) &= \begin{pmatrix} I_{N\times N} & & &\\ & \xi I_{N\times
N} & &
\\ & & \xi^2I_{N\times N} & \\
& & & \ddots \label{g}
\end{pmatrix} \,,\\ g^{\prime}(\xi) &=
\begin{pmatrix} I_{N^{\prime}\times N^{\prime}} & & & \\
& \xi I_{N^{\prime}\times N^{\prime}} & &
\\ & & \xi^2I_{N^{\prime}\times N^{\prime}} & \\
& & & \ddots\label{g'}
\end{pmatrix} \,,
\end{align}
where $\xi$ is the generator $\exp(2\pi i/k)$ of $\ZZ_k$. The
embedding of the $\ZZ_k$ orbifold group in the R-symmetry is given
by
\begin{align}
h(\xi) = e^{i\pi \sigma^3/k} \,,
\end{align}
where $h$ belongs to $SU(2)_R$. The field theory describing the
D3-D$3^{\prime}$ intersection at the orbifold is then obtained
from that of the D3-D$3^{\prime}$ intersection in flat space by
projecting out fields which are not invariant under the orbifold
action. The result is an $SU(N)^k \times SU(N^{\prime})^k$ gauge
theory with $(4,0)$ supersymmetry and $SU(2)_L \times U(1)$
R-symmetry.

In $(2,2)$ superspace, the orbifold acts on superspace coordinates as
\begin{align}\label{supe}
&\theta^- \rightarrow \xi\theta^- \,,
\end{align}
but trivially on $\theta^+$. On the $(2,2)$ superfields the orbifold
acts as
\begin{align}
&{\rm  D3}:  &&V\rightarrow g(\xi)Vg^{\dagger}(\xi)\,,
&&\Sigma \rightarrow \xi^{-1}g(\xi)\Sigma g^{\dagger}(\xi)\,,&&\nonumber \\
&&&Q_1 \rightarrow \xi g(\xi)Q_1g^{\dagger}(\xi)\,,  &&Q_2
\rightarrow g(\xi)Q_2g^{\dagger}(\xi) \,,&&\nonumber \\
&&&\Phi \rightarrow g(\xi)\Phi g^{\dagger}(\xi)\,,&&\nonumber\\
&{\rm D3}^{\prime}:&&{\cal V}\rightarrow
g^{\prime}(\xi){\cal V}g^{\prime\dagger}(\xi)\,,  &&\Omega
\rightarrow \xi^{-1}g^{\prime}(\xi)\Omega
g^{\prime\dagger}(\xi)\,,&&\nonumber\\
&&&S_1 \rightarrow \xi g^{\prime}(\xi)S_1 g^{\prime\dagger}(\xi)\,,
&&S_2 \rightarrow g^{\prime}(\xi)S_2g^{\prime\dagger}(\xi)\,,
&&\nonumber \\
&&&\Upsilon \rightarrow g^{\prime}(\xi)\Upsilon
g^{\prime\dagger}(\xi)\,,
&&\nonumber\\
&{\rm D3-D3^{\prime}}:&&B\rightarrow
g(\xi)Bg^{\prime\dagger}(\xi)\,, && \tilde B \rightarrow
g^{\prime}(\xi) \tilde B g^{\dagger}(\xi)\,, \label{orbac}
\end{align}
with $g$, $g'$ as in (\ref{g}), (\ref{g'}).
Starting with the action (\ref{action1}) -
(\ref{defectaction}) and projecting out the degrees of freedom
which are not invariant under (\ref{orbac})  will give a $(4,0)$
supersymmetric action with manifest $(2,0)$ supersymmetry.

To illustrate how the orbifold acts on components, we consider the
action (\ref{orbac}) on the $(2,2)$ twisted superfield $\Sigma$. On the
bosonic components, this corresponds to
\begin{align}
\sigma &\sim \xi^{-1} g(\xi) \sigma g^{\dagger}(\xi)\,,\quad
F_{01} \sim g(\xi) F_{01} g^{\dagger}(\xi) \,.
\end{align}
This is consistent with the fact that the field $\sigma$
characterises fluctuations transverse to both D3-branes, i.e.\
fluctuations in the orbifold directions. This field is naturally
associated with fluctuations in the $w= X^8+ i X^9$ directions
which satisfy the orbifold condition $w \sim \xi^{-1} w$.   Upon
projecting out the parts which are not invariant under the
orbifold, $\sigma$ becomes a set of $k$ bifundamentals in the
representations $( \cdots N, \bar N, \cdots)$ of $SU(N)^k$.  These
bifundamental fields are written as $\sigma_{j,j+1}$,  where $j=1
\cdots k$ and the first(second) index labels the gauge group with
respect to which the field is a fundamental (anti-fundamental).
Fields which are adjoints with respect to one of the factors will
be written with a single index.

\subsubsection*{Quiver action in two-dimensional $(2,0)$ superspace}

Since the $(4,4)$ supersymmetry of the action (\ref{action1}) -
(\ref{defectaction}) is broken down to $(4,0)$ by the orbifold, an adequate
formulation of the corresponding quiver gauge theory is best given in
$(2,0)$ superspace. In order to project out the degrees of freedom which are
not invariant under the orbifold, we rewrite the parent action using
manifest
$(2,0)$ supersymmetry. To this end, we decompose the $(2,2)$ superfields
under
$(2,0)$ supersymmetry.  The decomposition of the $(2,2)$
superfields is as follows (see for instance \cite{Witten,Garcia}):\\

i) $(2,2)$ vector $\rightarrow$ $(2,0)$ vector $+$
$(2,0)$
chiral,

ii) $(2,2)$ chiral $\rightarrow$ $(2,0)$ chiral $+$
$(2,0)$ Fermi.\\

\noindent These $(2,0)$ superfields have the following component
decomposition:  \\

\begin{minipage}{14.5cm}
\noindent i) $(2,0)$ vector superfield $V$:  two gauge connections
$A_0, A_1$ and one fermion $\chi_-$,

\noindent ii) $(2,0)$ chiral superfields  $\Phi$: one complex
scalar $\phi$ and a fermion $\psi_+$,

\noindent iii) $(2,0)$ Fermi superfields $\Lambda$: one chiral
fermion $\lambda_-$.  The full expansion of this anticommuting superfield
contains an auxiliary field and a holomorphic function of $(2,0)$ chiral
superfields. \\
\end{minipage}

\noindent For the theory given by the action (\ref{action1}) -
(\ref{defectaction}), the decomposition of the $(2,2)$ superfields of the
D3-D3 intersection in flat space gives the following $(2,0)$ superfields (we
shall henceforward write $(2,2)$ superfields in boldface):
\begin{align}
&{\rm D3}:  &&{\bf Q}_1 \rightarrow Q_1, \Lambda^{Q_1}, && {\bf
Q}_2 \rightarrow
  Q_2,\Lambda^{Q_2}, && {\bf\Phi} \rightarrow \Phi, \Lambda^\Phi, &&
  {\bf V} \rightarrow V, \Theta_V\,,&&\nonumber
  \\
&{\rm D3}^{\prime}: &&{\bf S}_1 \rightarrow S_1, \Lambda^{S_1}, &&
{\bf S}_2 \rightarrow S_2,
  \Lambda^{S_2} \,,&& {\bf\Upsilon} \rightarrow \Upsilon,
  \Lambda^\Upsilon, &&
  {\bf {\cal V}} \rightarrow {\cal V}, \Theta_{\cal V}\,,&&\nonumber\\
&{\rm D3}-{\rm D3}: &&{\bf B} \rightarrow B, \Lambda^B, && \tilde {\bf
B} \rightarrow \tilde
  B, \Lambda^{\tilde B}\,.
\end{align}
Since we wish to obtain the action for the D3-D3 intersection at
the $\CC^2/\ZZ_k$ singularity in $(2,0)$ superspace, we
write the orbifold action (\ref{supe}), (\ref{orbac}) in $(2,0)$
superspace. In terms of the $(2,0)$ decomposition, the orbifold
acts as follows:
\begin{align} \label{orbifoldconstraints}
&{\rm D3}:&& Q_1 \rightarrow \xi g(\xi)
Q_1g^{\dagger}(\xi)\,,\quad &&\Lambda^{Q_1} \rightarrow g(\xi)
\Lambda^{Q_1} g^{\dagger}(\xi)\,, &&\nonumber
\\
&&&Q_2 \rightarrow g(\xi) Q_2 g^{\dagger}(\xi)\,,\quad &&\Lambda^{Q_2}
\rightarrow \xi^{-1}g(\xi)
\Lambda^{Q_2} g^{\dagger}(\xi)\,, &&\nonumber\\
&&&\Phi \rightarrow g(\xi)\Phi g^{\dagger}(\xi)\,,\quad
&&\Lambda^{\Phi}
\rightarrow \xi^{-1} g(\xi) \Lambda^{\Phi} g^{\dagger}(\xi)\,, &&\nonumber\\
&&&V \rightarrow g(\xi) V g^{\dagger}(\xi)\,,\quad &&\Theta_V \rightarrow
\xi^{-1} g(\xi)\Theta_V g^{\dagger}(\xi)\,,&&\nonumber\\
\nonumber\\
&{\rm D3}^{\prime}:&& S_1 \rightarrow \xi g'(\xi)
S_1g'^{\dagger}(\xi)\,,\quad &&\Lambda^{S_1} \rightarrow g'(\xi)
\Lambda^{S_1} g'^{\dagger}(\xi)\,,
&&\nonumber\\
&&&S_2 \rightarrow g'(\xi) S_2 g'^{\dagger}(\xi)\,,\quad
&&\Lambda^{S_2} \rightarrow \xi^{-1}g'(\xi) \Lambda^{S_2}
g'^{\dagger}(\xi)\,, &&\nonumber\\
&&&\Upsilon \rightarrow g'(\xi)\Upsilon g'^{\dagger}(\xi)\,,\quad
&&\Lambda^{\Upsilon}
\rightarrow \xi^{-1} g'(\xi) \Lambda^{\Upsilon} g'^{\dagger}(\xi)\,,
&&\nonumber\\
&&&{\cal V} \rightarrow g'(\xi) {\cal V} g'^{\dagger}(\xi)\,,\quad
&&\Theta_{\cal V}\rightarrow \xi^{-1} g'(\xi)\Theta_{\cal V}
g'^{\dagger}(\xi)\,,&&\nonumber\\
\nonumber\\
&{\rm D3}-{\rm D3}^{\prime}:&& B \rightarrow g(\xi) B
g'{}^\dagger (\xi) \,,\quad &&\Lambda^{B} \rightarrow \xi^{-1}
g(\xi) \Lambda^{B}
g'{}^\dagger(\xi)  \,,\quad  &&\nonumber\\
&&&\tilde B \rightarrow g'(\xi) \tilde B g{}^\dagger (\xi) \,,\quad
&&\Lambda^{\tilde B} \rightarrow \xi^{-1} g'(\xi) \Lambda^{\tilde B}
g^\dagger(\xi)  \,.&&
\end{align}
Each component of a $(2,0)$ superfield transforms under the orbifold
action in the same way as the $(2,0)$ superfield itself. Note that
this was not the case for $(2,2)$ superfields.  The degrees of freedom
which are invariant under (\ref{orbifoldconstraints}) together with
their $SU(N)^k \times SU(N^{\prime})^k$ gauge transformation
properties are summarized by the quiver diagram in
Fig.~\ref{superquiver}. The quiver consists of an inner and an outer
ring. Each of them is equivalent to the quiver diagram shown in
Fig.~\ref{quiver1} which provides the field content for the
(de)construction of the six-dimensional $(2,0)$ superconformal field
theory. We will see below that the spokes in the diagram, which
connect both rings, represent the degrees of freedom for the
(de)construction of a $\N=2, d=4$ field theory located at the M5-M5
intersection.

\begin{figure}[!ht]
\vspace{0.5cm}
\begin{center}
\includegraphics[height=19cm,clip=true,keepaspectratio=true]{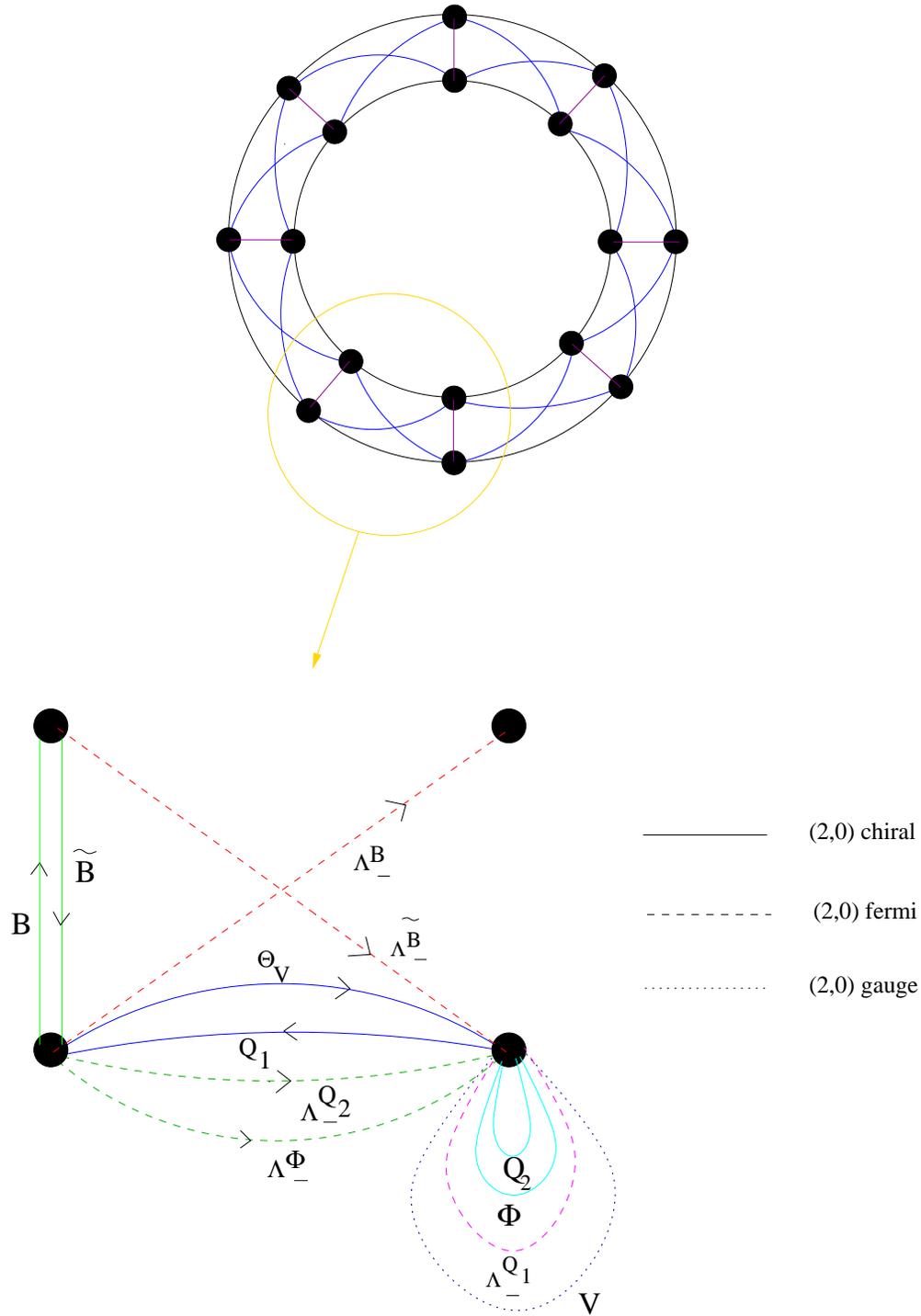}
\caption{Quiver diagram for intersecting D3-branes at a
$\CC^2/\ZZ_k$ orbifold (with $k$=8).  The nodes of the inner and
outer circle are associated with the $SU(N')^k$ and $SU(N)^k$
gauge groups respectively.  The parts which have not been drawn in
the detailed ``close-up'' are easily inferred from the $\ZZ_k$
symmetry and by swapping D3 degrees of freedom with D$3^{\prime}$
degrees of freedom.}\label{superquiver}
\end{center}
\end{figure}

We do not need the full action of the $(4,0)$ quiver theory. For now we just
give the $(2,0)$ term analogous to a superpotential, which will be all
that we require for most purposes.  Superpotentials of $(2,0)$ theories have
the generic structure
\begin{align}
W= \int d\theta^+ \sum_a \Lambda^a J_a(\Phi_i)|_{\theta^+ = 0} \,,
\end{align}
where $J_a(\Phi_i)$ is a holomorphic function of the chiral
superfields.  For
the D3-D3 intersection at a $\CC^2/\ZZ_k$ orbifold, this term descends
from the superpotential of the D3-D3 intersection in flat space which
is presented in $(2,0)$ superspace in App.~\ref{appSuperpot}. Upon
projecting out the degrees of freedom which are not invariant under
the orbifold (\ref{orbifoldconstraints}), one obtains the $(2,0)$
superpotential
\begin{align}
W=W_{\rm D3} + W_{\rm D3'} + W_{\rm D3-D3'}\,, \label{super}
\end{align}
where
\begin{align}
W_{\rm D3}=  \int d^2x d\theta^+ {\rm tr}_{N\times N}& \left(g
\Lambda^\Phi_{j,j+1}  ( Q^2_{j+1} Q^1_{j+1,j} \,- \,
   Q^1_{j+1,j} Q^2_{j})\right. \label{super1}\\
+ &\,\Lambda^{Q^1}_j[\partial_{\bar x}+g\Phi_{j},Q^2_{j}]\nonumber\\
+ &\, \left.\left.\Lambda^{Q^2}_{j,j+1}( - \partial_{\bar
x}Q^1_{j+1,j} - g Q^1_{j+1,j} \Phi_j
+ g \Phi_{j+1} Q^1_{j+1,j}) \right)\right\vert_{\bar\theta^+ = 0}
\,,\nonumber
\end{align}
\begin{align}
W_{\rm D3'}= \int d^2y
d\theta^+ {\rm tr}_{N^{\prime}\times N^{\prime}}& \left(g
\Lambda^\Upsilon_{j,j+1}  ( S^2_{j+1} S^1_{j+1,j} \,- \,
   S^1_{j+1,j} S^2_{j})\right. \label{super2} \\
+ &\,\Lambda^{S^1}_j[\partial_{\bar x}+g\Upsilon_{j},S^2_{j}]\nonumber\\
+ &\,\left.\left. \Lambda^{S^2}_{j,j+1}( - \partial_{\bar
x}S^1_{j+1,j} - g S^1_{j+1,j} \Upsilon_j
+ g \Upsilon_{j+1} S^1_{j+1,j}) \right)\right\vert_{\bar\theta^+ = 0}
\,,\nonumber
\end{align}
\begin{align}
W_{\rm D3-D3'} = g \int d\theta^+ {\rm tr}_{N\times N}& \left(
\Lambda^B_{j,j+1}(\tilde B_{j+1}Q^1_{j,j+1} - S^1_{j+1,j}
\tilde B_j) + \Lambda^{Q^1}_j B_j\tilde B_j \right) \label{superpotential}
\\
+ {\rm tr}_{N^{\prime} \times N^{\prime}} & \left.\left(
\Lambda^{\tilde B}_{j,j+1} (Q^1_{j+1,j}B_j - B_{j+1} S^1_{j+1,j})
- \Lambda^{S^1}_j \tilde B_j B_j \right)\right\vert_{\bar\theta^+
= 0} \,.\nonumber
\end{align}

\vspace{0.5cm}\mbox{}
\newpage

In order to see that this theory has indeed $(4,0)$ supersymmetry, we record
the basic structure of the $(4,0)$ multiplets which appear. These are as
follows:
\begin{itemize}
\item[i)] $(4,0)$ hypermultiplets composed of two $(2,0)$
chiral multiplets: There are five multiplets of this type
containing the pairs $(B,\tilde B), (\Phi,Q_2), (\Theta_V, Q_1),
(\Upsilon, S_2)$
and $(\Theta_{\cal V}, S_1)$. 

\item[ii)] $(4,0)$ vector multiplets composed of one $(2,0)$
vector multiplet and one $(2,0)$ Fermi multiplet: There are two
multiplets of this type containing the pairs $(V,
\Lambda^{Q_1})$ and $({\cal V}, \Lambda^{S_1})$.

\item[iii)] $(4,0)$ Fermi multiplets composed of
one\footnote{There is no need to add degrees of freedom to make a
$(4,0)$ Fermi multiplets out of a $(2,0)$ Fermi multiplet \cite{Garcia}.}
$(2,0)$ Fermi multiplet:  There are six multiplets of this type
corresponding to the $(2,0)$ Fermi multiplets $\Lambda^B$,
$\Lambda^{\tilde B}$, $\Lambda^{\Phi}$, $\Lambda^{Q_2}$,
$\Lambda^{\Upsilon}$ and
$\Lambda^{S_2}$. 
\end{itemize}

\noindent The transformation properties under the $(4,0)$
$SU(2)_L$ R-symmetry are readily obtained from Tab.~\ref{rsym} on page
\pageref{rsym}.  Note that the $SU(2)_L$ R-symmetry acts on the
degrees of freedom of either the inner or outer ring of the quiver
diagram as the $SU(2)$ R-symmetry of the associated ${\cal N} =2,
d=4$ theory.

In the following section, we shall make use of this superpotential to
discuss the (de)con\-struction of the M5-M5 intersection.

\subsubsection{(De)constructing the M5-M5 intersection}

The inner and outer circle of the quiver diagram in
Fig.~\ref{superquiver} are each separately equivalent to the quiver
diagram of Fig.~\ref{quiver1}, which (de)constructs the $(2,0)$ theory
upon taking the appropriate large $k$ limit
\cite{Arkani-Hamed}. The new twist here is that there are degrees
of freedom connecting the inner and outer rings. These are localized
at the intersection of the D3-branes, and it is natural to expect that
in the large $k$ limit, these correspond to the tensionless strings
localized at the intersection of M5-branes.  One reason to expect this
follows from a trivial extension of an argument given in
\cite{Arkani-Hamed} based on the IIB string theory embedding. As
discussed in Sec.~\ref{secString}, in the $k
\rightarrow\infty$ limit, the $\CC^2/\ZZ_k$ orbifold appears as a flat
$S^1 \times \RR^3$ geometry sufficiently far away from the orbifold
point (or sufficiently far out on the Higgs branch).  For intersecting
D3-branes, T-dualizing and lifting to M-theory on this space gives
rise to intersecting M5-branes wrapping a torus of fixed
dimensions. The strings stretched between the orthogonal D3-branes
then become membranes stretched between M5-branes, as shown in
Fig.~\ref{deconstruction}. In the following, we shall focus on the
field theoretic origins of the tensionless strings at the
intersection.

\begin{figure}[!h]
\begin{center}
\includegraphics[keepaspectratio, scale=0.9]{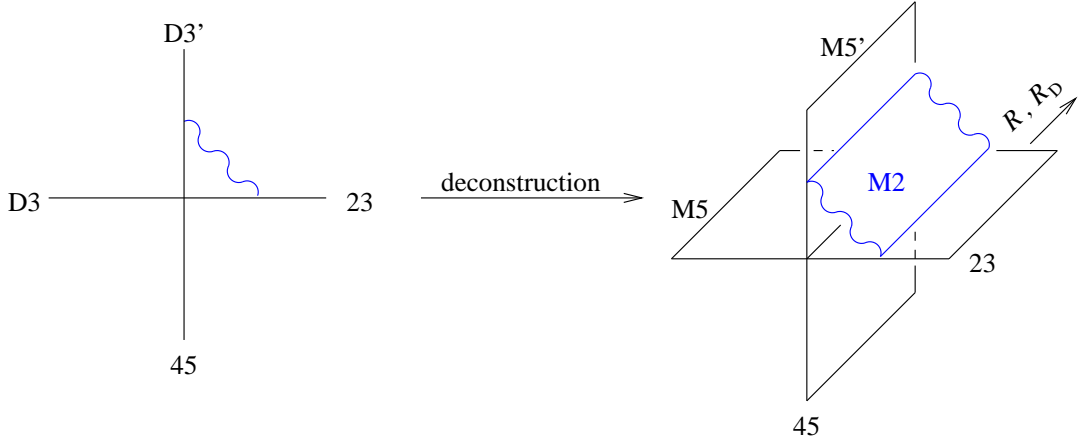}
\caption{(De)construction of two extra dimensions along the torus
with radii $R_5$ and $R_6$. The common directions $x^0$ and $x^1$ as
well as the four orbifold directions are suppressed.}\label{deconstruction}
\end{center}
\end{figure}

\subsubsection*{(De)constructing the $(2,0)$ theory}

Before discussing the strings localized at the intersection, we shall briefly
review the field theoretic arguments behind the (de)construction of the
six-dimensional $(2,0)$ theory discovered in \cite{Arkani-Hamed}. The quiver
diagram of the deconstructed theory is that of Fig.~\ref{quiver1},
which describes a superconformal ${\cal N} =2, d=4$ gauge theory with gauge
group $SU(N)^k$. The hypermultiplets described by the double lines stretched
between adjacent nodes contain two complex scalars in bifundamental
representations.  The quiver diagram can be viewed as a discretization of an
extra circular spatial dimension if one takes all the bifundamental scalars to
have the same non-zero expectation value.  At this point on the Higgs branch
the gauge symmetry is broken from $SU(N)^k$ to the diagonal $SU(N)$.

To make closer contact with our work,  we show how the extra
dimensions arise from the
${\cal N} =2, d=4$ theory using the language of two-dimensional
$(2,0)$ superspace.  Consider the term $W_{\rm D3}$ in the
superpotential (\ref{super}),  which involves only fields on the
outer ring of the quiver diagram. Deconstructing the
six-dimensional $(2,0)$ theory involves going to a particular
point on the Higgs branch of the ${\cal N} =2, d=4$ theory
described by the outer ring.  At this point $\langle q^1_{j+1,j}
\rangle = vI$ for all $j$, where $v$ is real and $q^1$ is the
scalar component of $Q^1$. One then has an effective
superpotential with the quadratic terms
\begin{align}
W_{\rm D3} = \int d^2x d\theta^+ {\rm tr}_{N\times N}& \left(g v
   \Lambda^\Phi_{j,j+1}(Q^2_{j+1}-Q^2_{j})
+  gv\Lambda^{Q^2}_{j,j+1}( \Phi_{j+1} - \Phi_{j} ) \right.
\nonumber \\
&\,+\left.\left.\Lambda^{Q^1}_j\partial_{\bar x}Q^2_{j} -
\Lambda^{Q^2}_{j,j+1}\partial_{\bar x}{\cal P}_{j+1,j}
\right)\right\vert_{\theta^+ = 0} \,, \label{three}
\end{align}
where\footnote{The field ${\cal P}_{j+1,j}$ can be interpreted as
part of a gauge connection in an extra spatial latticized
direction. Terms other than the superpotential must also be
included to see this.} ${\cal P}_{j+1,j} = v- Q^1_{j+1,j}$. The
first and second terms in (\ref{three}) can be viewed as kinetic
terms on a lattice with $k$ sites and lattice spacing $a =
1/{gv}$.  The bosonic kinetic terms arise upon integrating out
auxiliary fields.   From a two-dimensional point of view, the
first two terms in (\ref{three}) give rise to a mass
matrix\footnote{Strictly speaking, we must also include the
contribution to the mass matrix coming from terms other than the
superpotential.  These terms are related to those of the
superpotential by $(4,0)$ supersymmetry, and modify an overall
factor in the mass matrix.  } with eigenvalues
\begin{align}
m_l^2 = (gv)^2 |e^{2\pi i l/k} - 1|^2 \,. \label{spectrum}
\end{align}
For sufficiently large $k$,  this becomes a Kaluza-Klein spectrum
$m_l^2 = (l/R_5)^2$ with $R_5= \frac{k}{gv}$.

Yet another compact dimension is generated due to the S-duality of
the ${\cal N}=2$, $d=4$ gauge theory.  Under S-duality $g\rightarrow
k/g$ and one therefore expects a spectrum of S-dual states with
masses
\begin{align}
M_l^2 = \left (\frac{kv}{g} \right)^2|e^{2\pi i l/k} - 1|^2\,.
\end{align}
For large $k$ and fixed $n$ there is a Kaluza-Klein spectrum on an
S-dual circle of radius
\begin{align}
2\pi R_6 = \frac{g}{v}\,.
\end{align}
The continuum limit is obtained by taking $k\rightarrow \infty$
with $R_5, R_6$ fixed.  This requires that one goes to strong
coupling $g\sim \sqrt{k}$ and that one goes far out onto the Higgs
branch $v\sim \sqrt{k}$.

\subsubsection*{A note on stability of the spectrum}

The existence of the continuum limit is actually more subtle than the previous
discussion suggests since it includes a strong coupling limit.  Although
string theory indicates the limit should exist, a field theoretical argument
would have to demonstrate the validity of the semiclassical spectrum
(\ref{spectrum}) at strong coupling, and $k\rightarrow\infty$ at fixed $n$.
Strictly speaking, this spectrum is not a BPS mass formula for finite $k$,
since the ``charge'' $n$ is defined modulo $k$ and is therefore not a central
charge.  Assuming the existence of a continuum limit with enhanced
supersymmetry, the spectrum is BPS with respect to this enhanced
supersymmetry.  In \cite{Csaki:2002fy}, an argument that the supersymmetry
enhancement is robust at low energies was given by studying the Seiberg-Witten
curve of the ${\cal N}=2, d=4$ quiver gauge theory.

A further argument in favor of the stability of the spectrum at finite $k$ is
as follows.  Although the first two terms in (\ref{three}) are lattice kinetic
terms, they appear in the $(2,0)$ superpotential, which has a
holomorphic structure and is protected against radiative
corrections. If we were to work with four-dimensional ${\cal N}
=1$ superspace,  we would also find that the lattice kinetic terms
arise in part from the effective superpotential on the Higgs
branch.  In ${\cal N}=1, d=4$ superspace,  the superpotential is
\begin{align}
W = g \sum_{j=1}^k {\rm tr}\left( \Phi^1_j \,\Phi^2_{j,j+1}
\Phi^3_{j+1,j} - \Phi^1_{j+1}\Phi^3_{j+1,j}\Phi^2_{j,j+1}\right) \,.
\end{align}
The effective superpotential corresponding to lattice kinetic
terms is obtained on the Higgs branch by setting $\Phi^2_{j,j+1} =
v + \Gamma^2_{j,j+1}$ and $\Phi^3_{j+1,j} = v + \Gamma^3_{j+1,j}$.
The non-renormalization of the effective superpotential and terms
related to it by supersymmetry is crucial for the stability of the
spectrum (\ref{spectrum}) at large $g$, and to the existence of a
continuum limit.

Note that the non-renormalization of the lattice kinetic terms is
somewhat akin to the non-renormalization of the metric on the
Higgs branch of four-dimensional ${\cal N} =2$ gauge theories.  
The latter non-renormalization can be argued, albeit in an unconventional
way, by writing the action in two-dimensional $(2,2)$
superspace. The kinetic terms for the ${\cal N}=2, d=4$
hypermultiplet then arise partially from a $(2,2)$ superpotential
of the form $\epsilon_{ij}Q_i \partial_x Q_j$ as in
(\ref{action1}).

\subsubsection*{Strings at the intersection}

Let us now consider the same $k\rightarrow \infty$ limit as above
for the case in which there are orthogonal intersecting stacks of
D3-branes.  We will initially take the Higgs branch moduli for the
${\cal N}=2, d=4$ theories on the inner and outer ring of the
quiver to be equal,  such that  $\langle s^1_{j+1,j} \rangle
= vI_{N\times N}$ and $\langle q^1_{j+1,j}\rangle = vI_{N^{\prime}
\times N^{\prime}}$. In this case,
the inner and outer rings of the quiver can be expected to
separately (de)construct the six-dimensional $(2,0)$ theory
compactified on tori with the same dimensions.  However one must
also consider the strings stretching between the D3-branes, i.e.\
the ``spokes'' which connect the inner and outer rings of the
quiver. We shall now argue that these (de)construct tensionless
strings living at a four-dimensional intersection of the two
six-dimensional world-volumes.

The ``spoke'' degrees of freedom correspond to the $(2,0)$ chiral
fields $B_j,\tilde B_j$ and Fermi fields ${\Lambda^B}_{j,j+1},
{\Lambda^{\tilde B}}_{j,j+1}$,  which describe strings stretched
between the two stacks of D3-branes. For $\langle s^1_{j+1,j} \rangle=
vI_{N\times N}$ and $\langle q^1_{j+1,j}\rangle = vI_{N^{\prime} \times
N^{\prime}}$,  the quadratic part of the effective superpotential
is
\begin{align}\label{latkin}
W_{\rm D3-D3'} = gv\int d\theta^+ {\rm tr} \left. \left[
\Lambda^{\tilde B}_{j,j+1}(B_j - B_{j+1}) + (\tilde B_{j+1} -
\tilde B_j)\Lambda^B_{j,j+1}\right]\right\vert_{\bar\theta^+ = 0} \,
\end{align}
which follows from (\ref{superpotential}).  This can also be
viewed as a lattice kinetic term.  The same mass matrix arises for
the fundamental degrees of freedom at the intersection as for
those on the inner and outer circles of the quiver. Therefore
these degrees of freedom also carry momentum in an extra dimension
of radius $R_5$. The full theory is again expected to exhibit
S-duality, based on its embedding in string theory. Thus there
should also be S-dual degrees of freedom at the intersection which
carry momentum in an extra dimension of radius $R_6$. Dyonic
states carry momenta in both extra directions.  The precise nature
of degrees of freedom which are S-dual to the fundamental degrees
of freedom $B, \Lambda^B, \tilde B$ and~$\Lambda^{\tilde B}$
remains an open question at the moment. However, assuming
S-duality, the $k\rightarrow \infty$ limit generates two
six-dimensional world-volumes intersecting over four dimensions
from a theory with two four-dimensional world-volumes intersecting
over two dimensions. Note that the inner and outer rings of the
quiver do not see independent extra directions, since the apparent
$\ZZ_k \times \ZZ_k$ symmetry is broken to $\ZZ_k$ by couplings to
the degrees of freedom at the intersection.

The spoke degrees of freedom should be interpreted as tensionless
strings wrapping the compact directions rather than particles. To
see this, it is helpful to move the orthogonal stacks of D3-branes
to different points in the orbifold.  This corresponds to going to
different points on the Higgs branches of theories described by
the inner and outer rings of the quiver. For the inner ring the
Higgs branch is characterised by vevs for $\sigma_{j,j+1}$ and
$\bar q^1_{j,j+1}$ which form a doublet $Y_{j,j+1}$ of the
$SU(2)_L$ R-symmetry. Similarly the Higgs branch for the outer
ring is characterised by vevs for $\omega_{j,j+1}$ and
$s^1_{j,j+1}$ which also form a doublet $Y^{\prime}_{j,j+1}$ of
$SU(2)_L$.   Consider the following point in the moduli space:
\begin{align}
Y_{j,j+1} = \begin{pmatrix} v+\Delta/2 \cr v + \Delta/2 \end{pmatrix}
\,,\qquad
Y_{j,j+1}^{\prime} = \begin{pmatrix} v-\Delta/2 \cr v - \Delta/2
\label{sep}
\end{pmatrix}\quad.
\end{align}
where $\Delta$ is real.   One might worry that the extra
dimensions seen by the degrees of freedom on the inner and outer
rings of the quiver are no longer the same, since  at different
points on the Higgs branches, $\langle Y \rangle \ne \langle
Y^{\prime} \rangle$, the radii are apparently different. However
we shall keep $v\Delta$ fixed in the $k\rightarrow \infty$ limit
with $v\sim \sqrt{k}$. In this limit the deconstructed radii are
the same and correspond to the same spatial directions:
\begin{align}
&\lim_{k\rightarrow\infty} \frac{k}{g(v+\frac{\Delta}{2})} =
\lim_{k\rightarrow\infty} \frac{k}{g(v-\frac{\Delta}{2})} = R_5
\,, \nonumber \\
&\lim_{k\rightarrow\infty} \frac{g}{v+\frac{\Delta}{2}} =
\lim_{k\rightarrow\infty} \frac{g}{v-\frac{\Delta}{2}} = R_6 \,.
\end{align}
At the point in moduli space given in (\ref{sep}), the quadratic
part of the effective $(2,0)$ superpotential is
\begin{align}
  W &= gv \int d\theta^+ {\rm tr} \left.\left[ B_j({\Lambda^{\tilde
          B}_-}_{j,j+1} - {\Lambda^{\tilde B}_-}_{j-1,j}) +
      {\Lambda^B_-}_{j,j+1}(\tilde B_{j+1} - \tilde
B_j)\right]\right\vert_{\bar \theta^+} \nonumber \\
  &+ \,g\Delta\int d\theta^+ {\rm tr} \left.\left[B_j({\Lambda^{\tilde
          B}_-}_{j,j+1} + {\Lambda^{\tilde B}_-}_{j-1,j}) +
      {\Lambda^B_-}_{j,j+1}(\tilde B_{j+1} + \tilde
      B_j)\right]\right\vert_{\bar \theta^+} \label{massterm} \,.
\end{align}
The second term in (\ref{massterm}) is a mass term from the point
of view of the lattice theory.  For large $k$ and fixed $l$,
diagonalizing the mass matrix for the fundamental spoke degrees of
freedom gives
\begin{align}
m_l^2 = (g\Delta)^2 + (l/R)^2\,,
\end{align}
where the integer $n$ is the lattice momentum obtained by Fourier
transforming with respect to the index $j$ labeling points on the
quiver. For simplicity let us set $l=0$, so that $m = g\Delta$. The
S-dual modes then have $M=\frac{k}{g}\Delta$. Since $m/M =
R_6/R_5$, the fundamental spoke degrees of freedom should be
interpreted as strings wrapping the cycle of radius $R_6$, while their
S-duals wrap the cycle of radius $R_5$ (see
Fig.~\ref{wrappedstrings}). The string tension is
\begin{align}
T = \frac{m}{2\pi R_6} = \frac{M}{2\pi R_5} = v\Delta \,.
\end{align}
\bigskip

\begin{figure}[!ht]
\begin{center}
\includegraphics[height=6cm,clip=true,keepaspectratio=true]{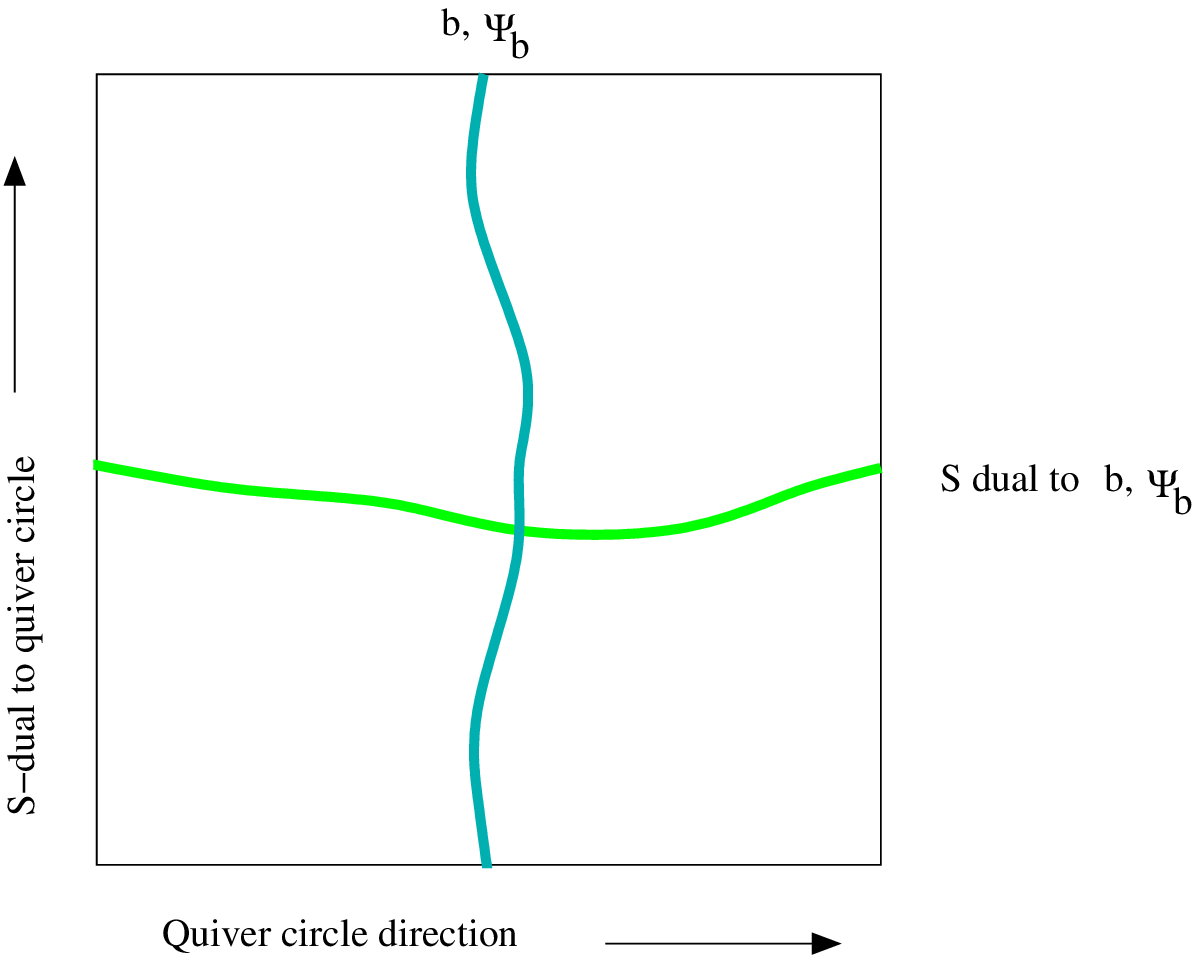}
\vspace{0.1cm}
\caption{Strings localized at the
intersection.}\label{wrappedstrings}
\end{center}
\vspace{-6.7cm} \hspace{4.0cm}
\rotatebox{90}{S-dual to quiver circle $\longrightarrow$} \\
\vspace{0.1cm}
\hspace{4.6cm} {Quiver circle direction $\longrightarrow$}

\vspace{-6.8cm} \hspace{6.9cm} $b, \psi_b$ \vspace{6.3cm}

\vspace{-4.1cm} \hspace{10.2cm} S-dual \\ \mbox{}\hspace{10.2cm} 
to $b, \psi_b$ 
\vspace{3.6cm}
\vspace{0.8cm}
\end{figure}

Note that the S-dual's to the fundamental degrees of freedom at
the intersection are strings wrapping the $S^1$ of the quiver
diagram. Thus it is tempting to speculate that they can built from
gauge invariant products of fundamental spoke degrees of freedom
which wrap the quiver.  An example of such an operator is ${\rm
tr\,} \Lambda^B_{1,2}\Lambda^{\tilde B}_{2,3} \cdots
\Lambda^B_{k-1,k}\Lambda^{\tilde B}_{k,1}$. On the other hand, one
expects the S-dual operators to be solitons without an expression
in terms of products of local operators, so this speculation is
probably not quite correct.

\subsubsection*{String condensation and M5-branes on a holomorphic curve}

When tensionless strings condense,  the M5-M5 intersection is
resolved to the holomorphic curve $xy =c$.  This can be seen very
explicitly from compactification on a torus.  In this case the low
energy theory is that of the D3-D3 intersection in flat space. In
Sec.~\ref{higgsbranch} we showed that the Higgs branch of the
corresponding $(4,4)$ dCFT can be interpreted as a resolution of
the intersection to the holomorphic curve $xy=c$. The resolved
intersection is also clearly captured by the $(4,0)$ dCFT.  At the
point in the moduli space for which extra dimensions are
generated,  the $(4,0)$ dCFT reduces to the $(4,4)$ dCFT at low
energies.  The potential is minimized by restricting to fields
with values independent of the quiver index $j$ and  satisfying
equations equivalent to (\ref{holomeq}).  The holomorphic curve
$xy=c$ arises when the fields $B_j$ and $\tilde B_j$ get
expectation values independent of $j$.  These fields correspond to
tensionless strings at the M5-M5 intersection.

\subsubsection*{Identifying R-symmetries and moduli}

The M5-M5 intersection has ${\cal N} =2, d=4$ supersymmetry with
$SU(2) \times U(1)$ R-symmetry.  We  would like to identify the
corresponding charges in the $(4,0)$ defect conformal field
theory.

For the $U(1)$ R-symmetry,  the identification is as follows. This
symmetry is manifest in both cases and corresponds to a
simultaneous rotation of the $x$ and $y$ planes, which are
transverse to one stack of parallel branes but not the orthogonal
stack.  In the $(4,0)$ dCFT, it is generated by $J_{23}-J_{45}$,
and the associated charges can be readily obtained from Tab.~\ref{rsym}.
Note that the other linear combination, $J_{23} + J_{45}$, is not
an R-symmetry, and acts trivially on the degrees of freedom
localized at the intersection.

We will now argue that the $SU(2)_L$ R-symmetry of the $(4,0)$ dCFT
should be identified with the $SU(2)$ R-symmetry of the ${\cal N} =2,
d=4$ theory of the M5-brane intersection. This matching is non-trivial
for the following reason. In order to generate the extra dimensions,
it was necessary to consider a point on the Higgs branch where the
$SU(2)_L$ doublets $\langle Y \rangle$ and $\langle Y^{\prime}
\rangle$ are non-zero, so that $SU(2)_L$ is spontaneously broken.  On
the other hand the $SU(2)$ R-symmetry of the M5-M5 intersection is
only broken when M5-branes are transversely separated. Nevertheless,
we shall find evidence that the identification makes sense.  This
suggests that when the M5-branes are not separated, the $SU(2)_L$
symmetry of the $(4,0)$ dCFT description is unbroken as far as the
non-trivial dynamics is concerned.

There are three directions transverse to both stacks of M5-branes,
corresponding to the moduli $\vec X$ and $\vec X^{\prime}$,  which
form a triplet under the $SU(2)$ R-symmetry.  This R-symmetry is
spontaneously broken if either $\vec X$ or $\vec X^{\prime}$ is
non-zero. However if all the eigenvalues of $\vec X$ and $\vec
X^{\prime}$ are the same,  then the symmetry breaking is due only
to trivial center of mass dynamics.  The string tension $T = |\vec
X- \vec X^{\prime}|$ vanishes at this point.

In the $(4,0)$ dCFT, the point in moduli space described by
(\ref{sep}) corresponds to a string tension $T= v\Delta$. If we
act with $SU(2)_L$,  we obtain another point in moduli space which
also deconstructs the same configuration of intersecting
M5-branes. The string tension can be written in an $SU(2)_L$
invariant way as the expectation value of $|Y^{\dagger}\vec\sigma
Y - {Y'}^\dagger \vec\sigma Y'|$ where $\vec \sigma$ are Pauli
matrices. We have dropped the $j,j+1$ subscript as we only
consider the zero momentum modes in the (de)constructed
directions.  On the other hand the string tension is related to
the moduli of the M5-M5 intersection by $T = |\vec X - \vec
X^{\prime}|$.  This motivates the proposal
\begin{align}
\vec X - \vec X^{\prime} \sim Y^{\dagger}\vec \sigma Y -
{Y'}^\dagger \vec\sigma Y' \,. \label{modreln}\end{align} Under
$SU(2)_L$, $Y^{\dagger}\vec \sigma Y- {Y'}^\dagger \vec\sigma Y'$
transforms as a triplet,  while $\vec X - \vec X^{\prime}$
transforms as triplet under $SU(2)$. This suggests that one should
identify the $SU(2)_L$ R-symmetry of the $(4,0)$ theory with the
$SU(2)$ R-symmetry of the M5-M5 intersection.

Thus far we have neglected a degree of freedom in the moduli space
which also contributes to the string tension. There are four
degrees of freedom in either $Y$ and or $Y'$, while only three are
characterised by $Y^{\dagger} \vec \sigma Y$ or ${Y'}^{\dagger}
\vec \sigma Y'$. Note that $Y^{\dagger} \vec \sigma Y$ is
invariant under $Y\rightarrow e^{i\theta} Y$, so the missing
degree of freedom is an angle.  The quantity
$|Y^{\dagger}\vec\sigma Y - {Y'}^\dagger \vec\sigma Y'| =
4vRe(\Delta)$ only gives the string tension for real $\Delta$. For
complex $\Delta$ the string tension is easily seen to be
$v|\Delta|$.  The imaginary part of $\Delta$ corresponds to the
additional angular degree of freedom. That the imaginary part is
an angle is evident from the orbifold condition $(u,\bar w) \sim
\exp(2\pi i/k) (u,\bar w)$ which for large $k$ gives $\Delta \sim
\Delta + 2\pi i v/k$.  By viewing the quiver action as an action
with only one extra discretized dimension (i.e.\ taking $R_6
\rightarrow 0$ with fixed $R_5$), one discovers that the angular
degree of freedom is a gauge connection in the compact discretized
fifth direction. If the associated Wilson lines differ for the two
intersecting branes, a mass term is generated for the degrees of
freedom localized at the intersection. In terms of the six
dimensional theories, this Wilson line may be interpreted as a
Wilson surface corresponding to the holonomy of the mysterious
non-abelian two-form on the torus.


 
\subsection{Deconstruction of M-theory on $T^3 \times A_{N-1}$} \label{Ch45}

The deconstruction of extra dimensions is also possible in theories including
gravity. Discrete gravitational extra dimensions were studied in
\cite{Arkani-Hamed:2003vb, Arkani-Hamed:2002sp, Schwartz:2003vj, Jejjala,
  Jejjala:2003qg}.  The basic idea is to consider \mbox{$d+1$}-dimensional
\mbox{Einstein} gravity as the low-energy effective theory of a
$d$-dimen\-sional gravitational theory with a discrete theory space.  The
continuum physics of the \mbox{$d+1$}-dimensional gravitational theory can be
reproduced correctly at energies parametrically higher than the
compactification scale. However, a peculiar UV/IR connection was found
forbidding the deconstruction all the way up to the $d+1$-dimensional Planck
scale.

As a special gravitational theory, it would be very interesting to deconstruct
M-theory itself.  In~\cite{Arkani-Hamed} it was proposed to deconstruct
M-theory on an $A_{N-1}$ singularity from a particular $(1, 0)$ little string
theory (LST).  This LST is defined as the decoupling limit of $N$ NS5-branes
at an orbifold singularity of the type $\CC^2/\ZZ_{k}$. A seventh dimension
arises on the Higgs branch of this theory. In a continuum ($k\rightarrow
\infty$) limit one expects to obtain a seven-dimensional gauge theory together
with its UV completion.  Exploiting string dualities for $k\rightarrow
\infty$, it was shown~\cite{Arkani-Hamed} that the stack of NS5-branes maps to
M-theory on $A_{N-1}$, which is a UV completion of the seven-dimensional gauge
theory.

A direct deconstruction seems to be impossible due to the obstructions to
finding a (conventional) Lagrangian description for LST.  Alternatively, one
could first deconstruct the NS5-brane theory out of the D3-brane theory at
$\CC^3/\ZZ_{N_5} \times \ZZ_{N_6}$ \cite{Arkani-Hamed}. One obtains a lattice action
for LST which could in principle be orbifolded again by projecting out degrees
of freedom which are not invariant under the (second) orbifold $\CC^2/\ZZ_k$.
Here one faces the problem that it is difficult, if not impossible, to find an
$SU(2)$ \mbox{R-symmetry} inside the lattice action into which the $\ZZ_k$
orbifold action can be embedded. Note that the $SO(4)$ \mbox{R-symmetry} of
the NS5-brane theory is only recovered in the continuum limit.  It is
therefore not obvious how to further discretize the latticized LST action.

In the following we apply the deconstruction method to M-theory following a
slightly different approach. We deconstruct M-theory directly from a
four-dimensional non-super\-symmetric quiver gauge theory with gauge group
$SU(N)^{N_4N_6N_8}$ and $N_{4,6,8}$ three positive integers. The corresponding
orbifold realization is given by a stack of D3-branes in type IIB string
theory placed at the origin of $\CC^3/\Gamma$, where the orbifold group
$\Gamma$ is the product of three cyclic groups $\ZZ_{N_4} \times \ZZ_{N_6}
\times \ZZ_{N_8}$. These groups generate three circular orbits in the
directions 468.  The quiver diagram is a three-dimensional body-centred cubic
lattice. At a certain point in the moduli space, each of the $\ZZ_{N}$ factors
generates a circular discretized extra dimension.  In an appropriate
$N_{4,6,8} \rightarrow \infty$ limit, the extra dimensions become continuous,
such that the theory appears to be seven-dimensional on the Higgs branch.
There is however a peculiarity in this deconstruction which suggests that the
strongly coupled Higgs branch theory is actually an eleven-dimensional
gravitational theory: The deconstructed seven-dimensional gauge theory has a
UV completion in terms of M-theory on $A_{N-1}$.  In the brane realization of
the present deconstruction, M-theory on $A_{N-1}$ arises naturally in the
continuum limit.


The Higgs branch of the quiver theory corresponds to the decoupling limit of
\mbox{D3-branes} a finite distance away from the orbifold singularity. In the
limit which we will consider the D3-branes probe an approximate \mbox{$\RR^3
  \times T^3$} geometry.  The generation of three extra dimensions along the
Higgs branch corresponds to T-dualizing along the three circular dimensions of
the three-torus $T^3$, giving D6-branes wrapped on $T^3$. The
seven-dimensional gauge theory living on the D6-branes does not decouple from
the bulk degrees of freedom, such that the deconstructed theory is not just a
seven-dimensional gauge theory. Due to a strong type IIA string coupling $g_s$
a better description is obtained by lifting to M-theory. The D6-branes in type
IIA string theory lift to M-theory on an $A_{N-1}$ singularity.  This suggests
the equivalence of M-theory on $A_{N-1}$ with the continuum limit of the
present quiver theory.


\subsubsection{D3-branes at $\CC^3/\ZZ_{N_4} \times \ZZ_{N_6} \times 
  \ZZ_{N_8}$: orbifold \mbox{realization} of a non-super\-sym\-metric quiver
  theory}

In this section we discuss the quiver theory, from which we deconstruct
M-theory. The quiver theory is a four-dimensional non-super\-symmetric field
theory with gauge group $SU(N)^{N_4N_6N_8}$. This theory describes the
decoupling limit of $NN_4N_6N_8$ D3-branes in type IIB string theory
placed at a $\CC^3/\Gamma$ orbifold singularity with $\Gamma \equiv \ZZ_{N_4}
\times \ZZ_{N_6} \times \ZZ_{N_8}$. 

The orbifold action on the complex coordinates $z_i=(h, v, n)$ parameterizing
$\CC^3$ is given by
\begin{align}  \label{orbaction}
h \rightarrow \xi_4^{\,2} h\,,\qquad
v \rightarrow \xi_6^{\,2} v\,,\qquad
n \rightarrow \xi_8^{\,2} n\,,
\end{align}  
where the generators of the groups $\ZZ_{N_a}$ are defined by $\xi_a =\exp(2\pi
i/N_a)$, $a=4, 6, 8$. Each of the factors $\ZZ_{N_a}$ ($a=4,6,8$) acts on one
of the three complex planes transverse to the stack of
D3-branes. The orbifold action can be embedded into the subgroup $U(1)^3$
of the rotational group $SO(6)$.

The fields of the quiver theory descend from the $\N=4, d=4$ vector multiplet
of the parent super Yang-Mills theory with gauge group $U(N N_4N_6N_8)$.
We project out degrees of freedom which are not
invariant under the orbifold group. The action of the product orbifold on
the gauge field $A_\mu$, the three scalars $\phi^i=(h,v,n)$, and
the four spinors $\psi^i=(\psi^h, \psi^v, \psi^n, \lambda)$ is given by
\begin{align}
  A_\mu &\rightarrow g(\xi) A_\mu g(\xi)^{-1} \,, \\
  \phi^i &\rightarrow \xi^{b^{(4)}_i}_4 \xi^{b^{(6)}_i}_6 \xi^{b^{(8)}_i}_8
            g(\xi) \phi^i g(\xi)^{-1} \,, \\
  \psi^i &\rightarrow \xi^{a^{(4)}_i}_4 \xi^{a^{(6)}_i}_6 \xi^{a^{(8)}_i}_8
            g(\xi) \psi^i g(\xi)^{-1} \,,
\end{align}
where $g(\xi)=g(\xi_4) \otimes g(\xi_6) \otimes g(\xi_8)$ is the
regular representation of the generator $\xi=\xi_4 \xi_6 \xi_8$ of $\Gamma$.
These transformation rules extend the invariance conditions given by
Eqns.~(\ref{transf1})-(\ref{transf3}).
For the vectors $a_i$ and $b_i$ we choose:
\begin{align}
 \ZZ_{N_4}:\quad a^{(4)}_i&=(-1,1,1,-1) \,,\quad b^{(4)}_i=(2,0,0) \,,
 \label{tuple1}\\
 \ZZ_{N_6}:\quad a^{(6)}_i&=(1,-1,1,-1) \,,\quad b^{(6)}_i=(0,2,0) \,,\\
 \ZZ_{N_8}:\quad a^{(8)}_i&=(1,1,-1,-1) \,,\quad b^{(8)}_i=(0,0,2) \,.
 \label{tuple3}
\end{align}

Each of the three pairs of vectors ($a_i, b_i$) gives rise to an orbifold
$\CC/\ZZ_N$ of the type given by Eq.~(\ref{nonsusy}). Together they define the
non-supersymmetric orbifold $\CC^3/\ZZ_{N_4} \times \ZZ_{N_6} \times
\ZZ_{N_8}$. The vectors $b_i$ determine the action (\ref{orbaction}) on the
coordinates $z_i=(h, v, n)$ via 
\begin{align}
z_i \rightarrow \xi^{b^{(4)}_i}_4 \xi^{b^{(6)}_i}_6 \xi^{b^{(8)}_i}_8 z_i\,.
\end{align}
The vectors $a_i$ give the corresponding action on the four fermions. The
invariant fermions $\psi^m_{i,j,k}$ transform under the gauge group
\begin{align}
  SU(N)^{N_4 N_6 N_8}
\end{align}
as $({\rm N}_{i,j,k}, \overline{\rm N}_{i\pm a^{(4)}_m, j\pm a^{(6)}_m, k \pm
  a^{(8)}_m})$, where N$_{i,j,k}$ ($\overline{\rm N}_{i',j',k'}$) denotes the
(anti-)funda\-mental representation of the gauge group $SU(N)_{i,j,k}$
($SU(N)_{i',j',k'}$). The invariant scalars $\phi^m_{i,j,k}$ transform as
$({\rm N}_{i,j,k}, \overline{\rm N}_{i\pm b^{(4)}_m, j\pm b^{(6)}_m, k\pm
  b^{(8)}_m})$. We summarize the field content of our quiver theory in the
Tab.~\ref{tab1}.

\begin{table}[!ht] 
\begin{center}
\begin{tabular}{|l|l||l|l|} 
\hline
  field & representation & field & representation \\
\hline
  $h_{i,j,k}$ & $({\rm N}_{i,j,k}, \overline{\rm N}_{i+2,j,k})$ &
  $\psi^h_{i,j,k}$  & $({\rm N}_{i,j,k}, \overline{\rm N}_{i+1,j-1,k-1})$\\
  $v_{i,j,k}$ & $({\rm N}_{i,j,k}, \overline{\rm N}_{i,j+2,k})$ &
  $\psi^v_{i,j,k}$  & $({\rm N}_{i,j,k}, \overline{\rm N}_{i-1,j+1,k-1})$\\
  $n_{i,j,k}$ & $({\rm N}_{i,j,k}, \overline{\rm N}_{i,j,k+2})$ &
  $\psi^n_{i,j,k}$  & $({\rm N}_{i,j,k}, \overline{\rm N}_{i-1,j-1,k+1})$\\
  $A^\mu_{i,j,k}$ & {\rm adjoint} &
  $\lambda_{i,j,k}$ & $({\rm N}_{i,j,k}, \overline{\rm N}_{i+1,j+1,k+1})$\\
 \hline
\end{tabular}
\caption{Fields in the quiver theory and their transformation behaviour
under the gauge group $SU(N)^{N_4 N_6 N_8}$.} \label{tab1}
\end{center}
\end{table}

\begin{figure}[!ht]
\begin{center}
{\includegraphics[scale=.75]{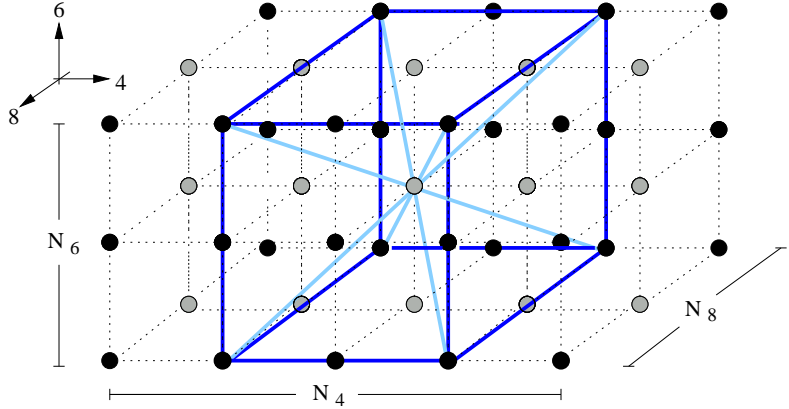}}
\caption{Theory space for the $\ZZ_{N_4} \times \ZZ_{N_6} \times
  \ZZ_{N_8}$ quiver theory. Dark and light (blue) lines in the basic cell
  represent bosonic and fermionic bifundamentals. Dotted lines are not
  physical and are just to guide the eye.} \label{box3}
\end{center}
\end{figure}

The theory space is a three-dimensional lattice with $N_4N_6N_8$ sites which
discretizes a three-dimensional torus $T^3$ as shown in Fig.~\ref{box3}.  Each
site represents one of the gauge groups $SU(N)_{i,j,k}$ and its associated
gauge boson~$A^\mu_{i,j,k}$. Link fields start at a site $i,j,k$, where they
transform in the fundamental representation ${\rm N}_{i,j,k}$, and end at a
site $i',j',k'$, where they transform in the anti-fundamental representation
$\overline{\rm N}_{i',j',k'}$. Fig.~{\ref{box3} shows the unit cell of the
  lattice spanned by the link fields. The bosonic bifundamentals $h_{i,j,k}$,
  $v_{i,j,k}$, $n_{i,j,k}$ (dark lines) form the edges of the unit cell,
  while the fermionic bifundamentals $\psi^h_{i,j,k}, \psi^v_{i,j,k},
  \psi^n_{i,j,k}, \lambda_{i,j,k}$ (light lines) connect the corners
  with the centre of the cell. Translating the unit cell in the lattice
we obtain a  body-centred cubic lattice which is invariant under the 48 element
octahedral symmetry group~$O_h$. Such bcc lattices were also studied in the
context of four-dimensional $\N=4$ super Yang-Mills theory on a
three-dimensional lattice \cite{Kaplan}.

Let us now construct the Lagrangian for our orbifold model which consists of
three parts,
\begin{align}
  {\cal L}={\cal L}_{\rm kin} + {\cal L}_{\rm Yuk} + {\cal L}_{\rm quartic}
  \,.  \label{action}
\end{align}
This Lagrangian follows from four-dimensional $\N=4$ super
Yang-Mills theory with gauge group $U(NN_4N_6N_8)$ upon projecting out
degrees of freedom which are not invariant under the orbifold
group. The kinetic terms have the form
\begin{align}
  {\cal L}_{\rm kin} \supset \frac{1}{2}{\rm Tr} (D_\mu
    \varphi_{i',j',k'})^\dagger D^\mu \varphi_{i,j,k} \,,
\end{align}
where the field $\varphi_{i,j,k}$ is one of the seven bifundamentals
transforming in the \linebreak $({\rm N}_{i,j,k}, \overline{\rm
  N}_{i',j',k'})$ as listed in Tab.~\ref{tab1}. The covariant derivative of
$\varphi_{i,j,k}$ is defined by
\begin{align}
  D_\mu \varphi_{i,j,k} = \partial_\mu \varphi_{i,j,k} - i g A^{i,j,k}_\mu
  \varphi_{i,j,k} + i g \varphi_{i,j,k} A^{i',j',k'}_\mu \,.
\end{align}

\begin{figure}
\begin{center}
{\includegraphics[scale=.75]{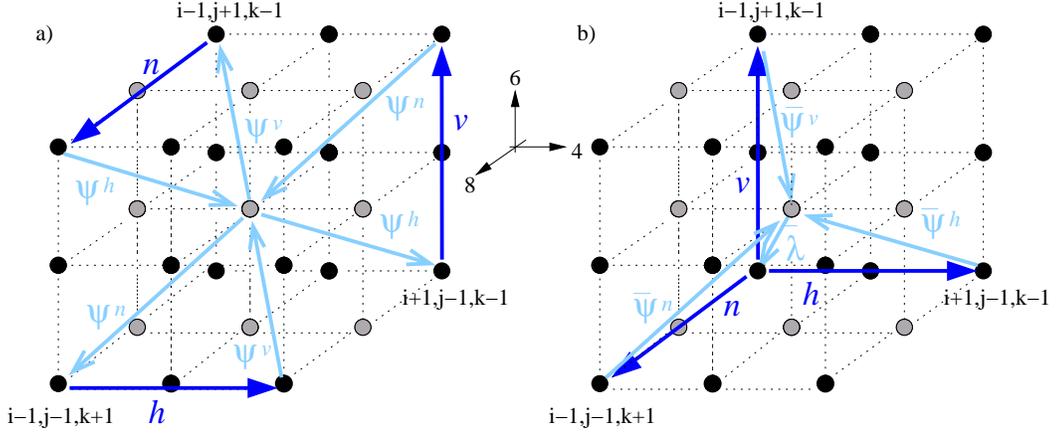}}
\caption{Oriented link fields in the basic cell. The site in the
          centre of the cell has labels $i, j, k$.  The triangles represent
          possible Yukawa couplings in the quiver action.} \label{box}
\end{center}
\end{figure}
We now consider the Yukawa and quartic scalar interactions ${\cal L}_{\rm
  Yuk}$ and ${\cal L}_{\rm quartic}$.  Fig.~\ref{box} shows six of twelve
possible triangles inside the basic cell.  These triangles consist of two
fermionic and one bosonic arrow and correspond to Yukawa couplings in the
quiver theory.  Each of the twelve triangles leads to a Yukawa term in the
action.  

For the Lagrangian ${\cal L}_{\rm Yuk}$ we thus find
\begin{align}
{\cal L}_{\rm Yuk}={\cal L}^1_{\rm Yuk}+{\cal L}^2_{\rm Yuk} \,,\label{Yukawa}
\end{align}
with
\begin{align}
  \!\!\!\!\!\!{\cal L}^1_{\rm Yuk}&= i\sqrt{2} g {\,\rm Tr}(\psi^v_{i,j,k}
  n_{i-1,j+1,k-1} \psi^h_{i-1,j+1,k+1}
  - \psi^n_{i,j,k} v_{i-1,j-1,k+1} \psi^h_{i-1,j+1,k+1} \nonumber\\
  &\,+ \psi^n_{i,j,k} h_{i-1,j-1,k+1} \psi^v_{i+1,j-1,k+1}
  - \psi^h_{i,j,k} n_{i+1,j-1,k-1} \psi^v_{i+1,j-1,k+1} \\
  &\,+ \psi^h_{i,j,k} v_{i+1,j-1,k-1} \psi^v_{i+1,j+1,k-1} - \psi^v_{i,j,k}
  h_{i-1,j+1,k-1} \psi^n_{i+1,j+1,k-1} + c.c.) \nonumber
\end{align}
and 
\begin{align}
  \!\!\!\!\!\!{\cal L}^2_{\rm Yuk}&= i\sqrt{2} g {\,\rm Tr}(\bar\lambda_{i,j,k} h_{i-1,j-1,k-1} \bar\psi^h_{i+1,j-1,k-1}
   - \bar \psi^h_{i,j,k} h_{i-1,j+1,k+1}  \bar \lambda_{i+1,j+1,k+1}\nonumber\\
  &+ \bar\lambda_{i,j,k} v_{i-1,j-1,k-1} \bar \psi^v_{i-1,j+1,k-1} 
   - \bar \psi^v_{i,j,k} v_{i+1,j-1,k+1} \bar\lambda_{i+1,j+1,k+1}  \\
  &+ \bar\lambda_{i,j,k} n_{i-1,j-1,k-1} \bar \psi^n_{i-1,j-1,k+1} 
   - \bar \psi^n_{i,j,k} n_{i+1,j+1,k-1} \bar\lambda_{i+1,j+1,k+1} + c.c.)
   \nonumber \,,
\end{align} 
where summation over $i,j,k$ is understood. The terms with a positive sign in
${\cal L}^{1}_{\rm Yuk}$(${\cal L}^{2}_{\rm Yuk}$) correspond to the triangles
in Fig.~\ref{box}a(b) (those with negative sign are not shown).  The Yukawa
couplings in ${\cal L}^{1}_{\rm Yuk}$ descend from the $\N=4$ superpotential
$[H,V]N$, while those in ${\cal L}^{2}_{\rm Yuk}$ come from the K\"ahler
potential.

 Quite analogously, squares in the quiver diagram represent quartic
scalar terms which are given by\footnote{There are some additional terms
  contributing to ${\cal L}_{\rm quartic}$: For brevity we did not list terms
  corresponding to degenerate squares coming from D-terms like $D\bar hh+D\bar
  vv+ D\bar nn$.}
\begin{align}
\!\!\!\!\!\!{\cal L}_{\rm quartic}  
&=g^2 {\,\rm Tr}(n_{i,j,k}h_{i,j,k+2}\bar{n}_{i+2,j,k+2}\bar{h}_{i+2,j,k}
 - h_{i,j,k}n_{i+2,j,k}\bar{h}_{i+2,j,k+2}\bar{n}_{i,j,k+2}\nonumber\\
&+h_{i,j,k}v_{i+2,j,k}\bar{h}_{i+2,j+2,k}\bar{v}_{i,j+2,k}
 -v_{i,j,k}h_{i,j+2,k}\bar{v}_{i+2,j+2,k}\bar{h}_{i+2,j,k}\\
&+v_{i,j,k}n_{i,j+2,k}\bar{v}_{i,j+2,k+2}\bar{n}_{i,j,k+2}
 -n_{i,j,k}v_{i,j,k+2}\bar{n}_{i,j+2,k+2}\bar{v}_{i,j+2,k})\nonumber
  \,. \label{scalarterms}
\end{align} 

This model belongs to the class of conformal non-supersymmetric orbifold
models studied in~\cite{Lawrence, Bershadsky}. In these models it is assumed
that the orbifold group $\Gamma \subset SU(4)$ acts solely on the transverse
space $\CC^3$ of $M$ parallel D3-branes. Kachru and Silverstein \cite{Kachru}
noticed that the orbifold group acts only on the $S^5$ factor of the near
horizon geometry \mbox{$AdS_5 \times S^5$}. In the AdS/CFT correspondence the
isometry group of the $AdS_5$ space is identified with the conformal group of
the field theory on its boundary.  This implies classical conformal invariance
of the world-volume theory on the D3-branes. In \cite{Lawrence} it was shown
that if $M$ is finite and the regular representation of $\Gamma$ is chosen,
the one-loop beta functions for the gauge couplings vanish in these theories.
In the large $M$ limit one can even prove the vanishing of the beta functions
to all orders in perturbation theory \cite{Bershadsky}. This holds for the
present quiver theory which we consider in the limit $N_{4,6,8} \rightarrow
\infty$ such that $M=N\,N_4N_6N_8 \rightarrow \infty$.\footnote{ The basic
  requirement for vanishing beta function, Eq.~(2.7) in \cite{Bershadsky}, is
  satisfied: Let $\gamma^a \in \Gamma \equiv \ZZ_{N_4} \times \ZZ_{N_6} \times
  \ZZ_{N_8}$ and $\gamma^a_{4,6,8} \in \ZZ_{N_{4,6,8}}$. Then
\begin{align} 
  {\rm tr}(\gamma^a)&=(1+\gamma_4^a+...+(\gamma_4^a)^{N_4-1})
  (1+...+(\gamma_6^a)^{N_6-1})(1+...+(\gamma_8^a)^{N_8-1})
  N \nonumber\\
  &= \left\lbrace \begin{tabular}{l} $N_4N_6N_8 N$ if $\gamma^1=1$\\
      $0$ $\forall \gamma^a$, $a \neq 1$ \end{tabular} \right. .  \nonumber
\end{align}
} Our non-supersymmetric quiver theory
has therefore quantum conformal invariance. As discussed in detail in
\cite{Arkani-Hamed} conformal invariance guarantees that the quiver theory remains in
the Higgs phase even at strong coupling.

A related question to that of conformal invariance is the stability of
the moduli space. Since the theory is not supersymmetric the potential
for the scalars is not necessarily protected against quantum
corrections.  This would change the moduli space of the classical
theory. However, as shown in \cite{Bershadsky} all Feynman diagrams in
the quiver theory are the same as in the $\N=4$ parent $U(M)$ gauge
theory up to possible $\frac{1}{M}$ corrections. In the large $M$
limit these corrections are suppressed and the potential remains
unchanged.\footnote{Note that it is not necessary to send $N
\rightarrow \infty$ in order to take $M=N N_4N_6N_8 \rightarrow
\infty$.} Although this points to a stable moduli space, we might still
have troubles with divergencies coming from the twisted closed string sector.

The question of stability of our model is highly non-trivial.  In contrast to
supersymmetric orbifold models there are {\em closed string tachyons} in the
twisted sector of non-supersymmetric orbifolds of the type $\CC/\ZZ_N$.  The
tachyon condensation leads to the decay of the orbifold as studied in
\cite{Adams}.  The initial effect of the tachyons is to smooth out the
orbifold singularity.  An RG flow is initiated by the tachyon and the orbifold
decays to flat space. If the initial state has been appropriately fine-tuned,
the orbifold decay can take place in a series of transitions $\CC/\ZZ_N
\rightarrow \CC/\ZZ_{N-2}$.  For finite $N$ the orbifold becomes flat in a
finite time. However, in the limit $N \rightarrow \infty$ the orbifold does
not decay in a {\em finite} amount of time, since the orbifold goes through
only finitely many transitions $\CC/\ZZ_N \rightarrow \CC/\ZZ_{N-2}$.  The
orbifold may however decay faster, e.g.~via transitions $\CC/\ZZ_N \rightarrow
\CC/\ZZ_{N-M}$ ($M>2$). If the quotient $M/N$ vanishes in the large $N$ limit
the orbifold remains stable.  In other words, the question is whether the
flattening of spacetime induced by the tachyon condensation outweighs the
extreme curvature at the singularity.  We believe it does not and presume that
our model is stable in the large $N_{4,6,8}$ limit. However, this issue
deserves some further investigation.

Another essential feature in the deconstruction of M-theory is S-duality of
the orbifold model.  In \cite{Lawrence} it was argued that in a conformal
(non-super\-sym\-metric) quiver theory the complex moduli $\tau_i$ are
inherited from the coupling $\tau_{\rm par}$ of the $\N=4$ parent theory
(recall $\tau_i=\tau_{\rm par}/\vert \Gamma\vert$). In the present quiver
theory the $N_4N_6N_8$ gauge couplings $\tau_{i,j,k}$ associated with the
gauge groups $SU(N)_{i,j,k}$ are all the same and related to $\tau_{\rm par}$
by
\begin{align}
 \tau \equiv \tau_{i,j,k}=\frac{\tau_{\rm par}}{N_4N_6N_8} \,.
\end{align}
The strong-weak duality $g_{\rm par} \rightarrow 1/g_{\rm par}$ thus
amounts to an $SL(2,\ZZ)$ S-duality \linebreak \mbox{$g \rightarrow N_4 N_6 N_8/g$} in
the quiver theory.

\subsubsection{Generation of three compact extra dimensions
  in the low-energy effective field theory}

We now show by studying the mass spectrum of the gauge bosons that the quiver
theory generates three circular extra dimensions at low energies.  On the
Higgs branch of the theory the scalar bifundamentals have expectation
values,
\begin{align} \label{diagvevs}
    \langle h_{i,j,k} \rangle=v_4\,, \qquad \langle v_{i,j,k} \rangle=v_6\,,
    \qquad \langle n_{i,j,k} \rangle=v_8 \,,
\end{align}
independent of $i,j,k$. These condensates break the gauge group
$SU(N)^{N_4N_6N_8}$ down to the diagonal subgroup $SU(N)$.  Upon substituting
the {\em vevs} $v_4, v_6, v_8$, the scalar kinetic terms inside ${\cal L}_{\rm
  kin}$ give rise to mass terms for the gauge bosons,
\begin{align}
  {\rm Tr} \vert D^\m h_{i,j,k} \vert^2 
  &=  g^2v^2_4 (A_\mu^{i,j,k} - A_\mu^{i+2,j,k})^2  
   \equiv A_\mu^{i,j,k} ({\cal M}_4)^2_{ii'} \delta_{jj'} \delta_{kk'} 
   A^\mu_{i',j',k'}\nonumber\\
 {\rm Tr} \vert D^\m v_{i,j,k} \vert^2
  &=  g^2v^2_6 (A_\mu^{i,j,k} - A_\mu^{i,j+2,k})^2 
   \equiv A_\mu^{i,j,k} \delta_{ii'} ({\cal M}_6)^2_{jj'} \delta_{kk'}
   A^\mu_{i',j',k'}\\
 {\rm Tr} \vert D^\m n_{i,j,k} \vert^2
  &=  g^2v^2_8 (A_\mu^{i,j,k} - A_\mu^{i,j,k+2})^2 
   \equiv A_\mu^{i,j,k} \delta_{ii'} \delta_{jj'} ({\cal M}_8)^2_{kk'} 
   A^\mu_{i',j',k'} \,.\nonumber
\end{align}
The matrices ${\cal M}_{4,6,8}$ have entries $2$ on the diagonal and $-1$ on
the second off-diagonal.  As in \cite{Arkani-Hamed:2001tb} diagonalization
of the mass matrices ${\cal M}_{4,6,8}$ yields the mass eigenvalues
\begin{align}
m^k_{4,6,8} = 2 g v_{4,6,8} \sin \frac{2 \pi k}{N_{4,6,8}} 
\approx 2 g v_{4,6,8}\frac{2 \pi k}{N_{4,6,8}} \quad{\rm for\,\,} 
k \ll N_{4,6,8}\,. 
\end{align}
For small enough $k$, this approximates the Kaluza-Klein spectrum of a
seven-dimen\-sional gauge boson compactified on a three-torus $T^3$ with radii
$R_{4,6,8}$.  The radii $R_{4,6,8}$ are fixed by the mass scales of the
lightest KK modes ($k=1$) which are given by
\begin{align}
m_{4,6,8} 
= \frac{1}{R_{4,6,8}} \,,\label{mass}
\end{align}
with
\begin{align} \label{radii}
2 \pi R_4 = N_4 a_4 = \frac{N_4}{2gv_4} \,, \quad
2 \pi R_6 = N_6 a_6 = \frac{N_6}{2gv_6} \,,\quad
2 \pi R_8 = N_8 a_8 = \frac{N_8}{2gv_8} \,
\end{align}
and $a_{4,6,8}$ the lattice spacings. 
In principle, this  KK spectrum is not protected and could receive quantum
corrections at strong coupling. In a similar context \cite{Arkani-Hamed} it was argued
that such quantum corrections are proportional to $\frac{1}{N_{4,6,8}}$ and
vanish in the large $N_{4,6,8}$ limit.  Provided this is true, we explicitly
deconstruct three compact extra dimensions with radii $R_{4,6,8}$. 

The Higgs mechanism does induce masses both for the gauge bosons as well
as for the bifundamental fermions and scalars. For instance, substituting {\em
  vevs} for the scalars inside the Lagrangian ${\cal L}_{\rm Yuk}$ leads to
fermionic mass terms. Such mass terms could in principle lead to a different
mass spectrum due to the non-supersymmetric nature of our model. Following
\cite{Csaki1} one can however verify that the fermionic mass spectrum is
identical to the gauge boson spectrum.  Both the bosonic as well as the
fermionic Kaluza-Klein spectra generate the same extra dimensions.

\subsubsection{M2-branes on the Higgs branch}

By studying the orbifold geometry we show in the next section that the Higgs
branch theory is equivalent to M-theory on an $A_{N-1}$ singular geometry. We
have seen that the Higgs branch theory contains seven-dimensional super
Yang-Mills theory with gauge group $SU(N)$.  In M-theory on $A_{N-1}$ the
seven-di\-men\-sional $SU(N)$ gauge symmetry arises from M2-branes wrapping
collapsed two-cycles at the singularity, see e.g.~\cite{Sen}. In other words,
the deconstructed 7d super Yang-Mills theory is part of M-theory on $A_{N-1}$.
However, \mbox{M-theory} contains more than just the 7d gauge theory. We have
to verify also the existence of M2- or M5-branes inside the quiver theory.

The states corresponding to M2-branes can be seen in the dyonic spectrum of
the quiver gauge theory.  The dyonic mass spectrum follows from that of the
gauge bosons by S-duality.  Substituting
\begin{align}
 g \rightarrow \frac{N_4N_6N_8}{g} \,
\end{align}
into Eq.~(\ref{mass}), we find for the lowest dyonic states the masses
\begin{align}
M_4 = 8\pi^3 R_6 R_8/g^2_7 \,,\qquad M_6 = 8\pi^3 R_8 R_4/g^2_7 \,,\qquad
M_8 = 8\pi^3 R_4 R_6/g^2_7 \,,
\end{align}
where the seven-dimensional coupling constant is $g_7^2 = a_4 a_6 a_8 g^2$.
These masses are identical to those of two-branes wrapping around
two-tori~$T^2$ inside the $T^3$. We can read off the tension of the
two-branes, {$T_{2}= 1/(2\pi)^2 g_7^2$}, which is identical to the tension of
M2-branes, $T_{\rm M2} = 1/(2\pi)^2 l_p^3$.  This gives field-theoretical
evidence that we really deconstruct M-theory.\footnote{We cannot see M5-branes
  in this way.  The theory we expect to deconstruct is \mbox{M-theory} on the
  geometry $\RR^{1,3} \times T^3 \times A_{N-1}$.  There are not enough
  compact dimensions inside this geometry which M5-branes could wrap around.}

\subsubsection{Summary of the field theory results}

Let us summarize the properties of the $SU(N)^{N_4N_6N_8}$ quiver
theory.  We have seen that three extra dimensions with fixed radii
$R_{4,6,8}$ are generated along the Higgs branch defined by
Eq.~(\ref{diagvevs}).  For finite lattice spacings $a_{4,6,8}$ our
four-dimensional quiver theory describes a seven-dimensional theory
with gauge coupling $g_7^2$ discretized on a three-dimensional
toroidal lattice. Seven-dimensional super Yang-Mills theory breaks
down at a certain cut-off $\Lambda_{\rm 7d}$ above which it requires a
UV completion.  The cut-off of the deconstructed theory is given by
the mass of the highest KK mode, $\Lambda=a^{-1}$ ($a=\max [a_4, a_6,
a_8]$). In the continuum limit $a_{4,6,8} \rightarrow 0$, which
requires \mbox{$g\rightarrow\infty$} while keeping the radii
$R_{4,6,8}$ and the seven-dimensional gauge coupling $g_7^2$ fixed,
$\Lambda$ becomes very large, $\Lambda \gg \Lambda_{\rm 7d}$.  In the
large $N_{4,6,8}$ limit we therefore expect to deconstruct not only 7d
super Yang-Mills theory but also its UV completion. This is shown
schematically in Fig.~\ref{energy}.  We show in the next section that
the UV completion is M-theory on $A_{N-1}$ with Planck length
$l_p^3=g_7^2$.

\begin{figure}[!ht]
\begin{center}
{\includegraphics[scale=0.9]{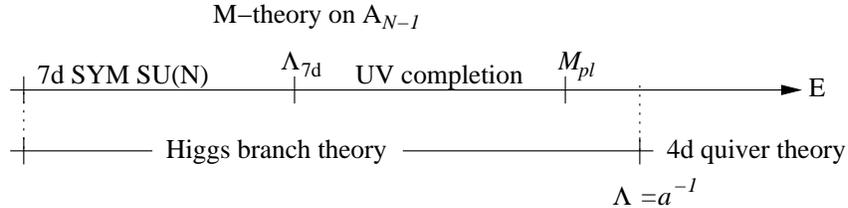}}
\caption{Various cut-offs in the deconstruction of M-theory.}
        \label{energy}
\end{center}
\end{figure}

M-theory on this geometry consists of two parts:~The seven-dimensional gauge
theory living on the singularity of the $A_{N-1}$ space couples to the
eleven-dimensional bulk degrees of freedom of M-theory. In the gauge boson
spectrum, we can therefore see only three of the seven string-theoretically
predicted extra dimensions.  It is not quite clear how the four dimensions of
the $A_{N-1}$ space are generated. However, we have found a spectrum of
massive dyons showing the presence of M2-branes in the deconstructed theory.
This supports our conjecture that the deconstructed theory is M-theory on
$A_{N-1}$.

\subsubsection{String-theoretical motivation for the equivalence} \label{sec25}
                
We now motivate the conjecture of the last section by a
string-theoretical analysis of the orbifold geometry. The discussion
will be similar to that in Sec.~\ref{secString}. We show that supersymmetry
with 16 supercharges is restored on the Higgs branch of our model,
where the theory becomes M-theory on a $T^3 \times A_{N-1}$
geometry. To this end, we consider the behaviour of the stack of
D3-branes on the Higgs branch.
                     
For this purpose, let us study the geometry of the orbifold in the vicinity
of the {D3-branes} which are located a distance $d$ away from the orbifold
singularity.  The product orbifold $\CC^3/\ZZ_{N_4} \times \ZZ_{N_6} \times
\ZZ_{N_8}$ can be parameterized by the complex coordinates 
($i=1,2,3$)~\cite{Brax}
\begin{align}
z_i = r_i \exp\left(i \frac{b^{(4)}_i}{N_4} \theta_1 + 
i \frac{b^{(6)}_i}{N_6} \theta_2 + 
i \frac{b^{(8)}_i}{N_8} \theta_3  \right) \,.
\end{align}
For orthogonal vectors $b^{(4)}_i, b^{(6)}_i, b^{(8)}_i$, the orbifold metric
$ds^2 = dz_i d\bar z_i$ \mbox{acquires} the form
\begin{align}
ds^2 = 
 d\vec r^{\,2} +   \frac{({\vec r} \cdot {\vec b}^{(4)})^2}{N_4^2} d\theta_1^2
+ \frac{({\vec r} \cdot {\vec b}^{(6)})^2 }{N_6^2} d\theta_2^2
+ \frac{({\vec r} \cdot {\vec b}^{(8)})^2 }{N_8^2} d\theta_3^2  \,,
\end{align}
where $\vec r=(r_1,r_2,r_3)$, $\vec b^{(4)}= (b^{(4)}_1,b^{(4)}_2,b^{(4)}_3)$,
etc. The orbifold has the geometry of a three-torus fibration over $\RR^3$: We
recover three circles $S^1$ parameterized by $\theta_{1,2,3} \in [0,2\pi]$.
For the particular choice of vectors $b_i$ as defined in
Eqns.~(\ref{tuple1})-(\ref{tuple3}), the circles $S^1$ have radii
$R_{S^1}= {l_s^2}/{R_{4,6,8}}$ with the radii $R_{4,6,8}$ given by
\begin{align}
R_{4}=\frac{N_{4} l_s^2}{2 d_4} \,,\qquad
R_{6}=\frac{N_{6} l_s^2}{2 d_6} \,,\qquad
R_{8}=\frac{N_{8} l_s^2}{2 d_8}\,.
\end{align}
Here the D3-branes were assumed to be located at $\vec r = (d_4,d_6,d_8)$.
Comparison with the radii defined in (\ref{radii}) yields relations between
the parameters of the quiver theory $g$, $v_{4,6,8}$, $N_{4,6,8}$ and the
string theory parameters $l_s$, $g_s$, $d_{4,6,8}=\vert z_{1,2,3} \vert$,
$N_{4,6,8}$:
\begin{align}
\frac{d_{4}}{N_{4} l_s^2} = \frac{2\pi gv_4}{N_{4}}  \,,\qquad
\frac{d_{6}}{N_{6} l_s^2} = \frac{2\pi gv_6}{N_{6}}  \,,\qquad
\frac{d_{8}}{N_{8} l_s^2} = \frac{2\pi gv_8}{N_{8}}  \,.
\end{align}
These relations show that giving {\em vevs} $v_4, v_6, v_8$ to the
scalar bifundamentals $h, v, n$ corresponds to moving the D3-branes a
distance $d =\sqrt{d_4^2+d_6^2+d_8^2}$ away from the singularity. The
continuum limit $a_{4,6,8} \rightarrow 0$ keeping $R_{4,6,8}$ fixed is
obtained if we take $l_s \rightarrow 0$, $N_{4,6,8}\rightarrow \infty$
with $g_s=g^2/N_4N_6N_8$ and $d_{4,6,8}/ N_{4,6,8} l_s^2$ fixed.
 
The orbifold may be visualized as a product of three cones.  In the large
$N_{4,6,8}$ limit each of the cones locally degenerates into a cylinder
$\RR\times S^1$ similar as in Sec.~\ref{secString}, see also
\cite{Arkani-Hamed, Brax}. The orbifold geometry in the vicinity of the
D3-branes becomes approximately $\RR^3 \times T^3$ with $T^3$ a three-torus.
Note that this induces a strong {\em supersymmetry enhancement} in the 
world-volume theory.  The $\N=4$ super Yang-Mills parent theory preserves 16
supercharges.  The orbifold projection reduces supersymmetry to $\N=0$. Now in
the large $N_{4,6,8}$ limit, the D3-branes probe the geometry $\RR^3 \times
T^3$, which in contrast to the orbifold, does not break supersymmetry. On the
Higgs branch the supersymmetry of the quiver theory is therefore enhanced
again to 16 supercharges.

In the large $N_{4,6,8}$ limit the radii of the three circles $S^1$
are sub-stringy, i.e.~$R_{S^1} \ll l_s$ if $R_{4,6,8} \gg l_s$, and
the appropriate description is obtained by T-dualizing along the three
directions 468 of the three-torus.  Details of the T-dualization are
shown in Tab.~\ref{TabMth}.

\begin{table}[!ht]
\hspace{1.8cm}
\begin{tabular}{|c|c|l|l|} 
\hline duality
  & D3 & $l_s \rightarrow 0$ & $g_s$ fixed \\
\hline
   T in $x^4$ & D4 & $l'_s=l_s \rightarrow 0$ 
              & $g'_s=g_s R_4/l_s \rightarrow \infty$\\
   T in $x^6$ & D5 & $l''_s=l_s \rightarrow 0$
              & $g''_s=g_s R_4R_6/l^2_s \rightarrow \infty$ \\
   T in $x^8$ & D6 & $l'''_s=l_s \rightarrow 0$
              & $g'''_s=g_s R_4R_6R_8/l^3_s \rightarrow \infty$ \\
\hline
  M-theory & M-theory & $l^3_p=(l'''_s)^3 {g'''_s}$& $R_{11} =l'''_s {g'''_s}$\\
    lift  & on $A_{N-1}$ & fixed &\\
\hline
\end{tabular}
\caption{T-dualization in $x^{4,6,8}$ and lift to M-theory.} \label{TabMth}
\end{table}

We started from D3-branes in the decoupling limit $l_s \rightarrow 0$, $g_s$
fixed.  These D3-branes T-dualize to D6-branes which wrap a three-torus $T^3$
with large radii $R_{4,6,8}$.  Due to a large string coupling $g'''_s$, a more
appropriate description is obtained by lifting the stack of $N$ D6-branes to a
singular $A_{N-1}$ geometry in M-theory.  The seven-dimensional gauge theory
located on the $A_{N-1}$ singularity has gauge coupling
$g_7^2=l_p^3=(l'''_s)^3 {g'''_s}=l^2_s R_{11}$. Since $R_{11}=g_s
R_4R_6R_8/l^2_s \rightarrow \infty$ and $l_s \rightarrow 0$, we can hold both
the gauge coupling $g_7$ and the eleven-dimensional Planck length $l_p$ fixed.
The seven-dimensional gauge theory therefore does not decouple from the bulk
gravity.

To conclude, string theory arguments suggest that our strongly coupled
non-super\-symmetric quiver theory on the Higgs branch describes M-theory on
$T^3 \times A_{N-1}$ in a large $N_{4,6,8}$ limit.  Since M-theory reduces to
eleven-dimensional supergravity at low-energies, we have deconstructed  a
gravitational theory!



\section{Summary and conclusion}
\label{conclusion}

 \setcounter{equation}{0}\setcounter{figure}{0}\setcounter{table}{0}
 \fancyhead{}
\fancyhead[LE,RO]{\bfseries\thepage}
\fancyhead[LO]{\bfseries Summary and conclusion}
\fancyhead[RE]{\bfseries Summary and conclusion}

In this chapter we summarize the results found in Ch.~2--4 and give 
an outlook on possible future developments.

\subsection{Results on dCFTs and their holographic duals}

In Ch.~\ref{cha:dCFT} we discussed extensively the D3-D$p$ brane
intersection as well as its dual probe-supergravity description. We have also
presented the action and some of the elementary properties of a defect
conformal field theory describing intersecting D3-branes, including some
aspects of the AdS/CFT dictionary.  There remain many interesting open
questions, of which we enumerate a few below.

The defect conformal field theory corresponding to the D3-D3 intersection
requires further field-theoretic analysis. One of the stranger features of
this theory is that it contains massless two-dimensional scalars with
(presumably) exactly marginal gauge, Yukawa, and scalar potential couplings.
It is not at all obvious that one can construct a Hilbert space corresponding
to operators with power law correlation functions, due to the logarithmic
correlators of the two-dimensional scalars.  It would be very interesting if
one could show this to all orders in perturbation theory.

As a precursor to including gravity into the holographic map, it
would be interesting to study the energy-momentum tensor of the
defect conformal field theory in detail. We did not find any
evidence of an enhancement of the two-dimensional $SO(2,2)$ global
conformal symmetry to a full infinite-dimensional conformal
symmetry on the two-dimensional defect. A study of the
energy-momentum tensor would allow us to address this question
conclusively at least from the field-theoretic side. For example,
if an enhancement did indeed occur it should manifest itself in
the form of a two-dimensional energy-momentum tensor which is
holomorphically conserved.

Another question concerns the light-cone open string vacuum for
D3-branes in the Penrose limit of the probe-supergravity
background which we have considered.  The operator proposed in
\cite{Skenderis} to correspond to the open string light-cone
vacuum is not really a chiral primary and gives negative
light-cone energy.   This operator is precisely the one given in
(\ref{wierdops}),  and contains the dimensionless scalars which
parameterize the Higgs branch.  One might instead propose the
operator ${\cal C}^{\m l}$, with $P_- = \Delta -l = 0$ as the dual
of the light-cone vacuum, however this is only $1/4$ BPS and is in
a non-trivial representation of the unbroken $SU(2)_L \times
SU(2)_R$ R-symmetry.  Presumably the subtleties regarding the
light-cone vacuum are related to the quantum spreading over
holomorphic embeddings $wy =c$ corresponding to the classical
Higgs branch of the defect CFT. While the origin of this spreading
is clear from the point of view of the dual defect CFT, and from
the difficulties in finding localized supergravity solutions for
intersecting D3-branes \cite{marolfpeet,peet,Gomberoff}, they are
not so clear from the point of view of a probe D3-brane embedded
in the plane-wave or $AdS$ backgrounds.

Although there is presumably no fully localized supergravity
solution for intersecting D3-branes, it would be surprising if
there is no closed string description,  in which both
stacks of D3-branes are replaced by geometry.  The problem of
finding a closed string description of the theory raises a closely
related question of how new degrees of freedom appear when $1/N$
(or $g_s$) corrections are taken into account in
probe-supergravity background which we have considered.  In
constructing the holographic dual of the defect CFT, we have fixed
the number $N^{\prime}$ of D3-branes in one stack, while taking
the number $N$ of D3-branes in the orthogonal stack to infinity.
In this limit, the degrees of freedom on one four-dimensional part
of the world volume of the defect become free. The remaining
coupled degrees of freedom live on a four-dimensional world volume
and a two-dimensional defect, which are the boundaries of $AdS_5$
and the embedded $AdS_3$ respectively. Because the defect degrees
of freedom are in the fundamental representation,  the genus
expansion of Feynman diagrams resembles an open string world-sheet
expansion. When $1/N$ corrections are taken into account, the
decoupled degrees of freedom must somehow reappear. The defect
degrees of freedom become bi-fundamental fields with respect to a
$SU(N) \times SU(N^{\prime})$ gauge group. The genus expansion for
Feynman diagrams of the theory can now be viewed as a closed
string world-sheet expansion, where a new branch of the target
space has opened up.\footnote{A similar although not directly
related picture has been discussed in \cite{OoguriVafa}.}

Finally, the string theory realization of the defect CFT leads one
to expect that it exhibits S-duality.  It would be very
interesting to find some field theoretic evidence for this.  In
particular one would need to find the S-duals of the fundamental
degrees of freedom localized at the intersection.

\subsection{Results on AdS/CFT with flavour}

In Ch.~\ref{cha:flavour} we have studied two non-supersymmetric gravity
backgrounds with embedded D7-brane probes, corresponding to Yang-Mills
theories with confined fundamental matter.  The AdS Schwarzschild black-hole
background which in the presence of the probe describes an ${\cal N}=2$
Yang-Mills theory at finite temperature, exhibits interesting behaviour such
as a bilinear quark condensate and a geometric transition which corresponds to
a first order transition in the gauge theory. The D7-brane embedding into the
Constable-Myers background is more QCD-like, showing a chiral condensate at
small quark mass as well as the accompanying pion (or large $N$
$\eta^{\prime}$).

In the Constable-Myers background with a D7-brane, there is a spontaneously
broken $U(1)$ axial symmetry.  A closer approximation to (large $N$) QCD would
require a spontaneously broken $U(N_f)_L \times U(N_f)_R$ chiral symmetry with
$N_f = 2$ or $N_f =3$.  Unfortunately, simply adding D7-branes does not
accomplish this in the Constable-Myers background.  Assuming that one were
able to find a background with the full chiral symmetry, it would be very
interesting to obtain a holographic interpretation of low-energy current
algebra theorems and make predictions for chiral-Lagrangian parameters based
on the non-abelian Dirac-Born-Infeld action. One challenge would be to obtain
such terms as the Wess-Zumino-Witten term, $\int\, d^5 x\, {\rm tr}(\Sigma d
\Sigma^{\dagger})^5$. Note that this term requires an auxiliary fifth
dimension, which appears naturally in the holographic context.

It is clearly important to look at more physical geometries. We chose these
backgrounds because they are particularly simple; the $S^5$ is left invariant
and hence embedding the D7 is straightforward and the RG flow depends only on
the radial direction. More complicated geometries, such as the Yang-Mills$^*$
geometry \cite{Babington:2002qt}, that include mass terms for the adjoint
scalars and fermions of ${\cal N}=4$, have extra dependence on the angles of
the $S^5$ and the resulting equations of motion are much less tractable.

In conclusion the results presented in this chapter represent another success
for the AdS/CFT correspondence. The results suggest that gravity duals of
non-super\-symmetric gauge theories may induce chiral symmetry breaking if
light quarks are introduced, just as is observed in QCD. This opens up the
possibility of studying the light meson sector of QCD using these new
techniques.

\subsection{Results on deconstructing extra dimensions}

\subsubsection{Summary of intersecting M5-branes}

In Ch.~\ref{Ch44}} we have presented a formulation of intersecting M5-branes
in terms of a limit of a $(4,0)$ defect conformal field theory.  We hope that
this will lead to an improved understanding of the low energy dynamics of
M5-branes although, as for the (de)construction of parallel M5-branes
\cite{Arkani-Hamed}, immediate progress is impeded by the fact that the
continuum limit is also a strong coupling limit.

At the moment we only have control of a few some simple properties
which are protected against radiative and non-perturbative
corrections. We were able to show the existence of tensionless strings
on the four-dimensional intersection of the two stacks of M5-branes.

It would be interesting to try to generalize the construction here
to more complicated intersections of branes in M-theory.  Such
generalizations might be of use in understanding the microscopic
origins of black hole entropy.

It would also be very interesting to find field theoretic
arguments in favour of the S-duality of the D3-D3 intersection,
either in flat space or at a $\CC^2/\ZZ_k$ orbifold.  We have only
assumed S-duality,  based on the S-duality of the
string theory background.  A starting point would be to find
solitons which are S-duals of degrees of freedom localized at the
intersection. This is clearly very important if one wishes to have
a better understanding of the degrees of freedom and dynamics of
intersecting M5-branes.

\subsubsection{Summary of the deconstruction of M-theory}

In Ch.~\ref{Ch45} we have deconstructed M-theory on a singular space of the
type $T^3\times A_{N-1}$ from a four-dimensional non-supersymmetric quiver
gauge theory with gauge group \linebreak $SU(N)^{N_4N_6N_8}$. This theory is
conformal in the large $N_{4,6,8}$ limit.  We have given some evidence for the
commutativity of the diagram shown in Fig.~\ref{fig3}, which summarizes the
deconstruction.

\begin{figure}[!ht]
\begin{center}
\input{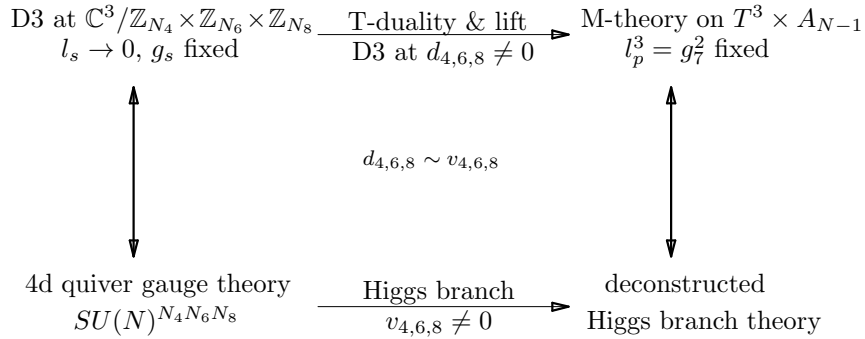} 
        \caption{Orbifold realization of the quiver gauge theory and 
          deconstruction of M-theory on $T^3 \times A_{N-1}$.}\label{fig3}
\end{center}
\end{figure}

The left hand side of the diagram shows the quiver gauge theory and its
realization in type IIB string theory as a stack of D3-branes placed at the
origin of an orbifold of the type $\CC^3/\ZZ_{N_4} \times \ZZ_{N_6} \times
\ZZ_{N_8}$. The right hand side represents the Higgs branch of the quiver
theory and its corresponding realization in M-theory.  Moving the D3-branes
away from the orbifold singularity corresponds to the Higgs branch of
the quiver theory. Exploiting string dualities, we map the geometry in the
vicinity of the D3-branes in type IIB string theory to a $T^3 \times A_{N-1}$
geometry in M-theory.  We conclude that the deconstructed Higgs branch
theory, in a particular strong coupling and large $N_{4,6,8}$ limit, is
equivalent to M-theory on $T^3 \times A_{N-1}$. This claim is further
supported by a dyonic spectrum in the quiver theory which corresponds to 
wrapped M2-branes. We thus provide a completely field-theoretical definition of
M-theory on $T^3 \times A_{N-1}$.

The deconstruction of M-theory does not suffer from the problems of the
deconstruction of pure gravitational theories discussed in
\cite{Arkani-Hamed:2003vb, Arkani-Hamed:2002sp, Schwartz:2003vj, Jejjala,
  Jejjala:2003qg}.  In quiver {\em gauge} theories  the  cut-off for the
higher-dimensional behaviour of the theory can be taken to infinity.

However it would be of interest to find further field-theoretical evidence for
the equivalence.  For example, one would like to see how eleven-dimen\-sional
supergravity is realized in the model. The quiver model is formulated in terms
of local fields of an ordinary four-dimensional Yang-Mills theory. An open
question is the relation between these fields and the metric tensor or higher
spin fields in M-theory.  It would also be exciting to find a
field-theoretical argument for the quiver theory to describe eleven dimensions
besides the string-theoretical argument given in this thesis.  There are some
indications that the Kaluza-Klein spectrum for gravitational extra dimensions
might arise from closed strings in the twisted sector of the orbifold which
has been ignored so far. This will be studied elsewhere.

The present 4d quiver theory provides a non-perturbative definition of
M-theory and might be an alternative to matrix models which describe the
discrete light-cone quantization (DLCQ) of M-theory \cite{Banks:1997vh,
  Susskind:1997cw}. It would be very interesting to find a relation between
both approaches. The matrix model for M-theory on $T^3 \times A_{N-1}$ is
given by a 4d $\N=2$ super Yang-Mills quiver theory with gauge group
$U(k)^N$~\cite{Ganor}. The parameter $k$ characterizes the discrete momentum
$P_-=k/R$ of states in the light-like direction.  The Coulomb branch of this
model describes the gauge theory located at the singularity of the geometry.
The Higgs branch encaptures the physics of the spacetime away from the
singularity.  This matrix model has to be compared with the Higgs branch
theory of our model, which has unbroken gauge group $SU(N)$ and preserves 16
supercharges in the continuum limit.  The matrix model describes a sector of
M-theory with fixed momentum $P_-$. In contrast, our model is not
restricted on a particular sector of M-theory. Note however that the continuum
limit $a_{4,6,8}\rightarrow 0$ requires one to consider the quiver theory at
strong coupling, impeding perturbative access to M-theory.

\subsection{General conclusions and perspectives}

As the conclusion shows, this thesis provides new techniques for studying
asymp\-totically-free Yang-Mills theories with fundamental matter in terms of
holographic supergravity duals.  The results presented in this thesis open up
the possibility of understanding non-per\-tur\-ba\-tive phenomena in the
strong coupling regime of QCD-like theories.  As an independent result, we
presented a way of understanding aspects of M-theory in terms of
deconstruction models.

There are several directions in which the present work can be extended.
We shall name a few.

As a future project we suggest to study the holographic dual description of
the chiral symmetry breaking $SU(N_f) \times SU(N_f) \rightarrow SU(N_f)$ with
$N_f$ the number of flavours.  One brane system which realizes this chiral
symmetry consists of two stacks of intersecting D7-branes in the background of
an orbifold \cite{Katz}. Each of the D7-branes gives rise to a $SU(N_f)$
symmetry giving $SU(N_f) \times SU(N_f)$ which can be broken down to $SU(N_f)$
by a blow-up of the orbifold.  We propose to study intersecting D7-branes in a
background which realizes this blow-up via an RG flow.

Another obvious extension of the present work is to go beyond the probe
approximation. Lattice gauge computations show that the ``quenched''
approximation in QCD, which corresponds to the probe approximation, is not a
very good one. In order to include the effect of the quarks on the gluons, one
has to take into account the backreaction of the branes onto the geometry. In
this context a challenge would be to find a holographic verification of the
Witten-Veneziano formula as given by Eq.~(\ref{witvene}), which is only
possible in a solution with backreaction.  We leave this as a subject of
further investigation.\footnote{Shortly after this work was completed, two
  papers were published \cite{Armoni:2004dc, Barbon:2004dq} which discuss the
  Witten-Veneziano formula within the AdS/CFT framework.}

\section*{Acknowledgements}

\noindent I wish to thank 
Johanna Erdmenger and Dieter L\"ust for giving me the opportunity to do my
Ph.D.\ at Humboldt University Berlin. I am especially indebted to Johanna
\mbox{Erdmenger} and Zachary Guralnik for their inspiration, stimulating
suggestions and encouragement which helped me during all the time of my
research.  I am also grateful to James Babington, Neil Constable, Nick Evans,
and Robert Helling for excellent collaboration on which part of this work is
based. My research was funded by the DFG (Deutsche Forschungsgemeinschaft)
within the Emmy Noether programme, grant ER301/1-2.



\begin{appendix}


\section{Defect conformal field theories}
\setcounter{equation}{0}

\subsection{Mass-dimension relation in $AdS_{d+1}/CFT_d$}

The mass of a $p$-form on an $AdS_{d+1}$ space is related to the
conformal dimension $\Delta$ of a $(d-p)$-form operator in the dual
CFT$_d$ by
\begin{align}
m^2=(\Delta-p)(\Delta + p-d) \,.
\end{align}
This formula can be inverted to give 
\begin{align} \label{massoprel}
 \Delta_\pm = ( d \pm \sqrt{(d-p)^2 + 4 m^2})/2 \,.
\end{align}

\subsection{Dirac-Born-Infeld action of an $AdS_{k+2}$-brane} \label{appDBI}

For the computation of the DBI action of the $AdS_{k+2}$-brane
we make use of the trace expansion of the determinant,
\begin{align}
\sqrt{\det (1+A)} = 1+ \frac{1}{2}{\rm Tr\,} A + \frac{1}{8}({\rm Tr\,} A)^2
- \frac{1}{4}{\rm Tr\,} A^2 + {\cal O}(A^3) \,,
\end{align}
and the identity
\begin{align}
\sqrt{-\det\left(\bar g + \omega \right)}=
\sqrt{-\det \bar g \left(1 + \bar g^{-1} \omega \right)}=
\sqrt{-\det \bar g}\sqrt{\det \left(1+ \bar g^{-1} \omega \right)} 
\end{align} 
which holds for any two matrices $\bar g$ and $\omega$.  The DBI
lagrangian can then be expanded to quadratic orders in fluctuations,
\begin{align}
{\cal L}_{DBI} &= \sqrt{-\det\left(g_{ab}^{PB} +  {\cal
F}_{ab}  \right) } = \sqrt{-\det\left(\bar g_{ab} + \partial_aZ^i\partial_bZ^jg_{ij} +2\pi l_s^2
{F}_{ab}  \right)}
 \nonumber\\ &\approx \sqrt{-\det \bar g_{ab}}
\left( 1+ \frac{1}{2} 
\bar g^{ab} \partial_aZ^i\partial_bZ^jg_{ij} +\frac{1}{4}(2\pi l_s^2)^2
F_{ab} F^{ab} \right) \,,  \label{DBIapprox}
\end{align} 
where we dropped terms involving $B_{ab}$ and $g_{ai}$.  Here the
$p+1$-dimensional metric $\bar g_{ab}$ is the reduction of the $AdS_5
\times S^5$ metric (\ref{AdS5metric}) to $p+1$ dimensions (set all
differentials $dZ^i =0$). Its determinant can be expanded,
\begin{align}
 \sqrt{-\det \bar g_{ab}}=\sqrt{\bar g_{p+1}} (\sin \phi_{k+1} \cdots
\sin \phi_{5})^k \approx \sqrt{\bar g_{p+1}} \left(1 - \frac{k}{2} 
\sum_{i=k+1}^{5}  {\phi'_i}^2  \right) \label{detexp}
\end{align}
with $\phi'_i \equiv\phi_i-\frac{\pi}{2}$ and $\bar g_{p+1}$ the
determinant of the $AdS_{k+2} \times S^k$ metric (\ref{AdSkmetric}).
Writing out Eq.~(\ref{DBIapprox}) and using the expansion (\ref{detexp}),
we find
\begin{align} 
{\cal L}= \sqrt{\bar g_{p+1}}
&\left[1+ \sum_{i=k+1}^{5} (\frac{1}{2}\partial_a\phi'_i\partial^a\phi'_i
  -\frac{k}{2} {\phi'}_i^2 ) \right.  \\
&\left. + \sum_{i=k+1}^3
\frac{1}{2u^2}\partial_a X^i \partial^a X^i
+\frac{1}{4}(2\pi l_s^2)^2
F_{ab} F^{ab} \right] \,, \nonumber
\end{align}
which is leads to  Eq.~(\ref{quadBI}).

\subsection{Sugra computation of one-point functions} \label{appA2}

The standard bulk-to-boundary propagator in $d$ dimensions is
given by \cite{FreedmanMathur}
\begin{align} \label{btobprop}
K_\D (w, \vec w, \vec z) =
\frac{\G(\D)}{\pi^{\frac{d}{2}} \G(\D-\frac{d}{2})}
\left(\frac{w}{w^2 + (\vec w-\vec z)^2} \right)^{\D} \,
\end{align}
with $\vec w, \vec z \in \RR^d$ and $w$ the radial coordinate in $AdS_{d+1}$.
Here we consider the bulk-to-boundary propagator in $AdS_5$ (i.e.~$d=4$).
Integration over the $AdS_{k+2}$ subspace yields the one-point function
(substitute $\vec z = (\vec x, \vec y)$ and $\vec w = (0,\vec w)$ in
Eq.~(\ref{btobprop})):
\begin{align}
\langle {{\cal O}}_\D(\vec x, \vec y) \rangle &=
\lambda^{(k-1)/2} \int \frac{dw
d\vec w^{k+1}}{w^{k+2}} \frac{\G(\D)}{\pi^2 \G(\D-2)}
\left(\frac{w}{w^2+\vec x^2 + (\vec w-\vec y)^2} \right)^{\D}
\end{align}
After rescaling $\vec w' =\vec y' + \sqrt{\vec x'^2 +
w'^2} \vec v$ and $w'=\vert \vec x' \vert u$ we get
\begin{align}
\langle {\cal O}_\D(\vec x, \vec y) \rangle &=\lambda^{(k-1)/2}
 \int \frac{du d \vec{v}^{k+1}}{\pi^2 u^{k+2}} \frac{\G(\D)}{\G(\D-2)}
\frac{ u^\D}{ \vert \vec x \vert^{\D}
(1+u^2)^{\D-(k+1)/2} (1+\vec v^2)^\D} \label{1point}
\end{align}
For the two integrals we find\footnote{use \cite{Gradshteyn} 
\begin{align}
\int_0^\infty (x+\beta)^{-\nu} x^{\mu-1} = \beta^{\mu-\nu} B(\nu-\mu,\mu)
\,,\quad B(a,b) \equiv \frac{\G(a)\G(b)}{\G(a+b)} \,,
\end{align} 
with $\beta=1$, $\nu=\D-(k+1)/2$, $\mu=(\D-k-1)/2$ in (\ref{int1}) and
 $\beta=1$, $\nu=\D$, $\mu=\D-(k+1)/2$ in (\ref{int2}).  }
\begin{align} \label{int1}
\int du \frac{u^{\D-(k+2)}}{(1+u^2)^{\D-(k+1)/2}} = \frac{1}{2} \int dx
\frac{x^{(\D/2-(k+1)/2)-1}}{(1+x)^{\D-(k+1)/2}} = 
\frac{\G(\frac{\D}{2}) \G(\frac{\D-(k+1)}{2})}{2 \G(\D-\frac{k+1}{2})}  \,,
\end{align}
and
\begin{align} \label{int2}
\int  \frac{d \vec{v}^{k+1}}{(1+ \vec{v}^2)^\D}&=\int dr \frac{2\pi^{(k+1)/2} 
r^k}{\G(\frac{k+1}{2})} \frac{1}{(1+r^2)^\D} = \frac{\pi^{(k+1)/2}}{\G(\frac{k+1}{2})} \int dx \frac{x^{(k+1)/2-1}}{(1+x)^\D} \nonumber\\
&= \frac{\G(\D-\frac{k+1}{2})}{\G(\D)} \pi^{(k+1)/2} \,.
\end{align}
Substituting these expressions into Eq.~(\ref{1point})
we finally get
\begin{align}
\langle {{\cal O}}_\D(\vec x, \vec y) \rangle
&=\lambda^{(k-1)/2} \frac{1}{\vert \vec x \vert^{\D}}
\frac{\G(\frac{\D}{2}) \G(\frac{\D}{2}-\frac{k+1}{2})}{2 \G(\D-2)} \pi^{(k+1)/2-2} \,.
\end{align}

\subsection{Conformal symmetry \label{confs}}

Here we review some basic implications of conformal symmetry in a
four-dimensional field theory with a $k+1$-dimensional defect \cite{Osborn}.

Consider four-dimensional Euclidean space with a $k+1$-dimensional
defect. The coordinates are given by $v_\mu=(\vec{z}, \vec{x})$
where $v_\mu$ are the four-dimensional coordinates, $\vec{z}_M$
are the $k+1$ defect coordinates and $\vec{x}_\alpha$ are the
coordinates perpendicular to the defect. The conformal
transformations which leave the defect invariant are given by
translations and rotations within the defect plane, rotations in
the plane perpendicular to the defect and by inversions $v_\mu
\rightarrow v_\mu/v^2$. The four-dimensional conformal group $SO(1,5)$
decomposes into $SO(1,k+2)\times SO(3-k)$. Under these transformations
we have for two points $v$, $v'$
\begin{equation}
(v-v')^2 \rightarrow \frac{(v-v')^2}{\Omega(v)\Omega(v')} \, ,
\quad \vec{x}_\alpha \rightarrow \frac{\vec{x}_\alpha}{\Omega(v)}
\,, \quad \vec{x}'_\alpha \rightarrow
\frac{\vec{x}'_\alpha}{\Omega(v')} \, .
\end{equation}
Hence there is a dimensionless coordinate invariant of the form
\begin{equation}
\xi = \frac{(v-v')^4}{(\vec{x}\cdot \vec{x}) (\vec{x}' \cdot
\vec{x}')} \,.
\end{equation}

There does not exist a $k+1$-dimensional conserved local
energy-momentum tensor. Only the four-dimensional energy-momentum
tensor of the combined bulk and defect action
contributions, given by
\begin{equation} \label{fourt}
T_{\mu \nu}(v) \, = \, T^{\rm bulk}{}_{\mu \nu }(v) \, + \, T^{\rm
def}{}_{MN} (\vec{z})\, \delta_{M(\mu} \delta_{\nu)N} \,
\delta^{(3-k)}(\vec{x})\, \, ,
\end{equation}
is conserved, $\partial_\mu T_{\mu \nu} = 0$. By integration over $x$
we obtain
\begin{gather}
{\cal T}_{MN}(z) = \int\! d^{3-k}x \, T^{\rm bulk}{}_{MN} (x,z) \, +
\, T^{\rm def}{}_{MN} (z) \, ,
\end{gather}
which is contained as a component in the $k+1$-dimensional super
current ${\cal J}_M$. ${\cal T}_{MN}(z)$ satisfies $\partial^z{}_M
{\cal T}_{MN} (z) = 0$. Nevertheless it is not a local traceless
$k+1$-dimensional energy-momentum tensor.

For a quasi-primary scalar operator of dimension $\Delta$ close to
the defect we have a one-point function
\begin{equation} \label{onepoint}
\langle {\cal O} (v) \rangle \, = \, \frac{A_{\cal
O}}{\vert \vec{x} \vert^{\Delta}} \, .
\end{equation}
Near to the defect we have a boundary operator expansion of the
bulk operators in terms of the defect operators, which reads
\begin{equation}
{\cal O}(v) \, = \, \sum\limits_{n} \frac{B_{{\cal O}, \hat {\cal O}_n
}}{ \vert \vec{x} \vert^{\Delta - \hat \Delta_n}} \, \hat
{{\cal O}}_n(\vec{z}) \, .
\end{equation}
This gives rise to a bulk-defect correlator
\begin{equation} \label{twopoint}
\langle {{\cal O}}(v) \hat {{\cal O}}_n(\vec{z}') \rangle \, = \,
\frac{B_{{\cal O}, \hat {{\cal O}}_n }}{ \vert \vec{x} \vert^{\Delta - \hat \Delta_n} (v-v')^{2\hat \Delta_n}}\, ,
\quad  (v - v')_\mu = (\vec{x}, \vec{z}-\vec{z}').
\end{equation}
For two operators of dimension $\hat \Delta_n$ on the defect, this
expression reduces to
\begin{equation}
\langle \hat {{\cal O}}_n(\vec{z}) \hat {{\cal O}}_n(\vec{z}')
\rangle \, = \, \frac{B_{{\hat {{\cal O}}_n}, \hat {{\cal O}}_n
}}{ (\vec{z}-\vec{z}')^{2\hat \Delta_n}}\,
\end{equation}
as expected.

\subsection{Multiplets in $(2,2), d=2$ superspace}\label{appA}

In this appendix we briefly summarize the component
expansions of the superfields in $(2,2), d=2$ superspace which can
be found in \cite{Hori}, for instance. We use chiral coordinates
$y^0, y^1, \theta^\pm, \bar \theta^\pm$ which are related to the
superspace coordinates $x^0, x^1, \theta^\pm, \bar \theta^\pm$ by
\begin{align}
y^M&=x^M + i \theta^+ \rho^M_{11} \bar \theta^+ + i \theta^-
\rho^M_{22} \bar
\theta^- \nonumber\\
  &=x^M + i \theta^+  \bar \theta^+ + (-1)^M i \theta^- \bar \theta^- \,,
\qquad M=0,1 \,,
\end{align}
where we use the Pauli matrices
\begin{align}\label{Pauli}
\rho^0\equiv\s^0=\begin{pmatrix} 1 & 0\\ 0 & 1 \end{pmatrix} ,\,
\rho^1\equiv\s^3=\begin{pmatrix} 1 & 0\\ 0 & -1 \end{pmatrix},\,
\rho^2\equiv\s^2=\begin{pmatrix} 0 & -i\\ i & 0 \end{pmatrix} ,\,
\rho^3\equiv\s^1=\begin{pmatrix} 0 & 1\\ 1 & 0 \end{pmatrix} .
\end{align}
Expansions of (2,2) superfields can be obtained by dimensional
reduction of $\N=1$, $d=4$ superfields in the $2$ and $3$ direction
and defining $\theta^+\equiv\theta^1= \theta_2$ and $\theta^-
\equiv \theta^2= -\theta_1$. In this way we find the expansions of
the chiral and the vector multiplet in Wess-Zumino gauge,
\begin{align}
  \Phi(y,\theta^\pm) &= \phi + \sqrt{2} \theta^+ \psi_+ +
  \sqrt{2} \theta^- \psi_- -2 \theta^+ \theta^- F \\
  V(y,\theta^\pm,\bar\theta^\pm) &=\,\, \theta^- \bar \theta^- (v_0 - v_1)+
  \theta^+\bar\theta^+ (v_0+v_1)
  - \theta^- \bar\theta^+ \s - \theta^+\bar\theta^- \bar \s\nonumber\\
  &\quad+i \sqrt{2} \theta^- \theta^+ (\bar\theta^- \bar\lambda_-+\bar 
\theta^+
  \bar\lambda_+)+i \sqrt{2}
  \bar\theta^+ \bar \theta^- (\theta^- \lambda_-+ \theta^+ \lambda_+) 
\label{Vexp}\\
  &\quad+2\theta^-\theta^+\bar\theta^+\bar\theta^- (D -i \partial^M v_M) \,.
  \nonumber
\end{align}
The scalar $\s$ is complex and is defined in terms of the
components $v_1$ and $v_2$ of the dimensionally reduced
four-vector $v_\m$ by $\s \equiv v_3+iv_2$. For the (abelian)
twisted chiral superfield $ \Sigma(y,\theta^\pm,\bar\theta^\pm)
\equiv \bar D_+ D_- V(y,\theta^\pm,\bar\theta^\pm) $ we find the
expansion
\begin{align}
\Sigma(y,\theta^\pm,\bar\theta^\pm) = & \,\sigma + i\sqrt{2}
\theta^+ \bar \lambda_+ - i\sqrt{2} \bar \theta^- \lambda_- + 2 \theta^+
\bar\theta^-  ( D - i f_{01}) + 2i \bar \theta^- \theta^-
( \pr_0  - \pr_1) \sigma \nonumber\\
& - 2 \sqrt{2} \bar \theta^- \theta^- \theta^+ (\pr_0 - \pr_1)
\bar \lambda_+ \,.
\end{align}

\subsection{Decomposing the $\N=2$, $d=4$ vector multiplet under $(2,2)$,
  $d=2$ supersymmetry}  \label{newappendix}

\lhead[\fancyplain{}{\bfseries\thepage}]%
{\fancyplain{}{\bfseries A.6 Decomposing the $\N=2$, $d=4$ vector multiplet}}
\rhead[\fancyplain{}{\bfseries\leftmark}]%
{\fancyplain{}{\bfseries\thepage}}

We start from the decomposition\footnote{IMPORTANT NOTE: In this section the
  $\N=2$, $d=4$ superspace is parametrized by ($x^0, ..., x^3$,
  $\theta_i^{\alpha},\bar\theta^{i}_{\dot \alpha}$) and the defect is placed
  at $x^1 = x^2=0$ in contrast with our convention in Sec.\ \ref{sec3}
  (defect at $x^2=x^3=0$).}  of the vector multiplet $\Psi$ under $\N=1$,
$d=4$ which is given by an expansion in $\theta_{(2)}$ \cite{Lykken},
\begin{align} \label{N=1expansion}
  \Psi(\tilde y,\theta_{(1)},\theta_{(2)})=
  \Phi'(\tilde y,\theta_{(1)}) + i \sqrt{2}
  \theta^\a_{(2)} W'_\a(\tilde y,\theta_{(1)}) + \theta_{(2)}
  \theta_{(2)} G'(\tilde y,\theta_{(1)})\,\,,
\end{align}
where $\Phi'$, $W'_\a$, and $G'$ are chiral, vector, and auxiliary
$\N=1$ multiplets, respectively. The superfield $\Psi$ is a
function of the coordinate $\tilde y^\m$ which is related to
$x^\m$ by
\begin{align}
  \tilde y^\m = x^\m &+ i \theta^+ \s^\m_{11} \bar \theta^+ + i \thetasl^-
  \s^\m_{21} \bar \theta^+ + i \theta^+ \s^\m_{12} \bar \thetasl^-
  +  i \thetasl^- \s^\m_{22} \bar \thetasl^- \nonumber\\
  &+ i \bar \theta^- \s^\m_{22}  \theta^- + i  \thetasl^+
  \s^\m_{12} \theta^- + i \bar \theta^- \s^\m_{21} \bar \thetasl^+
  +  i \thetasl^+ \s^\m_{11} \bar \thetasl^+ \,,
\end{align}
where $\s^0$ is the identity matrix and $\s^a$ ($a=1,2,3$) are the
Pauli matrices.

Our goal is to find an expression for $\Psi$ in terms of $(2,2)$,
$d=2$ multiplets,
\begin{align} \label{expansion}
\Psi \equiv \Psi \vert_{\thetasl=\bar\thetasl=0} + \thetasl^+
(\Dsl_+ \Psi ) \vert_{\thetasl=\bar\thetasl=0} + \thetasl^-
(\Dsl_- \Psi ) \vert_{\thetasl=\bar\thetasl=0} +  \thetasl^+
\thetasl^- (\Dsl_+ \Dsl_- \Psi) \vert_{\thetasl=\bar\thetasl=0}
\,,
\end{align}
with $\thetasl=(\thetasl^+,\thetasl^-)$. In order to find the
coefficients of this expansion, we substitute the component
expansions of $\Phi'$, $W'_\a$ and $G'$ in (\ref{N=1expansion})
and use the coordinates \mbox{($\tilde y$, $\theta^\pm$,
$\thetasl^\pm$, $\bar \theta^\pm$, $\bar \thetasl^\pm$)} as
defined in (\ref{coordinates}). We find\footnote{ conventions:
$(\psi_1, \psi_2)=(\psi_+,\psi_-)$; $\psi^+=\psi_-,
\psi^-=-\psi_+$}
\begin{align}
  \Psi =&\, \phi' + \sqrt{2} \theta^+ \psi'_+ + \sqrt{2} \thetasl^- \psi'_-
  - 2 \theta^+ \thetasl^- F' \nonumber\\
  &+ i \sqrt{2} \bar \theta^- \left( - i \lambda'_- + \theta^+ D' + \theta^+ 
(
    f'_{12} -i f'_{03} ) + \thetasl^- (f'_{02}-f'_{32} +i f'_{10} - i 
f'_{13})
  \right. \nonumber\\
  & \left. \qquad\qquad - 2 \theta^+ \thetasl^- (\pr_1 \bar \lambda'^+ + i
    \pr_2 \bar \lambda'^+ + \pr_0 \bar
    \lambda'^- - \pr_3 \bar \lambda'^-) \right) \nonumber\\
  &+ i \sqrt{2} \thetasl^+ \left( - i \lambda'_+- \thetasl^- D' - \theta^+
    (f'_{02} + f'_{32} - i f'_{10} - i f'_{13}) + \thetasl^-
    (f'_{21} + i f'_{03}) \right. \\
  &\left.\qquad\qquad - 2 \theta^+ \thetasl^- (\pr_1 \bar \lambda'^- - i 
\pr_2
    \bar \lambda'^- + \pr_0
    \bar \lambda'^+ + \pr_3 \bar \lambda'^+) \right) \nonumber\\
  &- 2 \thetasl^+ \bar \theta^- \left(F'^* - i \sqrt{2} \theta^+ (\pr_1 \bar
    \psi'^- - i \pr_2 \bar \psi'^- + \pr_0 \bar \psi'^+ + \bar \pr_3 
\psi'^+)
    + 2 \theta^+ \thetasl^- \square_4 \phi'^* \right) \,. \nonumber
\end{align}
Note that all fields are functions of $\tilde y$ and we have to
expand this expression such that all fields become functions of
the chiral coordinates $y^M=x^M + i \theta^+  \bar \theta^+ + (-1)^M i
\theta^- \bar \theta^- \, (M=0,3)$.  Evaluating $\Psi$, $\Dsl_+ \Psi$, and
$\Dsl_- \Psi$ at $\thetasl^+={\thetasl^-=0}$ we obtain
\begin{align}
  \Psi \vert_{\thetasl=0} &= - i  \Sigma \,,\nonumber\\
  \quad (\Dsl_+ \Psi ) \vert_{\thetasl=0} &= \bar D_+ \left( \bar \Phi -
    \partial_{\bar x} V \right)\,, \quad (\Dsl_- \Psi )
  \vert_{\thetasl=0} =  D_- \left( \Phi - \partial_{x}
    V \right) \,, \label{coeff}
\end{align}
where $\partial_x \equiv\partial_1+i\partial_2$ is the derivative transverse
to the defect.  Here we defined the (unprimed) components of the (2,2)
superfields $\Sigma$, $\Phi$, and $V$ in terms of the (primed) components of
the $\N=1$, $d=4$ superfield $\Phi'$ and $W'_\a$ by
\begin{align}
&\sigma \equiv i \phi', \quad \bar \lambda_+ \equiv \psi'_+, \quad \lambda_-
\equiv - \lambda'_-, \quad D \equiv \frac{1}{\sqrt{2}} ( D' + f'_{12}),
\quad
f_{03} \equiv \frac{1}{\sqrt{2}} f'_{03}   \,, \nonumber\\
&\phi{} \equiv \frac{1}{\sqrt{2}} (v'_1+iv'_2),\quad \bar \psi_+
\equiv \lambda'_+, \quad \psi_- \equiv \psi'_-,\quad F{} \equiv F'
\label{redefinition}
\end{align}
If we substitute the coefficients (\ref{coeff})  back into
the expansion (\ref{expansion}) of $\Psi$, we find the
decomposition (\ref{decomposition}).

The appearance of $f'_{12}$ in the definition (\ref{redefinition})
of the auxiliary field $D$ is required by $(2,0) \subset (2,2)$
supersymmetry. Consider the (2,0) supersymmetry transformation
rules for the spinor component $\lambda'^+$, the auxiliary field $D'$,
and the component $f'_{12}$ given by
\begin{align}
  \delta_{\eps}\lambda'^{+} &= i \eps^+ (D'+ f'_{12} -i f'_{03})\nonumber \\
  \delta_{\eps} D' &= \bar \eps^+ (\pr_0-\pr_3) \lambda'^+ - \eps^+
  (\pr_0-\pr_3) \bar \lambda'^+ - \bar\eps^+(\pr_1-i\pr_2)\lambda'^- - 
\eps^+
  (\pr_1-i\pr_2) \bar\lambda'^-
  \label{susyvec} \nonumber\\
  \delta_{\eps} f'_{12} &= \eps^+ (\pr_1-i\pr_2) \bar\lambda'^- + \bar 
\eps^+
  (\pr_1-i\pr_2) \lambda'^- \,.
\end{align}
Of particular interest in Eq.~(\ref{susyvec}) are the non-standard
terms appearing in the variations of $\lambda'^{+}$ and $D'$
involving transverse derivatives, $\pr_2$ and $\pr_{1}$.  Note
that in dimensional reduction these terms would have simply been
set to zero.  The susy variation of $f'_{12}$ in $\delta_{\eps} D
\equiv \textstyle\frac{1}{\sqrt{2}}\delta_{\eps}(D'+ f'_{12})$
precisely cancels the non-standard terms in the variation of the
auxiliary field $D'$. This leads to the familiar (2,0) susy
variation for $D$,
\begin{align}
\delta_{\eps} D &= \bar \eps^+ (\pr_0-\pr_3) \lambda^+ - \eps^+
(\pr_0-\pr_3)
  \bar \lambda^+  \,.
\end{align}

\subsection{Defect action in component form} \label{appendixC}
\lhead[\fancyplain{}{\bfseries\thepage}]%
{\fancyplain{}{\bfseries\rightmark}}
\rhead[\fancyplain{}{\bfseries\leftmark}]%
{\fancyplain{}{\bfseries\thepage}}

In this appendix we derive the component expansion of the defect
action in the decoupling limit which is given by
\begin{align} \label{impaction}
  S^{\rm dec}_{\rm D3-D3^{\prime}} \equiv &\, S_{\rm kin} + S_{\rm superpot}
  \nonumber\\
  = &\int d^2z d^4 \theta \left(\bar B e^{gV} B + {\tilde B} e^{-gV} \bar
    {\tilde B} \right) +\, \frac{ig}{{2}} \int d^2z d^2\theta (\tilde B Q_1 
B)
  + c.c.
\end{align}
with $d^4\theta= \frac{1}{4} d\theta^+ d\theta^- d\bar\theta^+
d\bar\theta^-$ and $d^2\theta = \frac{1}{2} d \theta^+ d
\theta^-$.  Using the following expansions for the (2,2)
superfields $B$, $\tilde B$, $Q_1$,
\begin{align}
B &= b + \sqrt{2} \theta^+ \psi^b_+ +
  \sqrt{2} \theta^- \psi^b_- -2 \theta^+ \theta^- F^b \nonumber\\
\tilde B &= \tilde b + \sqrt{2} \theta^+ \psi^{\tilde b}_+ +
  \sqrt{2} \theta^- \psi^{\tilde b}_- -2 \theta^+ \theta^- F^{\tilde b}\\
Q_1 &= q_1 + \sqrt{2} \theta^+ \psi^{q_1}_+ +
  \sqrt{2} \theta^- \psi^{q_1}_- -2 \theta^+ \theta^- F^{q_1} \nonumber
\end{align}
as well as Eq.~(\ref{Vexp}) for $V$, the defect action can be
expanded as
\begin{align}
  S_{\rm kin}&= \int d^2 z \big( \bar F^b F^b - \vert D_M b \vert^2 + i \bar
  \psi^b_- (D_0 + D_1) \psi^b_- + i\bar \psi^b_+ (D_0 - D_1) \psi^b_+
  \nonumber\\
  &\quad- \frac{g}{2} (\bar\psi^b_- \s \psi^b_+ + \bar \psi^b_+ \bar \s
  \psi^b_-) + \frac{ig}{{2}} (b \bar \psi^b_+ \bar \lambda_- - b \bar
  \lambda_+ \bar \psi^b_- - \bar b \lambda_- \psi^b_+ + \bar b \lambda_+
  \psi^b_- )
  \nonumber\\
  &\quad+ \frac{1}{2}(g D - \frac{1}{2} g_{YM}^2 \bar \sigma \sigma ) \bar b 
b
  \big)
  + (B \rightarrow \tilde B, g \rightarrow -g) \\
  S_{\rm superpot}&= \frac{ig}{{2}} \int d^2z \big( \tilde b F^{q_1} b +
  \tilde b \psi^{q_1}_- \psi^b_+ + {\psi}^{\tilde b}_+ \psi^{q_1}_- b +
  F^{\tilde b} q_1 b + \psi^{\tilde b}_- \psi^{q_1}_+ b +
  \psi_-^{\tilde b} q_1 \psi^b_+ \nonumber\\
  &\quad+ \tilde b q_1 F^b + \psi_+^{\tilde b} q_1 \psi_-^b + \tilde b
  \psi_+^{q_1} \psi_-^b \big) +c.c.
\end{align}
where we used the covariant derivative $D_M = \pr_M +
\textstyle\frac{i}{2} g v_M$ $(M=0,1)$.

For the ambient action we have the standard component expansion of
$\N=4$, $d=4$ SYM. Some of the components of the $\N=4$ ambient
vector field, which we gather in the (2,2) fields $V$ and $Q_1$,
couple to the defect. The components of $V$ and $Q_1$ are
related to the components of the $\N=1$, $d=4$ superfields $V'$,
$\Phi'$, $\Phi'_1$, and $\Phi'_2$, which form the $\N=4$ vector
multiplet, by
\begin{align}
&\sigma \equiv i \phi', \quad \bar \lambda_+ \equiv \psi'_+, \quad \lambda_-
\equiv - \lambda'_-, \quad D \equiv \frac{1}{\sqrt{2}} ( D' + f'_{32}),
\quad
f_{01} \equiv \frac{1}{\sqrt{2}} f'_{01}   \,, \nonumber\\
&q_i \equiv \phi'_i, \quad \psi^{q_i}_\pm = \psi'^{\phi_i}_\pm,
\quad F^{q_i} \equiv F'^{\phi_i} \qquad(i=1,2) \,.
\end{align}


\section{Mesons in AdS/CFT}
\setcounter{equation}{0}

\subsection{Mesons in the AdS black hole geometry} \label{appB1}

In the AdS black-hole background (\ref{BHmetric}), we consider
solutions of the type
\begin{align} \label{w5fluct}
 w_5 = f(\rho) \sin (\vec k \cdot \vec x) \,,\qquad w_6 = w_6(\rho) \,,
\end{align}
where $w_6(\rho)$ is a solution of Eq.~(\ref{eqnmot}). The 
equations of motion for the fluctuations~$w_5$,
\begin{align} \label{euler}
\frac{d}{d\rho} \frac{d{\cal L}}{d(\partial_\rho w_5)}
+ \frac{d}{d x} \frac{d{\cal L}}{d(\partial_x w_5)} 
- \frac{d {\cal L}}{d w_5}= 0 \,,
\end{align}
follow from the DBI Lagrangian (\ref{DBIBH}). The three terms are
\begin{align}
&\frac{d}{d\rho} \frac{d{\cal L}}{d(\partial_\rho w_5)}=
 {d \over d\rho} \left[ {\cal G}(\rho, w_5, w_6)
\sqrt{ 1 \over 1 +  w'_5{}^2+  w'_6{}^2} { d w_5
\over d \rho} \right] 
\,, \\
&\frac{d}{dx } \frac{d{\cal L}}{d(\partial_x w_5)} =
 {\cal G}(\rho,w_5,w_6) \sqrt{ 1 \over 1 +  w'_5{}^2+ w'_6{}^2}
 \frac{4}{4(\rho^2+w_5^2+w_6^2)^2+b^4} \frac{d^2 w_5}{d^2 x}
 \,, \\ 
&\frac{d {\cal L}}{d w_5}= \sqrt{1 +  w'_5{}^2+w'_6{}^2} \frac{d}{d w_5}  
{\cal G}(\rho,w_5, w_6) = \sqrt{1 + w'_5{}^2+ w'_6{}^2} {b^8 \rho^3
w_5 \over 2 ( \rho^2 + w_5^2+w_6^2)^5} \,,
\end{align}
where $' \equiv \partial_\rho$. Here we ignored the overall factors
$\mu_7$ and $\varepsilon_3$.  We linearize these equations in the
fluctuations $w_5$ by setting $w_5^2 \approx 0$ and $w'_5{}^2 \approx
0$. Upon substituting Eq.~(\ref{w5fluct}), we obtain the equation of
motion (\ref{meep}).

\subsection{Mesons in the Myers-Constable geometry} \label{appB2}

\subsubsection*{Fluctuations in $w_5$}

In the Myers-Constable background (\ref{mcmetric}) we consider again solutions
of the type (\ref{w5fluct}). Now the three terms in the Euler-Lagrange
equation (\ref{euler}) are given by
\begin{align}
&\frac{d}{d\rho} \frac{d{\cal L}}{d(\partial_\rho w_5)}=
 {d \over d\rho} \left[ \frac{e^\phi {\cal G}(\rho, w_5, w_6)}
{\sqrt{ 1 +  w'_5{}^2+  w'_6{}^2 + C (\partial_x w_5)^2}} { d w_5
\over d \rho} \right] 
\,, \\
&\frac{d}{dx} \frac{d{\cal L}}{d(\partial_x w_5)}=
 {d \over dx} \left[ \frac{e^\phi {\cal G}(\rho, w_5, w_6)}
{\sqrt{ 1 +  w'_5{}^2+  w'_6{}^2 + C (\partial_x w_5)^2}} C{ d w_5
\over d x} \right] 
 \,, \\ 
&\frac{d {\cal L}}{d w_5}= \sqrt{1 + w'_5{}^2+w'_6{}^2 + C
 (\partial_x w_5)^2} \frac{d}{d w_5} \left[ e^\phi {\cal G}(\rho,w_5,
 w_6)\right] \,,
\end{align}
where the factor ${C}$ is defined by
\begin{align}
{C} \equiv g^{xx}g_{55} = H \left( \frac{w^4+b^4}{w^4-b^4} 
\right)^{(1-\delta)/2} \frac{w^4-b^4}{w^4} \,,\quad w^2=\rho^2+w_5^2
+w_6^2 \,.
\end{align}
As in App.~\ref{appB1}, we linearize these equations in the fluctuations
$w_5$, and substitute the ansatz (\ref{w5fluct}). We finally obtain
Eq.~(\ref{lin}).

\subsubsection*{Fluctuations in $w_6$}

We now consider solutions of the type
\begin{align} \label{w6fluct}
 w_5 = 0 \,,\qquad w_6 = w^{(0)}_6 + \delta w_6\,,\qquad
 \delta w_6 =h(\rho) \sin (\vec k \cdot \vec x) \,,
\end{align}
where $w^{(0)}_6$ are embedding solutions (\ref{eommc}). The three terms in
the Euler-Lagrange equation for $w_6$,
\begin{align}
\frac{d}{d\rho} \frac{d{\cal L}}{d(\partial_\rho w_6)}
+ \frac{d}{d x} \frac{d{\cal L}}{d(\partial_x w_6)} 
- \frac{d {\cal L}}{d w_6}= 0 \,, 
\end{align}
are given by
\begin{align}
\frac{d}{d\rho} \frac{d{\cal L}}{d(\partial_\rho w_6)}&=
 {d \over d\rho} \left[ \frac{e^\phi {\cal G}(\rho, w_6)}
{\sqrt{ 1 +  w'_6{}^2 + C (\partial_x w_6)^2}} { d w_6
\over d \rho} \right] 
\,, \label{extraterm}\\
\frac{d}{dx} \frac{d{\cal L}}{d(\partial_x w_6)}&=
 {d \over dx} \left[ \frac{e^\phi {\cal G}(\rho, w_6)}
{\sqrt{ 1 +  w'_6{}^2 + C (\partial_x w_6)^2}} C{ d w_6
\over d x} \right] \nonumber\\
& \approx C \frac{e^\phi {\cal G}(\rho, w_6)}
{\sqrt{ 1 +  w'_6{}^2 }} M^2 \delta w_6 
 \,, \\ \frac{d {\cal L}}{d w_6} &= \sqrt{1 +w'_6{}^2 + C (\partial_x
 w_6)^2} \frac{d}{d w_6} \left[ e^\phi {\cal G}(\rho, w_6)\right] \,,
\end{align}
with ${C}\equiv g^{xx}g_{66} = g^{xx}g_{55}$ as above and $\frac{d}{d w_6}
\left[ e^\phi {\cal G}(\rho, w_6) \right]$ as in Eq.~(\ref{potliketerm}). The
term (\ref{extraterm}) is the extra term (\ref{extraterm2}).  As explained in
the text, we substitute the ansatz (\ref{w6fluct}) and solve numerically the
equations of motion in its non-linear form.

\section{Deconstruction of extra dimensions}
\setcounter{equation}{0}

\subsection{Gauge transformation properties}  \label{transformations}

The gauge transformation properties under the residual gauge group
$SU(N)^k\times SU(N')^k$ in $(2,0)$ superspace are as follows:\
\begin{align} \label{transform}
  \tilde B_i &\rightarrow e^{-i \lambda'_i} \tilde B_i e^{i \lambda_i}, \quad
  \Lambda^{\tilde B}_{-,i} \rightarrow e^{-i \lambda'_{i}}
  \Lambda^{\tilde B}_{-,i} e^{i\lambda_{i-1}}, \nonumber \\
  B_i &\rightarrow e^{-i \lambda_i} B_i e^{i \lambda'_{i}},\quad
  \Lambda^{B}_{-,i} \rightarrow e^{-i \lambda_{i}} \Lambda^{B}_{-,i}
  e^{i \lambda'_{i-1}}   \,, \nonumber\\
  Q^1_i &\rightarrow e^{-i \lambda_i} Q^1_i e^{i \lambda_{i+1}},\quad
  \Lambda_{-,i}^{Q^1} \rightarrow e^{-i \lambda_i} \Lambda_{-,i}^{Q^1}
  e^{i \lambda_i} , \\
  Q^2_i &\rightarrow e^{-i \lambda_i} Q^2_i e^{i \lambda_{i}},\quad
  \Lambda_{-,i}^{Q^2} \rightarrow e^{-i \lambda_i} \Lambda_{-,i}^{Q^2} e^{i
    \lambda_{i+1}} ,
  \nonumber\\
  \Theta^V_i &\rightarrow e^{-i \lambda_{i}} \Theta^V_i e^{i \lambda_{i-1}},
  \quad
  e^{V_i} \rightarrow e^{-i \lambda^\dagger_i} e^{V_i} e^{i \lambda_i},  \nonumber\\
  S^1_i &\rightarrow e^{-i \lambda'_i} S^1_i e^{i \lambda'_{i+1}},\quad
  \Lambda_{-,i}^{S^1} \rightarrow e^{-i \lambda'_i} \Lambda_{-,i}^{S^1} e^{i
    \lambda'_i} ,
  \nonumber\\
  S^2_i &\rightarrow e^{-i \lambda'_i} S^2_i e^{i \lambda'_{i}},\quad
  \Lambda_{-,i}^{S^2} \rightarrow e^{-i \lambda'_i} \Lambda_{-,i}^{S^2} e^{i
    \lambda'_{i+1}} ,
  \nonumber\\
  \Theta^\V_i &\rightarrow e^{-i \lambda'_{i}} \Theta^\V_i e^{i
    \lambda'_{i-1}}, \quad e^{{\cal V}_i} \rightarrow e^{-i
    {\lambda'}^\dagger_i} e^{{\cal V}_i} e^{i \lambda'_i} \nonumber\,,
\end{align}
where $i=1,\dots k$. These gauge transformations lead to the
quiver diagram of Fig.~\ref{superquiver}. Strictly speaking, the
transformation laws for the Fermi multiplets hold only at $\bar \theta^+=0$.
Note that Fermi multiplets in superpotentials act effectively  as chiral
multiplets.

\subsection{Superpotential in manifest $(2,0)$ language} \label{appSuperpot}

The conformal field theory corresponding to the D3-D3 intersection
placed at an orbifold singularity is $(4,0)$ supersymmetric. For an
adequate formulation of the parent defect theory in flat space, we use
$(2,0)$ superspace. For our purposes it is sufficient to give the
superpotential of the parent theory, which we now express in terms of
$(2,0)$ superfields. Writing the full $(4,4)$ supersymmetric action
(\ref{action1}) - (\ref{defectaction}) in $(2,0)$ superspace is
straightforward, but we do not give the result here.

We decompose the $(2,2)$ defect multiplets $\bf B$ and
$\bf \tilde B$ as well as the ambient multiplets $\bf Q_1$, $\bf Q_2$, $\bf
\Phi$, and $\bf V$ under $(2, 0)$ supersymmetry. The reduction of 
$(2,2)$ multiplets to $(2,0)$ multiplets is discussed in \cite{Witten,
Garcia}.

In general, a $(2,2)$ chiral multiplet $\bf \Phi$ reduces to two
$(2,0)$ multiplets, a chiral multiplet $\Phi$ and a
Fermi multiplet $\Lambda_-$, according to
\begin{align} \label{chiralexp}
  \mathbf{\Phi}(y,\theta^\pm)=\Phi(y,\theta^+,\bar\theta^+)
  \vert_{\bar\theta^+=0} + \sqrt{2} \theta^-
\Lambda_-(y,\theta^+,\bar\theta^+)
  \vert_{\bar\theta^+=0} \,
\end{align}
with $y^M=x^M + i \theta^+ \bar \theta^+ + (-1)^M i \theta^- \bar \theta^-$
($M=0,1$).  The chiral $(2,0)$ multiplet satisfies ${\cal \bar D}_+
\Phi= 0$ and can be expanded as
\begin{align}
\Phi = \phi+\sqrt{2}\theta^+ \lambda_+
- i \theta^+ \bar \theta^+ (D_0 + D_1) \phi \,
\end{align}
with covariant derivatives $D_M=\partial_M +\frac{ig}{2} v_M$.
The Fermi multiplet expansion is given by
\begin{align}
  \Lambda_- = \psi_- + \sqrt{2} \theta^+ F
  -i \theta^+ \bar \theta^+ (D_0 + D_1) \psi_- - \sqrt{2} \bar \theta^+ E
\,.
\end{align}
$\Lambda_-$ satisfies ${\cal \bar D}_+ \Lambda_- = \sqrt{2} E$.  In the reduction
of the above $(2,2)$ chiral superfield $\bf \Phi$, the function $E$ is $E= i
\sqrt{2} T^a
\Theta^a_V \Phi$, where $T_a$ are the generators of the gauge group. Here
$\Theta_V$ is another chiral superfield defined by
\begin{align}
\Theta_V \equiv \mathbf{\Sigma}
  \vert_{\theta^-=\bar\theta^-=0}=\s + i \theta^+ \bar \lambda_+
  - i \theta^+ \bar \theta^+ (D_0 + D_1) \s \,,
\end{align}
where $\bf \Sigma$ is the gauge invariant field strength of the
(2,2) gauge multiplet $\bf V$.

We can now write the superpotential $W^{\rm par}
=W^{\rm par}_{\rm D3} + W^{\rm par}_{\rm D3} +W^{\rm par}_{\rm D3-D3'}$
of the parent theory, by substituting the following expansions
into the action (\ref{action1})-(\ref{defectaction}),
\begin{align}
  \mathbf{Q}_i&= (Q_i+\sqrt{2}\theta^- \Lambda^{Q_i}_-
  )\vert_{\bar\theta^+=0} \,,\qquad
  \mathbf{B}=(B+\sqrt{2}\theta^- \Lambda^{B}_-)\vert_{\bar\theta^+=0}
  \,, \\
  \mathbf{\Phi}&= (\Phi+\sqrt{2}\theta^- \Lambda^{\Phi}_-
  )\vert_{\bar\theta^+=0} \,,\qquad\,\,\,\,
  \mathbf{\tilde B}=(\tilde B +\sqrt{2}\theta^- \Lambda^{\tilde B}_-)
  \vert_{\bar\theta^+=0} \,.
\end{align}
For the superpotential $W^{\rm par}_{\rm D3}$ associated with one stack
of D3-branes, we
find
\begin{align}
  W^{\rm par}_{\rm D3} = & \int d^2x d^2 \theta\, \epsilon_{ij} {\rm\, tr\,}
  \mathbf{Q}_i
  [\partial_{\bar x} + g \mathbf{\Phi}, \mathbf{Q}_j] + c.c \\
  =& \int d^2x d \theta^+ {\rm tr\!}\left.\left( \Lambda_-^{Q_1} [
      \partial_{\bar x} + g \Phi, Q_2] -\Lambda_-^{Q_2} [ \partial_{\bar x}
+
      g \Phi, Q_1] + g \Lambda_-^{\Phi} [Q_2, Q_1] \right)
  \right\vert_{\bar\theta^+=0} \!+\! c.c.  \nonumber
\end{align}
A similar expression holds for $W^{\rm par}_{\rm D3'}$, while the defect
action has the $(2,0)$ super\-potential
\begin{align} \label{sp}
  W^{\rm par}_{\rm D3-D3'}= \frac{ig}{2}\int d\theta^+ {\rm\, tr} &\left(
      B \tilde B \Lambda_-^{Q_1} + \Lambda^{B}_- \tilde B Q_1 + B
      \Lambda^{\tilde B}_- Q_1 \right. \nonumber\\
&-  \tilde B B \Lambda_-^{S_1} - \Lambda^{\tilde B}_- B S_1 - \tilde B
      \Lambda^{B}_- S_1\left.
\big) \right\vert_{\bar \theta^+=0} +c.c.
\end{align}
The parent superpotential $W^{\rm par}$ leads to the superpotential
(\ref{superpotential}) under the orbifold projection.


\end{appendix}


\bibliographystyle{unsrthep}

\newpage
\addcontentsline{toc}{section}{References}
\bibliography{referencesFdP}


\end{document}